\documentclass[12pt]{report}
\usepackage{amssymb}
\usepackage{epsfig}
\usepackage{color}
\newcommand{\qsq}{\mbox{$Q^2$}}
\newcommand{\s}{\mbox{$s$}}

\newcommand{\x}{\mbox{$x$}}
\newcommand{\y}{\mbox{$y$}}
\newcommand{\R}{\mbox{$R$}}

\newcommand{\rh}{\mbox{$\rho(R)$}}
\newcommand{\rhR}{\mbox{$\rho(R;\varepsilon_0)$}}
\newcommand{\rhRz}{\mbox{$\rho(R=0;\varepsilon_0)$}}
\newcommand{\rhRi}{\mbox{$\rho(R=\infty;\varepsilon_0)$}}
\newcommand{\srhR}{\mbox{$\sigma_{\rho,R}$}}

\newcommand{\srhRzsq}{\mbox{$\sigma^2_{\rho,R=0}$}}

\newcommand{\srhRisq}{\mbox{$\sigma^2_{\rho,R=\infty}$}}
\newcommand{\gev}{\mbox{\rm GeV}}

\newcommand{\pbinv}{\mbox{${\rm pb^{-1}}$}}

\newcommand{\rpv}{{\mbox{$\not\hspace{-.55ex}{R}_p$}}}

\def\ra{\rightarrow}

\def\be{\begin{equation}}
\def\ee{\end{equation}}
\def\bea{\begin{eqnarray}}
\def\eea{\end{eqnarray}}
\def\rp{R_p}
\def\se{\tilde{e}}
\def\su{\tilde{u}}
\def\sd{\tilde{d}}
\def\sq{\tilde{q}}

\def\lp{\lambda^\prime}

\begin{document}
\parskip 5pt
\pagestyle{empty}
\begin{flushright}
LAL 00-57
\end{flushright}
\vspace*{2.5cm}
\centerline {\Large\bf New Insights into the Proton Structure}
\vspace{0.2cm}
\centerline {\Large\bf with $ep$ Collider HERA}
\vspace{2cm}
\centerline {\sc Zhiqing Zhang}
\vspace{0.5cm}
\centerline {\it Laboratoire de l'Acc\'el\'erateur Lin\'eaire}
\centerline{\it IN2P3 - CNRS et Universit\'e de Paris-Sud}
\centerline {\it B\^at.\ 200, BP 34}
\centerline {\it F-91898, Orsay Cedex}
\centerline {\it France}
\centerline {E-mail: zhangzq@lal.in2p3.fr}
\vspace{2cm}
\centerline{\bf Abstract}
Since its commissioning in 1991, the $ep$ collider HERA has been running
successfully for almost a decade without stopping improving its performance. 
In this report, the inclusive cross section and structure function measurements
for the deep inelastic scattering of neutral and charged current processes
in the full HERA kinematic domain are reviewed. The results
are compared with the Standard Model expectations for the deep inelastic
scattering processes. The new insights into the proton structure and on
the underlying strong and electroweak interactions are discussed.\\
\centerline { }
\centerline { }
\centerline { }
\centerline { }
\centerline { }
\centerline {\sl Habilitation defended on Dec.\ 1, 2000}

\pagestyle{plain}
\pagenumbering{roman}
\tableofcontents
\chapter{Introduction} \label{chap:intro}
\pagenumbering{arabic}
\setcounter{page}{1}
Scattering experiments have proven to be a very effective technique for
probing the structure of matter. The large angle scattering of $\alpha$
particles (a few MeV) off a gold foil led Rutherford to conjecture 
around 1910 that the bulk of the mass of the atom must be contained 
in a very small, positively charged nucleus of the atom. 
The energy available in such a collision is related to a certain wavelength. 
The higher the energy, the shorter the wavelength and thus the smaller the
distances which can be resolved.
To investigate atomic nuclei
with higher accuracy it is therefore necessary to bombard them with 
particles of higher energy.
In the 1950s and 1960s, starting with electron scattering experiments at
Stanford~\cite{slac50s}, experiments were performed at Darmstadt, Orsay, Yale,
CEA and DESY. At the energies available, the experiments were however 
restricted to elastic scattering or excitation of the low-lying resonances.
As higher energy electron beams (up to $16\,{\rm GeV}$) became available 
at SLAC in the late 1960s, inelastic scattering experiments could be 
performed. These experiments~\cite{slac68,slac69} together with those
at DESY~\cite{desy69} showed that at large four-momentum transfers, 
the inelastic nucleon structure functions were independent of any 
dimensional quantity, a phenomenon known as scaling. This result was 
interpreted as evidence for the existence of point-like constituents 
in the nucleons. These constituents are now known as quarks.
Subsequent fixed-target lepton-nucleon scattering experiments\footnote{The
use of high energy leptons (electrons, muons and neutrinos, along with
their antiparticles) as probes of nucleon structure is unique because
the leptons do not interact strongly, so that they are able to penetrate
the nuclear surface; their short wavelength implies that the leptons collide
with individual charged or weakly interacting constituents~\cite{mishra89}.}
at still higher energies have established the existence of violations of 
the scaling behavior~\cite{sv_slac,sv_fermi} which has been one of the most 
dramatic successes of perturbative Quantum Chromodynamics (QCD).

Apart from the important roles deep inelastic scattering (DIS) has played 
in understanding nucleon structure, it also contributed to our understanding
in other ways (see e.g.\ \cite{mishra89}). Neutral-current phenomena, 
which provided the first
demonstration of the $SU(2)\times U(1)$ unification of weak and
electromagnetic interactions, were discovered~\cite{hasert73} and 
corroborated~\cite{barish74} in
neutrino-nucleon experiments; interference between neutral weak and
electromagnetic propagators was demonstrated first in an electron-nucleon
inelastic scattering experiment~\cite{prescott79}. 
The most precise measurement 
of the weak mixing angle $\theta_W$, parametrizing the weak neutral current
coupling, used to be measured from neutrino-nucleon data.
Today, the most precise DIS data are being widely used to determine the
momentum distributions of partons in hadrons allowing to predict
cross sections in high energy hadron collisions.

The fixed-target experiments, though precise, are limited in their
kinematic range. The ``Hadron Elekton Ring Anlage'' (HERA), the first
electron proton collider ever built (Sec.\ref{sec:hera}), 
running at a center-of-mass energy of up to $\sqrt{s} \simeq 320\,{\rm GeV}$, 
equivalent to a fixed-target experiment with a lepton beam of 50\,TeV,
has significantly extended the kinematic domain explored so far.
The proton structure are being probed at 10 times 
smaller distances (down to $\sim 10^{-16}\,{\rm cm}$, which is 
one per mill of the proton radius) than previously accessible. Partons
can be studied down to very small fractional proton momenta, Bjorken $x\sim 
10^{-6}$. Two general purpose experiments, H1 and ZEUS, are dedicated to the
study of this physics.

Already the first 0.02\,pb$^{-1}$ of
luminosity delivered by HERA in 1992 has brought striking results 
which have opened new interest in QCD. Since then the luminosity has been
increased by three orders of magnitude and the latest results achieved
are comparable in precision with fixed-target experiments but in a different 
kinematic regime. At high momentum transfers, based on an early data sample of
1994-1996, an excess of events with respect to the expectation of the standard
DIS has been reported by both H1~\cite{h19496} and ZEUS~\cite{zeus9496}. 
This has again initiated a considerable amount of theoretical interest in
possible explanations within both the Standard Model(SM) and models
beyond the SM~\cite{thpapers}. Results obtained using full data taken from
1994 to 2000 will be reviewed in this report. The emphasis will however 
be put on the experimental understanding of the detector performance 
by showing in some detail how some of the measurements have been carried out.

The report is organized as follows. In Chapter \ref{chap:exp}, the
experimental aspects concerning the HERA machine, the H1 detector and the
kinematic reconstruction are described. In Chapter \ref{chap:theory},
the theoretical framework is briefly reviewed.
In Chapter \ref{chap:lowq2}, the structure function results at low momentum
transfers and the interpretation within QCD are given.
In Chapter \ref{chap:hiq2}, events at high momentum transfers are studied.
Also discussed are their impact both within the SM for 
valence quark distributions and electroweak tests and 
beyond the SM for new physics search.
A summary and an outlook are presented in Chapter \ref{chap:summary}.

\chapter{Experimental setups and techniques} \label{chap:exp}
Deep inelastic scattering (DIS) experiments can roughly be regrouped into three
generations. The first generation of fixed-target experiments
between 1968 and the mid-1970s has played an essential role in establishing
the point-like substructure of the nucleons and in discovering the underlying
weak neutral current interactions. The second generation of fixed-target
experiments of higher energies and precision has helped in
establishing and testing the quantum chromodynamics (QCD) and in providing
data for extracting a first set of parton distribution functions.
A list of selected DIS experiments is briefly presented in Sec.\ref{sec:fte}
after having introduced the kinematics in Sec.\ref{sec:kine}. The
third generation experiments at HERA $ep$ collider are described 
in more detail in the later sections.

\section{Kinematics of deep inelastic scattering} \label{sec:kine}
\begin{figure}[htb]
\begin{center}
\begin{picture}(50,130)
\put(-320,-240){\epsfig{file=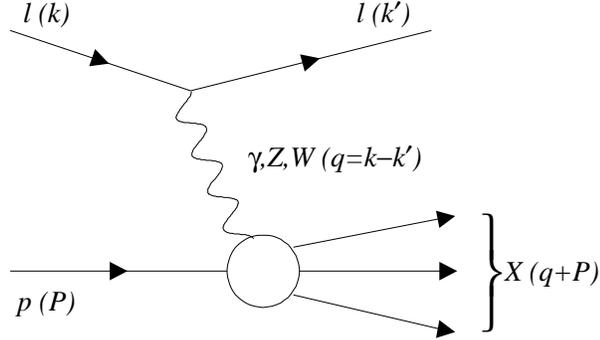,width=200mm}}
\end{picture}
\end{center}
\caption{\sl Schematic diagrams of lepton ($l$) and proton ($p$) scattering
via photon and $Z$ exchange for NC and $W$ exchange for CC. The four
momentum vectors of the particles or particle systems are given in
parentheses.}
\label{fig:kinedia}
\end{figure}

The kinematics of the inclusive DIS processes for neutral current
(NC) interaction $lp\rightarrow lX$ and charged current (CC) interaction
$ep\rightarrow \nu X$ or $\nu p\rightarrow lX$ (Fig.\ref{fig:kinedia})
at the given center-of-mass energy squared, $s=(k+P)^2$, is determined by two 
independent Lorentz invariant variables, which could be any of following
variables:
\begin{eqnarray}
Q^2=-q^2=(k-k^\prime)^2\,,& & Q^2\in \left[0,s\right],\label{eq:q2}\\
x=\displaystyle \frac{Q^2}{2P\cdot q}\,,& &
x\in\left[0,1\right],\\
y=\displaystyle \frac{q\cdot P}{k\cdot P}\,,& & y\in\left[0,1\right],\\
W^2=(q+P)^2\,,& & W^2\in\left[M^2_p,s\right].\label{eq:w2}
\end{eqnarray}
The square of the four momentum transfer (the invariant mass of the exchanged
vector boson), $q^2<0$, is space-like and determines the hardness of 
the interaction, or in other words, the resolving power of the interaction. 
The Bjorken variable $x$
is interpreted as the fraction of the proton momentum carried by the struck
parton in the quark parton model (QPM). The variable $y$ measures the
inelasticity of the interaction and its distribution reflects the spin
structure of the interaction in the rest frame of the target.

The variable $\nu$
\begin{equation}
\nu=\frac{q\cdot P}{M}\, \label{eq:nu}
\end{equation}
is often used in fixed-target experiments where $\nu=E-E^\prime$ and $M$
is the target mass.

\newpage
\section{The fixed-target experiments} \label{sec:fte}
Fixed-target DIS experiments may be divided into two classes
depending on the nature of the probe used, which in turn determines
the force involved.
In electroproduction, electrons or muons are scattered off
the target nucleon and the force involved is electromagnetic.
The leading process of the scattering is that of single-photon ($\gamma$)
exchange. The second class of processes
is called neutrinoproduction and, in this, neutrinos are scattered off the
target nucleons by the weak force. The leading process is that of
single-$W$-boson exchange for charged current interactions and of
single-$Z^0$-boson exchange for neutral current interactions. 

The earlier electroproduction experiments used electrons as probe. However
due to their large synchrotron radiation, the largest electron energy used
for scattering experiments was limited to about 12\,GeV (Cornell) and 26\,GeV
(SLAC). The kinematic range was extended by muon-nucleon scattering
to higher energies ($20<E<500$\,GeV).

The neutrinoproduction experiments complement the charged lepton scattering
experiments and comparisons between neutrino and charged lepton deep
inelastic experiments provide tests of universality of the parton density
functions. In addition, the neutrinoproduction experiments are unique as
they are able to distinguish quark flavors within target nucleons 
(Sec.\ref{sec:qpm}).
On the other hand, these experiments were in practice difficult to
realize as the electrically neutral, weakly interacting neutrinos cannot be
directed by electric and magnetic fields, as can electrons; and building a
usable neutrino beam is a complicated process. First, a primary beam of
protons is accelerated to high energy and is made to collide with a
stationary target such as a piece of iron. From these collisions, a host of
secondary particles, mainly mesons, will emerge in the general direction of
the incident proton beam, although with somewhat smaller energies. These
secondary mesons can then decay into neutrinos or antineutrinos and various
other particles by decays such as $\pi^\pm \rightarrow \mu^\pm + \nu_\mu
(\overline{\nu}_\mu)$. Because the muon-decay mode of the mesons is
generally the most common, it is mainly muon-type neutrinos which make up
the beam. Finally, the neutrinos are isolated by guiding the secondary beam
through a barrier of steel and rock equivalent of, perhaps, 0.5\,km of earth. 
Only the weakly interacting neutrinos can pass through this amount of matter 
and so the beam emerging from the far side of the barrier is a pure neutrino 
beam with a typical intensity of about $10^9$ particles per second. 
Two types of high energy beams are commonly used: narrow band beams with
a momentum and charge selection of the secondary beam, and wide band beams 
without such selection. The narrow band beams provide a measurement of 
the incident neutrino energy and flux. Furthermore, the beams are either 
almost purely neutrino or almost purely antineutrino, with little cross 
contamination. Wide band beams provide an order of magnitude more neutrinos; 
however, there is no direct check on the event neutrino energy and no direct 
measure of the flux.

A few selected fixed-target DIS experiments are listed in Table \ref{tab:fte}.
\begin{table}[tb]
\begin{center}
\begin{minipage}{14cm}
\caption{\sl A list of
selected fixed-target DIS experiments. Only the main targets used are
indicated with $p$, $d$, and Fe standing respectively for proton, deuterium,
and iron targets (the use of heavy nuclear targets have the advantage 
of high rate).}
\label{tab:fte}
\end{minipage}
\begin{tabular}{|l|l|l|c|l|}
\hline
Experiment & Year & Reaction & Beam Energy (GeV) & Reference \\\hline
SLAC-MIT & 1968 & $ep,ed$ & 4.5-20 & \cite{slac68,slac69,tayler69,slac72} 
\\\hline
DESY & 1969 & $ep$ & 6 & \cite{desy69,desy69b} \\\hline
Gargamelle$^\ast$ & 1971-1976 & $\nu_\mu (\overline{\nu}_\mu)p(d)$ & $<10$ &
\cite{hasert73} \\\hline
SLAC & 1974-1975 & $\mu$ Fe & 56.3, 150 & \cite{sv_slac} \\\hline
BEBC$^\ast$ & 1975-1983 & $\nu_\mu (\overline{\nu}_\mu)p(d)$ & $<200$ & 
\cite{bosetti78} \\\hline
Fermilab & 1977-1978 & $\mu p$ & 47, 96, 219 & \cite{sv_fermi} \\\hline
CDHS & 1979-1990 & $\nu_\mu (\overline{\nu}_\mu)$ Fe & $<200$ 
& \cite{abramowicz82} \\\hline
CCFR & 1979-1988 & $\nu_\mu (\overline{\nu}_\mu)$ Fe & $\leq 600$ & 
\cite{ccfr97} \\\hline
BCDMS & 1981-1985 & $\mu p, \mu d$ & 100,120,200,280 & \cite{bcdms89} \\\hline
NMC & 1986-1989 & $\mu p, \mu d$ & 90,120,200,280 & \cite{nmc97} \\\hline
E665 & 1987-1992 & $\mu p, \mu d$ & 470 & \cite{e665} \\\hline
\multicolumn{5}{l}{\begin{minipage}{13.2cm}{$^\ast$ The Gargamelle and BEBC 
(Big European Bubble Chamber) experiments used bubble chambers while 
other cited experiments used electronic detectors. The bubble chambers have
opted for wide-band beams that maximize neutrino flux.}\end{minipage}}
\end{tabular}
\end{center}
\end{table}                                                
The early fixed-target experiments between 1968 and the mid-1970s have
discovered a wealth of information on the structure of the proton, to name a
few:
\begin{itemize}
\item {\bf Point-like constituents:} The approximate independence of 
the measured structure functions of $Q^2$~\cite{slac68,slac69,desy69,slac72} 
indicated scattering off point-like constituents analogous to the classic 
Rutherford experiment on atomic structure.
\item {\bf Quark spin:} The near vanishing of $F_L$ in electron-nucleon
scattering~\cite{tayler69,desy69b} supported the assignment of half-integer
spin for the quarks.
\item {\bf Fractional charge:} The comparison of structure functions in 
electron and neutrino scattering reactions supported the assignment of 
fractional charges to the quarks.
\item {\bf Scaling violations and QCD:} 
The muon scattering experiments at high energies provided first evidences 
for scaling violations~\cite{sv_slac,sv_fermi}. The scaling violations of the 
DIS data have helped in establishing the theory of QCD (Sec.\ref{sec:qcd}).
\item {\bf Gluon:} The momentum sum rules in both electron- and neutrino-proton
scattering~\cite{perkins75} suggested that quarks carry only about half of
the total proton momentum. The other half was thought to be carried by
neutral gluons, the quanta of the interquark field force.
\end{itemize}
The neutrino scattering experiments have played other particular roles:
\begin{itemize}
\item {\bf Discovery of neutral currents:} The first observations of neutral
 currents were reported by the Gargamelle Collaboration~\cite{hasert73}.
 This was the first experimental success of the electroweak theory of the
 Standard Model.
\item {\bf Precise measurement of $\sin\theta_W$:} Before the advent of LEP
 and SLC, the electroweak mixing parameter $\sin\theta_W$ was best measured 
 by the neutrino scattering experiments with a first result 
 ($\sin\theta_W=0.28\pm 0.05$)~\cite{brisson76} already in 1976 leading to 
 a $W$ mass value of 70\,GeV~\cite{gayler95} well before the direct 
 observation of the $W$ boson in early 1980s.
\item {\bf First experimental hint of charm:} Neutrino productions of
 opposite-sign dimuon $(\mu^-\mu^+)$ in subprocesses $\nu_\mu+(d,s)
 \rightarrow \mu^- + c +X$ and $\overline{\nu}_\mu+(\overline{d},\overline{s})
 \rightarrow \mu^+ + \overline{c} +X$ (the second muon arises from the
 semileptonic decay of the charmed hadrons emerged from the charged quark $c$)
 provided a first experimental hint of charm~\cite{rubbia}.
\item {\bf Strange quark sea and evidence for an $SU(3)$ asymmetric flavor 
 sea:} The subprocess $\overline{\nu}_\mu + \overline{s}\rightarrow \mu^+ +
 \overline{c}+X$ offered a unique probe to measure the strange component of
 the nuclear sea. The subprocess $\nu_\mu+(d,s) \rightarrow \mu^- + c +X$
 provided a measure of the momentum fraction carried by the strange sea quarks
 relative to that carried by the non-strange sea
 $s/(\overline{u}+\overline{d})$ revealing an asymmetric strange sea.
\item {\bf Unique direct measurement of $|V_{cd}|$:} The Cabibbo suppressed 
 subprocess $\nu_\mu+d\rightarrow \mu^-+c+X$ has been the only direct 
 measurement of the Kobayashi-Maskawa matrix element
 $|V_{cd}|$~\cite{abramowicz82}.
\end{itemize}

During the last decade, several precise data have been published and a few 
old data have been reanalyzed. 
A number of previously observed discrepancies between different experiments
have been resolved. For a recent review, see Ref.\cite{cooper97}.

\newpage
\section{The HERA accelerator} \label{sec:hera}
\begin{figure}[htb]
\begin{center}
\begin{picture}(50,175)
\put(-155,-280){\epsfig{file=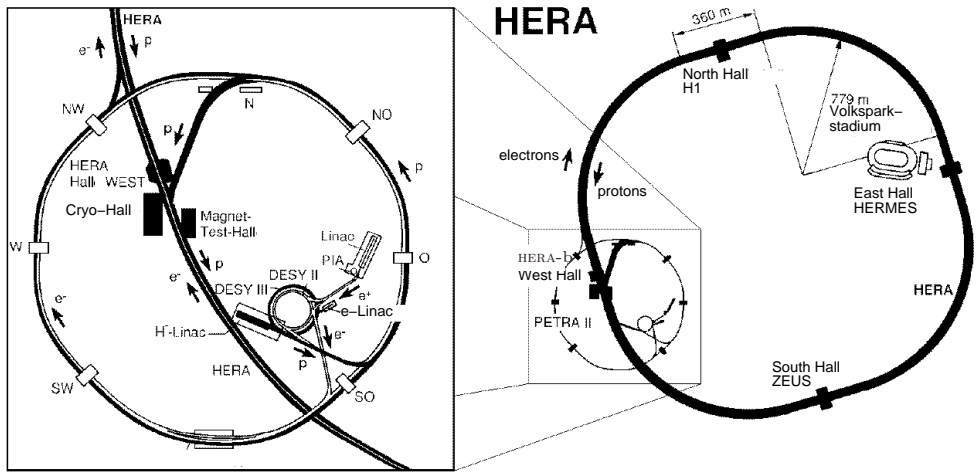,bbllx=0pt,bblly=0pt,bburx=594pt,
bbury=842pt,width=130mm}}
\put(43.6,70){\fcolorbox{white}{white}{\textcolor{white}{\tiny aaa}}}
\put(43.6,70){\tiny {\sc hera}-b}
\end{picture}
\end{center}
\caption{\sl The layout of the HERA accelerator and its preaccelerator 
facilities.}
\label{fig:hera}
\end{figure}

The HERA project was authorized in April 1984 and the construction was
completed late 1990 in accordance with the original time schedule.
The layout of the HERA accelerator and its preaccelerator facilities is
shown schematically in Fig.\ref{fig:hera}.
The electrons and protons are injected into HERA from PETRA (a previous 
$e^+e^-$ collider) with $14\,{\rm GeV}$ and $40\,{\rm GeV}$ respectively.
HERA is a double ring collider.
The electron ring is made of supercoducting cavities and normal conducting 
magnets for accelerating the electrons of 14\,GeV up to the nominal electron 
energy of $30\,{\rm GeV}$.
The electron ring provides either electron or positron beam\footnote{In the
following, the generic name electron is used for electrons as well as for
positrons unless stated otherwise.}. The year-dependent
lepton-beam charges and energies are shown in Table \ref{tab:hera_year}.
The actual maximum electron beam energy of $27.5\,{\rm GeV}$
is limited by the maximum available radio frequency ({\sc rf}) voltage.
The proton ring consists of conventional {\sc rf} cavities and superconducting 
magnets of 4.68\,Tesla magnetic field for accelerating the protons from
40\,GeV to 820\,GeV, maintaining them at this energy and keeping them on orbit.
The quality of the magnets is such that it is possible to rise the strength of
the magnetic field up to 5.8\,Tesla, which would correspond to 1\,TeV proton
beam. 
\begin{table}[tb]
\begin{center}
\begin{minipage}{10.3cm}
\caption{\sl The year-dependent lepton-beam charges and energies and the
corresponding center-of-mass energy $\sqrt{s}$ of the HERA collider.}
\label{tab:hera_year}
\end{minipage}
\begin{tabular}{|c|c|c|c|c|}
\hline
Year & $e^\pm$ beam & $E_e$ (GeV) & $E_p$ (GeV) & $\sqrt{s}$ (GeV) \\\hline
1992-93 & $e^-$ & 26.7 & 820 & 296 \\\hline
1994 & $\begin{array}{c} e^- \\ e^+ \end{array}$ & 27.5 & 820 & 300 \\\hline
1995-97 & $e^+$ & 27.5 & 820 & 300 \\\hline
1998 & $e^-$ & 27.5 & 920 & 320 \\\hline
1999 & $\begin{array}{c} e^- \\ e^+ \end{array}$ & 27.5 & 920 & 320 \\\hline
2000 & $e^+$ & 27.5 & 920 & 320 \\\hline
\end{tabular}
\end{center}
\end{table}                                                

HERA has four straight sections (interaction points) spaced evenly 
around its $6.3\,{\rm km}$ circumference.
The electron and proton beams collide head-on in two of these points
occupied by two general purpose detectors H1~\cite{h1det} (Sec.\ref{sec:h1det})
and ZEUS~\cite{zeusdet}.
Two fixed-target experiments HERMES~\cite{hermesweb} and 
HERA-B~\cite{herabweb} are located at the other
two interaction regions. At HERMES spin rotator provide the longitudinally 
polarized electron beam in collision with a polarized gas target for studying
spin dependent structure functions, while HERA-B uses the halo of the
proton beam with wire targets in an attempt to detecte CP violation 
in the $B$ system.

\begin{figure}[htb]
\begin{center}
\begin{picture}(50,350)
\put(-200,-150){\epsfig{file=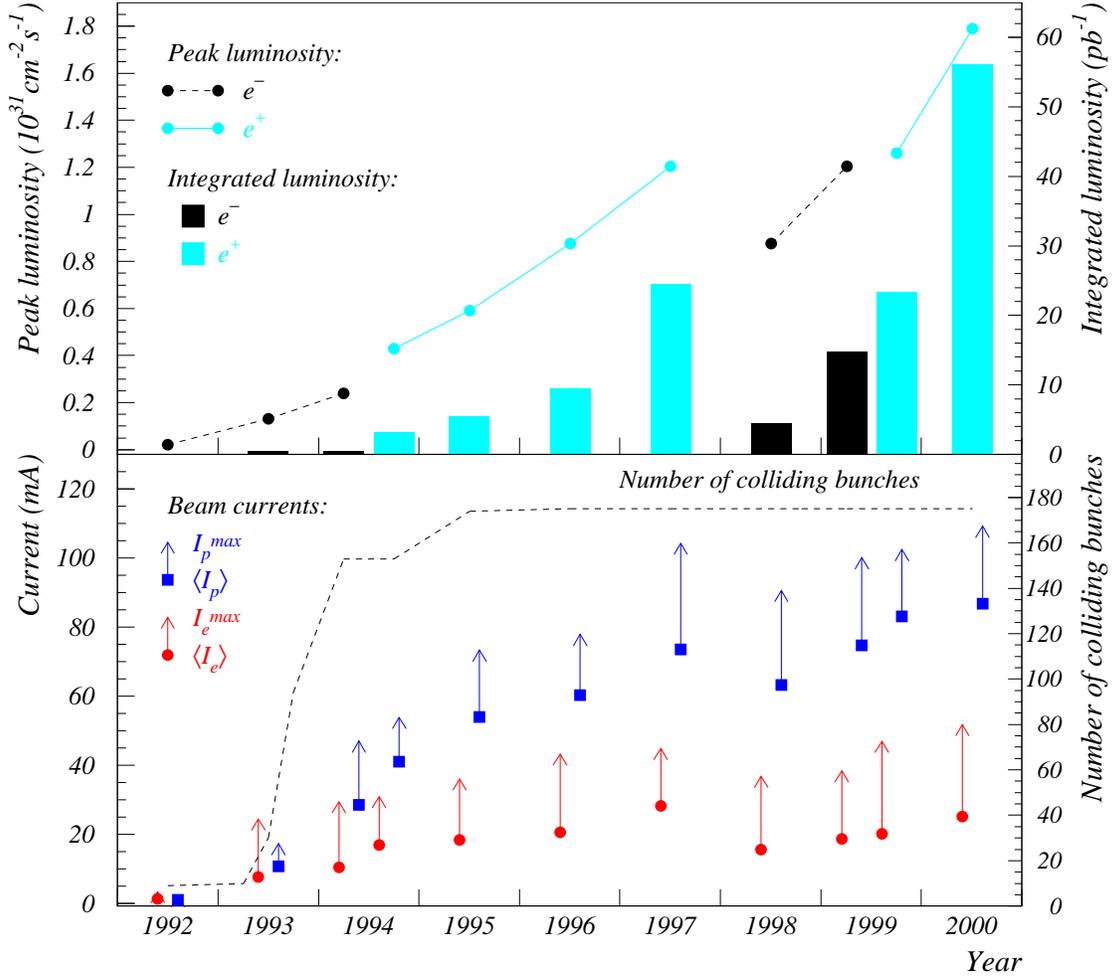,bbllx=0pt,bblly=0pt,
bburx=594pt,bbury=842pt,width=16cm}}
\end{picture}
\end{center}
\caption{\sl The time evolution of the following information: (1) the peak luminosity, (2) the
integrated luminosity collected by H1, (3) the mean as well as maximum
lepton and proton beam currents, and (4) the number of colliding bunches.}
\label{fig:heralumi}
\end{figure}

HERA has made steady progress since 1992. 
This can be seen from Fig.\ref{fig:heralumi} in which the time evolution of
the following information is shown: (1) the peak luminosity, (2) the
integrated luminosity collected by H1, (3) the mean as well as maximum
lepton and proton beam currents, and (4) the number of colliding bunches.
A comparison of the best achieved and design values of the main HERA
parameters is shown in Table \ref{tab:hera_para}.
\begin{table}[tb]
\begin{center}
\begin{minipage}{7.8cm}
\caption{\sl The main HERA parameters achieved so far compared with the
design values.}
\label{tab:hera_para}
\end{minipage}
\begin{tabular}{|l|r|r|}
\hline
Parameter & Design & Achieved \\\hline
$I_e$ (mA) & 58 & 51.8 \\\hline
$I_p$ (mA) & 160 & 109 \\\hline
$\sharp$ bunches & 210 & 189 \\\hline
$\sigma^p_x$ at IP ($\mu$m) & 280 & 179 \\\hline
$\sigma^p_y$ at IP ($\mu$m) & 50 & 48 \\\hline
$\sigma^p_z$ at IP (cm) & 11 & 11 \\\hline
${\cal L}_{\rm inst}$ (10$^{31}$ cm$^{-2}$s$^{-1}$) & 1.5 & 1.8 \\\hline
\end{tabular}
\end{center}
\end{table}
The improvement of the luminosity as a function of time can be understood
from the formula:
\begin{equation}
{\cal L}\propto \frac{I_eI_p}{\sigma_x\sigma_y}\,.
\end{equation}
The improvement in 1993 with respect to 1992 was mainly due to the increases
in the beam currents; from 0.94 to $10.8\,{\rm mA}$ for protons and from
1.33 to $7.7\,{\rm mA}$ for electrons. This was achieved mainly by
increasing the number of colliding bunches from 9 in 1992 up to 84 in 1993.
During the summer of 1994, the electron beam was replaced by a
positron beam which has considerably improved the beam lifetime\footnote{The
electron lifetime was severely limited due to presumably its interaction with
positively ionized impurities in the beam pipe. To cure the problem, 
the original ion getter pumps of the electron ring have been replaced 
in the 1997/1998 shutdown by passive non-evaporating getter pumps 
(adsorption pumps without high voltage that do not accelerate dust particles 
into the beam vacuum).} at high currents.
The number of colliding bunches have increased subsequently to 153 in 1994
and then to about 175 since 1995 with a maximum number of 189.
In addition to these colliding bunches, some
bunches are left unpaired (so-called pilot bunches, i.e. the corresponding
bunch in the other beam is empty). These pilot bunches are used for
evaluating the beam related backgrounds both for the luminosity determination
and for physics analyses. Two successive bunches are separated in time by $96\,{\rm ns}$.
This is to be compared with the collision frequency of $22\,\mu{\rm s}$ and
25\,ns respectively at LEP\footnote{The number corresponds to four bunches
per $e^\pm$ beam during the early (late) run of 1989-1992 (1996-2000), the
collision frequency was reduced by a factor of two or more by increasing
the number of bunches for the years between 1992 and 1995.} and LHC.

Other improvements include 
\begin{itemize}
\item the full exploitation in the aperture
margin of the machine in order to 
squeeze the beam cross sections at the collision point down
to $179\,\mu{\rm m}\times 48\,\mu{\rm m}$, which are two times smaller 
than the design value,
\item significant progress in the proton beam intensity
by improving controls and beam handling in the injector chain. The 
maximum proton beam current achieved in $ep$ collisions is now
$109\,{\rm mA}$ (or $\sim 8\times 10^{10}$ protons per bunch).
\end{itemize}
The maximum electron current obtained is 51.8\,mA. 
The beam current is limited by the available {\sc rf} power.
Therefore, although the beam currents still fall short of expectations, the
maximum peak luminosity achieved $1.8\times 10^{31}\,{\rm cm}^{-2}{\rm s}^{-1}$
exceeds the design goal of $1.5\times 10^{31}\,{\rm cm}^{-2}{\rm s}^{-1}$
by focusing very tightly the beams at the interaction points.

An important ungrading program has been prepared and has started since
September 2000. After the upgrade, an increase in the luminosity of 
more than a factor of 5 is expected~\cite{hera_upgrade}. 

\newpage
\section{The H1 detector} \label{sec:h1det}
\begin{figure}[htbp]
\begin{center}
\begin{picture}(50,270)
\put(-160,-15){\epsfig{file=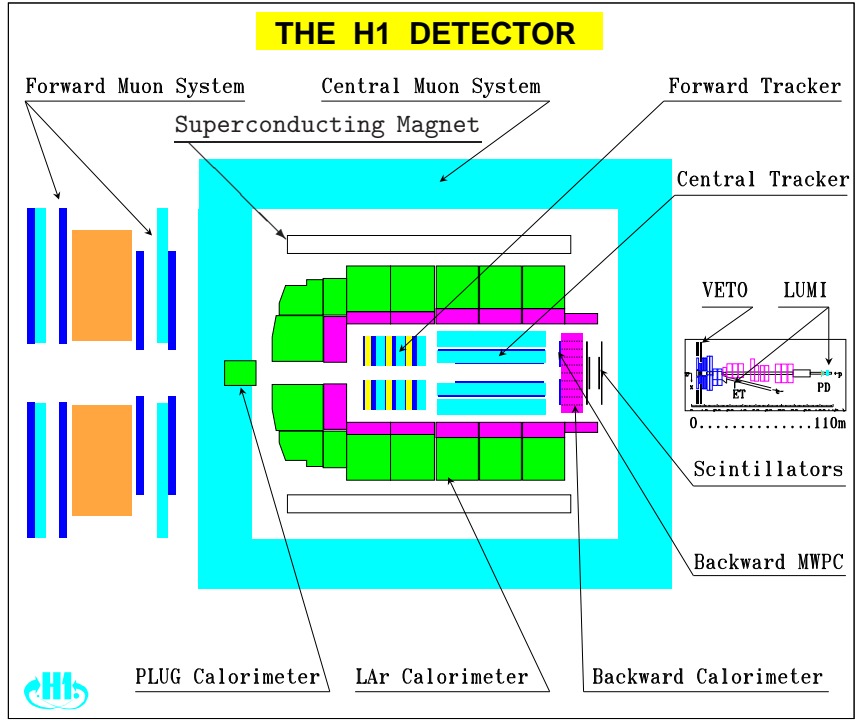,bbllx=0pt,bblly=0pt,bburx=594pt,
bbury=842pt,width=100mm,angle=90}}
\put(-60,215){\footnotesize {\tt Superconducting Magnet}}
\put(-60,211.5){\line(1,0){116}}
\put(-55,211.5){\vector(1,-1){37}}
\end{picture}
\end{center}
\caption{\sl An overview of the H1 detector.}
\label{fig:h1det}
\end{figure}

The H1 detector~\cite{h1det} 
(Fig.\ref{fig:h1det}) is nearly hermetic 
multi-purpose apparatus built to investigate $ep$ interactions at HERA.
The main components are the tracking detectors (Sec.\ref{sec:tracking}),
the calorimetry (Sec.\ref{sec:calorimetry}), the superconducting magnet, 
the muon detectors (Sec.\ref{sec:muon}), the very forward detectors 
(Sec.\ref{sec:fwddet}) and the luminosity detectors and various electron 
taggers (Sec.\ref{sec:lumisyst}).
In many aspects, the H1 detector\footnote{The ZEUS detector~\cite{zeusdet} 
is rather similar to the H1 detector in many aspects. The main difference 
lies in the main calorimeter; as described in Sec.\ref{sec:calorimetry}, 
the liquid-argon calorimeter of H1 has a fine granularity and 
a good resolution for electromagnetic objects whereas the uranium-scintillator 
calorimeter of ZEUS is compensating, giving equal response to hadrons and 
electrons, and has a good resolution for hadrons 
($\sigma(E)/E\simeq 35\%/\sqrt{E}\oplus2\%$). 
The ZEUS calorimeter has in addition a good time resolution, 
which is better than 1\,ns for energy deposits greater than 4.5\,GeV 
thus providing a fast time information for triggers and background rejections. 
The location of the superconducting coil providing magnetic field for trackers
is different as well; the coil of H1 is placed outside of the calorimeter 
thus minimizing the amount of dead material in front of the calorimeter, 
while ZEUS had placed it between the calorimeter and the tracker.}
does not differ strongly from the detectors
at $e^+e^-$ or $p\overline{p}$ colliders. Specific at HERA is however the
imbalance in the energy of the two colliding beams, which requires an
asymmetric detector.

The coordinate system convention for the experiment defines $x$ pointing to
the center of the HERA ring, $y$ the upward direction, and the forward,
positive $z$ direction as being that of the proton beam. The polar angle
$\theta$ is defined relative to this axis such that pseudo-rapidity,
$\eta=-\ln \tan\theta/2$, is positive in the forward region.

\subsection{The tracking detectors} \label{sec:tracking}

The tracking detectors consist of central jet chambers (CJC1, CJC2), central
trackers for measuring the $z$ coordinate (CIZ, COZ), central multiwire
proportional chambers for fast triggering (CIP, COP), forward tracking detector
(FTD), backward tracking detector (BPC, BDC), and central and backward silicon
microvertex detectors (CST, BST).

The CJC1 and CJC2, covering a polar angular range from 15$^\circ$ to
165$^\circ$, are two large, concentric drift chambers. The inner chamber,
CJC1, has 24 layers of sense wires arranged in 30 phi cells, while CJC2 has
32 layers of sense wires in 60 phi cells. The cells are at a 30$^\circ$
angle to the radial direction. The point resolution is 170\,$\mu$m in the
$r-\phi$ direction. The $z$ coordinate is measured by charge division and
has an accuracy of 22\,mm. A superconducting solenoid, which surrounds both
the tracking system and the liquid argon (LAr) calorimeter, provides a
uniform magnetic field of 1.15\,Tesla. The momentum of charged particles
may be determined from their track curvature in the magnetic field with a
transverse momentum resolution of $\sigma_{p_T}/p_T< 0.01 p_T$\,GeV.
The $dE/dx$ resolution for a well measured track is better than 7\%.

Two thin cylindrical drift chambers CIZ and COZ have sense wires perpendicular
to the beam axis, and therefore complement the accurate $r-\phi$ measurement
provided by the CJC with precise $z$ coordinates. The CIZ is
located at 18\,cm inside the CJC1, while COZ is located at 47\,cm
between CJC1 and CJC2. These two chambers deliver track elements with
typically 300\,$\mu$m resolution in $z$. To each of the $z$ chambers a
proportional chambers (CIP/COP) is attached for triggering.

The FTD, covering an angular range from 7$^\circ$ to 25$^\circ$, are 
integrated assemblies of three supermodules, each including, in
order of increasing $z$, three different orientations of planar wire drift
chambers (each rotated by 60$^\circ$ to each other in azimuth), a multiwire
proportional chamber (FPC), a transition radiation
detector and a radial wire drift chamber.

The backward proportional chamber BPC, located just in front on the backward
calorimeter, is made of four planes of wires with vertical, horizontal and
$\pm 45^\circ$ orientations. The wires are strung every 2.5\,mm, and signals
from two wires are fed to one preamplifier. Three out of four planes are
required in coincidence in order to reconstruct a space point with a spatial
resolution of about 1.5\,mm in the transverse plane.
The BPC provided an angular measurement of the electron in the
range from 155$^\circ$ to 174$^\circ$, together with the vertex given by
the main tracking detectors with a precision better than 1\,mrad. 
This detector has been replaced in the 1994/95 shutdown by an eight layer drift
chamber BDC with an extended acceptance between 155.1$^\circ$ and
177.5$^\circ$.

\subsection{Calorimetry} \label{sec:calorimetry}
The calorimetry system consists of the LAr calorimeter, the backward
electromagnetic calorimeter (BEMC/SPACAL), the forward calorimeter (PLUG) and 
the outer calorimeter, the so-called tail-catcher.

\begin{figure}[htbp]
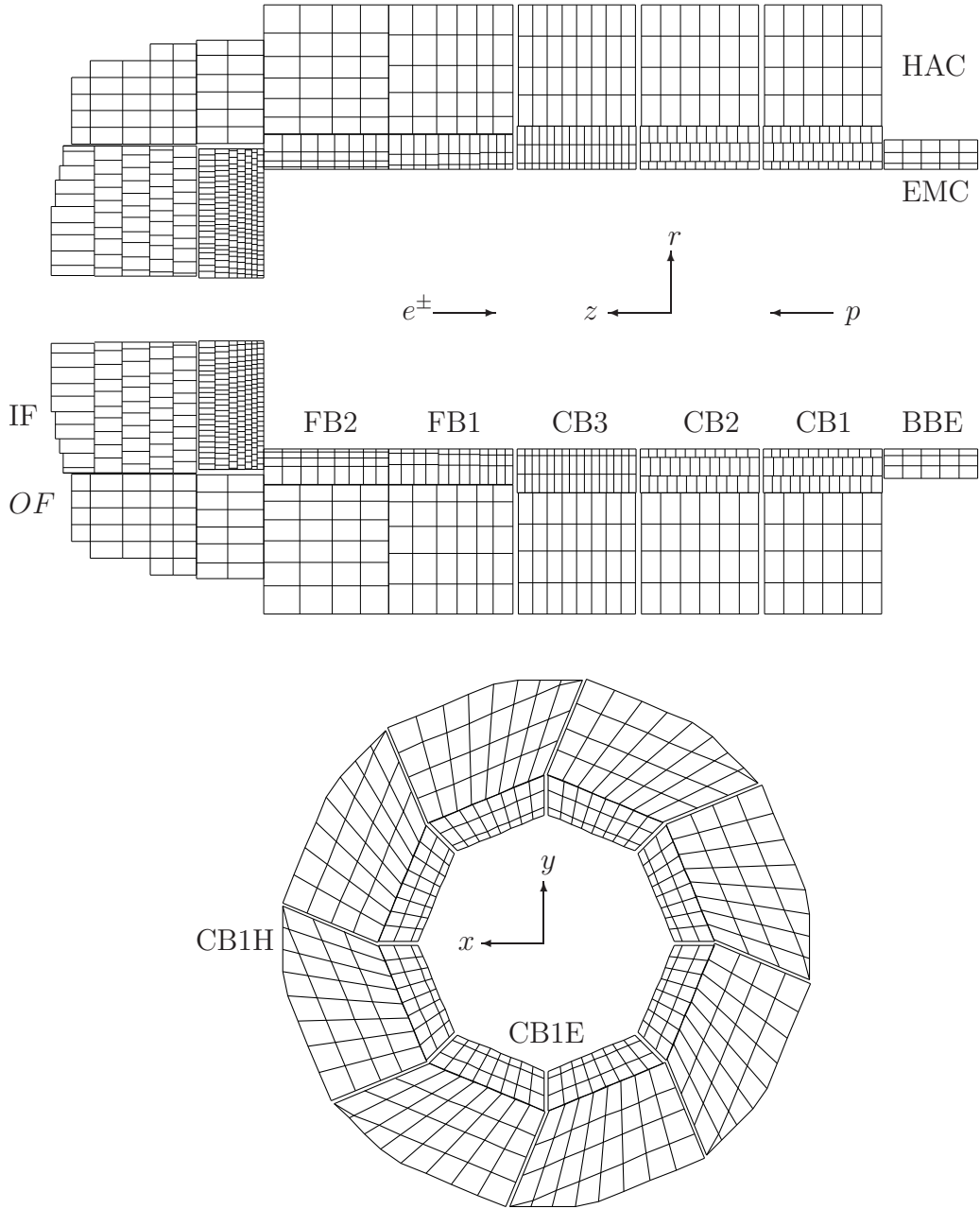

\begin{center}
\begin{picture}(50,470)
\put(-190,150){\epsfig{file=larcalo.ps,bbllx=0pt,bblly=0pt,bburx=594pt,
bbury=842pt,width=50mm}}
\put(95,350){\vector(0,1){25}}
\put(95,350){\vector(-1,0){25}}
\put(0,350){\vector(1,0){25}}
\put(160,350){\vector(-1,0){25}}
\put(60,347.5){$z$}
\put(94,377.5){$r$}
\put(-12.5,347.5){$e^\pm$}
\put(165,347.5){$p$}
\put(-170,305){IF}
\put(-170,270){$OF$}
\put(-52,302.5){FB2}
\put(-2.5,302.5){FB1}
\put(47.5,302.5){CB3}
\put(100,302.5){CB2}
\put(145,302.5){CB1}
\put(187.5,302.5){BBE}
\put(187.5,395){EMC}
\put(187.5,445){HAC}
\put(-250,-150){\epsfig{file=calo_phi.eps,bbllx=0pt,bblly=0pt,
bburx=594pt,bbury=842pt,width=350mm}}
\put(44.2,97.5){\vector(0,1){25}}
\put(44.2,97.5){\vector(-1,0){25}}
\put(10,95){$x$}
\put(43,127.5){$y$}
\put(-95,95){CB1H}
\put(30,58){CB1E}
\end{picture}
\end{center}
\caption{\sl A schematic $r-\phi$ view of the pad structure in different
wheels (top) and a transverse view of the pad structure in 8 octants in CB1.}
\label{fig:larcell}
\end{figure}

The emphasis is put on the electron recognition and energy measurement. This
led to placing the LAr inside the magnetic coil in order to minimize the
passive material. LAr was chosen because of its good stability, ease of
calibration, possibility of fine granularity and homogeneity of response. 
The LAr calorimeter covers the polar angle range between 3$^\circ$ and 
155$^\circ$. The calorimeter along the beam axis is segmented into 8 
``wheels'', with each wheel being further segmented into octants in 
$\phi$ (Fig.\ref{fig:larcell}). 
The structure of the electromagnetic section (EMC) consists of a pile of
G10-Pb-G10 sandwiches separated by spacers defining the LAr gaps.
The hadronic section (HAC) is made of stainless steel absorber plates
with independent readout cells inserted between the plates.
The orientation of the plates varies with $z$ such that
particles always impact with angles greater than 45$^\circ$.
The granularity ranges from 10
$\rightarrow$ 100\,cm$^2$ in the EMC section and to 50 $\rightarrow$
2000\,cm$^2$ in the HAC section. Longitudinal segmentation is 3-4 layers in
the EMC over 20-30 radiation lengths ($X_0$) and 4-6 layers in the HAC.
The total depth of both sections varies between 4.5 and 8 interaction lengths
($\lambda$). The most backward part of the LAr calorimeter is a smaller
electromagnetic calorimeter (BBE) which covers the polar angle ranging from
146$^\circ$ to 155$^\circ$. The LAr calorimeter has a total of
$45\,000$ readout cells. The noise per cell ranges from 10 to 30\,MeV.
The resolution measured in the test beam is $0.12/\sqrt{E({\rm GeV})}\oplus
0.01$ for an electromagnetic shower and $0.5/\sqrt{E({\rm GeV})}\oplus
0.02$ for a hadronic shower (the symbol $\oplus$ denotes addition in
quadrature). The electromagnetic and hadronic energy scale uncertainty is
respectively 1-3\% and 2-4\%. The calorimeter is non-compensating, with the
response to hadrons about 30\% lower than the response to electron of the
same energy. An offline weighting technique is used to equalize the response
and provide the optimal energy resolution~\cite{gayler85,h1nim336_93b}.

In the backward region, the LAr was complemented in the angular range from
151$^\circ$ to 176$^\circ$ by the BEMC, a conventional lead-scintillator
sandwich calorimeter. The BEMC was replaced in the 1994/95 shutdown by 
a lead/scintillating fiber calorimeter (SPACAL). The calorimeter BEMC 
is made of 88 stacks with a size of
$16\times 16\,{\rm cm}^2$ and has a depth of 21.7\,$X_0$, or approximately
1\,$\lambda$, which on average contains 45\% of the energy of a hadronic
shower. The four inner stacks around the beam pipe are of triangle shape.
A 1.5\,cm spatial resolution of the lateral shower position is achieved
using four photodiodes which detect the wavelength shifted light from each
of the scintillator stacks.
The electromagnetic energy resolution is
$\sigma/E=10\%/\sqrt{E({\rm GeV})}\oplus 1.7\%$.
A scintillator hodoscope (TOF) situated behind the BEMC is used to veto
proton induced background events based on their early time of arrival
compared with nominal $ep$ collision.
The new calorimeter SPACAL has both electromagnetic and hadronic sections.
The electromagnetic energy resolution is $7.5\%/\sqrt{E({\rm GeV})}\oplus 2.5\%$, and the
hadronic section has 2\,$\lambda$ and an integrated timing function to veto
proton beam induced background interactions (the old TOF system could thus
be removed). The angular region covered is
extended (from $153^\circ$ to $177.5^\circ$) compared to the BEMC,
and the calorimeter has a very high
granularity (1192 cells) yielding a spatial resolution of about 2\,mm.

The LAr and BEMC/SPACAL calorimeters are surrounded by the iron return yoke,
which is instrumented with 16 layers of limited streamer tubes (LST).
Eleven of the 16 layers are equipped with readout electrodes (pads). From
ionization energies of particles passing through the chambers the energy of
tails of hadronic showers leaking out of the LAr are measured in the analog
readout system. Therefore the system is also called tail-catcher (TC). 
The energy resolution is $\sigma/E\thickapprox 100\%/\sqrt{E({\rm GeV})}$. 

In the forward direction around the beam pipe, the angular range
from 0.3$^\circ$ to 3.3$^\circ$ is covered by the PLUG, 
a sampling calorimeter consisting of nine copper absorber plates interleaved
with eight sensitive layers of large area silicon detectors.

\subsection{Muon detectors} \label{sec:muon}
Recognition of muons is very important in the study of heavy quarks, heavy
vector mesons, $W$-production and in the search for exotic physics.

Muons in the central region are identified by looking for particles
penetrating the calorimeter and coil and leaving signals in the TC.
Three of the 16 instrumented LST layers are located before the first
iron plate, and three after the last iron plate. There is a double layer
after four iron plates, and eight single layers in the remaining gaps
between the iron sheets. A minimum muon energy of 1.2\,GeV is needed to
reach the first LST, while 2\,GeV muons just penetrate the iron.

In the very forward direction, a spectrometer composed of drift chambers
surrounding a toroidal magnet with a field of 1.6\,Tesla is used to measure
muons. This spectrometer measures muons in the momentum range between 5 to
200\,GeV. The lower limit is determined by the amount of material traversed,
while beyond the upper limit the muon charge can no longer be measured
unambiguously.

\subsection{Very forward detectors} \label{sec:fwddet}
H1 has spectrometers downstream of the main detectors in the proton beam
direction to measure high energy protons, as well as calorimeters at zero
degrees to measure high energy neutrons. These are used in the study of
diffractive scattering as well as in the study of leading-particle production.

A forward proton spectrometer (FPS) has been installed since 1995 at 81 and
90\,m away from the interaction point, which detects leading protons in the
momentum range from 580 to 740\,GeV and scattering angles below 1\,mrad.
The FPS has been extended with stations at 80 and
63\,m since 1997. In all stations the protons are detected
with scintillating fiber hodoscopes. The detector elements are mounted
inside plunger vessels, so called Roman Pots, which are retracted during
injection and are brought close to the beam after stable luminosity
conditions are reached.

The forward neutron calorimeter (FNC) is located at 107\,m downstream of the
interaction point. The calorimeter consists of interleaved layers of lead
and scintillating fibers. The calorimeter has a total depth of
9.5\,$\lambda$ and has an acceptance $>90\%$ for neutrons with a production
angle below 1\,mrad.

\subsection{Luminosity detector and electron taggers} \label{sec:lumisyst}
\begin{figure}[htb]
\begin{center}
\begin{picture}(50,110)
\put(-167.5,-32.5){\epsfig{file=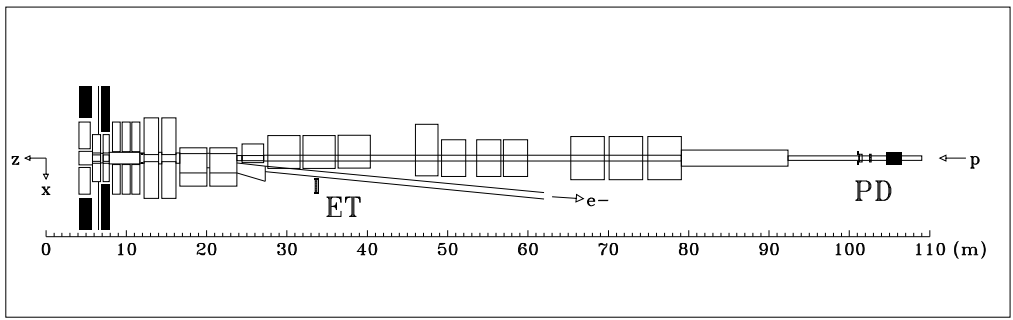,width=50mm,angle=90}}
\end{picture}
\end{center}
\caption{\sl The layout of the H1 luminosity system consisting of an
electron tagger (ET=etag) and a photon detector (PD).
Note that the $z$ axis coincides with the proton beam direction.}
\label{fig:detlumi}
\end{figure}

At HERA, the luminosity is determined from the rate of the bremsstrahlung
(Bethe-Heitler) process $ep\rightarrow ep\gamma$ (Sec.\ref{sec:rad_proc})
using a small luminosity system 
(the example from H1 is indicated schematically in Fig.\ref{fig:h1det} 
and shown in a larger view in Fig.\ref{fig:detlumi}).
The process has a large and precisely known cross section.
In the H1 area, the final state electron, deflected by a set of low-beta
quadruples and a bending magnet located in the region
$-23.8\,{\rm m}<z<-5.8\,{\rm m}$, passes an exit window at $z=-27.3$\,m and
hits the electron tagger (etag) at $z=-33.4$\,m. The etag is located beside
the electron beam pipe and is made of 49 crystals covering a total area of
$154\times 154$\,mm$^2$. The photon leaves the proton beam pipe through
an exit window at $z=-92.3$\,m where the proton beam pipe bends upwards.
The photon detector (PD), situated at $z=-102.9$\,m on the $z$ axis, is
built out of 25 crystals with a total surface of $100\times 100$\,mm$^2$
and a depth of 22 radiation lengths ($X_0$). 
A $2X_0$ Pb filter followed by a $1X_0$ water 
\v{C}erenkov veto counter (VC) in front of the PD protects the detector from 
the high synchrotron radiation flux. The VC eliminates events with photons
interacting in the filter. The whole system provides a fast
relative luminosity measurement with a statistical precision of about 2\%
at nominal beam conditions.

The luminosity system serves in addition several other purposes. It provides
\begin{itemize}
\item electron beam monitoring for the HERA machine, 
\item absolute luminosity measurement in the interaction
region with an accuracy of better than $5\%$, 
\item tagging of photoproduction events,
\item energy measurement for electrons scattered under small angles and 
\item energy measurement for photons from initial state radiation
(Sec.\ref{sec:radc}).
\end{itemize}

The etag at 33\,m measures electrons in the energy range
$0.25E_e<E_e^\prime <0.75E_e$ with an average acceptance of 48\%. The
acceptance to higher energy ($0.76E_e<E_e^\prime <0.96E_e$) is covered by a
new tagger at 44\,m installed since 1995. The lower energy is covered by
yet another tagger at 4\,m.

\subsection{Trigger and event reconstruction} \label{sec:trigrec}
The purpose of the trigger system is a fast separation of the interesting
physics events from background events. The main background sources, common
as those presented to other accelerator experiments, are synchrotron
radiation from the electron beam, proton gas interaction in the beam pipe
vacuum of about $10^{-9}$\,hPa and stray protons, which produce particle
showers by hitting the beam tube and other materials around the accelerator
(beam-gas and beam-wall). Beam halo muons and muons from cosmic radiation
also contribute. The rate of the background is up to $10^4$ times higher than
the rate of a variety of physics processes under study in $ep$ collisions
(Table \ref{tab:xs_rates}). The rate of physics processes extends from 
photoproduction, where the visible $ep$ cross section of
several $\mu$b implies an event rate of 20-30\,Hz at design luminosity (see
Table \ref{tab:hera_para}), towards $W$ production expected to occur a few
times per week.
\begin{table}[tbh]
\begin{center}
\begin{minipage}{12cm}
\caption{\sl A comparison of cross sections and/or rates between the main
background sources and various physics processes at design luminosity.}
\label{tab:xs_rates}
\end{minipage}
\begin{tabular}{|l|r|r|}
\hline
Source/process & Cross section & Rate \\\hline
Beam-gas(wall) interaction & & 50\,kHz \\
Cosmic $\mu$ in barrel & & 700\,Hz \\\hline
Tagged $\gamma p$ & 1.6\,$\mu$b & 25\,Hz \\
$c\overline{c}$ total & 1\,$\mu$b & 15\,Hz \\
DIS low $Q^2$ & 150\,nb & 2.2\,Hz \\
DIS high $Q^2$ (e in LAr) & 1.5\,nb & 1.4\,min$^{-1}$ \\
Charged current DIS ($p_T>25$\,GeV) & 50\,pb & 3\,h$^{-1}$ \\
$W$ production & 0.4\,pb & 0.5\,d$^{-1}$ \\\hline
\end{tabular}
\end{center}
\end{table}                                                

The trigger system is based on 4 levels in order to filter the intersting
physics events, followed by the offline reconstruction of the kept events.

The trigger level 1 (L1) system makes a decision within 2\,$\mu$s on whether to
accept or to reject an event using information provided by different
subdetectors (``trigger elements (TE)''). The central trigger logic (CTL)
combines these trigger elements into 128 ``subtrigger (ST)''. Not all
subdetectors can provide this information fast enough to make a decision
after each bunch crossing (BC) immediately. Therefore the information is
sent into pipelines where it is kept until all relevant subdetectors have
provided their trigger elements. The delay of 2\,$\mu$s (24\,BC) is
necessary as some subdetectors are relatively slow: the CJC takes 11\,BC 
due to the longest drift time of 1\,$\mu$s and the LAr takes 13\,BC due to 
long integration time of the preamplifiers\footnote{The typical drift time
of an ionized electron in the LAr gap is about 200\,ns/mm, a gap of 2.35\,mm 
between two absorber plates results already in about 500\,ns, i.e.\ about
5\,BC.}. If any of the ST conditions is
fulfilled by the event, the pipeline is stopped immediately and the signal
is passed to the next trigger level\footnote{Some subtriggers are prescaled
such that not every event fulfilling the subtrigger conditions is kept.}

The level 2 (L2) and level 3 (L3)\footnote{This trigger has not yet been
used in H1.} triggers operate during the primary dead
time of the readout of about 1.5\,ms. They work on the same data as the L1,
and reach a decision within 20\,$\mu$s and 800\,$\mu$s, respectively. While
at the L1, only minimum correlation between different subdetectors (e.g.
between the MWPC and LAr) is used, the L2 and L3 make full use of the
detailed, high granularity trigger data of most subsystems. The L2 system
includes a complex topological correlation and a neural network approach.

The level 4 (L4) trigger is based on full event reconstruction in MIPS
R3000 based processor boards. Algorithms similar to the ones used for the
offline analysis are used to select valid events. The events accepted by L4
are written to tapes with a rate of about 15\,Hz.

The accepted raw data or those simulated Monte Carlo data are then fully
reconstructed and assigned into different physics event classes. The data
information is written in a compressed format to Data Summary Tapes (DST)
which are the basis for physics analyses.

\section{Monte Carlo technique and detector simulation} \label{sec:simulation}
Monte Carlo (MC) technique has proven to be indispensable for the extraction
of physical quantities from the measurements. For the cross section
measurement, MC programs can be used to determine corrections for acceptance,
efficiencies, background contamination, and resolution effects of 
the detector system. Some of these corrections can be obtained directly from 
the data though often with a limited statistical precision. 
The MC, which provides in principle unlimited event samples, 
allows therefore to model the data with a better precision.
The simulated MC events are also very useful in defining variables and tuning cuts 
for selecting signal events from various background contributions.

More information concerning Monte Carlo generators used in different analyses
will be given later together with the analyses.
Once an event (either a signal or a background event) is generated, 
the H1 detector response to the particles generated in the event is simulated
in detail using the H1 simulation package, {\sc h{\small 1}sim}, 
which makes use of
the {\sc geant} program~\cite{geant}. The parameters used by this program
were determined in test beam measurements and optimized during $ep$ data
taking. For the simulation of the energy response of the calorimeter a fast
parametrization is used for the development of electromagnetic and hadronic 
showers to save computing time. These simulated events are then subject to 
the same reconstruction program (Sec.\ref{sec:trigrec}) as the data and 
the same analysis chain.

\newpage
\section{Kinematics reconstruction and coverage}
\subsection{Reconstruction of kinematic variables} \label{sec:kinerec}
For NC events, the kinematics is over-constrained as the
HERA experiments measure both the scattered electron and the hadronic
final state. Here are a few commonly used methods:
\begin{description}
\item [Electron method] uses the energy, $E_e^\prime$, and the polar
angle, $\theta_e$, of the scattered electron measured relative to 
the proton beam direction:
\begin{eqnarray}
& & Q^2_e=4E_eE_e^\prime\cos^2\left(\frac{\theta_e}{2}\right)
\label{eq:q2e} \\
& & y_e=1-\frac{E_e^\prime}{E_e}\sin^2\left(\frac{\theta_e}{2}\right) 
\label{eq:ye} \\
& & x_e=\frac{Q^2_e}{sy_e} \label{eq:xe} \,.
\end{eqnarray}
Since the incident beam energy $E_e$ appears both in $Q^2_e$ and in $y_e$, 
this method is thus sensitive to the initial state radiation (the effective
beam energy after radiation is smaller than $E_e$, see Sec.\ref{sec:radc},
and Eq.(\ref{eq:eb_eff}) in Sec.\ref{sec:isr}).
\item [Double angle (DA) method~\cite{kineda}] uses the electron polar angle,
$\theta_e$, and the inclusive hadronic polar angle, $\theta_h$,
which is the polar angle of the scattered quark in the QPM (Sec.\ref{sec:qpm})
with massless quarks
\begin{eqnarray}
& & Q^2_{\rm DA}=\frac{E_e}{E_p}\frac{s}{\alpha_e(\alpha_e+\alpha_h)}\,,
\label{eq:q2da} \\
& & y_{\rm DA}=\frac{\alpha_h}{\alpha_e+\alpha_h}\,, \\
& & x_{\rm DA}=\frac{Q^2_{\rm DA}}{sy_{\rm DA}}\,,
\end{eqnarray}
with
\begin{eqnarray} 
& \displaystyle \alpha_e\equiv \tan\!\left(\frac{\theta_e}{2}\right)=\frac{\Sigma_e}{p_{T,e}}\,, &
\alpha_h\equiv
\tan\!\left(\frac{\theta_h}{2}\right)=\frac{\Sigma_h}{p_{T,h}}
\label{eq:alpha_eh}\\
& \displaystyle \Sigma_e\equiv E_e^\prime-p_{z,e}\,, &
p_{T,e}\equiv E_e^\prime\sin\theta_e \label{eq:epzpt_e} \\ 
& \displaystyle \Sigma_h\equiv \sum_i E_i-p_{z,i}\,, &
p_{T,h}\equiv \sqrt{(\sum_i p_{x,i})^2+(\sum_i p_{y,i})^2}\label{eq:epzpt_h}
\end{eqnarray}
where the summations are over all particles of the hadronic final state.
The DA method is also sensitive to photon emission of the primary electron. 
On the other hand, it is, to 
a good approximation, insensitive to the energy scale uncertainties. It has
thus been widely used in various analyses to check and improve the energy
scales of the scattered electron and of the hadronic system.
\item[Sigma ($\Sigma$) method~\cite{kinesigma}] is constructed such that no
electron beam energy $E_e$ is used directly in $Q^2$ and $y$:
\begin{eqnarray}
& &
Q^2_\Sigma=\frac{E^{\prime 2}_e\sin^2\theta_e}{1-y_\Sigma}\,,\label{eq:q2s}\\
& & y_\Sigma=\frac{\Sigma_h}{\Sigma_e+\Sigma_h}\,,\label{eq:ys}\\
& & x_\Sigma=\frac{Q^2_\Sigma}{sy_\Sigma}\,.\label{eq:xs}
\end{eqnarray}
This method is thus less sensitive to the radiative effects.
\end{description}
The fact that different methods can be used to reconstruct the event kinematics
has two consequences:
\begin{itemize}
\item The kinematic domain is fully explored; for example the electron method is 
good at high $y$ while the $\Sigma$ method can be used at sufficiently low $y$.
\item Important cross-checks of systematic effects are possible between
different methods.
\end{itemize}

For a CC event, however, the kinematics can only be reconstructed from the
hadronic final state:
\begin{eqnarray}
& & Q^2_h=\frac{E^2_{T,h}}{1-y_h}\,,\\
& & y_h=\frac{\Sigma_h}{2E_e}\,,\label{eq:yh} \\
& & x_h=\frac{Q^2_h}{sy_h}\,.
\end{eqnarray}
The understanding of the hadronic energy scale and its scale uncertainty is
thus crucial for all CC measurements.

\subsection{Coverage of the kinematic phase space}
\begin{figure}[htb]
\begin{center}
\begin{picture}(50,370)
\put(-190,-20){\epsfig{file=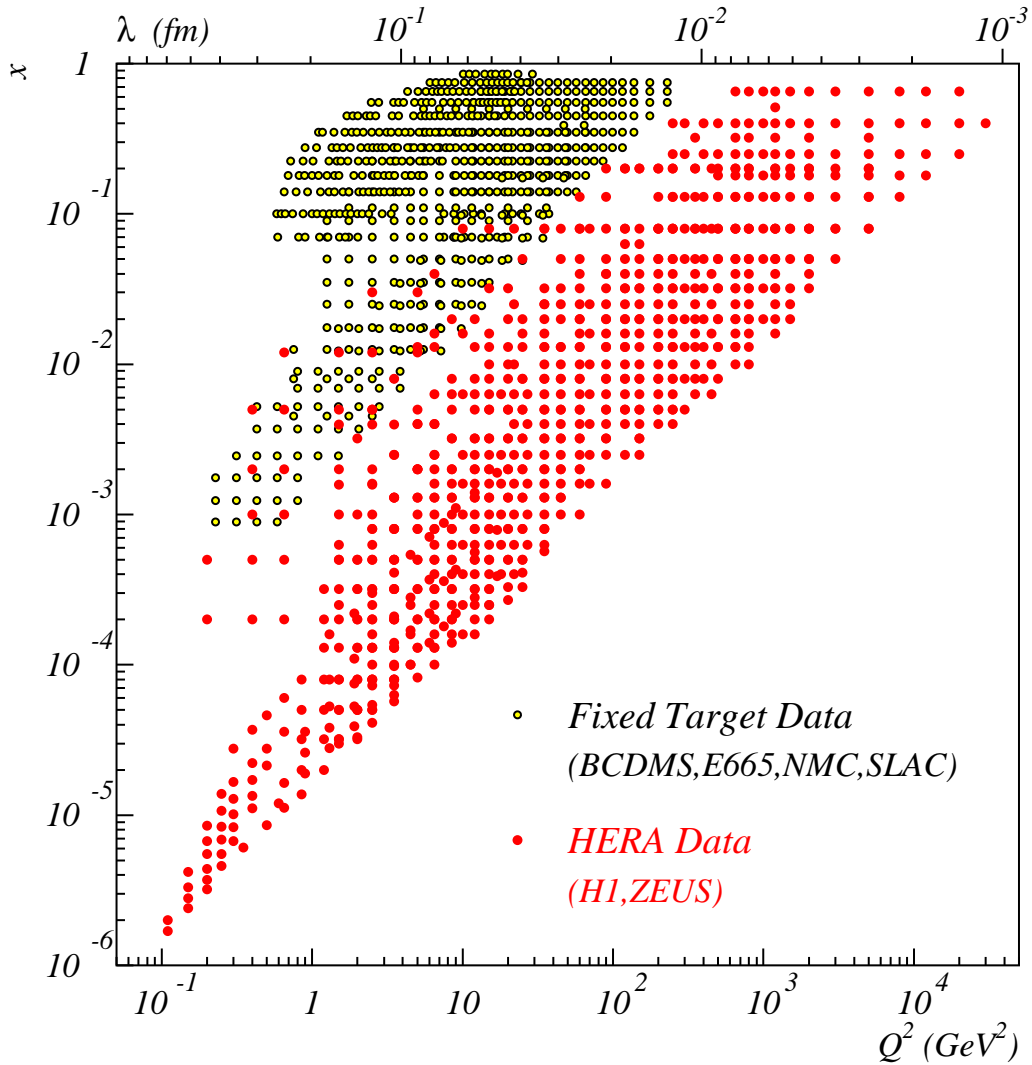,width=150mm}}
\end{picture}
\end{center}
\caption{\sl Coverage in the kinematic plane $x-Q^2$ of the HERA experiments,
H1 and ZEUS, compared with that of the fixed-target experiments from
BCDMS~\cite{bcdms}, NMC~\cite{nmc97}, E665~\cite{e665}, and
SLAC~\cite{slac92}. The corresponding scale $(\lambda)$ probed by a virtual 
boson is also indicated. The points indicate the location where 
cross sections or structure functions have been measured.}
\label{fig:kine_xq2}
\end{figure}

The HERA measurements in DIS cover a vast kinematic range as shown in \
Fig.\ref{fig:kine_xq2}. The HERA measurements extend those of fixed-target
experiments by more than 2 orders of magnitudes in both $x$ and $Q^2$. 
At the highest values of $Q^2$, the measurements are still statistically 
limited (Chapter \ref{chap:hiq2}). 
The measurements for $Q^2\lesssim 4$\,GeV$^2$ 
have been made possible by using different techniques and by upgrading
experimental apparatus:
\begin{itemize}
\item Using data in which the interaction point is shifted from its nominal
position to the forward direction by $\sim 70$\,cm (so-called shifted vertex
events). For the same backward calorimeter, the
angular acceptance is increased when the interaction vertex is moved away
from the nominal interaction point.
\item Using data with hard initial state radiation. A large fraction ($\sim
30\%$) of the photons from events with hard initial state radiation are
measured in the photon calorimeters of the luminosity system
(Sec.\ref{sec:lumisyst}). The resultant $ep$ collisions therefore occur at 
lower center-of-mass energies, and lower $Q^2$ and higher $x$ values can be
attained. As an example, the H1 analysis is shown in some detail 
in Sec.\ref{sec:isr} to illustrate how the measurement is realized
in an extended kinematic region.
\item Upgraded apparatus. The angular acceptance of the scattered electron 
has been significantly extended in the shutdown period 1994-95 by replacing 
the old backward calorimeter and proportional chamber, BEMC and BPC, with the
new calorimeter and drift chamber, SPACAL and BDC (Sec.\ref{sec:h1det}). 
The ZEUS experiment has added a small angle electron calorimeter, 
the beam pipe calorimeter, in the same shutdown period.
\end{itemize}

\chapter{Theoretical framework} \label{chap:theory}
\section{Deep inelastic scattering and the quark parton model}
\label{sec:disqpm}
\subsection{Cross sections and structure functions} \label{sec:xs_sf}
The neutral current cross section for charged leptons scattering off nucleons
$l^\pm N \rightarrow l^\pm X$ studied in all fixed-target experiments and
in the low $Q^2$ part of the HERA kinematic domain arises to a good 
approximation from the photon exchange only. Taking the proton as an example,
the cross section is directly proportional to the amplitude
squared~\cite{book_esw}
\begin{equation}
\displaystyle \frac{d^2\sigma^{l^\pm p}}{dxdy} \propto |{\cal A}|^2
= \frac{1}{q^4}L_{\mu\nu}W^{\mu\nu}\label{eq:em_xs}
\end{equation}
with
\begin{equation}
{\cal A}=\frac{1}{q^2}j_\mu J^\mu
\end{equation}
where the term $1/q^2$ corresponds to the propagator of the exchanged photon,
$j_\mu$ and $J^\mu$ are respectively the lepton and proton currents.
In Eq.(\ref{eq:em_xs}), $L_{\mu\nu}$ and $W^{\mu\nu}$ are respectively the
leptonic and hadronic tensors. The leptonic part is completely determined by
QED, whereas the hadronic one contains all the information about the
interaction of the electromagnetic current $j_\mu$ with the proton target.
By writing down the most-general possible combinations of all the momenta
appearing in the reaction in $L_{\mu\nu}$ and $W^{\mu\nu}$ and then 
simplifying the result using general theoretical 
assumptions
such as parity 
and time-reversal invariance, the cross section can finally be expressed
as
\begin{equation}
\displaystyle \frac{d^2\sigma^{l^\pm p}}{dxdy}= \frac{4\pi\alpha^2 (s-M^2)}
{Q^4}\left[\left(1-y-\frac{M^2xy}{s-M^2}\right)F_2+\frac{y^2}{2}2xF_1\right],
\label{eq:xsnc_lowq}
\end{equation}
where $\alpha$ is the fine structure constant, $M$ is the proton target mass,
and $F_2(x,Q^2)$ and $2xF_1(x,Q^2)$ are so-called structure functions,
to be determined by the deep inelastic experiments (Chapter \ref{chap:lowq2}).
There are two structure functions because there are two essential degrees of
freedom. The exchanged virtual photon can be transversely or longitudinally
polarized. If one defines $\sigma^{\gamma^\ast p}_{\rm tot}$ as the total 
$\gamma^\ast p$ cross section, then it can be decomposed into 
the longitudinal part $\sigma_L$ and the transverse one
$\sigma_T$ as
\begin{equation}
\sigma^{\gamma^\ast p}_{\rm tot}=\sigma_L +\sigma_T\,, \label{eq:xs_tot}
\end{equation}
with
\begin{eqnarray}
\sigma_T \!\!&=&\!\! \frac{2\pi^2\alpha}{xMK}2xF_1\,,\\
\sigma_L \!\!&=&\!\! \frac{2\pi^2\alpha}{xMK}F_L\,.
\end{eqnarray}
For virtual photon, the flux factor $K$ cannot be unambiguously defined.
The longitudinal structure function $F_L$ is a combination of $F_2$ and $2xF_1$:
\begin{equation}
F_L=\left(1+\frac{4M^2x^2}{Q^2}\right)F_2-2xF_1\,. \label{eq:fl}
\end{equation}
The measurement is carried out either directly on $F_L$ or on $R$, the 
cross-section ratio of the longitudinally polarized over transversely
polarized virtual photon:
\begin{equation}
R=\displaystyle
\frac{\sigma_L}{\sigma_T}=\left(1+\frac{4M^2x^2}{Q^2}\right)\frac{F_2}{2xF_1}-1\,,
\label{eq:r}
\end{equation}
so that the arbitrary flux factor $K$ does not enter.

At HERA, high $Q^2$ regime ($Q^2\gtrsim M_Z^2$) is reached, where the
contribution from pure $Z^0$ exchange and its interference with $\gamma$
exchange becomes increasely important. As the weak neutral current does not
conserve parity, an additional term ($x\tilde{F}_3(x,Q^2)$) has
to be taken into account in Eq.(\ref{eq:xsnc_lowq}):
\begin{eqnarray}
\displaystyle \frac{d^2\sigma^{l^\pm p}}{dxdQ^2}
\!\!\!&=&\!\!\!\frac{4\pi\alpha^2}{xQ^4}
\left[(1-y)\tilde{F}_2+\frac{y^2}{2}2x\tilde{F}_1\mp \left(y-\frac{y^2}{2}\right)
x\tilde{F}_3\right] \nonumber \\
\!\!\!&=&\!\!\!\frac{2\pi\alpha^2}{xQ^4}\left[Y_+\tilde{F}_2-y^2\tilde{F}_L\mp
Y_-x\tilde{F}_3\right],\label{eq:xsnc}
\end{eqnarray}
where $Y_\pm=1\pm(1-y)^2$ are helicity functions and the term proportional to
$M^2$ has been neglected.
Note that the structure functions $F_2(x,Q^2)$, $2xF_1(x,Q^2)$, and
$F_L(x,Q^2)$ have been replaced respectively by $\tilde{F}_2(x,Q^2)$,
$2x\tilde{F}_1(x,Q^2)$, and $\tilde{F}_L(x,Q^2)$
to account for the contributions from the $\gamma Z^0$ interference and $Z^0$ 
exchange terms (Eqs.(\ref{eq:f2nc_qpm}),(\ref{eq:gf2nc})).

The cross section for charged current processes,
\begin{eqnarray}
\begin{array}{l} \nu_\mu N\rightarrow \mu^- X \\
                      \overline{\nu}_\mu N\rightarrow \mu^+ X \end{array}
& & (\mbox{Fixed-target experiments})\\
\begin{array}{l} e^- p \rightarrow \nu_e X \\
                     e^+ p \rightarrow \overline{\nu}_e X \end{array} 
& & (\mbox{HERA experiments})
\end{eqnarray}
can be derived in a similar way and reads
\begin{eqnarray}
& \displaystyle \frac{d^2\sigma^{\rm CC}}{dxdQ^2}= &
\frac{G^2_F M^4_W(s-M^2)}{2\pi xs}\left(\frac{1}{Q^2+M^2_W}\right)^2 \times
\label{eq:xscc} \\
& & \left[\left(1-y-\frac{M^2xy}{s-M^2}\right)F^{\rm CC}_2+
\frac{y^2}{2}2xF^{\rm CC}_1 \pm
\left(y-\frac{y^2}{2}\right)xF^{\rm CC}_3\right],\nonumber
\end{eqnarray}
where $M_W$ is the mass of the exchanged $W$ boson. Again since the weak force 
does not respect parity invariance, a similar but different weak structure 
function $xF^{\rm CC}_3(x,Q^2)$ has to be introduced (compare Eqs.(\ref{eq:f3nc}) and
(\ref{eq:f3cce-}),(\ref{eq:f3cce+})).

\subsection{Structure functions in the quark parton model} \label{sec:qpm}
Two ideas, both put forward in 1969, have played an important role in the 
development of the experiments and in our understanding of them. 
The two ideas are those of the parton model by Feynman~\cite{feynman69} and of
scaling by Bjorken~\cite{bjorken69}.

The parton model is simply a formal statement of the notion that the nucleon
is made up of smaller constituents, the partons. No initial assumptions
about the partons are necessary as it is the purpose of the experiments to
determine their nature.

The scaling prediction states that when the momentum carried by the probe 
becomes very large, the dependence of the cross section on parameters such as 
the energy $\nu=E-E^\prime$ and momentum squared $q^2$, transferred by 
the photon, becomes simple. In the parton model, the onset of this simple 
scattering behavior has a straightforward interpretation. 
The complicated scattering
of the probe off a nucleon of finite spatial extent has been replaced by the
scattering of the probe off a point-like parton. The photon ceases to
scatter off the nucleon as a coherent object and, instead, scatter off the
individual point-like partons incoherently.

When it became clear that the hypothesized point-like constituents, partons
had properties characteristic of quarks, the parton model came to be called
the Quark Parton Model (QPM). 

While the cross sections shown in Eqs.(\ref{eq:xsnc_lowq}),(\ref{eq:xsnc}),
(\ref{eq:xscc}) make no assumptions about the underlying structure of the
hadron involved in the interaction, in the QPM, the structure functions 
$\tilde{F}^{l^\pm p}_2, x\tilde{F}^{l^\pm p}_3$ 
can be expressed in terms of parton distribution functions: 
\begin{eqnarray}
\displaystyle \tilde{F}^{l^\pm p}_2(x,Q^2)=\sum_i
A_i(\lambda,Q^2)x\left[q_i(x)+\overline{q}_i(x)\right], & &
\label{eq:f2nc_qpm} \\
x\tilde{F}^{l^\pm p}_3(x,Q^2)=\sum_i
B_i(\lambda,Q^2)x\left[q_i(x)-\overline{q}_i(x)\right], & & \label{eq:f3nc}\\
\tilde{F}_L=\tilde{F}_2-2x\tilde{F}_1=0\,, & & \label{eq:callan-gross}
\end{eqnarray}
where the sum runs over all quark flavors with
$xq(x)$ specifying the probability of finding a parton $q$
carrying a momentum fraction $x$ of the proton's momentum
in a frame where the proton's momentum is large. The last relation is so
called Callan-Gross relation~\cite{callan-gross} and is a direct consequence of
quark spin $1/2$ due to the fact that a spin-$1/2$ quark cannot absorb 
a longitudinally polarized vector boson. The couplings of the fermions
to the currents depend on the lepton-beam polarization $\lambda$:
\begin{eqnarray}
& & A_i(\lambda,Q^2)=
\frac{1-\lambda}{2}A^L_i(Q^2)+\frac{1+\lambda}{2}A^R_i(Q^2)\,,\\
& & B_i(\lambda,Q^2)=
\frac{1-\lambda}{2}B^L_i(Q^2)+\frac{1+\lambda}{2}B^R_i(Q^2)\,,
\end{eqnarray}
with
\begin{eqnarray}
& & A^{L,R}_i(Q^2)=
e^2_i-2e_i(v_l\pm a_l)v_iP_Z+(v_l\pm a_l)^2(v_i^2+a^2_i)P^2_Z\,,\label{eq:a_lr}\\
& & B^{L,R}_i(Q^2)=
\mp 2e_i(v_l\pm a_l)a_iP_Z\pm 2(v_l\pm a_l)^2v_ia_iP^2_Z\,,\label{eq:b_lr}
\end{eqnarray}
for $l^-p$ scattering, the corresponding structure functions for $l^+p$
scattering are obtained by swapping $L\rightarrow R$, $R\rightarrow L$ in
Eqs.(\ref{eq:a_lr}),(\ref{eq:b_lr}).
The vector and axial-vector couplings of the fermions are given by
\begin{equation}
v_f=T_{3f}-2e_f\sin^2\theta_W\,,\hspace{1cm} 
a_f=T_{3f}\,,
\end{equation}
where the definition holds for both changed leptons and quarks, $T_{3f}$ 
is the third component of weak isospin, and $\theta_W$ is the Weinberg angle.
The $Z^0$ propagator appears in the quantity $P_Z$ as:
\begin{equation}
P_Z=\frac{Q^2}{Q^2+M^2_Z}\frac{1}{\sin(2\theta_W)}\,.
\end{equation}

\begin{figure}[htbp]
\begin{center}
\begin{picture}(50,390)
\put(-200,-135){\epsfig{file=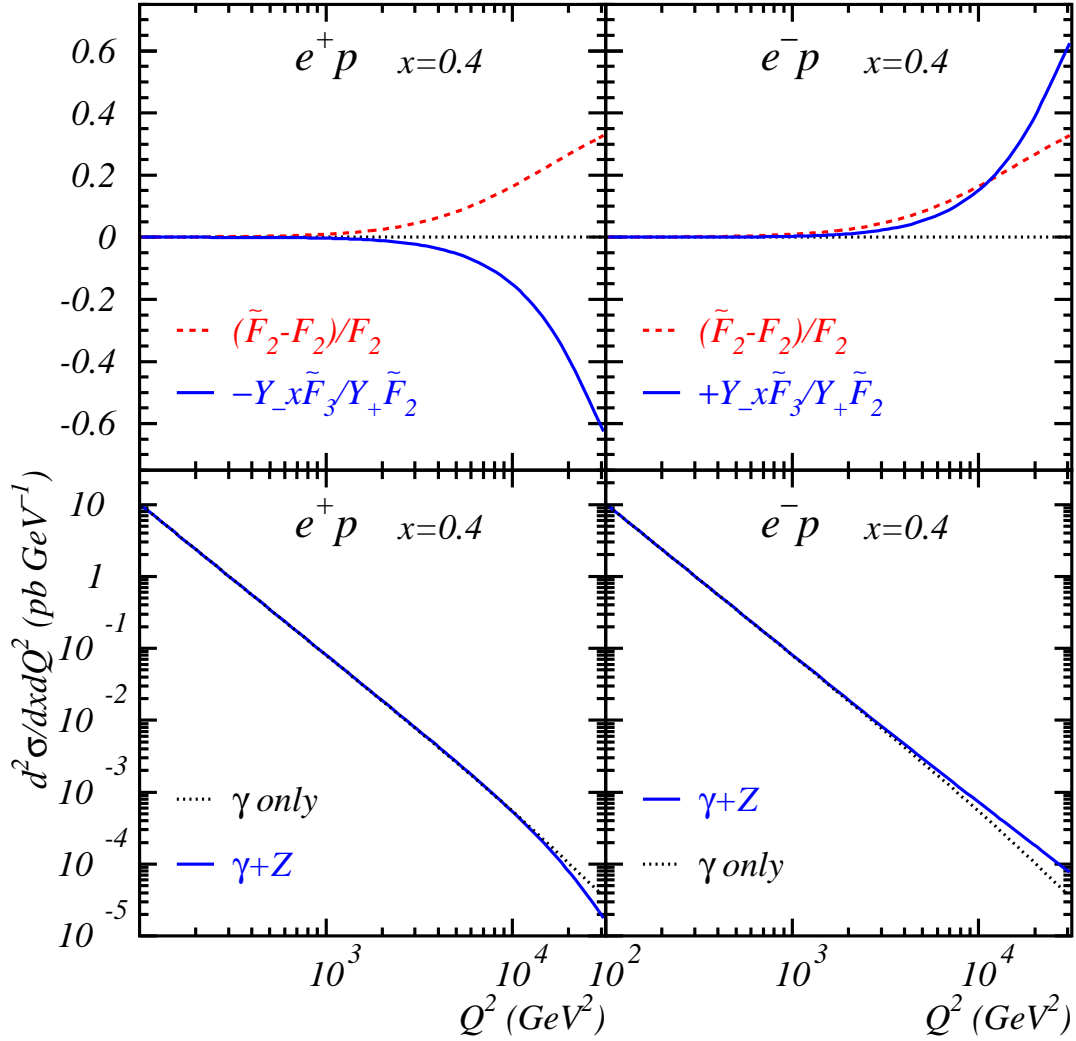,bbllx=0pt,bblly=0pt,
bburx=594pt,bbury=842pt,width=16.5cm}}
\end{picture}
\end{center}
\caption{\sl Comparisons of the generalized structure function
$\tilde{F}_2$ with the electromagnetic structure function $F_2$
and of the structure function $x\tilde{F}_3$ with $\tilde{F}_2$
for $e^+p$ and $e^-p$ collisions, and of the full cross sections 
with that from pure photon exchange.}
\label{fig:weak_cor}
\end{figure}

For unpolarized lepton beams, Eqs.(\ref{eq:f2nc_qpm}) and (\ref{eq:f3nc}) can
be expressed\footnote{Here we have taken the convention that
$v_{e^+}\equiv v_{e^-}=-1/2+2\sin^2\theta_W$ and 
$a_{e^+}\equiv a_{e^-}=-1/2$.} as
\begin{eqnarray}
& & \tilde{F}_2=F_2-v_lP_ZF^{\gamma
Z}_2+(v^2_l+a^2_l)P^2_ZF^Z_2 \approx F_2+ a^2_lP^2_ZF^Z_2 
 \label{eq:gf2nc}\\
& & x\tilde{F}_3=-a_lP_ZxF^{\gamma Z}_3+
2a_lv_lP^2_ZxF^Z_3 \approx -a_lP_ZxF^{\gamma Z}_3 \label{eq:gf3nc}
\end{eqnarray}
with $F_2$, $F^{\gamma Z}_2$ and $F^Z_2$ 
\begin{equation}
\left[F_2, F^{\gamma Z}_2, F^Z_2\right]=\sum_i\left[e^2_i, 2e_iv_i,
v^2_i+a^2_i\right]x\left(q_i+\overline{q}_i\right)
\end{equation}
standing for contributions respectively from pure photon exchange, $\gamma
Z^0$ interference and pure $Z^0$ exchange. The terms $xF^{\gamma Z}_3$ and 
$xF^Z_3$
\begin{equation}
\left[xF^{\gamma Z}_3, xF^Z_3\right]=\sum_i\left[2e_ia_i,
2v_ia_i\right]x\left(q_i-\overline{q}_i\right) \label{eq:xf3_nc}
\end{equation}
have similar meanings. The dominant correction to $F_2$ arises from the
$Z^0$ exchange term (the $F^Z_2$ term in Eq.(\ref{eq:gf2nc})) since $v_l$ is
small with respect to $a_l$. For the same reason, the dominant contribution
to $x\tilde{F}_3$ is from the $\gamma Z^0$ term (the $xF^{\gamma Z}_3$ term in
Eq.(\ref{eq:gf3nc})). As shown in Fig.\ref{fig:weak_cor}, the generalized
structure function $\tilde{F}_2$ is always larger than the electromagnetic
structure function $F_2$ and does not depend on the charge of the incident
lepton beam, whereas the structure function $x\tilde{F}_3$ has a negative
(positive) interference for a positron (electron) beam and its contribution
is large than that from the $F^Z_2$ term. As a result, the full cross
section is expected to be smaller (larger) than that from pure photon
exchange in the HERA kinematic range.

The structure functions for charged current processes $\overline{\nu}_\mu 
p\rightarrow \mu^+ X$, $e^-p\rightarrow \nu_eX$ with left handed
polarization are defined as
\begin{eqnarray}
F^{\rm CC}_2(x)=2xF^{\rm CC}_1(x)=
2x\left[u(x)+c(x)+ \overline{d}(x)+\overline{s}(x)\right], & &
\label{eq:f2cce-}\\
xF^{\rm CC}_3(x)=
2x\left[u(x)+c(x)-\overline{d}(x)-\overline{s}(x)\right], & & \label{eq:f3cce-}
\end{eqnarray}
and $F^{\rm CC}_2=2xF^{\rm CC}_1=F^{\rm CC}_3=0$ for $e^-p\rightarrow \nu_eX$
with right handed polarization.
The corresponding structure functions for processes $\nu_\mu
p\rightarrow \mu^- X$, $e^+p\rightarrow \overline{\nu}_e X$ with right handed
polarization are given by
\begin{eqnarray}
F^{\rm CC}_2(x)=2xF^{\rm CC}_1(x)=
2x\left[d(x)+s(x)+ \overline{u}(x)+\overline{c}(x)\right], & &
\label{eq:f2cce+}\\
xF^{\rm CC}_3(x)=
2x\left[d(x)+s(x)-\overline{u}(x)-\overline{c}(x)\right], & & \label{eq:f3cce+}
\end{eqnarray}
and $F^{\rm CC}_2=2xF^{\rm CC}_1=F^{\rm CC}_3=0$ for $e^+p\rightarrow
\overline{\nu}_e X$ with left handed polarization.

Comparing Eqs.(\ref{eq:f2cce-}),(\ref{eq:f2cce+}) with Eq.(\ref{eq:f2nc_qpm}), 
one notices an important feature of the charged current interactions: they are
able to distinguish between the partons and the antipartons of the target
nucleon. This is because the space-time structure of the weak interaction
ensures that target partons of differing helicities are affected
differently. In the relativistic limit, in which the rest mass of a particle
is regarded as being negligible, the parton and antiparton helicities are
opposite, so they will interact with the $W$-boson probe differently. Also,
because the $W$-boson probe is electrically charged, the target parton must
be able to absorb the charge. This rules out the participation of some types
of parton, making the weak interaction a more selective probe of the
nucleon's structure than the photon probe. 

\newpage
\section{Quantum Chromodynamics and parton density evolution equations} 
\label{sec:qcd}
The QPM has been successful in describing the early DIS data,
a few serious problems need, however, to be solved. The first is the absence
of free quark, which may be ascribed to some confinement mechanism which
requires very strong binding forces between the quarks, to prevent them
getting out. The second has to do with the seemingly violated Pauli principle
for having two identical $u$ valence quarks existing in a same proton.
Finally, the successful description of lepton scattering by
nucleons in terms of elastic scattering by quasi-free quarks equally
requires very weak binding forces between them, apparently inconsistent with
the first problem. These problems are naturally understood in the framework of
the non-Abelian gauge theory of the strong interactions of colored quarks and
gluons, the Quantum Chromodynamics (QCD), one of the components of
the $SU(3)\times SU(2)\times U(1)$ Standard Model, with the $SU(2)\times U(1)$
for the electroweak sector of the gauge theory.

The gauge bosons, gluons, are massless like the photons in the
electromagnetic interaction but in contrast to the photons, they interact
among themselves. This results in a strong scale dependence of the coupling
strength $\alpha_s$, one fundamental constant of QCD that must be determined
from experiment (Sec.\ref{sec:alphas}). 
At small distance (large energy scale) the
coupling between the quarks and gluons is small, this is known as
``asymptotic freedom''. At large distance the coupling increases which leads
to confinement. Taking $Q^2$ as the energy scale, the coupling strength in
leading order (LO) and next-to-leading order (NLO) is given respectively by
\begin{eqnarray}
& & \alpha_s^{\rm LO}(Q^2)=\frac{1}{\beta_0\ln (Q^2/\Lambda^2)}\,,\\
& & \alpha_s^{\rm NLO}(Q^2)=\frac{1}{\beta_0\ln (Q^2/\Lambda^2)}
\left[1-\frac{\beta_1}{\beta_0}\frac{\ln\ln(Q^2/\Lambda^2)}{\ln(Q^2/\Lambda^2)}\right]\,,
\end{eqnarray}
where $n_f$ is the number of quarks with mass less than the energy scale
$Q^2$, $\Lambda$ represents the scale at which the coupling
would diverge~\cite{book_esw}, and $\beta_i$ are functions which control the
renormalization scale dependence of the coupling:
\begin{equation}
\mu\frac{\partial \alpha_s}{\partial \mu}=-\frac{\beta_0}{2\pi}\alpha^2_s -
\frac{\beta_1}{4\pi^2}\alpha^2_s - \frac{\beta_3}{64\pi^3}\alpha^3_s - {\cal
O}(\alpha^4_s)
\end{equation}
with
\begin{eqnarray}
& & \beta_0=11-\frac{2}{3}n_f \\
& & \beta_1=51-\frac{19}{3}n_f\\
& & \beta_2=2857-\frac{5033}{9}n_f+\frac{325}{27}n^2_f
\end{eqnarray}
where contrary to $\beta_0$ and $\beta_1$, $\beta_2$ is scheme dependent
(see below).

\subsection{Structure functions in QCD}
\begin{figure}[htb]
\begin{center}
\begin{picture}(50,130)
\put(-275,-250){\epsfig{file=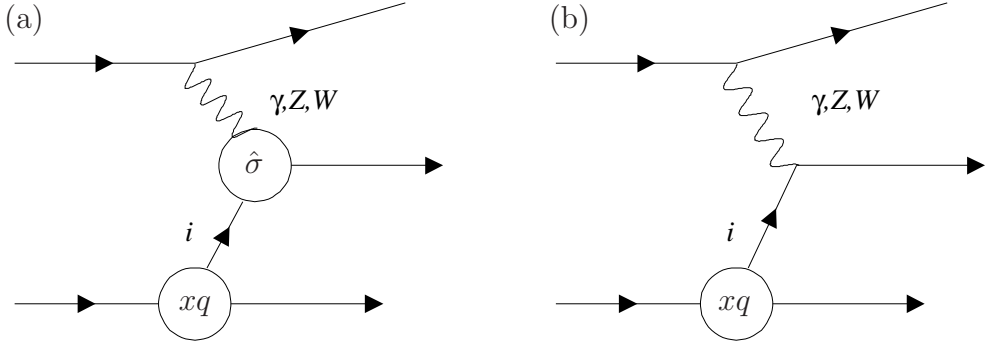,width=200mm}}
\put(-74,50){$\hat{\sigma}$}
\put(-99,-2){$xq$}
\put(105,-2){$xq$}
\put(-165,105){(a)}
\put(41,105){(b)}
\end{picture}
\end{center}
\caption{\sl (a) Schematic diagram showing QCD corrections and factorization 
theorem in DIS, (b) diagram in the QPM.}
\label{fig:qcd_qpm}
\end{figure}

In QCD, the structure function definitions of the QPM are modified to
accommodate strong interactions between the partons and to include mass
effects. Using the factorization theorem, the structure function can be
expressed as
\begin{equation}
\tilde{F}_2(x,Q^2)=\sum_i \hat{\sigma}_i \left(\frac{x}{z},
\frac{Q^2}{\mu^2},\frac{\mu^2_F}{\mu^2},\alpha_s(\mu^2)\right)
\otimes xq_i(z,\mu_F,\mu^2)\,, \label{eq:sf_qcd}
\end{equation}
where, as illustrated in Fig.\ref{fig:qcd_qpm}(a), $i$ is the parton label
and $f\otimes g$ denotes a convolution integral
\begin{equation}
f\otimes g=\int^1_x\frac{dy}{y}f(y)g\left(\frac{x}{y}\right)\,,
\label{eq:conv_int}
\end{equation}
of the hard vector-boson-parton cross section $\hat{\sigma}$ and the parton
distribution function $xq$.
The hard cross section $\hat{\sigma}$ can be calculated in perturbative QCD:
\begin{equation}
\hat{\sigma}\left(\frac{x}{z},\frac{Q^2}{\mu^2},
\frac{\mu^2_F}{\mu^2},\alpha_s(\mu^2)\right)=
\hat{\sigma}_0\delta\left(1-\frac{x}{z}\right)+
\alpha_s(\mu^2)\hat{\sigma}_1\left(\frac{x}{z},\frac{Q^2}{\mu^2},
\frac{\mu^2_F}{\mu^2}\right)+
{\cal O}(\alpha_s^2)\,, \label{eq:sigma}
\end{equation}
where $\hat{\sigma}_0$ is the contribution in the QPM 
(Fig.\ref{fig:qcd_qpm}(b)). 
In performing calculations beyond leading order,
various divergences arise and the renormalization scale $\mu$ is introduced
to regulate these divergences. The factorization scale $\mu_F$ serves to
define the separation of short-distance from long-distance scale. Roughly
speaking, any propagator that is off-shell by $\mu_F$ or more will
contribute to $\hat{\sigma}$ while it is grouped into $xq$ below this scale.
Since the physical structure function $\tilde{F}_2(x,Q^2)$ is
independent\footnote{This is true only when the result is summed to all
orders. Truncating the perturbative series at a given order spoils the
perfect compensation and introduces an artificial dependence on the choice
of the scale.}
of the scheme and scale dependence, these dependences of $\hat{\sigma}$
must be compensated by corresponding dependences of the parton distribution
functions $xq$. In the inclusive DIS analysis, these scales are usually
chosen to be $Q^2$, the characteristic large momentum scale of the process,
to avoid logarithms of large ratios $\ln(Q^2/\mu^2)$ and $\ln(\mu^2_F/\mu^2)$.

The two most commonly used schemes are the modified minimal subtraction 
$\overline{\rm MS}$~\cite{msbar} and DIS.
The $\overline{\rm MS}$ scheme is appealing for its theoretical elegance and
calculational simplicity. The DIS scheme~\cite{altarelli79}, on the other hand,
is appealing for its close correspondence to experiment. In this
scheme, one demands that, order-by-order in perturbation theory, all
corrections to the structure functions $\tilde{F}_2$ be absorbed into the
distributions of the quarks and antiquarks, such that the higher order formula
for $\tilde{F}_2$ is the same as the leading-order formula. On the other hand,
both $\tilde{F}_L$ and $x\tilde{F}_3$ acquire nontrivial order $\alpha_s$ 
corrections. For example the structure function $\tilde{F}_L(x,Q^2)$ reads
\begin{equation}
\tilde{F}_L(x,Q^2)=\frac{\alpha_s(Q^2)}{\pi} \int^1_x \frac{dz}{z} 
\left(\frac{x}{z}\right)^2 \left[ \frac{4}{3} \tilde{F}_2(z,Q^2)+
2c\left(1-\frac{x}{z}\right)zg(z,Q^2)\right], \label{eq:fl_qcd}
\end{equation}
where $c=N_f$ for neutrino scattering and $c=\sum_ie^2_i$ for
charged lepton scattering.
Higher order corrections and possibly non-perturbative contributions are
sizable~\cite{zijlstra92}.
At small $x$ ($x\lesssim 10^{-3}$) the second terms dominates. In fact for
$z\simeq 2.5x$, a measurement of $\tilde{F}_L$ is almost a direct measure of 
the gluon distribution~\cite{cooper88}.

\subsection{Evolution of structure functions}
\begin{figure}[htb]
\begin{center}
\begin{picture}(50,160)
\put(-275,-215){\epsfig{file=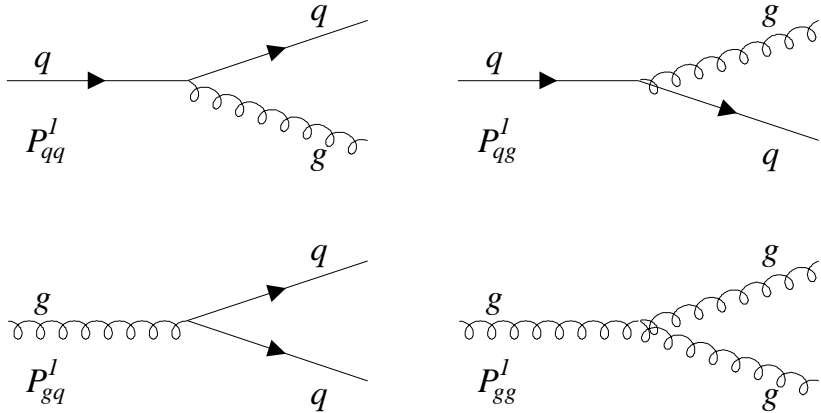,width=200mm}}
\end{picture}
\end{center}
\caption{\sl Leading-order diagrams contributing to the splitting function
$P^1_{ij}$.}
\label{fig:splitf}
\end{figure}

The most powerful quantitative prediction of perturbative QCD is the
breaking of Bjorken scaling in DIS.
Although the parton distribution functions in the hadron cannot be calculated
from first principle, their $Q^2$ dependence can be calculated within
perturbative QCD.
The scale dependence of the parton distribution functions in QCD has its
origin in the interactions of the quarks and gluons via such
elementary processes (Fig.\ref{fig:splitf}) as gluon emission from quarks,
$q\rightarrow qg$, the creation of quark-antiquark pairs by gluons,
$g\rightarrow q\overline{q}$, and gluon emission by gluons, $g\rightarrow gg$.
In describing the way in which scaling is broken in QCD, it is convenient to
define nonsinglet and singlet quark distributions:
\begin{eqnarray}
& & q^{NS}=q_i-q_j\,,\\
& & q^{S}=\sum_i(q_i+\overline{q}_i)\,.
\end{eqnarray}
It is understood that the parton distribution functions $q^{NS}$ and $q^S$
are functions of $x$ and $Q^2$.
The nonsinglet structure functions have nonzero values of flavor quantum
numbers such as isospin or baryon number.
The variation with $Q^2$ of these and the gluon distribution function
$g(x,Q^2)$ is described by the so-called DGLAP
(Dokshitzer-Gribov-Lipatov-Altarelli-Parisi) equations~\cite{dglap}, valid
to all orders in $\alpha_s$:
\begin{eqnarray}
& & \frac{\partial q^{NS}}{\partial \ln Q^2}=\frac{\alpha_s(Q^2)}{2\pi}
P_{qq}\otimes q^{NS}\,, \label{eq:qns_evol}\\
& & \frac{\partial}{\partial\ln Q^2}\left(\begin{array}{c} q^S \\
g\end{array}\right)=
\frac{\alpha_s(Q^2)}{2\pi}\left(\begin{array}{cc} P_{qq} & P_{qg} \\ P_{gq}
& P_{gg} \end{array}\right)\otimes \left(\begin{array}{c} q^S \\
g\end{array}\right).\label{eq:qs_g_evol}
\end{eqnarray}
The splitting functions $P_{ij}(x)$, representing the probability of a parton
$j$ emitting a parton $i$ with a fraction $x$ of the parent's momentum,
is calculable in perturbative QCD:
\begin{equation}
\frac{\alpha_s(Q^2)}{2\pi}P_{ij}(x,Q^2)=\frac{\alpha_s(Q^2)}{2\pi}P^1_{ij}(x)+
\left(\frac{\alpha_s(Q^2)}{2\pi}\right)^2P^2_{ij}(x)+\cdots\,.
\label{eq:splitf}
\end{equation}
The splitting functions in leading order $P^1_{ij}$~\cite{altarelli77}
correspond to contributions shown in Fig.\ref{fig:splitf}.
The truncation after the first two terms in Eq.(\ref{eq:splitf}) defines the
NLO evolution. The splitting functions at NLO have
been calculated since long. The splitting functions at next-to-NLO are however 
only known partially~\cite{neerven00}.

As we will see in later chapters, the conventional DGLAP evolution
equations have been working very successful in describing the HERA structure
function data. However, due to the approximation in which only terms
involving $\left(\alpha_s\ln Q^2\right)^n$ are summed to all orders in $n$ in
LO and $\alpha_s\left(\alpha_s\ln Q^2\right)^n$ in NLO, it is believed that
it may not be applicable at very low $x$ when $\alpha_s\ln(1/x)\sim 1$.
Indeed, in BFKL (Balitsky-Fadin-Kuraev-Lipatov) equation~\cite{bfkl} terms
involving $\left(\alpha_s\ln(1/x)\right)^n$ are summed instead.  
The solution of the equation gives a functional form for the gluon distribution
as
\begin{equation}
xg\sim x^{-\omega} \label{eq:xg_bfgl}
\end{equation}
with $\omega=12\ln 2 \alpha_s/\pi\simeq 0.5$ for $\alpha_s=0.19$. Recently,
the next-to-leading logarithmic (NLL) corrections have been
obtained~\cite{bfkl_nll} giving 
$\omega=2.65\alpha_s-16.3\alpha_s^2\simeq -0.12$ for the same $\alpha_s$ value.
Unfortunately, neither LL nor NLL
value is compatible with the data. For the latest development, see 
\cite{salam,thorne00} and references therein.

When $Q^2$ is large and $x$ is small, namely $\alpha_s\ln Q^2\ln(1/x)\sim 1$
but $\alpha_s\ln Q^2$ and $\alpha_s\ln(1/x)$ are both small, terms
involving $\left(\alpha_s\ln Q^2\ln(1/x)\right)^n$ have to be summed to all
orders in $n$ to give the so-called double leading logarithmic (DLLA)
approximation. Theoretical research has been trying to find a more general
approximation scheme resulting in an equation that is applicable in these
different regions. For example, the CCFM equation~\cite{ccfm} gives BFKL at
small-$x$ and DGLAP at large $x$.

Understanding the dynamics of the small $x$ region is one of the fundamental
problems of QCD. For a relatively low transverse scale of $p_T$ in a process
at a future high energy hadron machine with a center-of-mass energy $\sqrt{s}$,
the value of $x=p_T/\sqrt{s}$ can be low. When $x$ is small enough, the
density of partons becomes very large so that partons start to interact and
overlap, the perturbative QCD will eventually fail not because the strong
coupling $\alpha_s$ is large but because the parton density is high.

\subsection{Higher twist} \label{sec:hw}
The predictions of QCD discussed so far is at leading twist (twist two). 
In QCD, the structure functions have higher-twist power corrections:
\begin{equation}
F_2(x,Q^2)=F^{(2)}_2(x,Q^2)(1+h(x)/Q^2+\cdots)\label{eq:hw}
\end{equation}
where $h(x)$ has a form $(1-x)^{-1}$ according to a phenomenological analysis
of the structure function data of BCDMS and SLAC~\cite{virchaux92}.
The higher twist contribution is therefore expected to be important at high
$x$ and low $Q^2$.

The higher twist contribution involves presumably reinteraction of the
struck quark with the proton remnant and thus a full calculation may have to
await a solution to the problem of confinement.

\newpage
\section{Parameterizations of parton distribution functions}\label{sec:qcdfit}
One of the main strands of interest in experiments on DIS is to determine the
parton distribution functions. This is related to the fact that the parton
distribution functions have a key feature, universality, i.e.\ they are
independent of the structure functions in which they appear, and of the
physical processes to which they are applied. Therefore, the parton
distribution functions can be extracted from a quantitative comparison of
experimental data from a wide range of physical processes with QCD master
equations, Eqs.(\ref{eq:sf_qcd}),(\ref{eq:qns_evol}),(\ref{eq:qs_g_evol}).
These can then be used in other applications to make predictions as well as
to provide stringent tests of the self-consistency of the perturbative QCD
framework itself or the Standard Model in general. Since any compelling
indications of inconsistency of the Standard Model are signs of new physics,
and since even direct search for new physics must rely heavily on
understanding of the background from conventional physics, the systematic
analysis of parton distributions is intimately tied to all these ventures.

\subsection{Global analysis of parton distribution functions}
The analysis of structure function data in extracting parton density 
functions has a long history, the earlier
parameterizations dated around the late 1970's. These together
with many new generation parameterizations are available in the package {\sc
pdflib}~\cite{pdflib}. There are many subtle differences among analyses by
different groups or even a same group at different times,
but the technique used can be broadly summarized as follows.
\begin{itemize}
\item {\bf\boldmath Initial scale $Q^2_0$:} The initial scale is arbitrary, 
 but should be large enough to ensure that $\alpha_s(Q^2_0)$ is small enough 
 for perturbative calculations to be applicable. 
\item {\bf Functional forms:} Functional forms for parton distributions 
 (valence, sea and gluon, or non-singlet, singlet and gluon) are assumed 
 to be valid at $Q^2_0$:
\begin{eqnarray}
xu_v(x)\!\!\!&=&\!\!\!A_ux^{B_u}(1-x)^{C_u}P_u(x)\,,\\
xd_v(x)\!\!\!&=&\!\!\!A_dx^{B_d}(1-x)^{C_d}P_d(x)\,,\\
xS(x)\!\!\!&=&\!\!\!A_Sx^{B_S}(1-x)^{C_S}P_S(x)\,,\label{eq:par_sea} \\ 
xg(x)\!\!\!&=&\!\!\!A_gx^{B_g}(1-x)^{C_g}P_g(x)\,,\label{eq:par_gluon}
\end{eqnarray}
 where $P_i(x)$ takes the function form $1+\epsilon_i\sqrt{x}+\gamma_i
 x$ for the MRS (Martin, Roberts, Stirling) group~\cite{mrst} and
 $1+\gamma_i x^{\epsilon_i}$ for CTEQ (the Coordinated
 Theoretical-Experimental project on QCD) group~\cite{cteq}.
 In some (earlier) analyses, the gluon distribution takes a simplified form
 $1+\gamma_g x$. While the difference in a given parton density using
 these two functional forms does usually not exceed ${\cal O}(1)\%$ level 
 over a large $x$ range when same precise data are used, large differences 
 are possible at the kinematic boundaries~\cite{ph9706470}.
\item {\bf Evolution:} With a particular value of $\Lambda$,\footnote{Depending
 on the analyses, the variable $\Lambda$ was either treated as a free 
 parameter or fixed using the precise value from other independent 
 measurements~\cite{pdg00}.}\ the DGLAP 
 equations (\ref{eq:qns_evol}) and (\ref{eq:qs_g_evol}) are then used to 
 evolve the parton distributions up to
 a different $Q^2$ value, where they are convoluted with coefficient
 functions, appropriate to the chosen renormalization scheme, in order to
 make predictions for various processes corresponding to the chosen 
 measurements.
\item {\bf Data sets and comparison between data and parameterizations
 through minimization:} Data sets from various measurements are selected and 
 compared with the corresponding predictions. The free parameters are
 obtained by a minimizing procedure (e.g.\ with the {\sc minuit}
 package~\cite{minuit}) and by taking into account the experimental errors.
 Some of the parameters are constrained by the flavor counting rules and 
 the momentum sum rule.
\end{itemize}

\begin{figure}[htbp]
\begin{center}
\begin{picture}(50,510)
\put(-185,-35){\epsfig{file=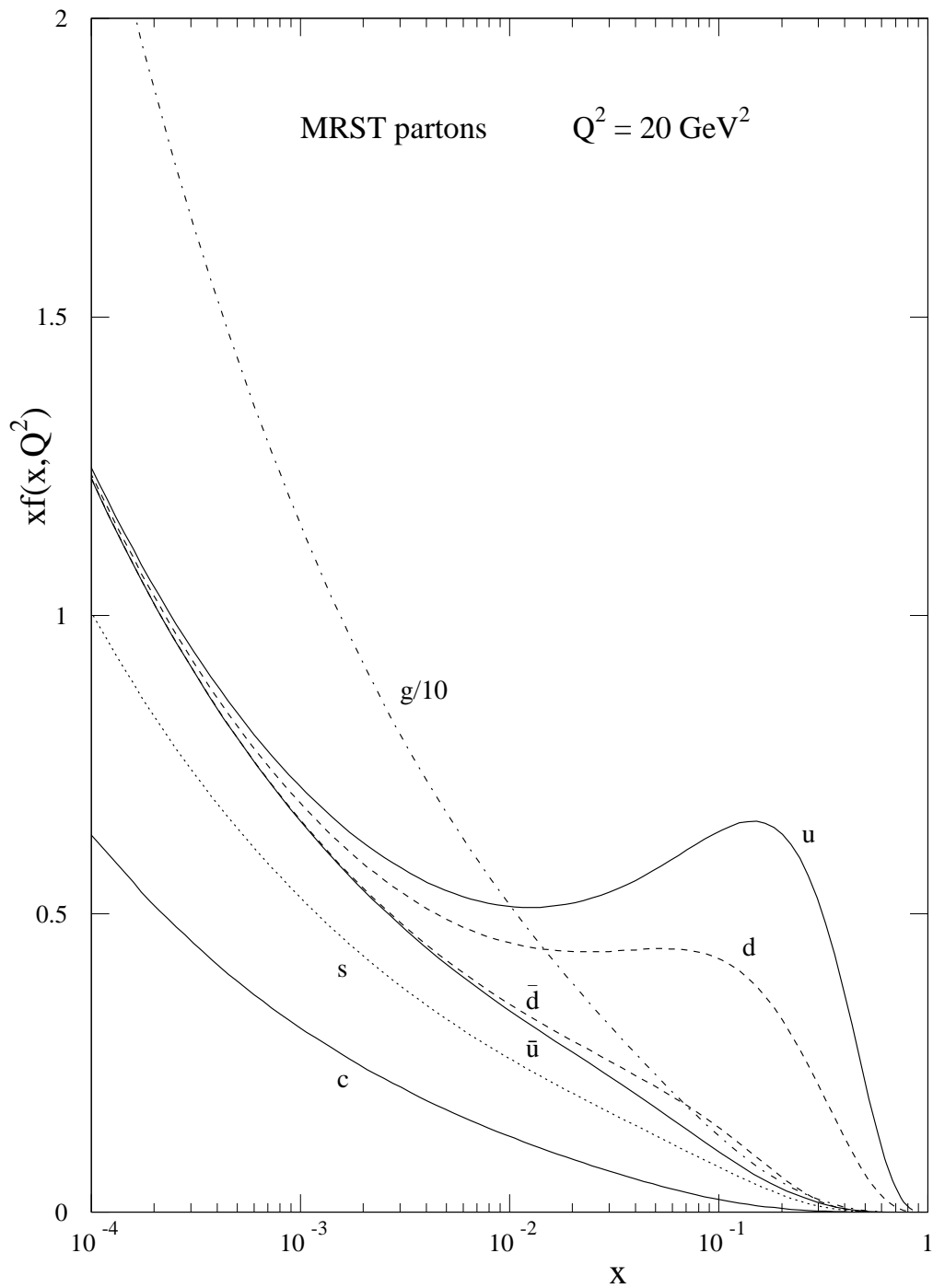,bbllx=0pt,
bblly=0pt,bburx=594pt,bbury=842pt,width=180mm}}
\end{picture}
\end{center}
\caption{\sl Parton distributions at $Q^2=20\,{\rm GeV}^2$ from the MRST
parameterizations~\cite{mrst}. The gluon density is scaled down by a factor of
10.}
\label{fig:pdf_mrst}
\end{figure}

A typical set of parton distributions at $Q^2=20\,{\rm GeV}^2$ obtained by the
MRS group~\cite{mrst} is shown in Fig.\ref{fig:pdf_mrst}. 
The core constraint for these parton density functions comes from the DIS
structure function data at low $x$ by the HERA experiments, as will see in
Chapter \ref{chap:lowq2}, and at high $x$ from fixed-target data.
As an example, four combinations $u+\overline{u}$, $d+\overline{d}$,
$\overline{u}+\overline{d}$, and $s$ can be derived from the following
structure functions\footnote{When deriving
Eqs.(\ref{eq:f2lp-f2ln})-(\ref{eq:xf3_vn}), we have neglected the charm
contribution and assumed $s=\overline{s}$.}
in the leading-order form:
\begin{eqnarray}
F^{l p}_2-F^{l n}\!\!\!&=&\!\!\!
 \frac{1}{3}x\left(u+\overline{u}-d-\overline{d}\right),\label{eq:f2lp-f2ln}\\
\frac{1}{2}\left(F^{l p}_2+F_2^{l n}\right)\!\!\!&=&\!\!\!
 \frac{5}{18}x\left(u+\overline{u}+d+\overline{d}+\frac{4}{5}s\right),\\
F_2^{\nu N}=F_2^{\overline{\nu} N}\!\!\!&=&\!\!\!
 x\left(u+\overline{u}+d+\overline{d}+2s\right),\\
\frac{1}{2}x\left(F^{\nu N}_3+F^{\overline{\nu} N}_3\right)\!\!\!&=&\!\!\!
 x\left(u-\overline{u}+d-\overline{d}\right), \label{eq:xf3_vn}
\end{eqnarray}
where $p$, $n$, $N$ stands respectively for proton, neutron, and isoscalar
targets.

The fact that $u$ valence shape is different from that of $d$ valence has been
known since the earliest days of neutrino scattering when neutrino and
antineutrino scattering data on protons and deuterium were
compared~\cite{parker84,allasia8485}. The recent data which fix these
valence shapes have come from taking the difference and the ratios of
$F_2^{\mu p}$ and $F_2^{\mu n}$ from NMC~\cite{nmc97}. At large $x$, when
only valence distributions are significant, one has in leading order
\begin{equation}
\frac{F_2^{\mu p}}{F_2^{\mu n}}=\frac{1+4d_v/u_v}{4+d_v/u_v}\,.
\end{equation}
The $W^\pm$ charge asymmetry at the Tevatron $p\overline{p}$ collider provides
additional information in the region of $x\sim 0.1$ and $Q^2\sim M^2_W$.
Because the $u$ quarks in the proton carry more momentum on average than
the $d$ quarks, the $W^+$ bosons tend to follow the direction of the incoming
proton and the $W^-$ bosons that of the antiproton. When $x\rightarrow 1$,
the behavior of the ratio $d_v/u_v$ is, however, largely unsettled with model
predictions varying between 0~\cite{field77} and 0.2~\cite{farrar75} 
(Sec.\ref{sec:ud}).

In the global analysis of the parton distribution functions, 
the sea quark is assumed to be the same as the anti-quark sea, 
i.e.\ $q_i=\overline{q}_i$. In the earlier analyses, the flavor symmetry 
$\overline{u}=\overline{d}$ is also assumed. 
In 1992 the NMC data~\cite{nmc92} together with the Gottfried sum
rule~\cite{gottfried} 
\begin{equation}
\int^1_0\frac{dx}{x}\left(F_2^p-F^n_2\right)=\frac{1}{3}\int^1_0dx(u_v-d_v)+
 \frac{2}{3}\int^1_0dx(\overline{u}-\overline{d})
\end{equation}
gave the first evidence that $\overline{d}>\overline{u}$.
The recent global analyses~\cite{mrst,cteq5} also use the asymmetry of
Drell-Yan production in $pp$ and $pn$ collisions first from NA51~\cite{na51}
for $x=0.18$ and then from E886~\cite{e886} for an extended $x$
range ($0.04<x<0.3$) to determine directly $\overline{u}-\overline{d}$.
The semi-inclusive DIS data from HERMES are expected to give confirmation or 
independent information on the $\overline{u}$ and $\overline{d}$ flavor
asymmetry.

In Fig.\ref{fig:pdf_mrst}, the strange quark sea $s$ is shown to be different 
from the quark seas $\overline{u}$ and $\overline{d}$. 
In fact in the recent global analysis of e.g.\ MRST, $s$ is assumed to 
have the same $x$ dependence as
$\overline{u}+\overline{d}$ but suppressed by 50\%. This is supported by
the CCFR dimuon data~\cite{ccfr_dimu}. The suppression is presumably due to
the mass difference between $s$ and $u$ and $d$ though they all
have been treated as massless quarks in the global analysis.

\begin{figure}[htb]
\begin{center}
\begin{picture}(50,115)
\put(-160,-255){\epsfig{file=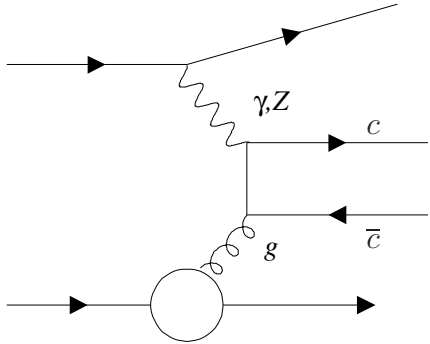,width=200mm}}
\put(90,60){$c$}
\put(90,17){$\overline{c}$}
\end{picture}
\end{center}
\caption{\sl Diagram of the boson-gluon fusion in deep-inelastic
lepton-proton scattering.}
\label{fig:bgf}
\end{figure}

The charm sea $c$ is further suppressed with respect to other light quark
seas. Until recently different groups have used
different procedures. One such procedure is to treat the charm quark
(similarly for the bottom quark) as infinitely massive below a threshold
$Q^2=m^2_c$, and as massless above the threshold thus evolving according to
the normal massless evolution equations. Up to NLO in $\alpha_s$ this
prescription guarantees that the correct results are obtained asymptotically,
but it is unsatisfactory near the threshold. An alternative
procedure considers the charm as being produced from the hard scattering
between the electroweak boson and a gluon, i.e.\ the boson gluon
fusion process (Fig.\ref{fig:bgf}). 
The latter treatment incorporates the correct threshold
behavior automatically but is unsuitable for $Q^2\gg m^2_c$ due to the
unsummed potentially large logarithm in $Q^2/m^2_c$.
The recent measurements of charm production at HERA~\cite{h1charm,zeuscharm} 
have emphasized the importance of having a consistent theoretical
framework for heavy flavor production in DIS. This has been achieved and
applied in the recent global analyses of parton distribution
functions~\cite{mrst, cteq5}. As a consequence, contrary to light quarks,
the charm quark density is determined by the other parton distributions and
no extra parameters are introduced apart from the charm quark mass.

At values of $Q^2$ far above that shown in Fig.\ref{fig:pdf_mrst},
the bottom quark sea becomes increasingly important and eventually all sea
quark distributions evolve to a common form, concentrated at small values of
$x$, since they are driven by $g\rightarrow q\overline{q}$ transitions.

\subsection{Dynamical parton distributions}\label{sec:grv}
Whereas the parameterizations of the global analyses depend crucially on the
non-perturbative input parameterizations at $Q^2_0$, the ones of the GRV
group~\cite{grv92,grv94,grv98} are constructed to be less dependent on 
their inputs.

The original idea behind these parameterizations is that at some very low
scale $Q^2=\mu^2$ with $\mu\simeq 3\Lambda$, the nucleon is assumed to 
consist only of constituent valence quarks.
As $Q^2$ increases, one generates the gluon and sea quarks in the nucleon
dynamically from the valence quarks, through the conventional DGLAP
equations for the processes $q\rightarrow qg$, $g\rightarrow q\overline{q}$.
The resulting predictions turned out, however, to be too steep in the small
$x$ region and subsequently the parameterizations for the sea quarks and 
gluons were modified to be valence like instead of being null. This
modification was made based on the argument that partonic quark
distributions should rather be identified with the current quark content of
hadrons instead of the constituent quarks.

In an earlier version in 1992~\cite{grv92}, only fixed-target DIS data at
high $x$ ($x>0.01$) were used in fixing the valence-like input parameters at
$\mu^2$. The resulting prediction at low $x$ was in qualitative agreement
with first structure function measurements at HERA (see
Sec.\ref{sec:1stf2_hera}). Quantitatively, however, it could still deviate
systematically from the later more precise data from HERA. The
parameterizations were thus updated first in 1994~\cite{grv94} and then in
1998~\cite{grv98} by including the HERA structure function data together
with other DIS and non-DIS data. Therefore, as far as the used data sets
are concerned, the GRV parameterizations do not differ from the other global
analyses. The only important remaining difference lies in the chosen initial
scale which is much smaller than in the other analyses. It is from
this low scale that the gluon and sea quarks acquire the sufficient large
radiative evolution length $\ln\left(\alpha_s(\mu^2)/\alpha_s(Q^2)\right)$
to change dynamically its behavior from valence-like ($B_i>0, i=g,S$, see
Eqs.(\ref{eq:par_sea}) and (\ref{eq:par_gluon})) to sea-like ($B_i<0$) found in
the other global analyses choosing $Q^2_0\sim {\cal O}(4)\,{\rm GeV}^2$.

\subsection{Other model parameterizations}\label{sec:para_other}
In addition to the parameterizations described in the previous subsections,
there are a number of other models or parameterizations~\cite{gotsman}. 
One example is the
model based on the Regge theory~\cite{regge}. The Regge theory has been very
successful in describing the energy dependence of the soft hadron
interactions:
\begin{equation}
\sigma\sim s^{\alpha_{\cal P}-1}
\end{equation}
where $\alpha_{\cal P}=1.08$ is the intercept of the soft pomeron (the so-called
Regge trajectory corresponding to the exchange of families of particles with
different spin). The connection between the small $x$ and the energy 
is made by $W^2$, the invariant mass of hadronic system defined
in Eq.(\ref{eq:w2}):
\begin{equation}
W^2=Q^2(1/x-1)\simeq Q^2/x\,. \label{eq:x_w2}
\end{equation}
In this approach, the gluon
distribution is expected to behave therefore as~\cite{dola}
\begin{equation}
xg\sim x^{1-\alpha_{\cal P}}\,. \label{eq:dola}
\end{equation}
When this description is applied to the HERA structure function data, it is
found unsuccessful as soon as $Q^2$ reaches ${\cal O}(1)\,{\rm GeV}^2$ 
(see Sec.\ref{sec:isr}). The model has been extended
recently to include a hard pomeron in addition to the soft
one to describe the HERA data~\cite{dola_new}. One interesting finding of
the new model is that while the hard pomeron is needed to describe the
strong rise of the $F_2$ as $x$ decreases (Sec.\ref{sec:1stf2_hera}), the
contribution of the soft pomeron at $Q^2=5\,{\rm GeV}^2$ dominates at high
$x$ and can be still important for $x$ down to $\sim 0.0002$, a behavior
of the higher-twist contribution (Sec.\ref{sec:hw}).

\newpage
\section{Radiative corrections and hard radiative processes} \label{sec:radc}
We have given explicitly in Sec.\ref{sec:xs_sf} the Born cross
sections corresponding to the contributions from the lowest order
lepton-nucleon scattering processes shown in Fig.\ref{fig:qcd_qpm}(b).
Experimentally, the inclusive cross sections that are measured include
higher order electroweak corrections. These corrections, in particular
the conventional electromagnetic (QED) bremsstrahlung corrections, 
can be very large ($>100$\%) depending on the kinematic phase space and 
the kinematic reconstruction method used~\cite{hollik87,hubert87,hubert+92}. 
These QED contributions do not contain additional information about 
the strong and weak interaction part of the complete theory, and are
generally removed from the measured cross sections. 

\subsection{Radiative corrections for neutral and charged current DIS processes
at HERA} \label{sec:rc}
\begin{figure}[htbp]
\begin{center}
\begin{picture}(50,230)
\put(-250,-300){\epsfig{file=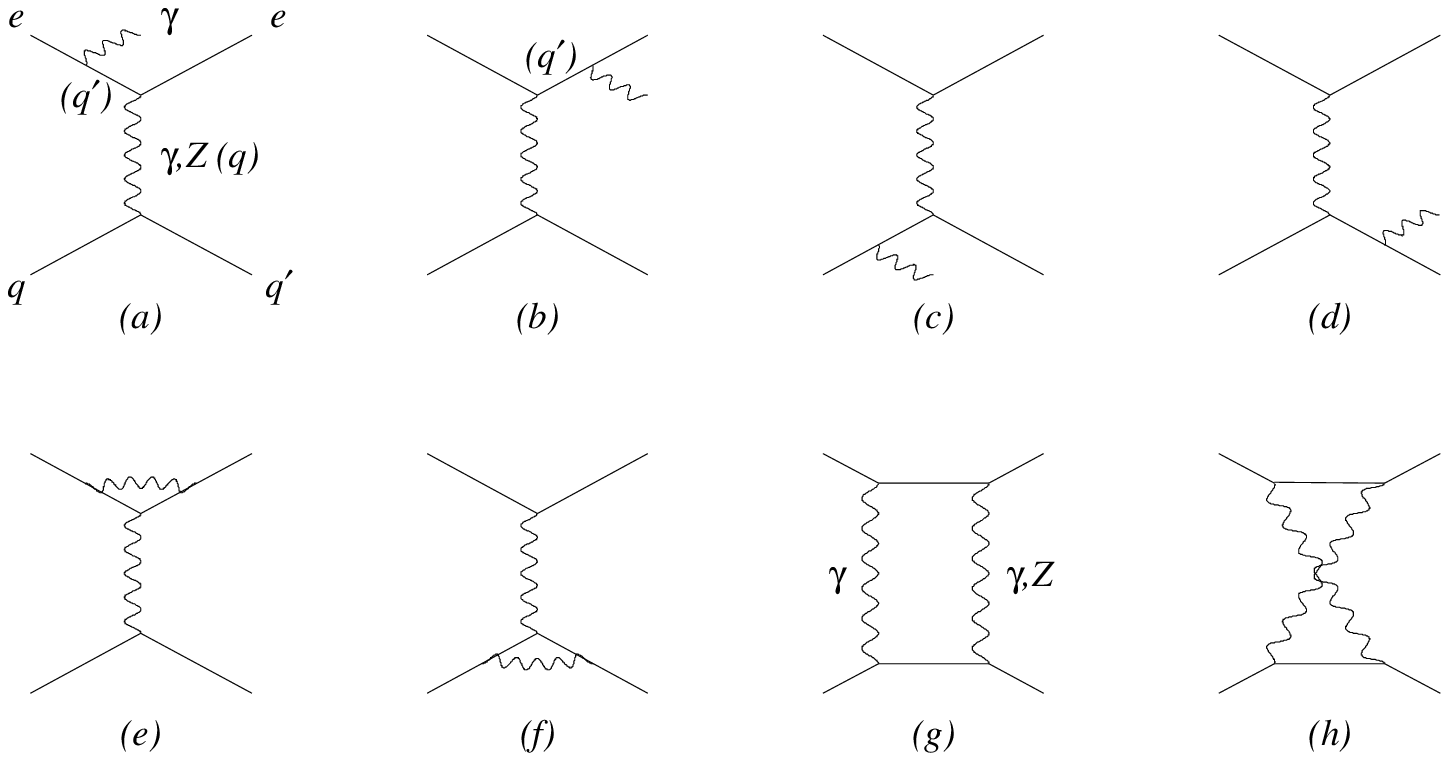,width=195mm}}
\end{picture}
\end{center}
\caption{\sl Diagrams of QED corrections to the NC DIS process.}
\label{fig:radc_em}
\begin{center}
\begin{picture}(50,230)
\put(-190,-300){\epsfig{file=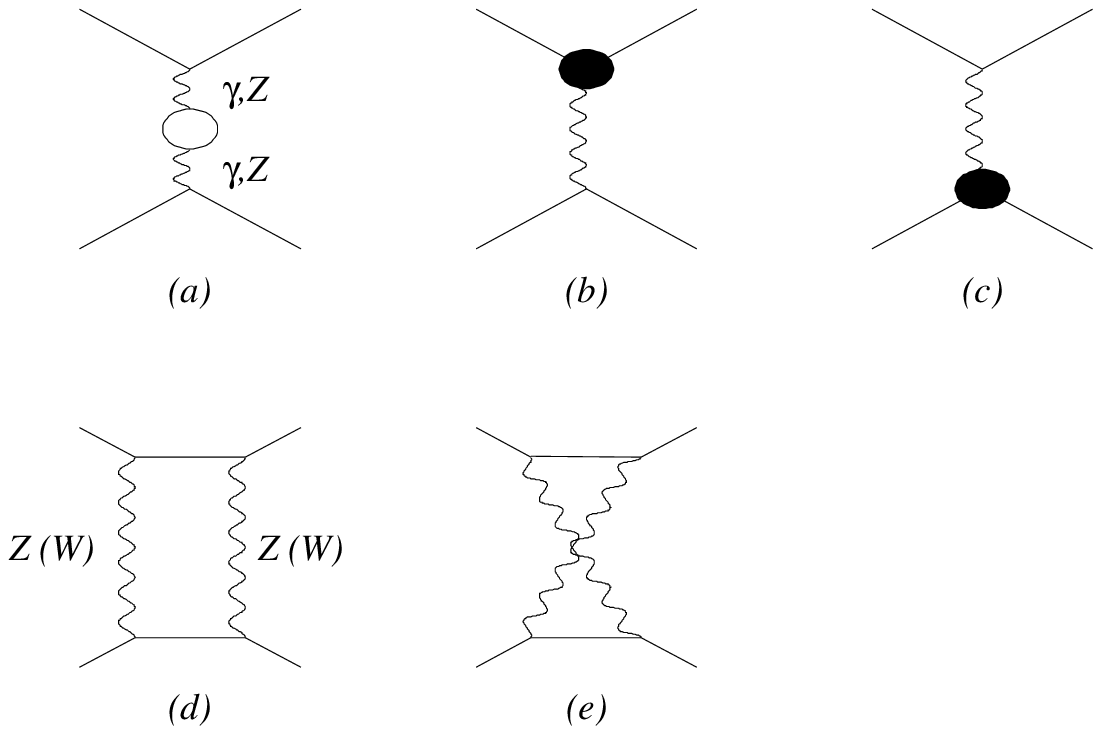,width=195mm}}
\end{picture}
\end{center}
\caption{\sl Diagrams of weak corrections to the NC DIS process.}
\label{fig:radc_weak}
\end{figure}

For the neutral current (NC) process, higher order electroweak contributions
can be separated into the QED and weak corrections as shown
respectively in Figs.\ref{fig:radc_em} and \ref{fig:radc_weak}.

The QED contributions can be further subdivided into
\begin{itemize}
\item the leptonic corrections described by diagrams containing an
additional photon attached to the electron line, i.e.\ the photon emission
from the electron line, Figs.\ref{fig:radc_em}(a) and \ref{fig:radc_em}(b),
and the photonic lepton vertex correction combined with the self energies of 
the external fermion lines, Fig.\ref{fig:radc_em}(e);
\item the quarkonic corrections represented by diagrams with an additional
photon at the quark line, Figs.\ref{fig:radc_em}(c), \ref{fig:radc_em}(d)
and \ref{fig:radc_em}(f);
\item the interference of bremsstrahlung from the electron and the quark line,
Figs.\ref{fig:radc_em}(a)-(d), and the box diagrams,
Figs.\ref{fig:radc_em}(g) and \ref{fig:radc_em}(h), which connect the
electron and the quark line by an extra virtual photon.
\end{itemize}

The leptonic corrections constitute the bulk of all radiative corrections and
represent the practically most important contribution. There are two reasons
why these corrections can be large. First, large logarithmic term of the form
$\alpha/\pi \ln(Q^2/m^2_e)$ are present due to the radiation of photon
collinear with the emitted lepton. Secondly, the emission of a very
energetic photon shifts the momentum in the propagator of the exchanged
photon to a value which is essentially smaller than determined from the
energy and momentum of the final electron. The second effect is primarily of
kinematical nature; it can be reduced in magnitude by applying suitable cuts
(Sec.\ref{sec:f293_sel}) and depends on the method used to reconstruct the
event kinematics (Sec.\ref{sec:kinerec}). The process in which the emission
of an energetic photon occurs in the incident electron line
(Fig.\ref{fig:radc_em}) can be further used to measure the structure
function $F_2$ in an extended kinematic region (Sec.\ref{sec:isr}) and the
longitudinal structure function $F_L$ (Sec.\ref{sec:fl_isr}).

The quarkonic corrections exhibit a different behavior: they are essentially
smaller in magnitude, typically a few percent, and are flat functions of
$Q^2$. The main reason for this difference is the absence of the kinematical
effect in the photon propagator, which dominates the radiation from the
lepton. The occurrence of quark mass singularities of the type $\alpha/\pi
e^2_q \ln(Q^2/m^2_q)$ can be absorbed into the quark distribution functions.

The interference corrections are not affected by either lepton or quark
masses and therefore free of the mass singularities. 

The weak corrections (Fig.\ref{fig:radc_weak}) are infrared finite and
numerically small (of the order of 1\%).

\begin{figure}[htbp]
\begin{center}
\begin{picture}(50,210)
\put(-250,-300){\epsfig{file=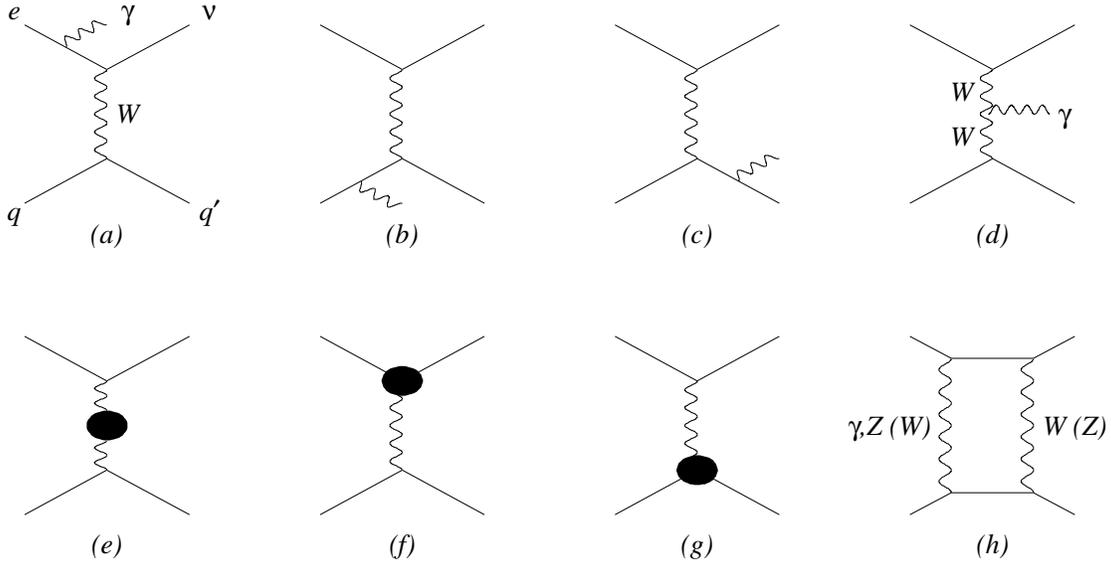,width=195mm}}
\end{picture}
\end{center}
\caption{\sl Diagrams of electroweak corrections to the CC DIS process. In
addition to (h), the corresponding crossed diagram also contributes.}
\label{fig:radc_cc}
\end{figure}

Diagrams of electroweak corrections for the charged current (CC) process is
shown in Fig.\ref{fig:radc_cc}.
In contrast to the NC process the subsets of the real
bremsstrahlung diagrams for the CC process are not gauge
invariant. Moreover the appearance of the non-abelian $\gamma WW$ vertex in
the real (Fig.\ref{fig:radc_cc}(d)) and virtual (Figs.\ref{fig:radc_cc}(f)
and \ref{fig:radc_cc}(g)) corrections, not present within the conventional
QED but typical for a non-abelian gauge theory, indicates that the
classification convenient for NC processes is less sensible in the CC case.
The radiative corrections for CC are flat functions of $Q^2$, in contrast to
the QED corrections in the NC case. This is a consequence of the absence of
the photon exchange diagram; the kinematical effect of the lowering the
exchanged $Q^2$ is less important in the $W$ propagator due to the presence
of the large $W$ mass.

\subsection{Hard radiative processes} \label{sec:rad_proc}
In the process, $ep\rightarrow e\gamma X$, where a real photon is emitted
(Fig.\ref{fig:radc_em}(a) and \ref{fig:radc_em}(b)), the virtuality of the
intermediate electron $Q^{\prime 2}$ can be defined
\begin{equation}
Q^{\prime 2}=-q^{\prime 2}=\left\{\begin{array}{lcl} -(k-K)^2 & &
\mbox{(Fig.\ref{fig:radc_em}(a))} \\ -(k^\prime+K)^2 & &
\mbox{(Fig.\ref{fig:radc_em}(b))}\,,\end{array}\right.
\end{equation}
in analogy to that of the exchanged virtual boson, $Q^2$, defined 
in Eq.(\ref{eq:q2}), where $k, k^\prime$ and $K$ are respectively the four 
momentum of the incident electron, the scattered electron, and 
the radiative photon.

The presence of the real photon introduces in the cross section formula an
additional propagator due to the intermediate virtual electron.
One has therefore
\begin{equation}
\frac{d^2\sigma}{dQ^2dQ^{\prime 2}}\sim
\frac{1}{Q^4}\frac{1}{\left(Q^{\prime 2}-m^2_e\right)^2}\,.
\end{equation}
Depending on the relative values of $Q^2$ and $Q^{\prime 2}$, one refers to 
the following processes:
\begin{itemize}
\item {\bf Bethe-Heitler and radiative photoproduction processes} correspond
respectively to the elastic and inelastic channel when
$Q^2\rightarrow 0$ and $Q^{\prime 2}\rightarrow 0$. The Bethe-Heitler
process has the largest cross section and has been used by both H1 and ZEUS
collaborations to provide the
luminosity measurement (Sec.\ref{sec:lumisyst}).
\item {\bf QED Compton process} corresponds to the case when $Q^2\rightarrow 0$ 
but $Q^{\prime 2}>0$. Since the radiative photon and the scattered 
electron can both be measured in the main detector, these events have been
used to check and calibrate the energy scale of the electromagnetic 
calorimeters (Secs.\ref{sec:isr_kinerec} and \ref{sec:em_scale}).
\item {\bf Radiative DIS process} corresponds to the case when $Q^2>0$ and 
$Q^{\prime 2}\rightarrow 0$. These events are of particular interest, see
Secs.\ref{sec:isr} and \ref{sec:fl_isr}.
\end{itemize}

\chapter{Measurement of structure functions and their interpretation}
\label{chap:lowq2}
\section{Pre-HERA results and expectations}
\begin{figure}[htb]
\begin{center}
\begin{picture}(50,320)
\put(-155,-30){\epsfig{file=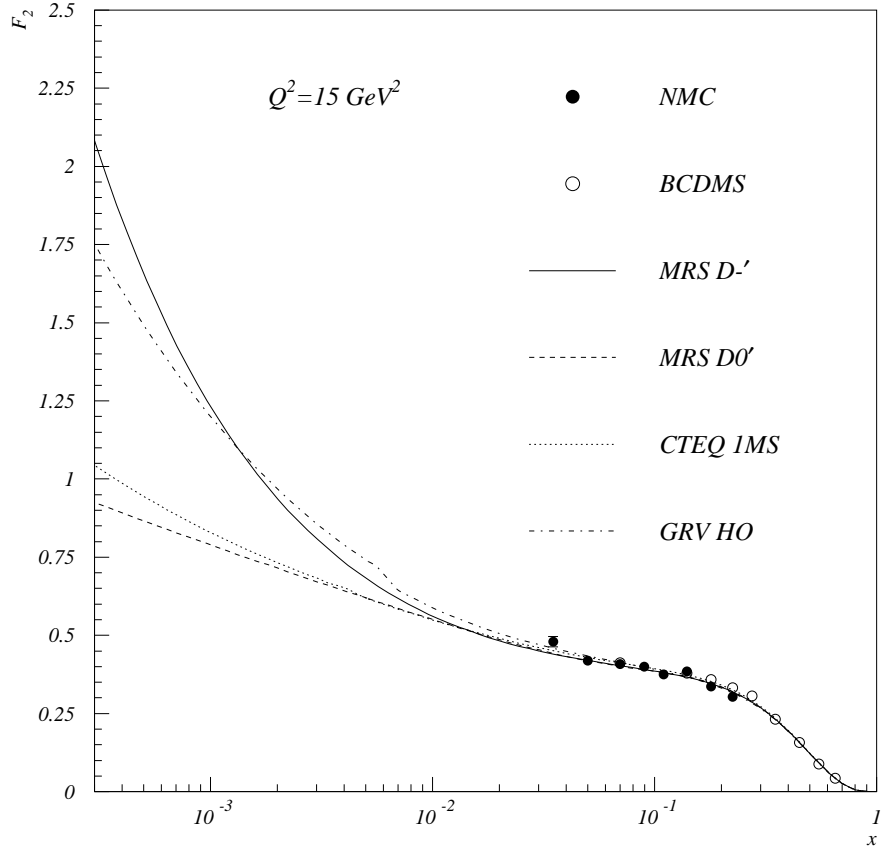,width=130mm}}
\end{picture}
\end{center}
\caption{\sl The proton structure function $F_2$ measured by the fixed-target
experiments NMC~\cite{nmc92} and BCDMS~\cite{bcdms} and compared with
various parameterizations: MRS D$-$$^\prime$, MRS D0$^\prime$~\cite{mrs92}, CTEQ
1MS~\cite{cteq1}, GRV~\cite{grv92}.}
\label{fig:f2p_fte}
\end{figure}

Before the advent of the HERA experiments, various structure functions have
been precisely measured by several fixed-target experiments for
$x\lesssim 0.01$ and values of $Q^2$ ranging from ${\cal O}(10)$\,GeV$^2$ to
about $100\,{\rm GeV}^2$ (Fig.\ref{fig:kine_xq2}). As an example, the proton
structure function $F_2$ measured by NMC~\cite{nmc92} and BCDMS~\cite{bcdms}
is presented in Fig.\ref{fig:f2p_fte}. These data have been used by various
groups to extract parton density functions.
A few parameterizations of the proton structure functions are also
shown in Fig.\ref{fig:f2p_fte}. These parameterizations, which all described
the then existing low energy data, differ at $x\simeq 10^{-4}$ by more than a
factor of four. The large uncertainty at small $x$ arises for two reasons
\begin{enumerate}
\item theoretically, there were concerns that within the perturbative QCD
framework, the occurrence of powers of $\ln (1/x)$ can spoil the
conventional (twist-2) formalism,
\item phenomenologically, even within the standard approach, the initial
parton distributions, which are needed in solving the evolution equations,
were largely unknown at $x\lesssim 10^{-2}$, because the existing data 
did not extend into this region, and the only constraint, the momentum sum 
rule, does not fix the shape of the distributions.
\end{enumerate}
All phenomenological analyses of parton distributions based on the usual QCD
formalism used certain assumed parameterizations of the initial distribution
functions that implicitly determined the extrapolated small-$x$ behavior.
For the MRS\,D~\cite{mrs92} parameterizations the
small $x$ evolution of the gluon density (at $Q^2_0=4\,{\rm GeV}^2$) was
singular ($B_g=-0.5$, see Eq.(\ref{eq:par_gluon})) for MRS\,D$-$$^\prime$ and
constant ($B_g=0$) for MRS\,D0$^\prime$. 
Similarly, for the CTEQ\,1MS~\cite{cteq1} parameterization the
gluon density was singular, but the sea quark density was not strongly coupled
to the gluon density, leading to a slower rise of $F_2$ with decreasing $x$.
For the GRV~\cite{grv92} parameterization small $x$ partons were dynamically
generated according to the DGLAP equations, starting from ``valence like''
quark and gluon density functions at $Q^2_0=0.3\,{\rm GeV}^2$.

\newpage
\section{First measurements at low $x$ from HERA} \label{sec:1stf2_hera}
\begin{figure}[htb]
\begin{center}
\begin{picture}(50,180)
\put(-190,-240){\epsfig{file=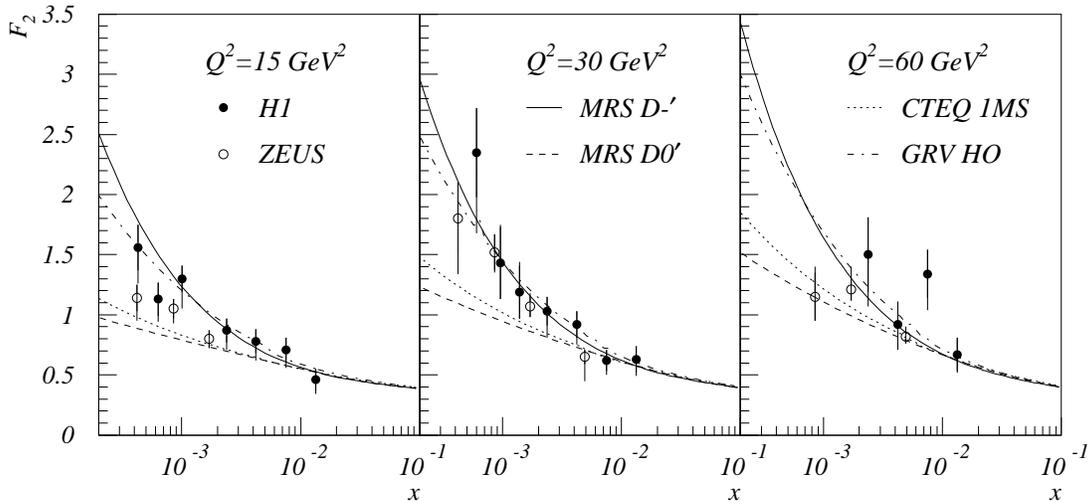,width=160mm}}
\end{picture}
\end{center}
\caption{\sl The first proton structure function $F_2$ measured at HERA by
H1~\cite{h1f292} and ZEUS~\cite{zeusf292} compared with
4 parameterizations: MRS\,D$-^\prime$, MRS\,D0$^\prime$~\cite{mrs92}, \
CTEQ\,1MS~\cite{cteq1}, GRV~\cite{grv92}.}
\label{fig:f2p_hera92}
\end{figure}

With about 25\,nb$^{-1}$ of data collected in 1992, the first year of the HERA
running, both H1 and ZEUS have made a first measurement~\cite{h1f292,zeusf292}
at small $x$ down to $0.5\times 10^{-4}$ at values of $Q^2$ comparable with 
the low energy fixed-target data~\cite{nmc92,bcdms}. The $x$ dependence of 
the measured $F_2$ for three selected $Q^2$ values is shown in 
Fig.\ref{fig:f2p_hera92}.

The data showed a significant rise in $F_2$ towards lower values of $x$.
This is a striking feature compared to all previous structure function data
obtained at fixed target experiments. These latter data can be successfully
described~\cite{dola92} by a single nonperturbative pomeron with intercept 
close to 1.08 (Sec.\ref{sec:para_other}). This is no longer possible for the
new data. These new
data, though limited in precision, have been used by several QCD analysis
groups to constrain the parton density functions at low $x$
resulting in better parameterizations of MRS\,H~\cite{mrsh} and
CTEQ\,2~\cite{cteq}.
The behavior of the strong $x$ dependence has been subsequently confirmed
with much improved precision already with 1993 data. The same data have
allowed a first measurement be made at values of $Q^2$ beyond those covered by
fixed-target experiments. This analysis is presented in the next section to 
illustrate as an example how these measurements were performed.

\newpage
\section{First $F_2(x,Q^2)$ measurement at values of $Q^2$ beyond those 
covered by fixed-target experiments} \label{sec:hiq2_93}
\begin{figure}[htb]
\begin{center}
\begin{picture}(50,325)
\put(-155,-25){\epsfig{file=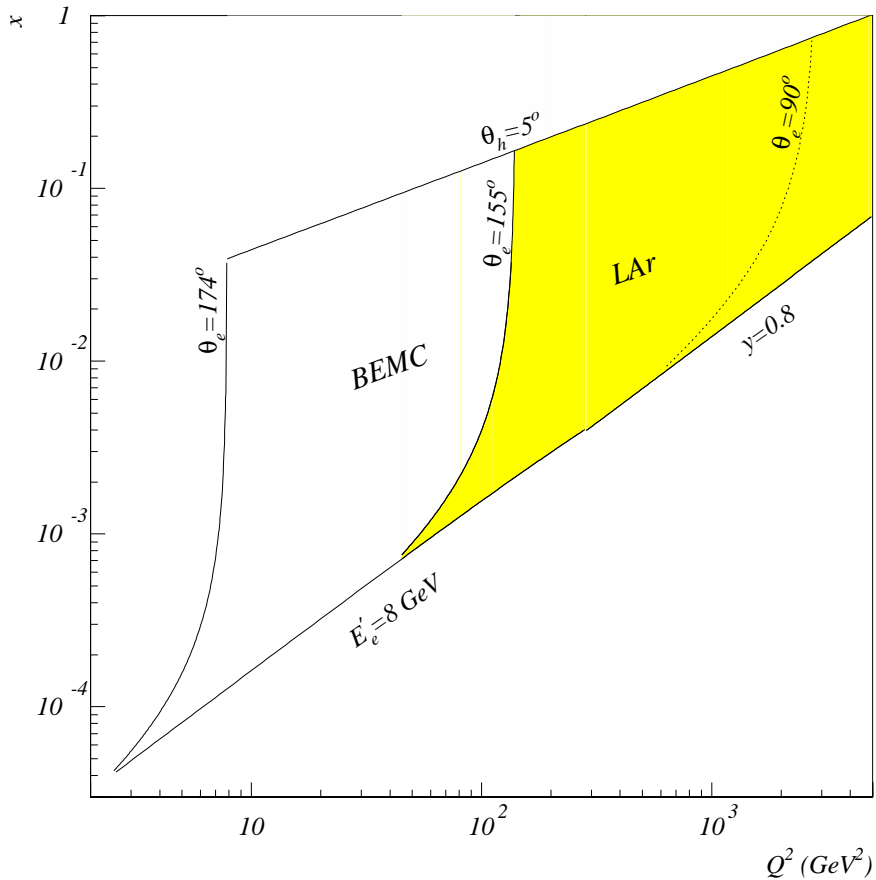,width=130mm}}
\end{picture}
\end{center}
\caption{\sl The kinematic region covered by 1993 data. It is limited on the
top by the angular acceptance of the hadronic final state
($\theta_h>5^\circ$), on the left by the angular acceptance of the scattered
electron, and from the bottom by the cuts on $E_e^\prime>8$\,GeV and
$y<0.8$. The shaded region, analyzed in the following, corresponds to 
the acceptance of the liquid argon (LAr) calorimeter, while the open area 
corresponds to the acceptance of the backward calorimeter (BEMC).}
\label{fig:kine_f293}
\end{figure}

In 1993, both H1 and ZEUS have collected a factor of 10 more data than in 1992.
The integrated luminosity collected by H1 was 0.271\,pb$^{-1}$.
The kinematic region covered by the 1993 data is shown in 
Fig.\ref{fig:kine_f293}.
The open region on the left corresponds to the acceptance of the backward
calorimeter (BEMC) in which the scattered electron is detected. The first
measurement of $F_2(x,Q^2)$ based on 1992 data was made in this region. The
increase of the luminosity has significantly extended the kinematic coverage
to the higher $Q^2$ region (shaded area in Fig.\ref{fig:kine_f293}) 
in which the scattered electron is measured in the liquid argon calorimeter 
(LAr). This kinematic region is covered in the following analysis.

\subsection{Event selection and background filters} \label{sec:f293_sel}
\begin{figure}[htbp]
\begin{center}
\begin{picture}(50,320)
\put(-210,-25){\epsfig{file=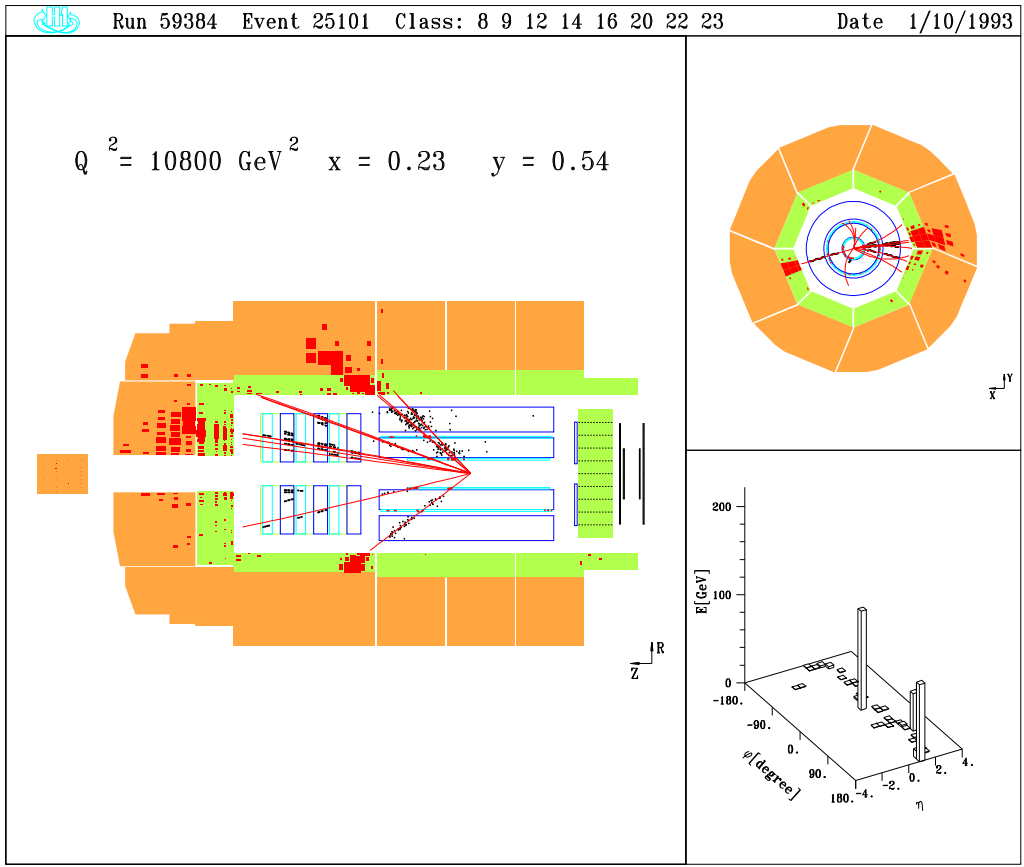,bbllx=0pt,bblly=0pt,
bburx=594pt,bbury=842pt,width=120mm,angle=90}}
\end{picture}
\end{center}
\caption{\sl A neutral current event at high $Q^2$ measured by the H1 detector
in 1993. The event kinematics is also shown.}
\label{fig:nc_r59384}
\end{figure}

The selection of DIS events at high $Q^2$ was based on an identified
scattered electron in the LAr calorimeter and additional requirements
for background rejection. One of the highest $Q^2$ neutral current events 
measured by the H1 detector is shown in Fig.\ref{fig:nc_r59384}.

The electron identification algorithm developed for this analysis used 
the salient
feature of the LAr calorimeter, namely the fine granularity in both lateral
and longitudinal direction. Requiring the longitudinal profile of an isolated
energy deposit (cluster) to be consistent with an electromagnetic object,
the cluster having the most compact lateral size was assigned as the scattered
electron resulting in a very good efficiency and low misidentification
probability~\cite{zhang94}. As far as the inclusive cross section and 
structure function measurement is concerned, the misidentification is more 
critical since it results in a wrong kinematic reconstruction for a DIS event 
and brings in additional background contribution from other events. 
The misidentification probability
was found to be $1.2\pm 0.5$\% in data which was in good agreement with the
simulation $1.5\pm 0.3$\%. These values corresponded to an identification 
using purely calorimetric information\footnote{The significant part of 
the inner central tracker was not operating in the whole 1993 data taking 
period due to the presence of broken wires, thus the track-cluster link 
could not be used in the electron identification.}. A study based on the
Monte Carlo events showed that if the track-cluster link were used, the
misidentification probability would have been reduced by a factor of two.

The main non-$ep$ background in this high $Q^2$ analysis was due to
muons traveling off axis parallel to the proton beam.
These muons are produced by proton beam halo interactions and
occasionally generate an electromagnetic shower in the LAr calorimeter.
Other important background sources included the cosmic ray events and 
beam gas/wall interactions. Most of these backgrounds were rejected 
by requiring 
\begin{itemize}
\item a reconstructed event vertex around the nominal interaction
point\footnote{The cut corresponds to about three times of the proton-bunch 
length ($\sigma^p_z=11$\,cm, see Table \ref{tab:hera_para}).}: \newline 
$|z_{\rm vtx}-z_0|<30$\,cm with $z_0=-5$\,cm and 
\item at least one reconstructed charged track having 
less than 2\,cm of the distance of closest approach between the track and 
the $z$ axis.
\end{itemize}
The remaining background events were rejected by a few topological background
filters developed for this analysis~\cite{zhang94}. For example, the halo
events were eliminated by searching for energy distributions in the LAr
calorimeter which are very
localized in the $r-\phi$ plane while having large spread in the $z$ direction.
These background filters have been corroborated and complemented later with 
other filters~\cite{ban_thesis,qbgfmar} and are being widely used in 
the current physics analyses~\cite{h1hiq9497}.

The dominant $ep$ background was from photoproduction in which the scattered
electron escaped the detector along the beam pipe and an energy cluster from
the hadronic final state faked a scattered electron. A sizable fraction of
the background, concentrated at low energies and high $y$, was suppressed by 
the following cuts:
\begin{eqnarray}
& & y_e<0.8 \label{eq:ycut}\\
& & E_e^\prime>8\,{\rm GeV} \label{eq:ecut} \\
& & \Sigma_e+\Sigma_h>30\,{\rm GeV}\,. \label{eq:epztot}
\end{eqnarray}
For $Q^2$ below $284\,{\rm GeV}^2$, the minimum energy
requirement is more restrictive than the $y_e$ cut, the latter becomes
effective at higher $Q^2$ (Fig.\ref{fig:kine_f293}).
The quantities $\Sigma_e$ and $\Sigma_h$ are defined respectively in
Eqs.(\ref{eq:epzpt_e}) and (\ref{eq:epzpt_h}).
Note that the sum $\Sigma_e+\Sigma_h$ is rather insensitive to energy loss
in the forward beam hole as the energy and the $z$ component of the momentum
essentially cancel, while it is very sensitive to energy loss (e.g.\ the
scattered electron of a photoproduction event or the initial state radiative
photon of a DIS event) in the backward beam hole.
While the sum for a DIS event should be around
$2E_e$,\footnote{The relation $\Sigma_e+\Sigma_h=2E_e$ is derived from the
energy and momentum conservation:
$\Sigma_e+\Sigma_h=E_e-P_{e,z}+E_p-P_{p,z}\simeq 2E_e$ by neglecting 
the electron and proton mass.}
a smaller value is expected for the photoproduction events as the contribution
from the scattered electron was not included in the sum.
The same cut also rejects DIS events with an energetic photon radiated
along the electron beam direction, thus reducing the radiative corrections
to the measurement. 

The final selected sample consisted of 1038 events. The residual non-$ep$
background was estimated to be smaller than 1.2\% with dominant contribution
from the cosmic ray event candidates and negligible other non-$ep$
background events by analyzing the pilot bunch data and a visual scan.
The overall photoproduction contribution was less than 1.5\% and 
the largest contribution at high $y$ did not exceed 10\%.

\subsection{Monte Carlo simulation}
For the high $Q^2$ analysis, three Monte Carlo samples have been generated for
the neutral current DIS interaction using the event generator
{\sc django}~\cite{django} and parton density distribution parameterizations
MRS\,D$-^\prime$, MRS\,D0$^\prime$, and MRS\,H. The {\sc django} program
is based on {\sc heracles}~\cite{heracles} for the electroweak interaction
and on {\sc lepton}~\cite{lepto} to simulate the hadronic final state.
{\sc Heracles} includes first order radiative corrections,
the simulation of real bremsstrahlung photons and the longitudinal structure
functions. The acceptance corrections were performed using the MRS\,H
parameterization, which is constrained to the HERA $F_2$ results of 1992.
To describe higher order QCD radiative processes {\sc lepto} uses 
the color dipole model~\cite{cdm} as implemented in {\sc
ariadne}~\cite{ariadne} which is in good agreement with data on the energy
flow and other characteristics of the final state as measured by
H1~\cite{h1had92} and ZEUS~\cite{zeushad92}. The program {\sc
jetset}~\cite{jetset} is
then used for the fragmentation of the resulting partons into hadrons, and
for their decay. {\sc Jetset} is based on the Lund string model of
fragmentation~\cite{lund}.

The ``soft'' vector meson contribution of the photoproduction interaction
was simulated using the {\sc rayvdm}~\cite{rayvdm} program, and the ``hard''
scattering part using the {\sc pythia}~\cite{pythia87} program. The relative
contributions of both were adjusted to agree with the total photoproduction
cross section analysis~\cite{gptot}.

\subsection{Kinematic reconstruction}
As mentioned in Sec.\ref{sec:kinerec}, the kinematics for a neutral current
DIS event can be redundantly reconstructed. Three methods have been studied
and compared for this analysis: the electron method, the double angle
method and the mixed method (Mixed). The first two methods have been defined 
already in Sec.\ref{sec:kinerec}. In the third method, the $Q^2$ is that of 
the electron method, i.e.\ $Q^2_{\rm Mixed}=Q^2_e$, while the $y$
is determined from the hadronic system, i.e.\ $y_{\rm Mixed}=y_h$.

For the electron method, the energy and the polar angle of the scattered
electron were used. For values of $Q^2\gtrsim 120\,{\rm GeV}^2$, the
scattered electron is detected in the LAr calorimeter. The polar angle
$\theta_e$ was defined with the cluster center in LAr with the vertex position.
With the 1993 data, a 5\,mrad accuracy and a 7\,mrad resolution were obtained
by using the Monte Carlo simulation and comparing data and the simulation
with the central tracking chambers~\cite{zhang94,h1f293}. 

Using the redundancy in the reconstruction of the kinematic variables 
by requiring $Q^2_e=Q^2_{\rm DA}$ with $Q^2_e$ and $Q^2_{\rm DA}$ being defined
respectively in Eqs.(\ref{eq:q2e}) and (\ref{eq:q2da}), the energy of the
scattered electron can be predicted
\begin{equation}
E_{e,{\rm DA}}=E_e\frac{\alpha_e+\alpha_e^{-1}}{\alpha_e+\alpha_h}
\label{eq:eda}
\end{equation}
where $\alpha_e$ and $\alpha_h$, defined in Eq.(\ref{eq:alpha_eh}), are
related to the angles of the scattered electron and of the hadronic final
state. The energy scale determined by test beam measurements~\cite{h1nim350_94}
was thus refined using this method. The resulting systematic
uncertainties in $E^\prime_e$ were smaller than 3\% for the barrel part of
the calorimeter and 5\% in the BBE region and the cracks~\cite{zhang94,h1f293}.

The mixed method needs the hadronic final state for reconstructing $y$.
The hadronic final state is measured in the LAr calorimeter\footnote{In fact,
this is only true for medium and low $y$. For very high $y$ hadrons are mainly
reconstructed in the BEMC, but these events were suppressed due to the
cuts on $E^\prime_e$ and $y$, see Eqs.(\ref{eq:ycut}) and (\ref{eq:ecut}).}.
The hadronic energy scale was known to 6\% as determined from
studies of the transverse momentum balance of DIS events. The test-beam data of
pions between 3.7\,GeV and 205\,GeV showed agreement on 3\% level with the
Monte Carlo description~\cite{h1nim336_93a}.

The hadronic angle $\theta_h$ was reconstructed according to
Eq.(\ref{eq:alpha_eh}) from the energy deposits in the calorimeter cells.
The Monte Carlo simulation showed that the angle was well measured except for
the small and large angles; at small angles it is sensitive to the calorimeter
noise effect (see Sec.\ref{sec:noise} for more explanation), while at large 
angles, hadrons were not well contained in the BEMC. 

\begin{figure}[htbp]
\begin{center}
\begin{picture}(50,400)
\put(-210,-35){\epsfig{file=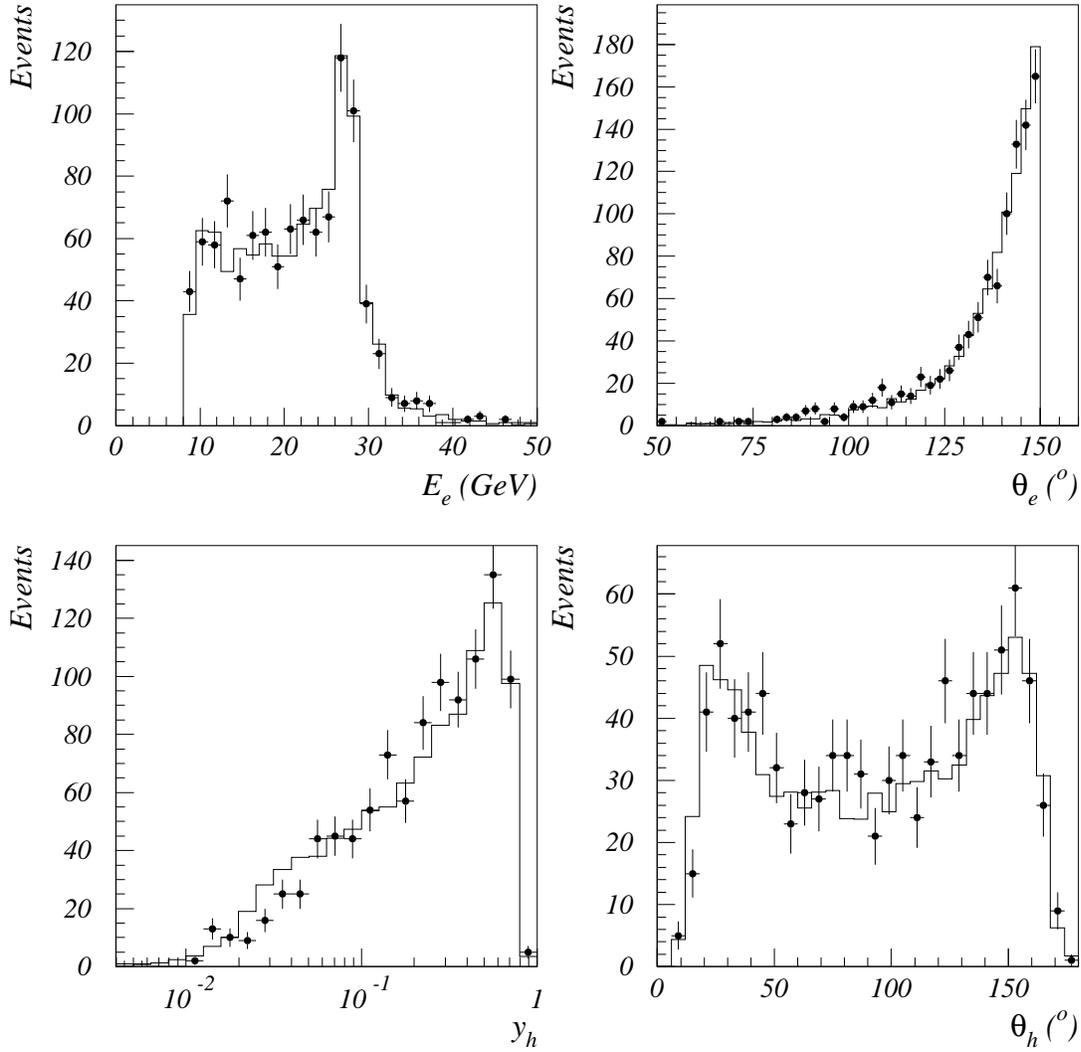,width=160mm}}
\end{picture}
\end{center}
\caption{\sl A comparison between data (points) and Monte Carlo
(histograms) of quantities relevant for the kinematics
reconstruction: $E_e^\prime$, $\theta_e$, $y_h$ and $\theta_h$.}
\label{fig:control_f293}
\end{figure}

The four quantities, $E^\prime_e, \theta_e, y_h$ and $\theta_h$, are compared 
in Fig.\ref{fig:control_f293} between data and the Monte Carlo simulation.
Given the precision, the agreement was reasonably good.

\begin{figure}[htb]
\begin{center}
\begin{picture}(50,200)
\put(-210,-35){\epsfig{file=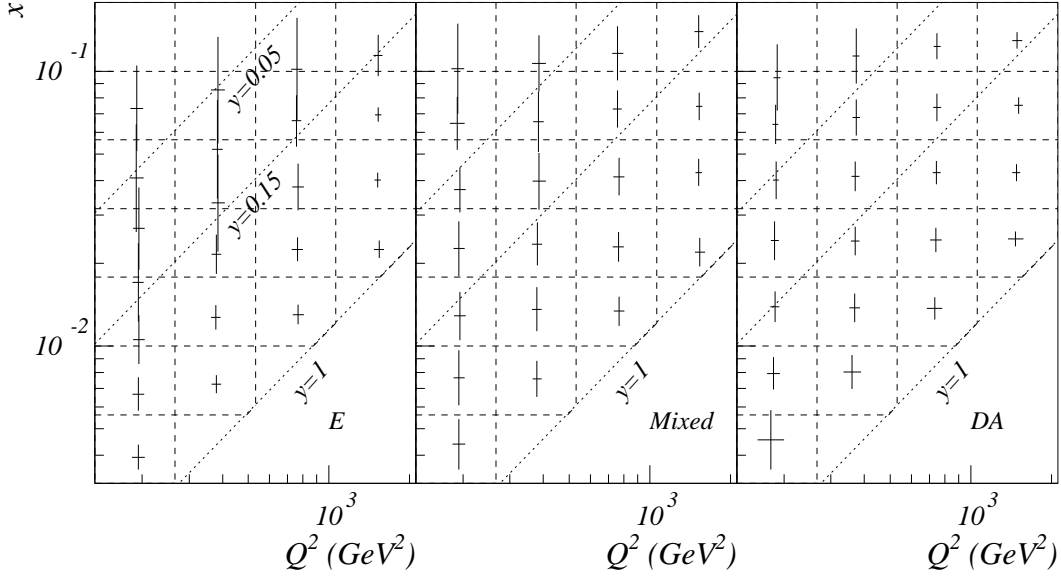,width=160mm}}
\end{picture}
\end{center}
\caption{\sl A comparison of the $x$ and $Q^2$ resolutions of the electron
method (E), the mixed method (Mixed) and the double angle method (DA). The
dashed lines indicate the bin boundary. Three $y$ values are shown with 
the dotted lines. The averaged true kinematics is around the center of
each bin. The location of the crosses with respect to the corresponding bin 
center indicates the magnitude of the migration of a method and 
the size of the crosses shows the resolution of the reconstructed kinematics.}
\label{fig:resoxq2}
\end{figure}

Since different quantities were used in the reconstruction of the kinematic
variables, the resulting precision could differ. This is illustrated in 
Fig.\ref{fig:resoxq2}.
While $Q^2_e$ resolution is the best over all the kinematic region studied,
the $x_e$ resolution deteriorates as $y$ decreases, this can be easily
understood by making the derivatives of $Q^2_e$ and $y_e$ defined
respectively in Eqs.(\ref{eq:q2e}) and (\ref{eq:ye}):
\begin{eqnarray}
& & \frac{\delta Q^2_e}{Q^2_e}=\frac{\delta
E_e^\prime}{E_e^\prime} \oplus
\tan\!\left(\!\frac{\theta_e}{2}\!\right)\delta \theta_e \label{eq:dq2e} \\
& & \frac{\delta y_e}{y_e}=\frac{1-y_e}{y_e}\left[\frac{\delta
E_e^\prime}{E_e^\prime} \oplus
\tan^{-1}\!\left(\!\frac{\theta_e}{2}\!\right)\delta
\theta_e\right]. \label{eq:dye}
\end{eqnarray}
The $x_h$ resolution is better at lower $y$ than that of $x_e$ but is still
worse than that of $x_{\rm DA}$. Overall the double angle method alone gives 
maximum coverage of the available kinematic range with reasonable precision.

\subsection{Measurement procedure of the structure function $F_2(x,Q^2)$}
\label{sec:method}
In this section, the method used to measure the inclusive cross section and
to extract the proton structure function $F_2$ is first introduced. 
Various efficiencies and their systematic uncertainties relevant for the 
measurement are then very briefly mentioned only as their precisions have
been superseded by high statistics data samples to be discussed in Chapter
\ref{chap:hiq2}.

\subsubsection{Method}\label{method}
What is measured experimentally is in fact the inclusive cross section. 
The measurement of the inclusive cross section and the
extraction of the structure function $F_2(x,Q^2)$ are performed 
in the following steps:
\begin{itemize}
\item {\bf Bin definition:} In commensurating with the available statistics and
the size of migration of the reconstructed kinematic variables due to the
finite detector resolution, a finite bin
size $\Delta x\Delta Q^2$ in $(x,Q^2)$ plane is defined.
\item {\bf Averaged cross section in a bin:} From the number of 
observed events in data
$N_{\rm data}$ and the number of estimated background events $N_{\rm bg}$
normalized to the integrated luminosity ${\cal L}$, an integrated cross
section over a defined bin $\Delta x\Delta Q^2$
is measured
\begin{equation}
\sigma|_{\Delta x\Delta Q^2}=\frac{1}{\epsilon}\frac{N_{\rm data}-N_{\rm
bg}}{\cal LA}
\end{equation}
where $\epsilon$ is an extra correction factor for those efficiencies which
are not simulated properly in Monte Carlo (see below). 
The variable ${\cal A}$ includes the efficiency (acceptance) and unfolding 
corrections, which in the simplest bin-by-bin unfolding method is
\begin{equation}
{\cal A}=\frac{N_{\rm MC}}{N_{\rm gen}}=\frac{N_{\rm MC}}{{\cal L}_{\rm gen}
\sigma_{\rm gen}|_{\Delta x\Delta Q^2}}\,,
\end{equation}
where $N_{\rm MC}$ and $N_{\rm gen}$ are respectively the number
reconstructed and generated events within the bin, and ${\cal L}_{\rm gen}$ is
the integrated luminosity of the generated Monte Carlo (MC).
Other sophisticated methods have been proposed~\cite{unfold} to unfold
the acceptance correction and the migration effects associated with the
finite detector resolution, the (strong) variation in the DIS cross section,
and the QED radiative corrections (Sec.\ref{sec:rc}). However, in practice,
since the Monte Carlo from which the unfolding matrix is generated describes
the data reasonably well (after e.g.\ the extra correction mentioned above
and the iteration on the input structure function, see below) 
the different methods give similar results for
the unfolded data~\cite{quadt_thesis}. \newline
To some extent, the migration effects depend on the shape of the structure
function, it is therefore important that the structure function
parameterization used in generating the input Monte Carlo cross section
$\sigma_{\rm gen}$ 
is not substantially different from the measured one. Otherwise (this is the
case for the earlier analyses), an iteration is often
necessary by reweighting\footnote{The reweighting is in practice performed on
an event-by-event basis using the true kinematics $x$ and $Q^2$.} the Monte
Carlo events by the measured cross section:
\begin{equation}
N_{\mbox{new}}=N_{\rm MC}
\frac{\sigma_{\rm new}|_{\Delta x\Delta Q^2}}{\sigma_{\rm gen}|_{\Delta
x\Delta Q^2}}\,. \label{eq:sf_reweight}
\end{equation}
\item {\bf Differential cross section:} A differential cross section at a 
quoted kinematic point $(x,Q^2)$ within a bin is obtained by applying a bin 
center correction
\begin{eqnarray}
\left.\frac{d^2\sigma(x,Q^2)}{dxdQ^2}\right|_{\rm meas}
\!\!&=&\!\!\frac{\sigma|_{\Delta x\Delta Q^2}}{\sigma_{\rm
(new)gen}|_{\Delta x\Delta Q^2}}
\frac{d^2\sigma_{\rm (new)gen}(x,Q^2)}{dxdQ^2} \\
\!\!&=&\!\!\displaystyle \frac{1}{\epsilon}\frac{N_{\rm data}-N_{\rm
bg}}{\cal L}\frac{{\cal L}_{\rm gen}}{N_{\rm (new)MC}}
\frac{d^2\sigma_{\rm (new)gen}(x,Q^2)}{dxdQ^2}\,. \label{eq:xs_meas}
\end{eqnarray}
\item {\bf Extraction of the structure function $F_2(x,Q^2)$:} Once the
inclusive cross section is measured, to extract the structure function
$F_2(x,Q^2)$ from the cross section formula (Eq.(\ref{eq:xsnc})):
\begin{eqnarray}
\frac{d^2\sigma}{dxdQ^2}\!\!\!&=&\!\!\!\frac{2\pi\alpha^2}{xQ^4}
 \left[Y_+\tilde{F}_2-y^2\tilde{F}_L+Y_-x\tilde{F}_3\right](1+\delta_{\rm
rc}) \\
 \!\!\!&=&\!\!\!\frac{2\pi\alpha^2}{xQ^4}
 \left\{\left[2(1-y)+\frac{y^2}{1+\tilde{R}}\right]\tilde{F}_2+
 Y_-x\tilde{F}_3\right\}(1+\delta_{\rm rc})
\end{eqnarray}
one needs to know the contribution of the $\gamma Z^0$ interference and $Z^0$
exchange to $\tilde{F}_2$ (Eq.(\ref{eq:gf2nc})), and the structure functions
$\tilde{R}(x,Q^2)$~\footnote{The structure function $\tilde{R}$ is
generalized from the $R$ defined in Eq.(\ref{eq:r}) when the $\gamma Z^0$
interference and $Z^0$ exchange contribution is taken into account.} and
$x\tilde{F}_3(x,Q^2)$, neither of which has been measured so far in
the kinematic region covered by HERA. However, according to
the QCD prediction, $\tilde{R}$ is small.
Due to the additional suppression factor $y^2$, its contribution was found
to be well below 1\% for most of the kinematic region considered and
reached up to 4.5\% at the high $y$ region~\cite{zhang94}.
The related issue will be discussed later in Sec.\ref{sec:fl}.
At values of $Q^2<1000\,{\rm GeV}^2$, the contribution from the $Z^0$
exchange and $\gamma Z^0$ interference is expected to be less than 1\% (see
Fig.\ref{fig:weak_cor}).
At higher $Q^2$, the structure function $x\tilde{F}_3$ becomes increasingly
larger and the structure function $\tilde{F}_2$ also deviates from $F_2$
for the one-photon exchange.
The largest correction was estimated to be 4.3\%~\cite{zhang94} for
the highest $Q^2=1600\,{\rm GeV}^2$ studied in this analysis.
The experimental evidence of the $x\tilde{F}_3$ contribution
at HERA will be discussed in Chapter \ref{chap:hiq2}.
In addition, a radiative correction $\delta_{\rm rc}$ should be applied 
in order to extract the structure function from the
measured cross section. First-order QED radiative effects were included
in the Monte Carlo simulation. The dominant contribution, partially
suppressed by the selection cut (Eq.(\ref{eq:epztot})), is originating
from the initial state radiation (Sec.\ref{sec:rc}). 
The corrections are highly correlated with
the method used in reconstructing the kinematics. 
Because of small separation angles between the final state radiative photon
aand the scattered electron, the photon is mostly not resolved from the
scattered electron by the calorimeter despite of its fine granularity. This
considerably reduces the kinematical migration effects caused by the final
state radiation and therefore the radiative effects of the final state
radiation are less important than those of the initial state radiation.
\end{itemize}

\subsubsection{Efficiencies and uncertainties}
In Eq.(\ref{eq:xs_meas}), an efficiency correction $\epsilon$ is explicitly 
shown to be applied. In principle, all efficiency corrections could be 
taken into account in the Monte Carlo simulation and if it is properly done, 
no additional correction is needed. 
In practice, this has to be verified and any systematic uncertainty 
has to be propagated into the measurement of the cross section or 
the structure function. In two cases, an additional correction will be needed:
\begin{enumerate}
\item when there should be a discrepancy between data and the Monte Carlo,
e.g.\ the correction of the vertex efficiency in Sec.\ref{sec:isr},
\item when a correction is determined after the Monte Carlo has been made,
e.g.\ the correction of the trigger efficiency in Secs.\ref{sec:isr} and
\ref{sec:cctrig}.
\end{enumerate}

All efficiencies were determined in this analysis from the data and 
compared with the Monte Carlo simulation. Agreement between the experimental 
and the simulated
values for the individual efficiencies (trigger, vertex) was found to
be better than 2\%. An overall error of 4\% was assigned due to the
imperfect description of the various efficiencies.

\subsection{Results}
\begin{figure}[htbp]
\begin{center}
\begin{picture}(50,380)
\put(-195,-40){
\epsfig{file=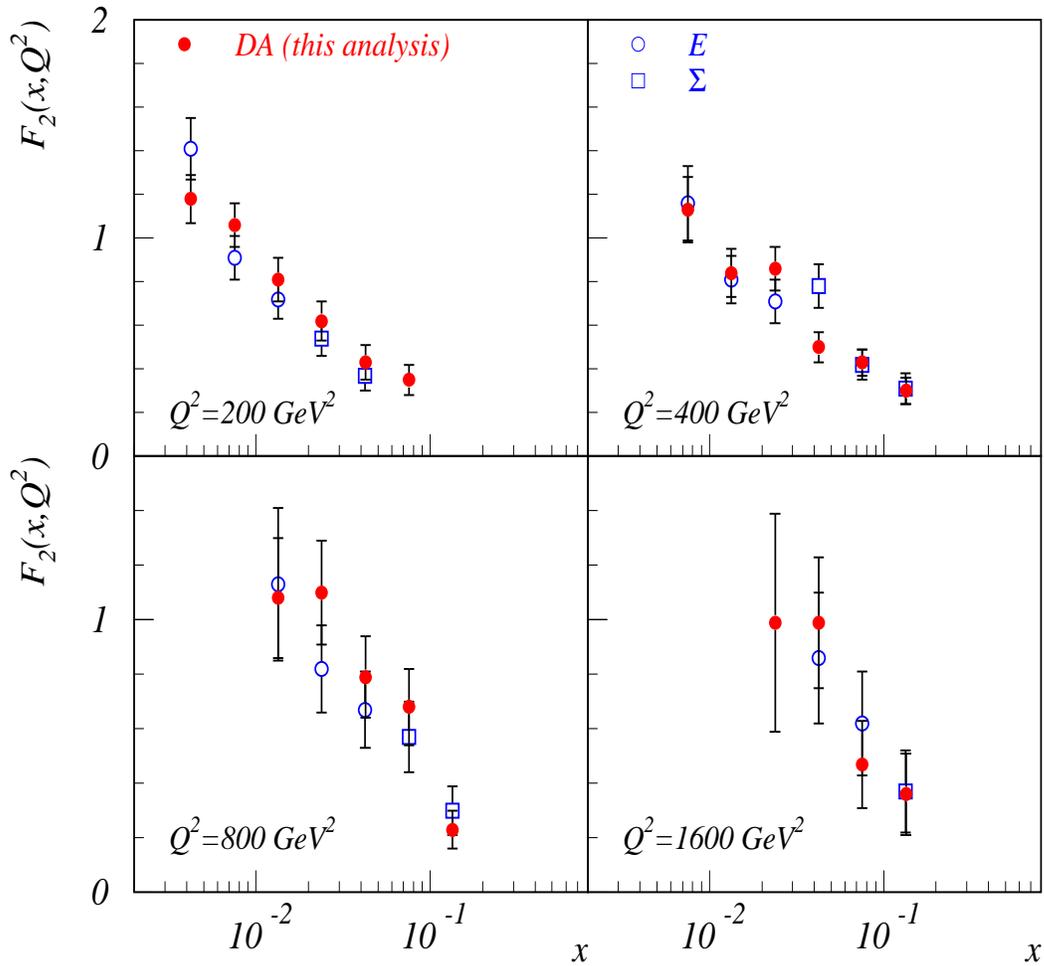,height=155mm,width=155mm}}
\end{picture}
\end{center}
\caption{\label{fig:f293_com}\sl {A comparison of the measured 
 $F_2(x,Q^2)$ from this analysis using the double angle (DA) method with 
 other analyses~\cite{leverenz} based on the electron (E) and sigma ($\Sigma$)
 methods. The error bars show the statistical errors only.
 }}
\end{figure}

\begin{figure}[htbp]
\begin{center}
\begin{picture}(50,410)
\put(-185,-45){\epsfig{file=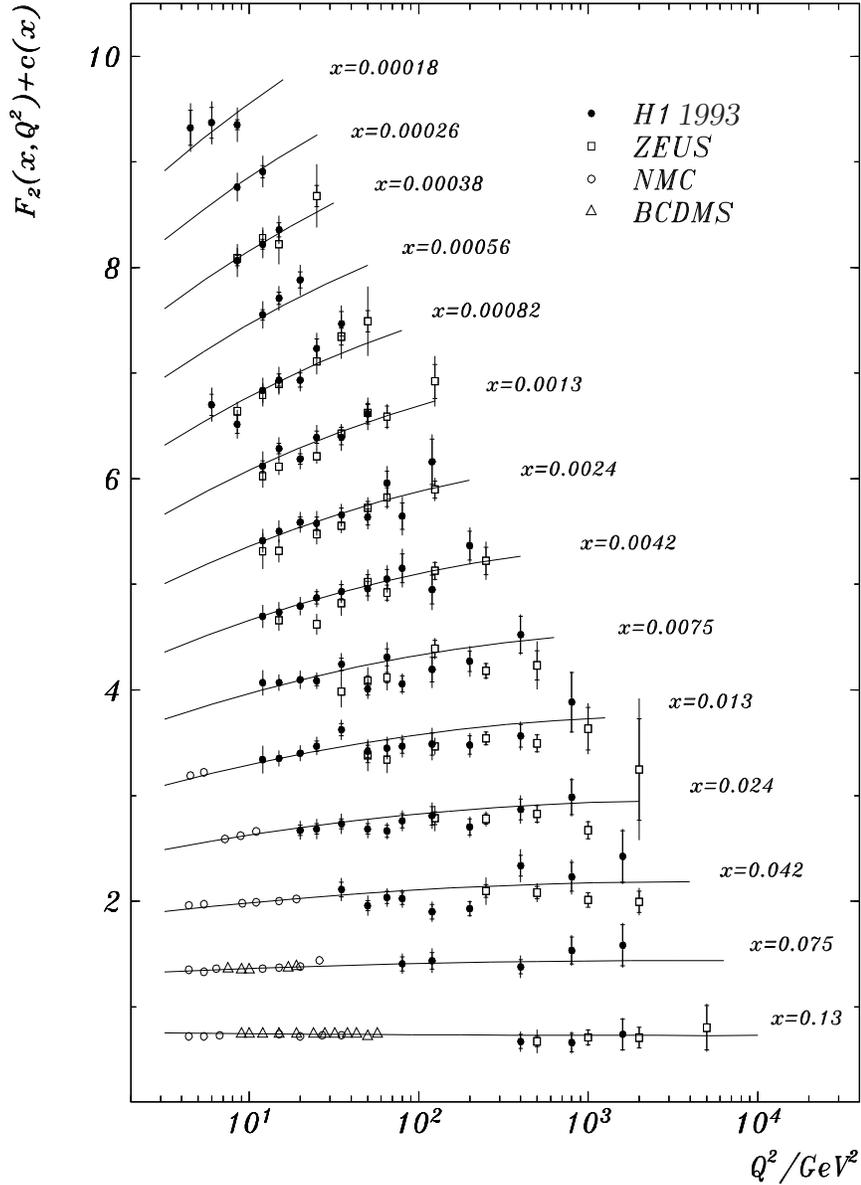,bbllx=0pt,bblly=0pt,
bburx=594pt,bbury=842pt,width=155mm}}
\put(103,393){\sl 1993}
\end{picture}
\end{center}
\caption{\label{fig:h1f293_q2}\sl {Measurement of the proton structure
function $F_2(x,Q^2)$ based on the 1993 $e^-p$ data from H1. The results of the
ZEUS collaboration are also shown with open squares (the
ZEUS $F_2$ data were shifted to the H1 $x$ values by using the
parameterization in \cite{zeusf293}). The curves represent a phenomenological 
fit to the H1, NMC and BCDMS data (see Ref.\cite{h1f293}). The $F_2$ values are
plotted with all but normalization errors in a linear scale adding a term
$c(x)=0.6(i_x-0.4)$ to $F_2$ where $i_x$ is the bin number starting at
$i_x=1$ for $x=0.13$.}}
\end{figure}

The proton structure function $F_2$ measured with the 1993 H1 data by
this analysis is compared in Fig.\ref{fig:f293_com} with other
analyses~\cite{leverenz} using the electron
method at low $x$ and the sigma method at high $x$.
Good agreement is observed with few exceptions. For this comparison,
only statistical errors are shown because the measurements were highly
statistically correlated although there were differences in the electron
identification and selection cuts,
which can result in a systematic difference in the $F_2$ measurements.
Additional systematic differences can also arise because the analyses use
different quantities in the kinematic reconstruction and do not have the
same radiative corrections.
Systematic uncertainties have been studied in
detail~\cite{zhang94,leverenz,h1f293}. Instead of giving the long list of 
considered systematic sources here, it is worth simply noting that most of the
uncertainties were limited either directly or indirectly by the low
statistics data sample available then. The cross section measurement and its
precision have been much improved with the increased integrated luminosity,
as we will see in 
Chapter \ref{chap:hiq2}.
Nevertheless, these measurements together with others at low
$Q^2$~\cite{h1f293}, shown in Fig.\ref{fig:h1f293_q2},
provided the basis for a QCD analysis~\cite{h1gluon93} in which 
a first determination of gluon density was performed (Sec.\ref{sec:fit}).
Also shown in Fig.\ref{fig:h1f293_q2} are the results from the ZEUS
experiment~\cite{zeusf293}, which are in good agreement with that of H1.

\newpage
\section{The structure function $F_2$ at low $Q^2$ from radiative events}
\label{sec:isr}
\subsection{Extended kinematic domain using radiative events}
When extracting the structure functions from a measured inclusive cross
section of DIS processes, one of the corrections that one has to make
is the radiative correction. The effect is dominated for the
electron scattering by the emission of the real energetic photons from the
incident electrons. At HERA, a significant fraction ($\sim 30\%$) 
of these initial state radiative (ISR) photons are detected 
in the photon detector of the luminosity system. 
The photon detector, situated at about 100\,m from the interaction 
point, has an angular acceptance of about 0.45\,mrad. 
These radiative photons can thus be considered as being emitted collinearly
with respect to the incident electrons. The energy of the electron after 
the radiation
\begin{equation}
E^{\rm eff}_e=E_e-E_\gamma \label{eq:eb_eff}
\end{equation}
available for the subsequent deep inelastic interaction has effectively been 
reduced and so does the center-of-mass energy
\begin{equation}
s^{\rm eff}=4(E_e-E_\gamma)E_p=\frac{E^{\rm eff}_e}{E_e}s\,,
\end{equation}
where $E_p$ and $s$ are respectively the proton beam energy and the nominal
center-of-mass energy squared.

\begin{figure}[htbp]
\begin{center}
\begin{picture}(50,325)
\put(-155,-25){\epsfig{file=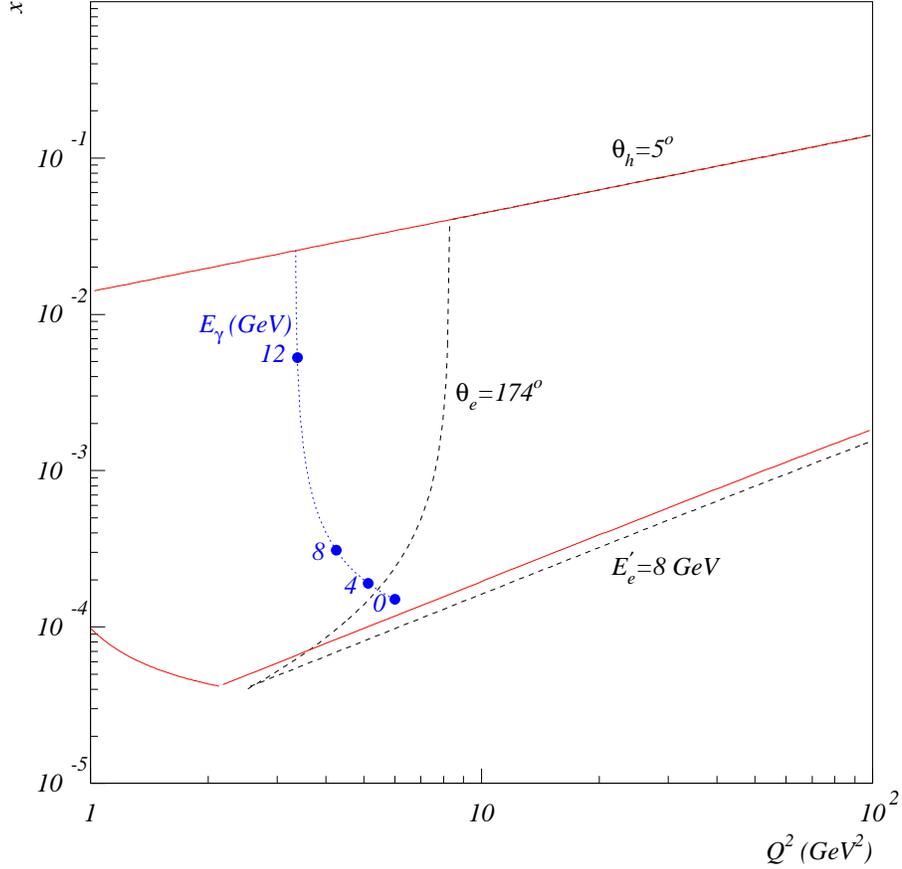,width=130mm}}
\end{picture}
\end{center}
\caption{\sl The extended kinematic domain from radiative events (the region
enclosed by the full lines from the top by the inclusive hadronic angle
$\theta_h>5^\circ$ and from the bottom by the energy threshold of
the scattered electron $E_e^\prime>8$\,GeV for $E_\gamma=4$\,GeV
(Eq.(\ref{eq:egacut})) compared with
the corresponding one from non-radiative events (the region enclosed by the 
dashed lines, in addition to the limits on $\theta_h$ and $E_e^\prime$, 
the main limit is on the angular acceptance of $\theta_e<174^\circ$).
The dotted curve and large dots illustrate how the kinematics is modified for
different photon energies $E_\gamma$.}
\label{fig:kine_isr}
\end{figure}

When replacing $E_e$ and $s$ respectively by $E^{\rm eff}_e$ and $s^{\rm eff}$
in Eqs.(\ref{eq:q2e})-(\ref{eq:xe}), the resulting kinematic values can be very
different from those of a non-radiative event depending on $E_\gamma$.
This is illustrated in Fig.\ref{fig:kine_isr} with a dotted curve originating
from $x=0.00015$ and $Q^2=6\,{\rm GeV}$ and continuing upward to the left
as the photon energy increases. As a consequence the limited kinematic region
for the non-radiative events due to mainly the angular acceptance
($\theta_e<174^\circ$) of the scattered electron in the backward calorimeter 
BEMC is significantly extended to lower $Q^2$ and high $x$.

A measurement of the structure functions in the extended kinematic region is
of basic interest:
\begin{itemize}
\item The uncertainty in the size of the radiative correction originates
from {\it a priori} unknown shape of the structure functions in the
unmeasured kinematical region rather than from the technical aspects of the
matrix element integration. The measurement of the cross section in the
extended kinematical region allows thus a direct control of this uncertainty.
\item The extended kinematical region covers an intermediate $Q^2$ region
between the photoproduction processes (at $Q^2\simeq 0$) and the
DIS regime ($Q^2>$ a few GeV$^2$). The measurement in this region may shed
light on the underlying dynamics of the transition.
\end{itemize}

\subsection{Event selection and background studies}\label{sec:sel_bg}
The analysis shown here is based on the data taken in 1994 by the H1 detector. 
The data correspond to an integrated luminosity of 2.7\, pb$^{-1}$, 
which is about a tenfold increase with respect to the earlier data collected 
in 1992 and 1993, upon which first experimental studies of the radiative 
process were performed~\cite{h1rad9293}.

\subsubsection{Event selection}
The selection of DIS events with hard photon emission collinear to the 
incident electron is based on an identification of the scattered electron in
the backward calorimeter BEMC,
a measured radiative photon in the photon detector and additional requirements
for background rejection. The non-$ep$ background events, dominated by 
interactions of beam protons with residual gas and beam line elements 
upstream of the H1 detectors, are efficiently rejected at the trigger level
using a time of flight system consisting of two scintillator planes
installed behind the BEMC.

The scattered electron, which is defined to be the most energetic cluster in
the BEMC, has to satisfy the following criteria:
\begin{itemize}
\item The energy of the cluster measured with the BEMC is larger than 8\,GeV.
\item The lateral size of the cluster is required to be smaller than 5\,cm,
as expected for the signature of an electron.
\item The cluster in the BEMC is required to be associated with at least one
reconstructed space point in the BPC by less than 4\,cm.
\end{itemize}

An energetic photon detected in the photon detector (PD) is required to have:
\begin{equation}
        E_{\gamma}=E_{\rm PD}+E_{\rm VC}>4\, {\rm GeV}\label{eq:egacut}
\end{equation}
where $E_{\rm PD}$ and $E_{\rm VC}$ are the energies deposited in the
photon detector and in the water \v{C}erenkov veto counter (VC).
This condition suppresses beam related background events and
cosmic rays which produce energetic showers in the BEMC.

The remaining non-$ep$ background events are further
rejected by requiring an event vertex, reconstructed from tracks in the
central and forward tracking chambers, within $\pm 35$\, cm from the nominal 
interaction point.
The vertex position together with the impact point of the scattered electron
measured in the BPC/BEMC\footnote{It is the position measured by the BPC 
that has been used as it has a better spatial resolution than the BEMC
because of the coarse granularity of the latter.}
also defines the polar angle $\theta_e$ of the scattered electron.

The main source of $ep$ background is pile-up events due to accidental 
coincidence of DIS and $\gamma p$ events with a Bethe-Heitler~\cite{bh} (BH) 
Bremsstrahlung event ($ep\rightarrow e\gamma p$, Sec.\ref{sec:rad_proc}) 
in a time window of $\pm 5\, {\rm ns}$. 
These pile-up events are efficiently rejected with the following cuts:
\begin{eqnarray}
 \label{eqndelta}
& & E_{\rm etag}<2\, {\rm GeV} \nonumber \\ 
& & \Delta=\frac{E_{\gamma}-E_{\rm miss}}{E_{\gamma}}<0.5 \label{eq:delta}
\end{eqnarray}
where $E_{e{\rm tag}}$ is the energy deposited in the electron tagger.
The quantity $\Delta$ compares the measured photon energy in the photon 
detector $E_{\gamma}$ with the measured missing energy $E_{\rm miss}$ based
on the main detector without including the photon detector:
\begin{equation}
        E_{\rm miss}=E_e(y_e-y_h)
\end{equation}
with $y_e$ and $y_h$ being defined respectively in Eqs.(\ref{eq:ye}) and
(\ref{eq:yh}).
      One expects for 
      radiative DIS events $\Delta=0$ ($E_{\gamma}=E_{\rm miss}$) while for 
      pile-up DIS+BH events $\Delta=1$ ($E_{\rm miss}=0$). More details 
      concerning these cuts and the remaining 
      background events are discussed in the following paragraphs.
      
To ensure a high trigger efficiency, two fiducial cuts
\begin{eqnarray}
& & \sqrt{x^2_{\rm BPC}+y^2_{\rm BPC}}>15\,{\rm cm} \label{eq:trig_fidu1}\\
& & |x_{\rm BPC}|+|y_{\rm BPC}|>18\,{\rm cm} \label{eq:trig_fidu2}
\end{eqnarray}
are applied to remove the region around the beam pipe where the efficiency
degrades (Fig.\ref{fig:efftg_xy}).

\subsubsection{Background studies}
After having applied the selection cuts described above, a
sample of 8229 events is selected. It consists mainly of the following 
types of events\footnote{Note that the angular
acceptance for photons from radiative DIS events is about 30\% while for BH 
events it is 
about 98\% due to the different angular distributions of photon emission.}:
\begin{enumerate}
\item Radiative DIS events alone or in random coincidence with BH events (with 
      the probability to be discussed below) where the radiated photon
      is detected in the PD,
\item Pile-up events due to radiative DIS with BH events where the radiated 
      photon from the DIS event is not detected,
\item Pile-up events due to non-radiative DIS with BH events,
\item Pile-up events due to $\gamma p$ with BH events or inelastic BH events 
      alone.
\end{enumerate}

Events of type 1 constitute the signal since the radiated photons are
collinear with respect to the direction of the incident electron. All
other events are background. The events of type 4 contribute because of
the $\pi^0\rightarrow \gamma\gamma$ decay in the hadronic 
final state, and of the electromagnetic nature of the BEMC making 
an unambiguous separation between electrons and hadrons at low energies 
difficult.

The probability for random coincidence depends on the minimum photon energy
($E^{\rm min}_{\gamma}$) of BH events. According to an 
analysis of BH events~\cite{sergej}, this probability is 5.6\% for 
$E^{\rm min}_{\gamma}=0.13\, {\rm GeV}$, which is consistent with the 
value determined using the $\Delta$ distribution described below.

\begin{figure}[htbp]
\begin{center}
\begin{picture}(50,510)
\put(-242.5,-30){
\epsfig{file=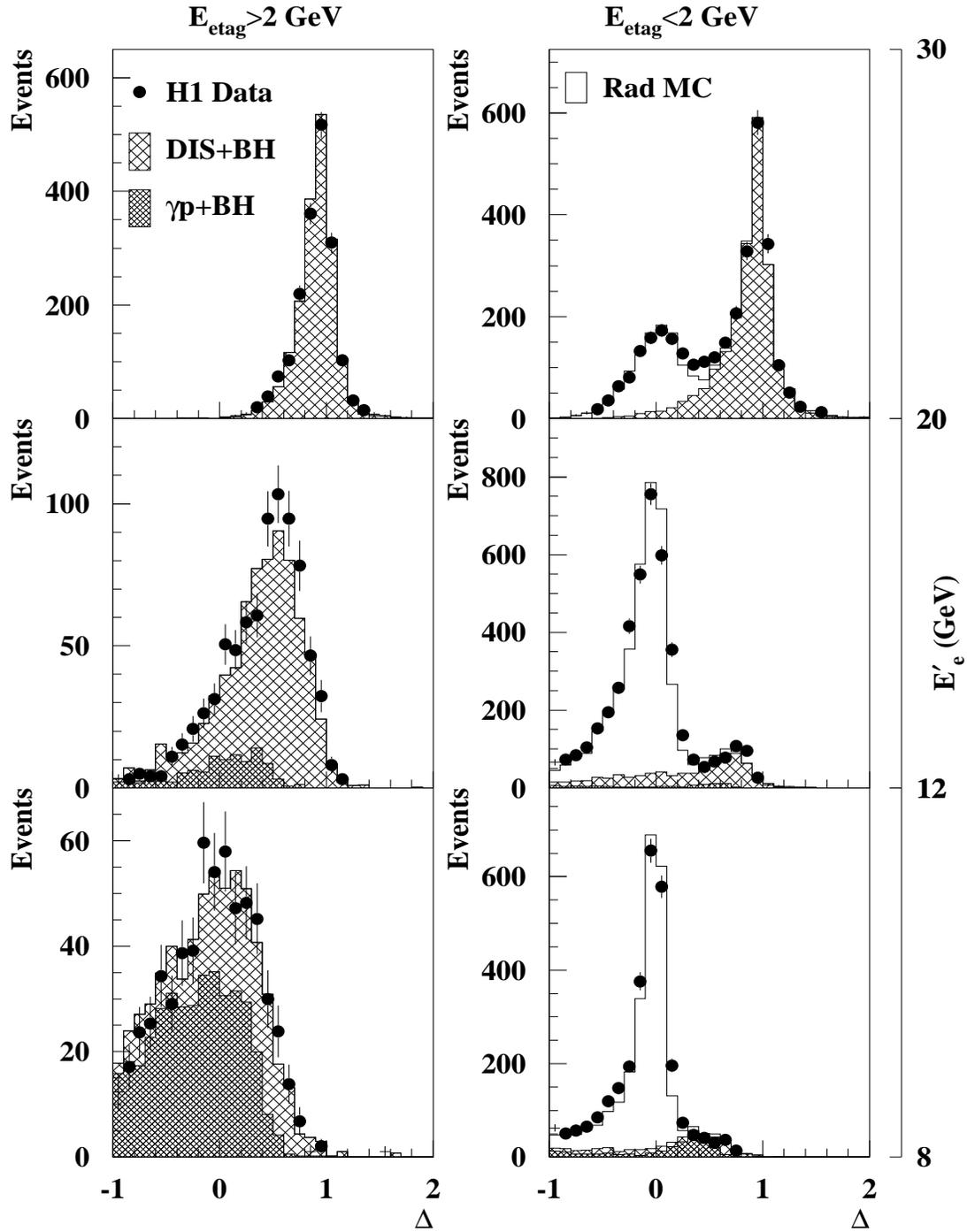,height=200mm,width=175mm}}
\end{picture}
\end{center}
\caption{\label{delta}\sl {Distribution of $\Delta$ (see
Eq.(\ref{eq:delta}) for definition) for tagged events (left plots) and
for non-tagged events (right ones) in three different electron energy ranges
as shown with the axis on the right side. The normalization for the
simulated DIS+BH events is fixed by the upper left plot and the
normalization for the $\gamma p$+BH events is fixed by the lower left plot.}}
\end{figure}

As mentioned above the variable $\Delta$ can be used to reject 
pile-up events. The distribution of $\Delta$ is shown in Fig.~\ref{delta} for 
six different event samples. The plots on the left show
those events which have activity in the electron tagger ($E_{\rm etag}>2\, 
{\rm GeV}$, tagged sample) while the plots on the right correspond to the 
non-tagged sample.
The dependence on $E_e^\prime$ is shown in three different energy ranges 
indicated with the axis on the right side. As expected, the peaks around 
one in the two upper plots are due to DIS+BH events.
The normalization of this background is fixed\footnote{With this method we 
found a probability for random coincidence
of 6.4\% for $E_{\gamma}^{\rm min}=0.1\, {\rm GeV}$, which corresponds
to 6.0\% for $E_{\gamma}^{\rm min}=0.13\, {\rm GeV}$, and is consistent with 
the value mentioned above. The large value of this method may be understood
since second order overlaps are included.} by the upper left plot
and confirmed with the upper right one for events at high $\Delta$ values
($\Delta>0.8$) where the expected signal (peaking around $\Delta=0$) does
not contribute. The difference (4\% ) is taken as the uncertainty in the
background subtraction of pile-up events of types 2 and 3.

The normalization for the events of type 4 is fixed by the lower left plot 
after the DIS+BH events have been subtracted. In this way, any possible 
background contribution from inelastic BH events (with photons being detected
in the PD) is also taken into account. In fact, among the estimated background
events of type 4, only about one third is expected from $\gamma p$ events
in random coincidence with BH events using the probability
given above. A 30\% uncertainty is assigned for this normalization due to 
the missing inelastic BH process in the MC and to the dependence on 
the low energy spectrum of the structure function.

The relative contribution of these background events after all selection
cuts is smaller than 10\% in most of the kinematic region to be measured and
the largest contribution at high $y$ does not exceed 30\%.

\subsection{Monte Carlo simulation}\label{mcsim}

For this analysis, neutral current DIS events (both radiative events and
non-radiative events) were generated using {\sc django}~\cite{django}.
The GRV~\cite{grv92} parton density parameterization was used for the generation
of events because this parameterization provided a reasonable description of
previous H1 and ZEUS measurements~\cite{h1f292,zeusf292} and was one of
few parameterizations that provided structure function parameterization 
at both low and high $Q^2$ needed for this analysis.

The $\gamma p$ events were simulated using the {\sc phojet} 
generator~\cite{phojet},
which generates the total $\gamma p$ cross section
by taking into account both soft and hard processes.

The BH events were generated according to the Bethe-Heitler 
approximation~\cite{bh} in which the proton recoil energy is neglected, 
i.e. $E_e^\prime+E_{\gamma} \simeq E_e$. The acceptance of the electron 
tagger was determined directly from the H1 data.

\subsection{Kinematic reconstruction} \label{sec:isr_kinerec}
A precise reconstruction of the kinematic variables relies crucially on
the measurement of the angle $\theta_e$ and the energy
$E_e^\prime$ of the scattered electron, of the energy $E_\gamma$ of
the radiative photon, and of the hadronic final state. Most of the measurement
is better studied with the high statistics sample of non-radiative
events~\cite{panitch_thesis}. This is the case for $\theta_e$,
$E_e^\prime$, and the measurement of the hadronic system. The key points
and the methods used for the measurement
of these quantities are briefly discussed here. The calibration of
$E_\gamma$, a quantity which is most relevant for this analysis, is then
described.

The angle of the scattered electron is defined as the straight line between
the reconstructed vertex and the impact point in the BPC. Several effects
contribute to the precision of the $\theta_e$ measurement:
\begin{enumerate}
\item the relative alignment of the BPC with respect to the central trackers,
\item the precision of the vertex reconstruction,
\item the amount of dead material the scattered electron encounters on its
way to the BPC affecting the multiplicity of the BPC hits and thus the
resolution of the impact point.
\end{enumerate}
With the 1994 data, a systematic precision of up to 1\,mrad is achieved.

The energy calibration and the resolution of the BEMC is originally determined
with test beams~\cite{bemc92}. Both the energy scale and the resolution are 
checked and improved based on the $ep$ data using the following three methods.
\begin{itemize}
\item {\bf The kinematic peak method} uses events in the pronounced peak of
quasi-elastically scattered electrons in the region close to the beam energy.
The absolute energy scale of the individual stacks is calibrated. For the
energy scale of the electron cluster, the effects of the presence of dead
material in front of the BEMC (up to 2\,$X_0$) and of cracks between the BEMC
stacks are taken into account by a Monte Carlo simulation.
\item {\bf The double angle method} uses the predicted electron energy
$E_{e,{\rm DA}}$ as defined in Eq.(\ref{eq:eda}) to 
an independent cross check of the energy scale of the cluster
and of its uniformity over the whole BEMC and allows to perform additional
corrections in the crack region, which is difficult to calibrate with
sufficient precision.
\item {\bf The QED Compton method} uses events from the physical process
corresponding to the Compton scattering of a quasi-real photon on an
incident electron (Sec.\ref{sec:rad_proc}), with the dominant contribution 
due to the elastic channel $(ep\rightarrow ep\gamma)$. 
The energy and the angle of the scattered
electron and of the photon are constrained by the QED theory, making this
process well suited for calibration of the BEMC~\cite{smain_thesis}.
\end{itemize}
The three methods are complementary and allow the energy linearity be studied.
After the recalibration, the systematic uncertainty on the energy scale is 
reduced from 1.7\% obtained from 1993 data to 1\%. The dead material 
description is also improved resulting in a comparable energy resolution 
between data and the Monte Carlo.

\begin{figure}[htb]
\begin{center}
\begin{picture}(50,220)
\put(-190,-260){
\epsfig{file=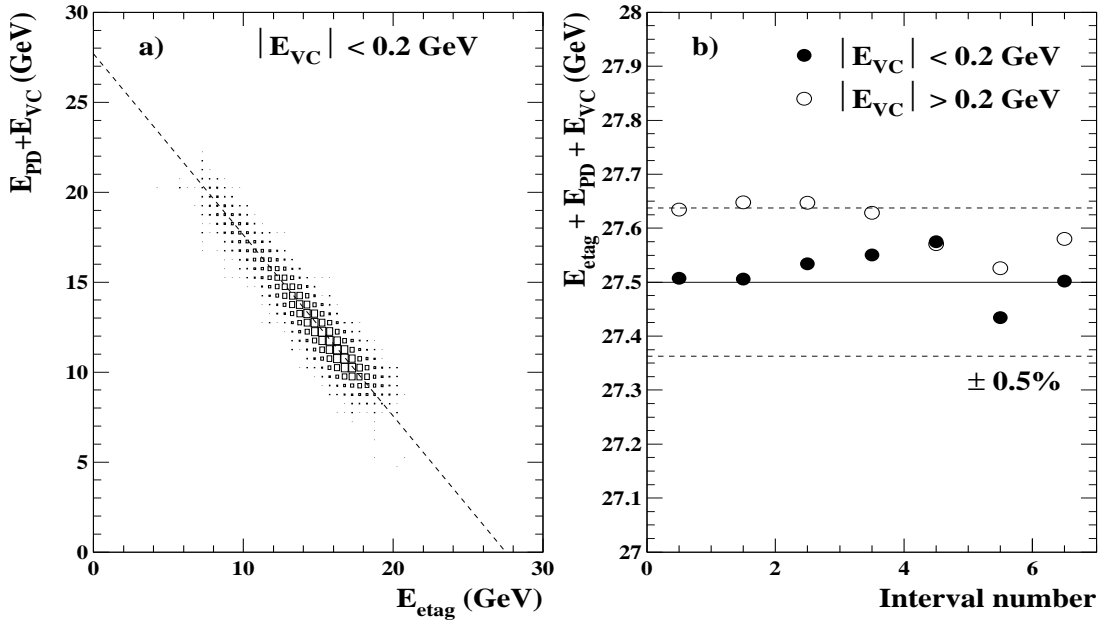,height=180mm,width=145mm}}
\end{picture}
\end{center}
\caption{\label{luminote}\sl {Correlation of the photon and electron 
 energies for Bremsstrahlung events in a). The energy sum is expected to be 
 $27.5\, {\rm GeV}$. In b) the mean values of the energy sum is plotted for 7 
 different run intervals. The calibration is stable within $\pm 0.5\%$.}}
\end{figure}

For a precise measurement of the energy of the radiated photon it is
important to know the energy calibration~\cite{sergej} of the photon detector 
with high precision. The off-line calibration of the photon detector 
(and also of the electron tagger) has been determined with a sample of 
BH events. These events were selected using the following requirements:
\begin{eqnarray}
 & &  E_{\rm etag} \ge 4\, {\rm GeV} \nonumber \\
 & &  E_{\gamma} \ge 4\, {\rm GeV} \nonumber \\
 & &  24\, {\rm GeV} \le E_{\rm etag} + E_{\gamma} \le 31\, {\rm GeV} \\
 & &  \left|x_{\rm etag}\right| \le 65\, {\rm mm} \quad\mbox{and}\quad 
      \left|y_{\rm etag}\right| \le 65\, {\rm mm} \nonumber 
\end{eqnarray}
The last condition, where $x_{\rm etag}$ and $y_{\rm etag}$ are the coordinates of 
the impact point with respect to the center of the electron tagger, was used 
to reject events in which a large amount of energy leaks over the transverse 
detector boundaries. The crystal calorimeters were calibrated with events
having $E_{\rm VC} \le 0.2\, {\rm GeV}$ making use of the kinematic constraint
$E_{\rm etag}+E_{\gamma} = E_e$, while the water \v{C}erenkov veto counter was
calibrated with those events having $E_{\rm VC} \ge 0.2\, {\rm GeV}$.
In Fig.~\ref{luminote}a) the correlation of $E_\gamma$ and $E_{\rm etag}$ is 
shown
for those Bremsstrahlung events with $E_{\rm VC} \le 0.2\, {\rm GeV}$. 
The stability of the calibration was checked by dividing the selected 
event sample into 7 subsamples. For each subsample a gaussian function was
iteratively fitted within $\pm 2\sigma$ to the energy sum 
$E_{\rm etag}+E_\gamma$.
The resulting mean values of the fits are given in Fig.~\ref{luminote}b),
separately for the two event samples with and without energy deposit in
the water \v{C}erenkov counter. For both samples the calibration could be
verified within $0.5\%$. The relative calibration of $E_\gamma$ and
$E_{\rm etag}$ is determined with a precision of $1.3\%$ from studies of the
$E_\gamma$ and $E_{\rm VC}$ dependence of the mean $E_\gamma + E_{\rm etag}$.
Taking into account a maximum nonlinearity of 1.3\% in the response of the
photon arm the global energy scale is known with a precision of 1.5\%
for $E_\gamma>4\, {\rm GeV}$.

The precision of the kinematic variables reconstructed with the electron
method is similar to that shown in Eqs.(\ref{eq:dq2e}) and (\ref{eq:dye})
except that, here, one has to take into account of an additional 
term $\delta E_\gamma/E_\gamma$ and
$(1-y_e)/y_e \delta E_\gamma/(E_e-E_\gamma)$ respectively for
$\delta Q^2_e/Q^2_e$ and $\delta y_e/y_e$ arising from the
energy resolution of the radiative photon.
Therefore, 
as for non-radiative events, $y_e$ degrades as $y_e$ decreases.
For this reason, the electron method is used only for $y>0.15$ and for
lower $y$, the $\Sigma$ method (Eqs.(\ref{eq:q2s})-(\ref{eq:xs})) is used
instead, which needs $\Sigma_h$ in addition to $\Sigma_e$ and $P_{T,e}$.

At lower $y$, the hadronic final state is measured by the LAr calorimeter.
Its energy scale is controlled by checking the transverse momentum
balance between $P_{T,h}$ and $P_{T,e}$. A systematic uncertainty of 4\% is 
obtained.

\subsection{Measurement of the structure function $F_2(x,Q^2)$}
The analysis follows essentially the same method as used in the previous
section. The only difference is that here, we measure the inclusive
radiative cross section which is part of the QED radiative corrections to 
the usual DIS cross section.
In the $Q^2$ range of the present analysis, the effect of $Z^0$ exchange is
negligible and the double differential cross section for single virtual
photon exchange in DIS, integrated over the solid angle within the
acceptance of the photon detector $\theta_\gamma<\theta_a\simeq 0.45$\,mrad
and from 0 to $z_1=0.85$ corresponding to $E_{\gamma}>4$\,GeV,
is given by~\cite{witek}
\begin{eqnarray}
\frac{d^2\sigma}{dxdQ^2}\!\!&=&\!\!\!\int _0^{z_1}\!\!dz\frac{d^3\sigma}{dxdQ^2dz}\\
\!\!&=&\!\!\frac{\alpha^3}{xQ^4}\left[2(1-y)+ \frac{y^2}{1+R}\right]F_2(x,Q^2)
(1+\delta_{ho})\int _0^{z_1}\!\!dzP(z) \label{eq:bornxs}
\end{eqnarray}
with 
\begin{eqnarray}
 & & z=\frac{E^{\rm eff}_e}{E_e}=\frac{E_e-E_{\gamma}}{E_e} \\
 & & P(z)=\frac{1+z^2}{1-z}\ln\!\left(\frac{{E_e}^2\theta_a^2}{m^2_e}\right)-
       \frac{2z}{1-z} \nonumber
\end{eqnarray}
where $R$, defined in Eq.(\ref{eq:r}), is related to the longitudinal
structure function, $F_L$.
The higher order (ho) correction to the first order radiative contribution
is given by $\delta_{\rm ho}$(see Sec.\ref{sec:horc}). 
In comparison with the usual DIS cross section, the
radiative cross section to be measured here is thus suppressed by a factor of
$\frac{\alpha}{2\pi}\int_0^{z_1}dzP(z)$.

\subsubsection{Efficiencies and their uncertainties}
The electron identification and most of the selection cuts described in
Sec.\ref{sec:sel_bg} are similar to the analysis on the sample dominated
by non-radiative events. Their efficiencies have
been studied with this high statistics sample and found to be well described
by the Monte Carlo simulation~\cite{beatriz_thesis,panitch_thesis}.
What are particular for this analysis are the trigger and vertex efficiencies.
The trigger efficiency differs from the other analysis since a different cut
on the minimum electron energy is applied (8\,GeV versus 11\,GeV). 
The hard radiation studied in this analysis also corresponds to a different
event topology, on which the vertex efficiency strongly depends: a scattered 
electron at large polar angle beyond the acceptance of the central tracker 
in the backward direction is unlikely to contribute to the vertex 
reconstruction and similarly the efficiency also drops when the hadronic 
system goes very forward beyond the acceptance in the other 
direction\footnote{The forward tracker
can be and has been used for the vertex reconstruction, its efficiency is
however worse than that of the central tracker due to the fact that the
passive material in front of it is relatively more important and that the
track density is higher because of the Lorentz boost in the proton beam
direction.\label{ft:effvtx_fwd}}.

The events studied in this analysis are triggered by requiring a local energy
deposit, or cluster of more than 4\,GeV in the BEMC. 
The efficiency of this trigger was determined as function of the electron 
energy and the impact position with an independent data sample
(dominated by non-radiative events) which were triggered with an independent 
tracker-related trigger. 

During the 1994 data taking period the four inner triangle stacks in the BEMC
did not contribute to the trigger for most of the
time (``closed triangle'' period). During the last weeks of data taking these
modules were included in the trigger (``open triangle'' period).
It turned out that for the open triangle period the trigger efficiency
did not depend on the impact position but on the electron energy only.
For electrons with an energy of 8 GeV an efficiency of $\simeq 80\%$ was
found and from 11 GeV on it was found to be $\simeq 100 \%$.

\begin{figure}[hbtp]
\begin{center}
\begin{picture}(50,380)
\put(-195,-10){
\epsfig{file=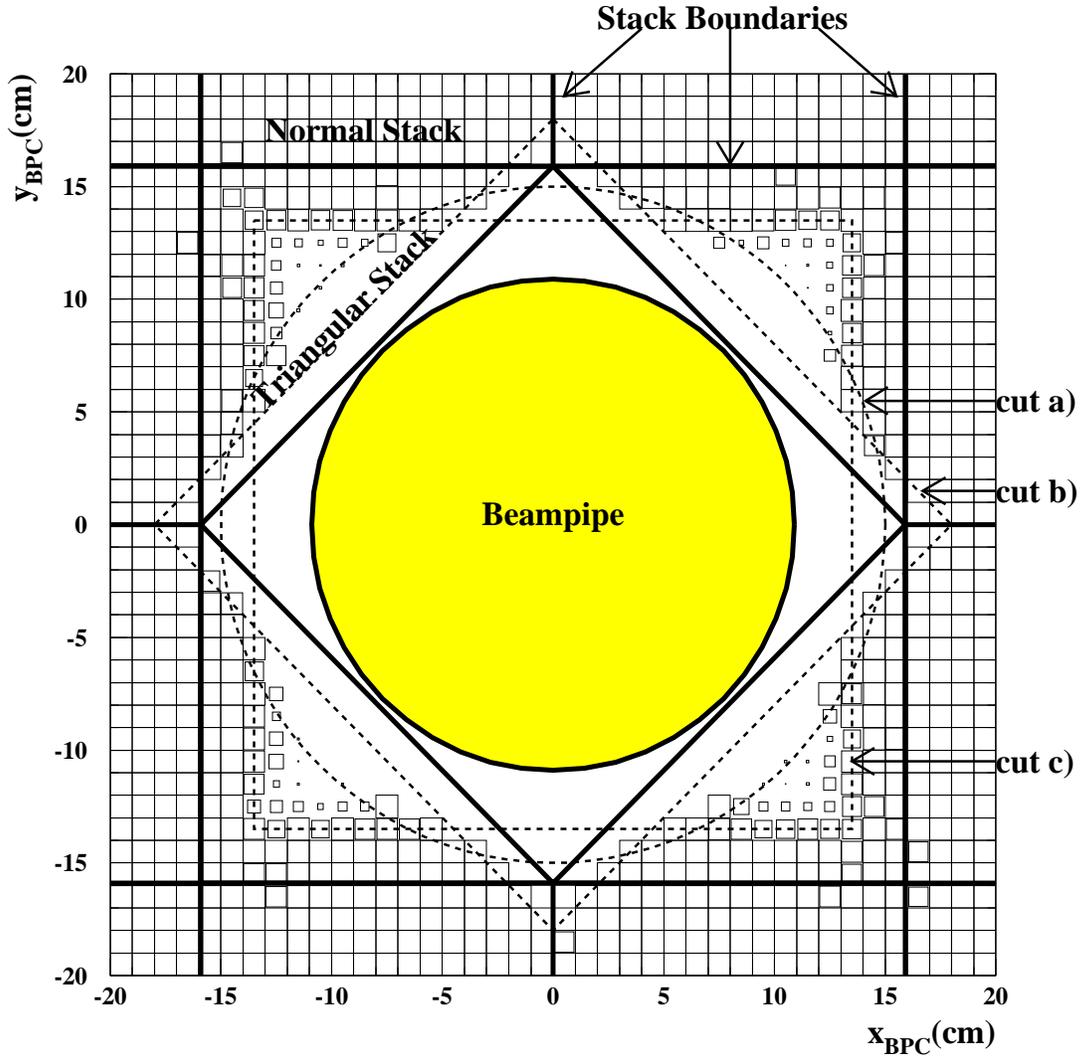,height=150mm,width=150mm}}
\end{picture}
\caption{\label{fig:efftg_xy}\sl {Trigger efficiency in the BEMC inner 
 region as functions of the impact position $(x_{\rm BPC},y_{\rm BPC})$ for 
 the closed triangle period. The efficiency is proportional to the box size
 with the full box size corresponding to 100\%. The electron energy was 
 required to be far above the trigger threshold to separate the spatial 
 dependence of the trigger efficiency from the energy dependence. The cuts a)
 and b) are always applied, see
 Eqs.(\protect{\ref{eq:trig_fidu1}}),(\ref{eq:trig_fidu2}). The cut c) 
 $\max\{\left|x_{\rm BPC}\right|,\left|y_{\rm BPC}\right|\}>13.5\,{\rm cm}$ 
 was applied for the closed triangle period only to ensure a high trigger 
 efficiency.}}
\end{center}
\end{figure}

For the closed triangle period the trigger efficiency strongly depends on
the impact position, as can be seen in Fig.~\ref{fig:efftg_xy}. To ensure the
trigger efficiency to be acceptably high an additional cut,
$\max\{\left|x_{\rm BPC}\right|,\left|y_{\rm BPC}\right|\}>13.5\, {\rm cm}$,
was applied for this period, as it is indicated in Fig.~\ref{fig:efftg_xy}. 
In the region
$\max\{\left|x_{\rm BPC}\right|,\left|y_{\rm BPC}\right|\}>17\, {\rm cm}$
the trigger efficiency depended on the electron energy only, while in the 
region $13.5\, {\rm cm}<\max\{\left|x_{\rm BPC}\right|,
\left|y_{\rm BPC}\right|\}<17\,{\rm cm}$ it was determined as function of 
the electron energy $E_e^\prime$ and $\max\{\left|x_{\rm BPC}\right|,
\left|y_{\rm BPC}\right|\}$. More details on the trigger efficiencies are
given in Refs.\cite{f2rad94,hutte_thesis}.

The vertex inefficiency has two contributions: either there is no
reconstructed vertex or there is a vertex but outside the vertex cut
($\pm 35$\,cm).
These contributions may be determined for radiative MC events in the following
way:
\begin{equation}
 \label{cipvtx1}
  \epsilon_{\exists\; z_{\rm vtx}}=
  \frac{N(\mbox{all cuts except $|z_{\rm vtx}-z_o|<35\, {\rm cm}$})}
  {N(\mbox{all cuts except $\exists\; z_{\rm vtx}$ and 
   $|z_{\rm vtx}-z_o|<35\, {\rm cm}$})}
\end{equation}
and
\begin{equation}
 \label{cipvtx2}
  \epsilon_{|z_{\rm vtx}-z_o|<35\, {\rm cm}}=\frac{N(\mbox{all cuts})}
  {N(\mbox{all cuts except $|z_{\rm vtx}-z_o|<35\, {\rm cm}$})}
\end{equation}
giving an overall value of respectively $82.6\pm 0.3\%$ and $94.4\pm 0.2\%$ 
for the selected radiative sample. Unfortunately this simple method 
when applied to the real data may bias the efficiency due to the residual 
background events contained in the denominators.

\begin{figure}[htb]
\begin{center}
\begin{picture}(50,230)
\put(-175,0){
\epsfig{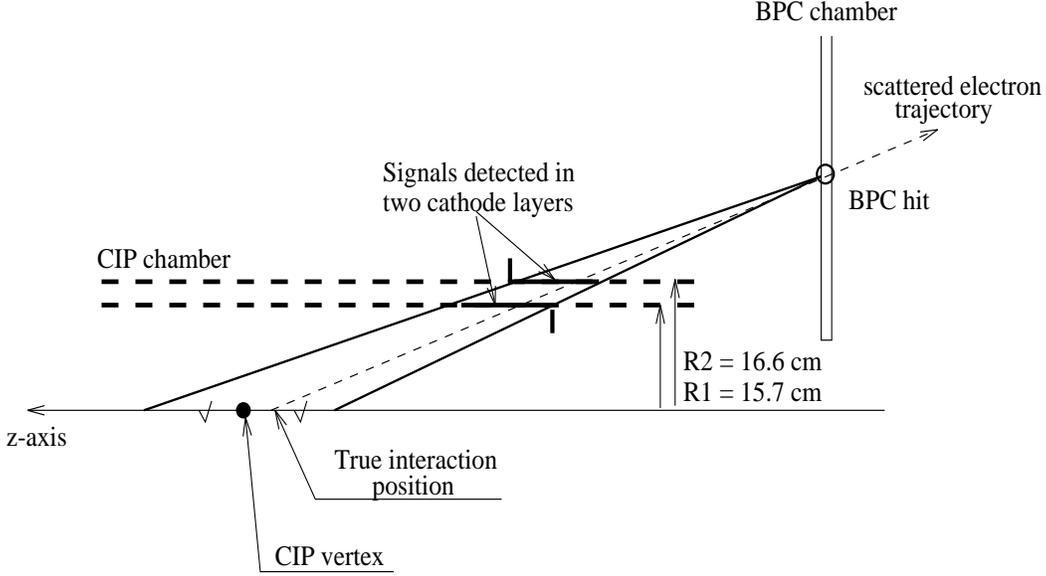}}
\end{picture}
\caption{\label{fig:cipvtx}\sl {Schematical presentation of the CIP
vertex definition, using the impact point in the BPC and two CIP hits in
coincidence (figure taken from Ref.\cite{panitch_thesis}).}}
\end{center}
\end{figure}

Alternative method exists~\cite{cipvtx}, which is schematically shown in
Fig.\ref{fig:cipvtx}.
The vertex $z_{\rm CIP}$ is defined in this method
as the crossing point between the $z$ axis and a straight
line formed with the impact point in the BPC and two CIP hits in coincidence
in the $r-z$ plane.
This method deffers from the standard method in that $z_{\rm vtx}$ is
still defined by charged particles from the hadronic final state 
when the electron track is inefficient in a DIS event or when
there is no such a track in a background event, while $z_{\rm CIP}$ can not
be defined if there is no CIP hits associated the electron track. Therefore
the CIP vertex is less sensitive to the photoproduction background.
For DIS events it has been checked that a good correlation was obtained 
when both vertices were defined~\cite{f2rad94}.

By applying the vertex cut using $z_{\rm CIP}$, namely 
$|z_{\rm CIP}-z_o|<35\, {\rm cm}$ 
to both the numerators and the denominators in Eqs.(\ref{cipvtx1}) and
(\ref{cipvtx2}), the vertex efficiencies are found to be consistent between
the data and the MC. Globally, one has $\epsilon_{\exists\; z_{\rm
vtx}}=82.8\pm 0.5\%$ and $\epsilon_{|z_{\rm vtx}-z_o|<35\, {\rm cm}}=94.2\pm
0.4\%$ in data to be compared with $\epsilon_{\exists\; z_{\rm vtx}}=83.8\pm
0.4\%$ and $\epsilon_{|z_{\rm vtx}-z_o|<35\, {\rm cm}}=95.2\pm 0.2\%$ in the 
MC. In principle, it is straightforward to compare the data and the MC more
differentially, e.g. in terms of $(x,Q^2)$ bins. However, due to the
limited CIP geometrical acceptance, part of the sample in which the electrons
scattered at large polar angles cannot be checked by this method.
Here an improvement in the experimental apparatus is highly desirable. 
The installation of a backward silicon tracker certainly helps in this 
direction.

\begin{figure}[htbp]
\begin{center}
\begin{picture}(50,180)
\put(-185,-220){
\epsfig{file=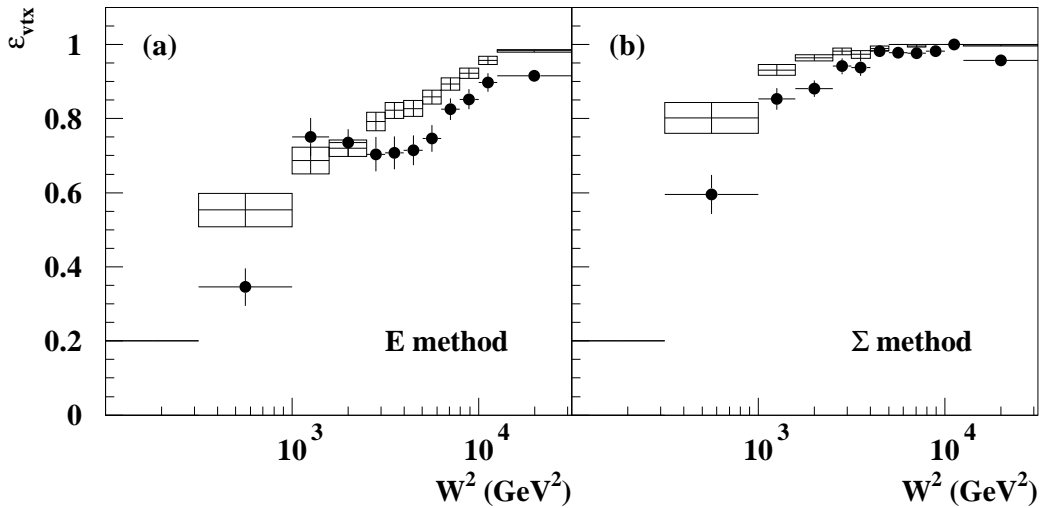,height=155mm,width=155mm}}
\end{picture}
\end{center}
\caption{\label{vtx_w2}\sl {Vertex efficiency as a function of $W^2$ 
 for radiative DIS events within the considered $(x,Q^2)$ bins in data(points)
 and in MC(open squares) for the $E$(a) and $\Sigma$(b) methods. }}
\end{figure}

Selecting events only at the considered $(x,Q^2)$ region,
a difference shows up between the data and the MC when the
efficiencies are plotted as a function of the hadronic invariant mass squared
($W^2$)\footnote{We have chosen $W^2$ on which the vertex efficiency depends
in a similar way as it depends on the inclusive hadronic angle $\theta_h$.
In fact, the smaller the $W^2$ is, the more forward the hadronic system
goes. The tracker covering the forward region has a smaller efficiency than 
the central tracker, see footnote \ref{ft:effvtx_fwd}.} in Fig.~\ref{vtx_w2}.
The difference is more pronounced using the $E$ method than using the $\Sigma$
method. One explanation is that the former is more sensitive to
higher order QED radiative contributions which are present in the data but are
missing in the MC. If we assume that the difference between the data and the
MC with the $\Sigma$ method is entirely due to the vertex efficiency, then
a correction of $\sim 6\%$ is needed at low $W^2$ values corresponding to low
$y$ values($<0.14$)\footnote{Using Eq.(\ref{eq:x_w2}), the inelasticity $y$ is
related to $W^2$ as $y\simeq W^2/s^{\rm eff}$.}. 
This correction has been applied for both $E$ and 
$\Sigma$ method in the low $y$ region with a systematic error of 10\% for the 
whole $W^2$ region using the $E$ method, and of 10\% and 4\% respectively for 
the low and high $W^2$ values using the $\Sigma$ method.

\begin{figure}[htbp]
\begin{center}
\begin{picture}(50,560)
\put(-205,-25){
\epsfig{file=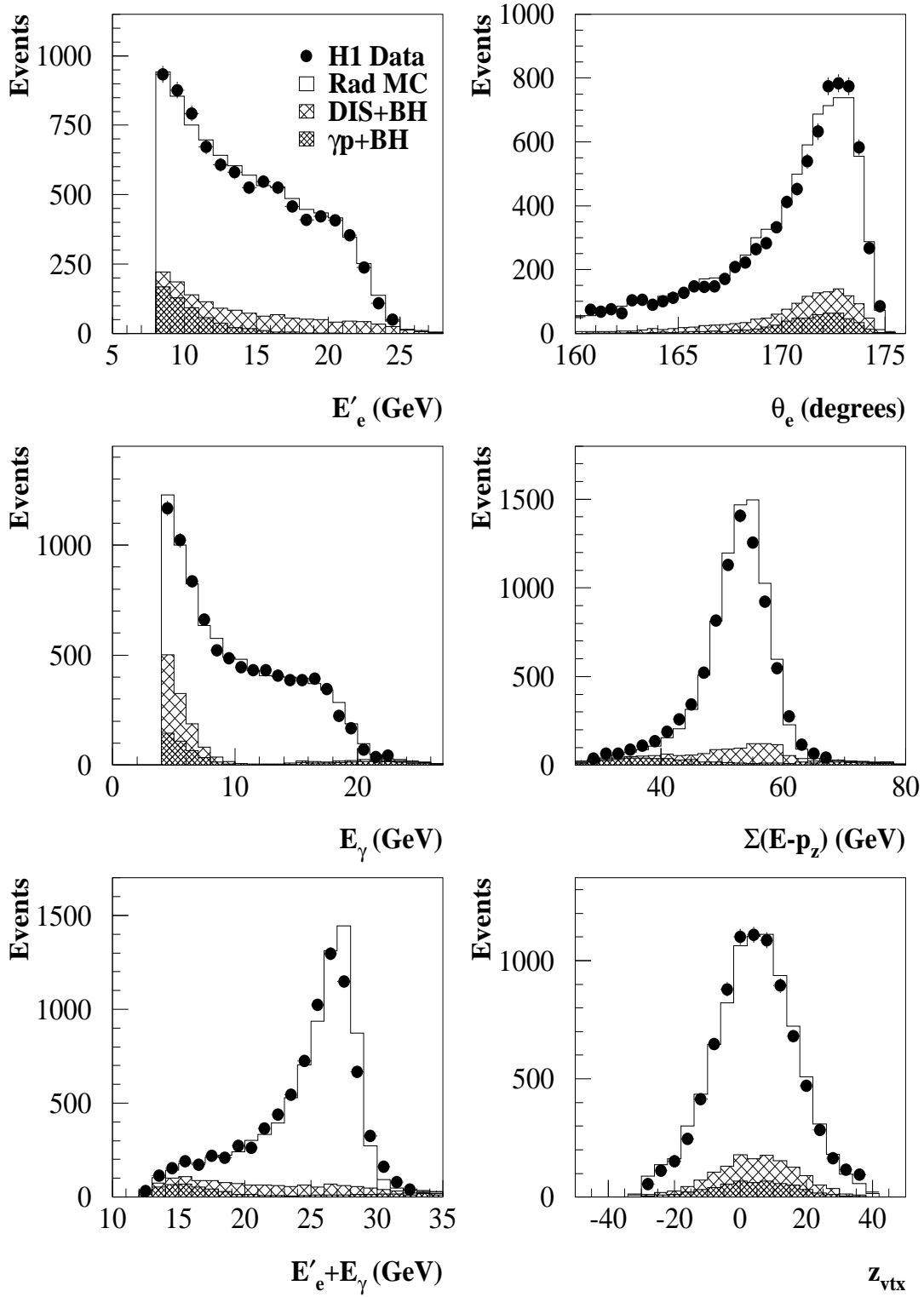,height=220mm,width=165mm}}
\end{picture}
\end{center}
\caption{\label{control}\sl {A comparison between data and MC for 
 various quantities relevant for the kinematics reconstruction.}}
\end{figure}

After having corrected for the trigger and vertex efficiency,
data are compared in Fig.~\ref{control}
with the MC for various relevant quantities used in the reconstruction
of the kinematics. The MC events generated with the GRV parameterization
have been reweighted according to Eq.(\ref{eq:sf_reweight}) by a
next-to-leading-order (NLO) fit\footnote{More precisely,
since the fit is done for $Q^2>5\, {\rm GeV^2}$, so for lower $Q^2$ values,
it is in fact a backward extrapolation of the fit.}
to the measured $F_2$ for $Q^2>5\, {\rm GeV}^2$ based on the H1 non-radiative 
DIS events taken also in 1994 (see Sec.\ref{sec:fit}).
The data distributions are found to be fairly well described by the MC
simulation.

\subsection{Higher order QED correction}\label{sec:horc}

As shown in Eq.(\ref{eq:bornxs}), we measure $F_2(x,Q^2)$ 
using events with collinear hard photon radiation in first order. 
The measured inclusive cross section  
has thus to be corrected for higher order radiative contributions.

\begin{figure}[htb]
\begin{center}
\begin{picture}(50,250)
\put(-215,-280){
\epsfig{file=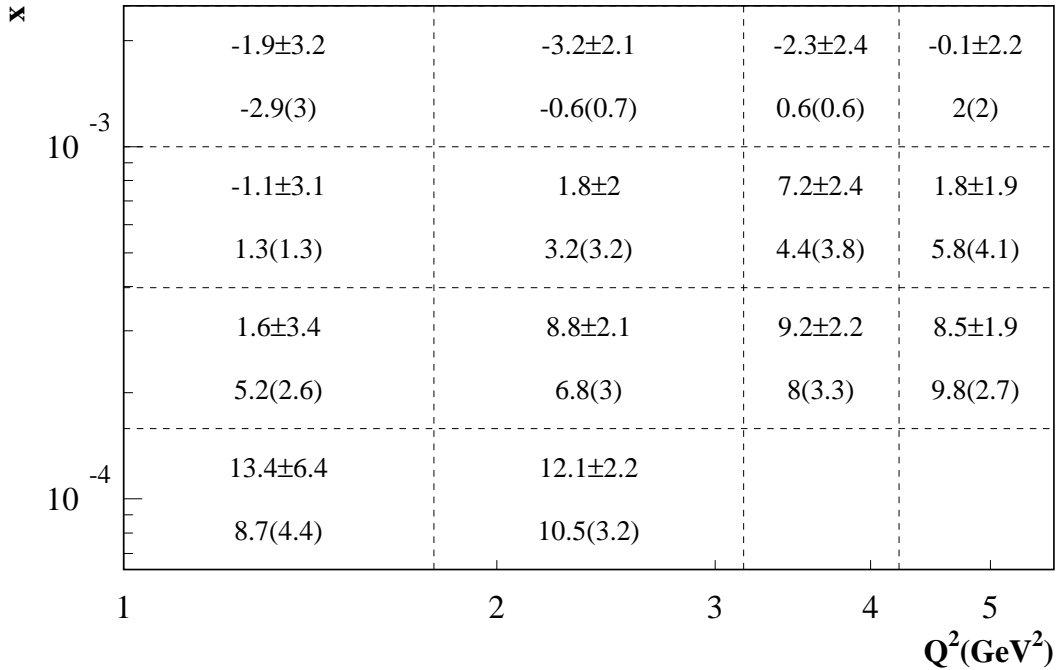,height=200mm,width=165mm}}
\end{picture}
\end{center}
\caption{\label{radcor}\sl {Higher order QED correction $\delta_{\rm ho}$
 (in \%) for the $E$ method estimated with {\sc lesko}~\cite{lesko} (upper 
 number, the error is statistical) and with an independent method based 
 on {\sc hector}~\cite{hector} (lower number). The numbers in brackets 
 are the assigned systematics errors.}}
\end{figure}

The effect of high order contributions, which is dominated by multiphoton
emission collinear to the incident electron, is estimated with the event 
generator {\sc lesko}~\cite{lesko}:
\begin{equation}
1+\delta_{ho}=\frac{\sigma^{YFS}}{\sigma^F}
\end{equation}
where $\sigma^F$ and $\sigma^{YFS}$ are the two options of the {\sc lesko} program.
The first option includes ${\cal O}(\alpha)$ QED radiative corrections, and 
the second one describes multiphoton leptonic radiation in a framework of the 
Yennie-Frautschi-Suura exclusive exponentiation procedure~\cite{YFS}.
The correction ($\delta_{\rm ho}$ in \%) for the $E$ method is shown with the 
upper number in each $(x,Q^2)$ bin (separated with the dashed lines) in 
Fig.~\ref{radcor}. 
The errors are statistical. The correction can be as large as
13\% in the lowest $x$ bins considered and becomes small for the high $x$ 
(or low $y$) region. The correction has been checked with an independent 
method~\cite{sasha} in which the higher order effect was simulated using 
the {\sc hector} package~\cite{hector} with the varying incident electron beam 
energy $E^{\rm eff}_e$ ($=E_e-E_{\gamma}$, $E_{\gamma}$ being the energy
spectrum of the radiated photon in first order). The results\footnote{A cut 
on $W^2>225\, {\rm GeV}^2$ has been applied as higher order contributions at 
lower $W^2$ values are unlikely to contribute according to Fig.~\ref{vtx_w2} 
when a reconstructed vertex is required.} are shown with the lower numbers in 
each bin in Fig.~\ref{radcor}. Reasonable agreement is seen and occasional 
large differences are due either to an intrinsic cut on $p_T>1\, {\rm GeV}$ in 
{\sc lesko}, which introduces some edge effect for the first $Q^2$ bin
at $1.5\, {\rm GeV}^2$, or to a statistical fluctuation in {\sc lesko}, e.g.
as a function of $Q^2$ for the $x$ bins in the second row from the top.
Since both methods give comparable results and the latter one has less 
fluctuation, this latter correction was used with systematic 
uncertainties given with the numbers (in \%) in brackets in Fig.~\ref{radcor}. 
The correction
for the $\Sigma$ method is small ($\lesssim 3\%$) and is rather uniform in the
$(x,Q^2)$ bins considered.

\subsection{$F_2$ results} \label{f2result}

\begin{figure}[htbp]
\begin{center}
\begin{picture}(50,410)
\put(-195,-20){
\epsfig{file=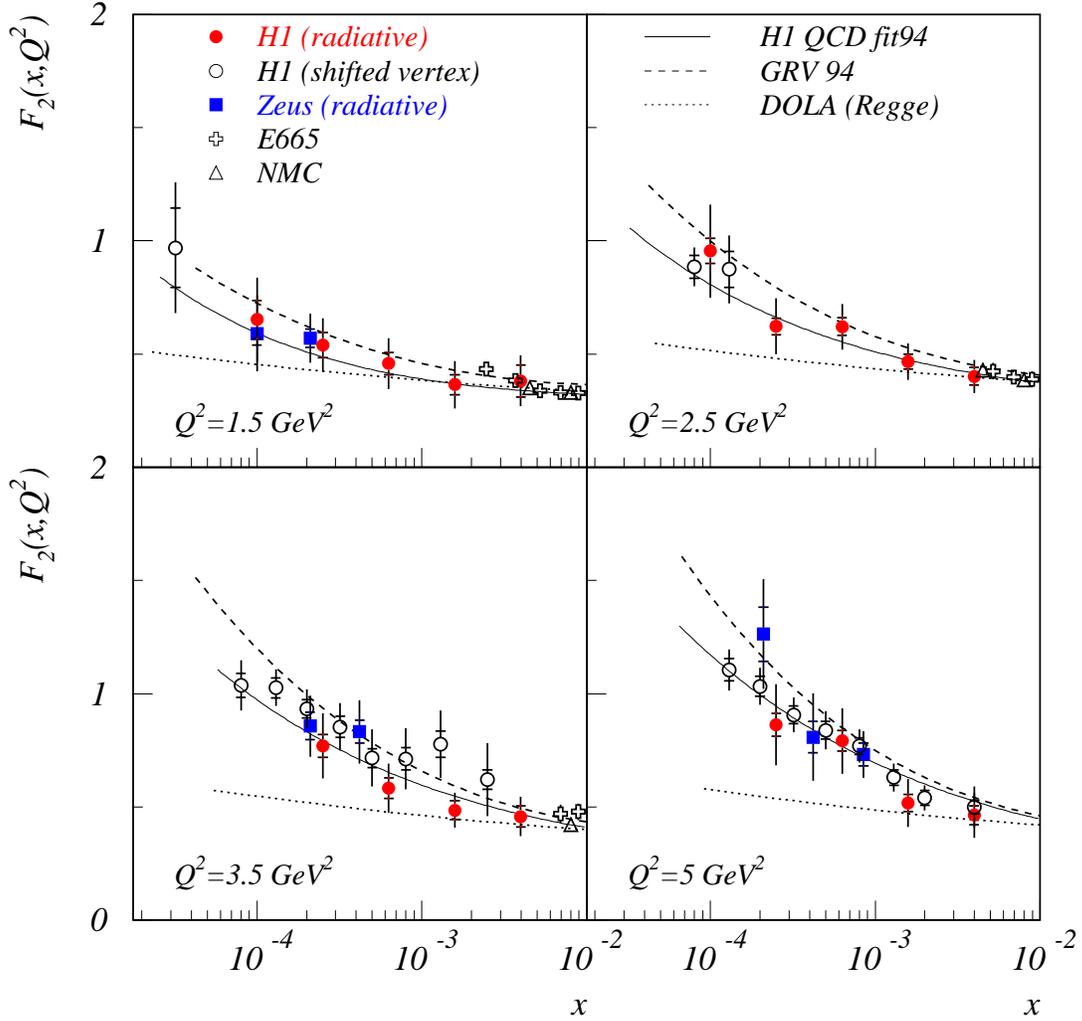,height=155mm,width=155mm}}
\end{picture}
\end{center}
\caption{\label{fig:f2rad94}\sl {A comparison of the measured $F_2(x,Q^2)$
 from this analysis with other measurements from H1 (based on
 non-radiative DIS events with shifted vertex)~\cite{h1f294}, from ZEUS
 (based on radiative DIS events)~\cite{zeusrad94}, from E665~\cite{e665},
 and from NMC~\cite{nmc97};
 also shown is an extrapolation to low $Q^2$ of the H1 QCD fit 
 94~\cite{h1f294} to the data at $Q^2\geq 5\,{\rm GeV}^2$, 
 and two model predictions: 
 GRV~\cite{grv94} and DOLA~\cite{dola}. The inner error bars show 
 the statistical errors only, while the outer ones represent statistical and 
 systematic errors added in quadrature.}}
\end{figure}

The final proton structure function $F_2$ combines
the $E$ method at high $y$ ($\gtrsim 0.15$) with the $\Sigma$ method at the
lower $y$ region such that the measurable kinematic
domain is significantly extended. The results of this
analysis~\cite{h1f294,f2rad94} have been cross checked by an independent
analysis~\cite{hutte_thesis}. The $x$ dependence of $F_2$ at four lowest
$Q^2$ values are shown in
Fig.\ref{fig:f2rad94} and compared with the measured $F_2$ from the
other analyses using non-radiative events with the interaction point shifted
in the proton beam direction by $\sim 70$\,cm (shifted vertex data), 
and the results of a similar analysis from the ZEUS 
collaboration~\cite{zeusrad94}.
The fixed-target experiment data from NMC~\cite{nmc95} and a few
parameterizations are also shown. Several observations can be made:
\begin{itemize}
\item This measurement has allowed to extend the HERA $F_2$ measurement
 for $Q^2$ down to $1.5\, {\rm GeV}^2$. The rise of $F_2$ with decreasing
$x$ observed in the previous HERA measurements for $Q^2\gtrsim 5\,{\rm
GeV}^2$~\cite{h1f292,zeusf292,h1f293,zeusf293} persists down to
$Q^2=1.5\,{\rm GeV}^2$. This extension to lower $Q^2$ values is relative and
remains true even after the angular acceptance of the BEMC is improved by
the new backward calorimeter SPACAL. Indeed, when the same analysis
technique was applied on the 1996 data collected with the SPACAL,
the $Q^2$ range has been further extended down to 0.2\,GeV$^2$~\cite{h1isr96}.
\item This measurement has allowed to fill the gap between the HERA
 measurements at low $x$ and that of fixed-target experiments at high $x$.
 In the common $x$ region, the measurement is in good agreement
 with other measurements with a tendency to be slightly lower than e.g.\
those measured with a data sample (58\,nb$^{-1}$) in which the interaction 
point was shifted to an averaged value of $+67$\,cm with respect to the
nominal interaction point in the proton beam direction (labeled ``shifted 
vertex'' in Fig.\ref{fig:f2rad94}). The shifted vertex data sample 
has a luminosity uncertainty of 3.5\% which is independent of the uncertainty 
of 1.5\% of this analysis.
\item All measurements are compatible with the GRV~\cite{grv94} calculation 
 except in the very low $x$ region where the measurements seem to be 
 (systematically) below the prediction.  
\item The parameterization DOLA~\cite{dola}, which is motivated by Regge theory
 and relates the structure function to Reggeon exchange phenomena, are seen
 to be ruled out by the measurement.
\item The measurement agrees well with an extrapolation to low $Q^2$ of a
 next-to-leading-order QCD fit~\cite{h1f294}, based on the H1
 1994 data and fixed target data at $Q^2\geq 5\,{\rm GeV}^2$.
\end{itemize}

\newpage
\section{Precision measurement at HERA} \label{sec:h1f29697}
With an increased data sample taken in 1996-1997, corresponding to
an integrated luminosity of about 20\,pb$^{-1}$, and the new backward 
apparatus, the H1 collaboration has recently published a precision measurement
on the inclusive cross section and structure functions covering
$1.5\leq Q^2\leq 150\,{\rm GeV}^2$
and $3\cdot 10^{-5}\leq x\leq 0.2$~\cite{h1f29697}.
The statistical accuracy of the measured cross section is better than 1\%
for a large part of the data. The systematic precision has reached 3\% apart
from the boundary of the covered region.

Contrary to the previous HERA measurements, the new measurements are presented 
in so-called reduced cross sections, which are related to the double 
differential cross sections defined in Eq.(\ref{eq:xsnc}):\footnote{In the 
considered kinematic region, the $\gamma Z^0$ interference and $Z^0$ exchange 
contributions can be safely neglected.}
\begin{equation}
\sigma_{\rm r}=\tilde{\sigma}=\frac{Q^4x}{2\pi\alpha^2Y_+}\frac{d^2\sigma}{dxdQ^2}
=F_2(x,Q^2)-\frac{y^2}{Y_+}F_L(x,Q^2)
\label{eq:redxs}
\end{equation}
with $Y_+=1+(1-y)^2$ being the helicity function. Thus the dominant 
$Q^2$ dependence in the double differential cross section due to the
propagator of one photon exchange is explicitly
suppressed in the reduced cross sections so that in most of the kinematic region
the relation $\sigma_{\rm r}=F_2$ holds to good approximation.
The only exception occurs at high $y$, 
where $\sigma_{\rm r}$  can be substantially different from
$F_2$ due to the sizable contribution from the longitudinal structure function
$F_L$. 

\begin{figure}[htbp]
\begin{center}
\begin{picture}(50,450)
\put(-180,-52.5){
\epsfig{file=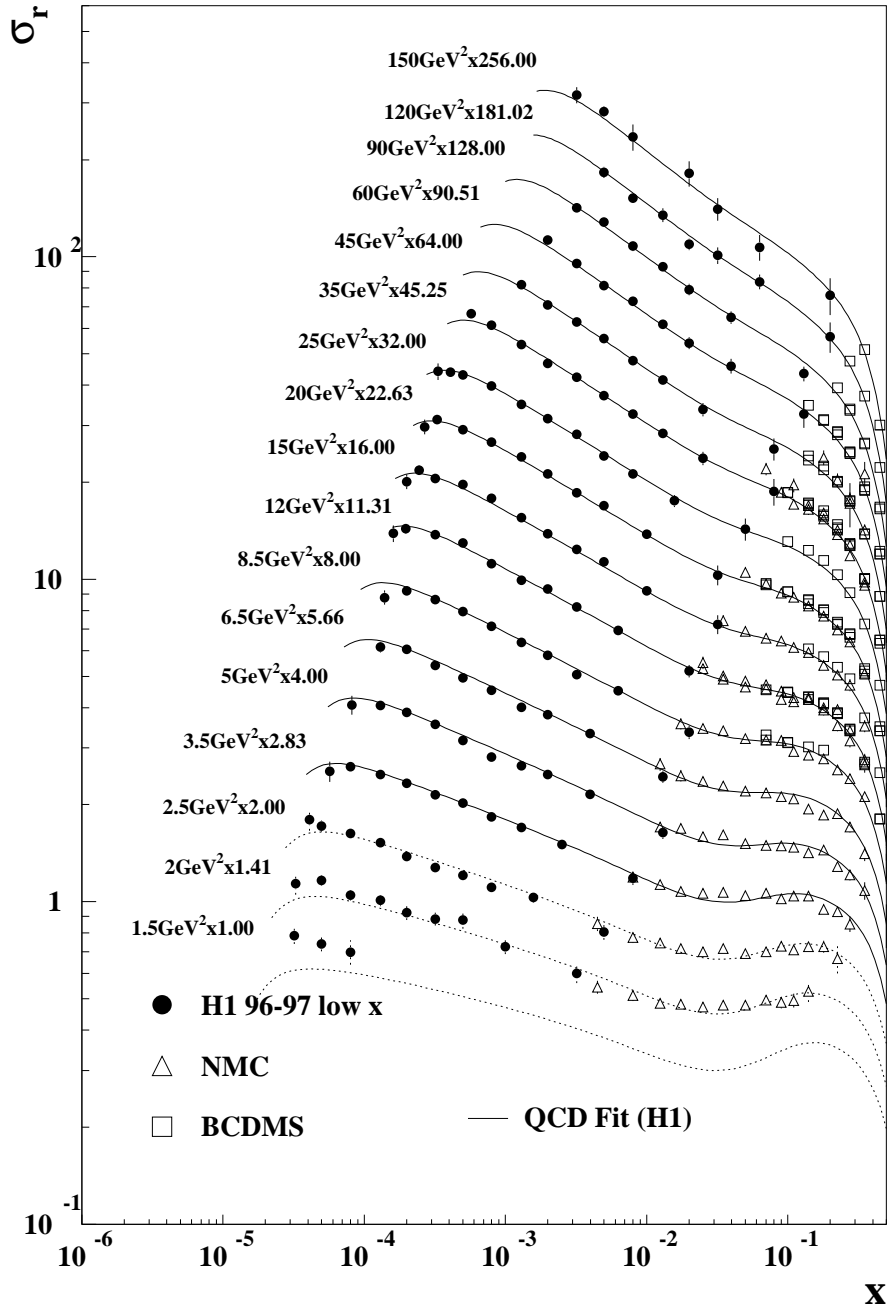,bbllx=0pt,bblly=0pt,bburx=594pt,
bbury=842pt,width=145mm}}
\end{picture}
\end{center}
\caption{\label{fig:h1f29697}\sl {The precision measurement of the
inclusive reduced cross sections from H1~\cite{h1f29697} in comparison with
the measurements from NMC~\cite{nmc95} and BCDMS~\cite{bcdms}. The full curves
represent the corresponding theoretical expectation based on a NLO QCD fit
to the H1 low $Q^2$ data and the published high $Q^2$ data taken in
1994-1997~\cite{h1hiq9497} for $3.5 \leq Q^2 \leq 3000\,{\rm GeV}^2$. 
The dashed curves show the extrapolation of the fit towards lower $Q^2$.}}
\end{figure}

The reduced cross sections are shown in Fig.\ref{fig:h1f29697} 
together with the $\mu p$ data by the NMC~\cite{nmc95} and \
BCDMS~\cite{bcdms} experiments, and a new H1 NLO QCD fit to the new low 
$Q^2$ data and the published $e^+p$ high $Q^2$ data taken in
1994-1997~\cite{h1hiq9497} (Sec.\ref{sec:hiq2_xs}), for 
$3.5\leq Q^2\leq 3000\,{\rm GeV}^2$.
The cross section rises at low $x$. This rise
is observed to be damped at the smallest values of $x$ which is attributed
to $F_L$, see Sec.\ref{sec:fl}.

The data at $Q^2<3.5\,{\rm GeV}^2$ seem to overshoot the backward
extrapolation of the QCD fit. A fit with $Q^2_{\rm min}=1.5\,{\rm GeV}^2$,
however, describes the low $x$ data well. Theoretically one expects higher
order logarithmic and power corrections to be larger for $Q^2\simeq 1\,{\rm
GeV}^2$ such that a NLO DGLAP treatment may be inadequate.
Further exploration of this interesting effect requires low $x$, high
precision data at low $Q^2\simeq 1\,{\rm GeV}^2$.

\begin{figure}[htbp]
\begin{center}
\begin{picture}(50,470)
\put(-295,-185){
\epsfig{file=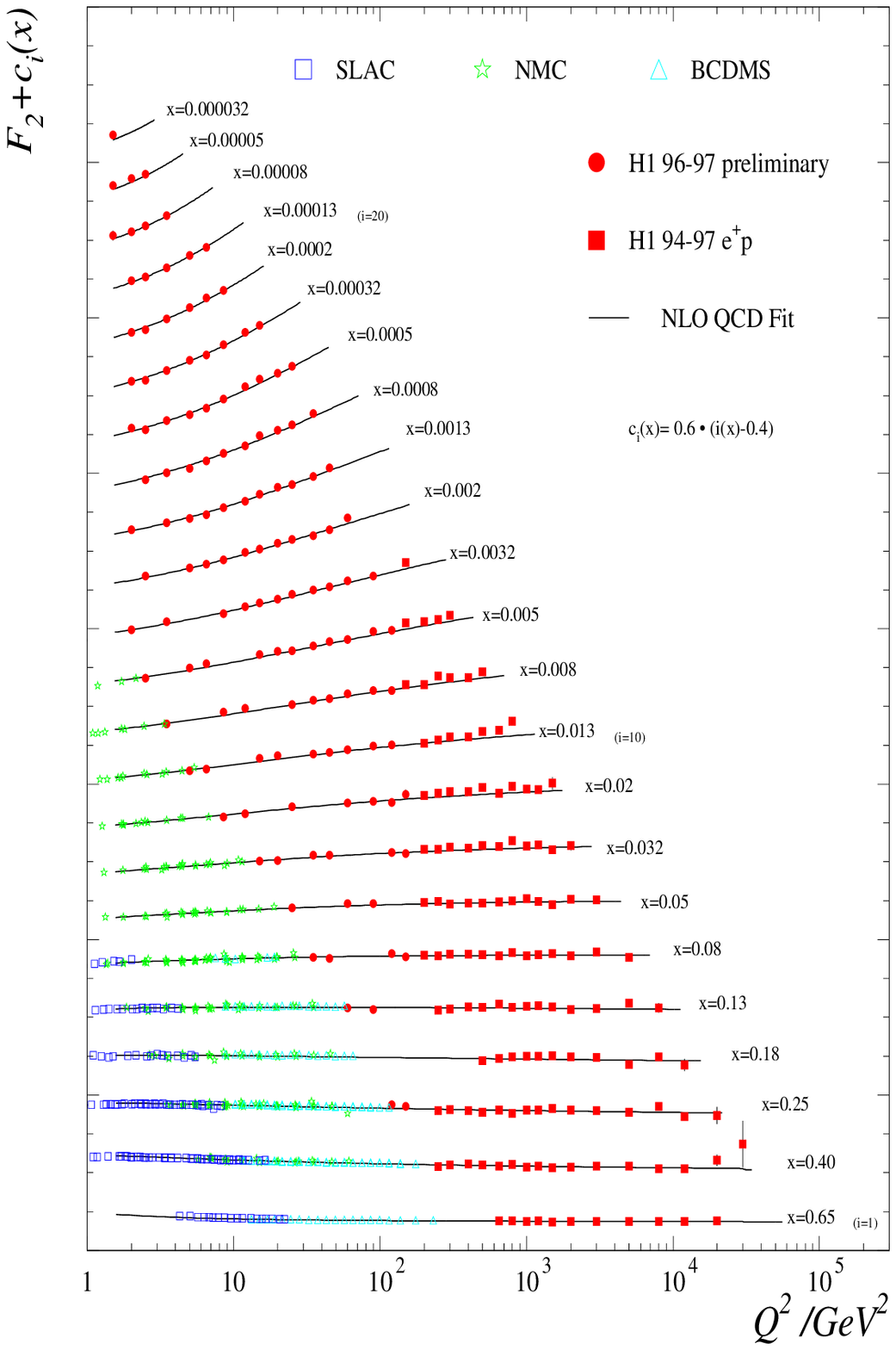,bbllx=0pt,bblly=0pt,bburx=594pt,
bbury=842pt,width=215mm}}
\end{picture}
\end{center}
\caption{\label{fig:f2_all}\sl {A compilation of the proton structure
function $F_2(x,Q^2)$ from the 1996-1997 H1 $e^+p$ data at low 
$Q^2 (\leq 150\,{\rm GeV}^2)$ and low $x$~\cite{h1f29697}, from the 1994-1997
H1 $e^+p$ data at high $Q^2 (\geq 200\,{\rm GeV}^2)$~\cite{h1hiq9497}
(Sec.\ref{sec:hiq2_xs}), and from the fixed-target data 
by SLAC~\cite{slac92},
NMC~\cite{nmc95}, and BCDMS~\cite{bcdms}.
The $F_2$ values are plotted in linear scale adding a constant 
$c(x)=0.6(i-0.4)$ where $i$ is the $x$ bin number starting at $i=1$ from 
$x=0.65$. 
The full lines correspond to the NLO QCD fit~\cite{h1f29697}.}}
\end{figure}

The proton structure function $F_2(x,Q^2)$ can be extracted from the
reduced cross sections by applying a (small) correction for the $F_L$
contribution in Eq.(\ref{eq:redxs}). The resulting $F_2$ is shown as a function
of $Q^2$ in Fig.\ref{fig:f2_all}.
Also shown are the structure functions obtained from H1 at high $Q^2 
(\geq 200\,{\rm GeV}^2)$ and from the fixed-target data by SLAC~\cite{slac92},
NMC~\cite{nmc95}, and BCDMS~\cite{bcdms} at low $Q^2$ and high $x$.
To extract the $F_2(x,Q^2)$ from the inclusive cross sections measured with
the 1994-1997 H1 $e^+p$ data at high $Q^2$~\cite{h1hiq9497} 
(Sec.\ref{sec:hiq2_xs}), the increasingly important contributions from 
the $\gamma Z^0$ interference and $Z^0$ exchange have been subtracted.

The H1 NLO QCD fit, based on the H1 data alone, provides a good overall
description to all data with a few exceptions at kinematic boundaries.
The strong scaling violation is clearly
displayed. It is this data feature and the precision of the data which
allows the strong coupling constant $\alpha_s$ and various parton density
distributions in particular the gluon distribution to be determined
within the QCD framework(Secs.\ref{sec:gluon}, and \ref{sec:alphas}).

\newpage
\section{Current knowledge of the gluon density} \label{sec:gluon}
\subsection{Impact of the HERA $F_2$ data on the determination of the gluon 
density at low $x$} \label{sec:fit}
About half of the proton's momentum is carried by gluons. Despite this, the
determination of the density of gluons $g(x,Q^2)$ in the proton has turned
out to be a difficult task.
Before the advent of the HERA experiments, our knowledge on the gluon density
distribution was very imprecise and limited to the region at
$x>0.01$~\cite{bcdmsgluon89,nmc93}.
The principle difficulty in measuring the gluon density in deep inelastic
scattering is that the gluons contribute only in high order processes
through gluon bremsstrahlung from quarks and quark pair creation from gluons 
(Fig.\ref{fig:splitf}). 
At small $x (<10^{-2})$ the latter process dominates the scaling
violation~\cite{cooper88}.
The HERA data on the proton structure function $F_2(x,Q^2)$ at low values of
$x$ down to $10^{-4}$ can thus be exploited to extract the gluon density in
the kinematic range.

Indeed, already with the first $F_2$ measurement in 1992, an
extraction has been performed by H1~\cite{h1gluon92} and
ZEUS~\cite{zeusgluon92} using the following approximation relation under the
assumption that the quark contribution at low $x$ is negligible~\cite{prytz93}
\begin{equation}
xg(x,Q^2)\thickapprox\frac{27\pi}{20\alpha_s(Q^2)}\frac{dF_2(x/2,Q^2)}{d\ln
Q^2}\,.
\end{equation}
It was derived that the gluon density rises strongly toward low $x$.

\begin{figure}[htb]
\begin{center}
\begin{picture}(50,210)
\put(-110,-35){
\epsfig{file=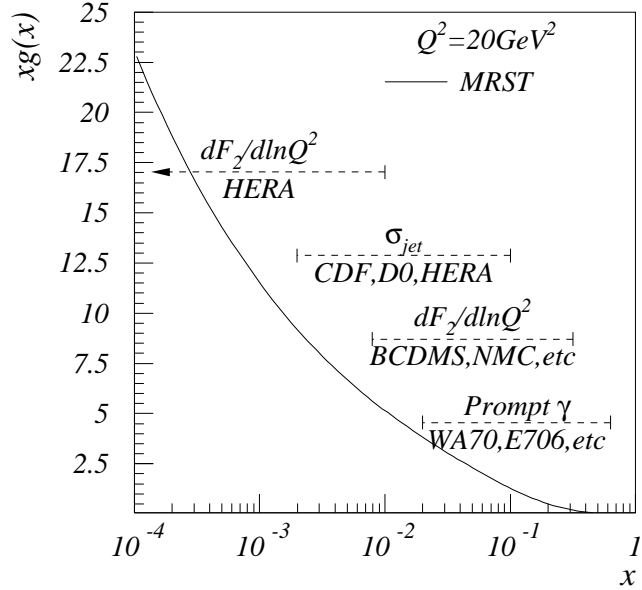,width=100mm}}
\end{picture}
\end{center}
\caption{\label{fig:gluon_constraints}\small\sl {Various experimental
measurements used by global QCD analyses to constrain the gluon density $xg$
at different $x$ range. The resulting $xg$ from the MRST group~\cite{mrst} 
at $Q^2=20\,{\rm GeV}^2$ is shown with the curve.
The ranges in $x$ for the
various measurements are only indicative since they are $Q^2$ dependent. For
instance, the HERA data reach to much large $x$ values than indicated at
higher $Q^2$ though the statistical precision of the data there is still 
limited.}}
\end{figure}

With the improved $F_2$ measurement in 1993, both H1 and ZEUS have performed
for the first time NLO QCD fits to $F_2$ in a similar manner as in the global
analyses by the MRS and CTEQ groups (Sec.\ref{sec:qcdfit}).
The emphasis is however different. While in a global
analysis, universal parton density functions are extracted, the analyses by
the experimental collaborations emphasize the determination of the gluon
density at low $x$. The result of the fit to $F_2$ is also used to have
a better parameterization of the cross section or the structure function than
the initial one used in the Monte Carlo generation.
For this reason, only a minimum number of data sets have been used
in the analyses; in
addition to the HERA $F_2$ data at low $x$, a few data sets at high $x$
from fixed-target experiments (e.g.\ BCDMS and NMC) were used.
This is in contrast with the global analyses where in addition to the
structure function data, other constraints from inclusive jet (and/or dijet) 
cross sections and the prompt photon data have also been used
as illustrated in Fig.\ref{fig:gluon_constraints}.

\begin{figure}[htbp]
\begin{center}
\begin{picture}(50,400)
\put(-187,-80){
\epsfig{file=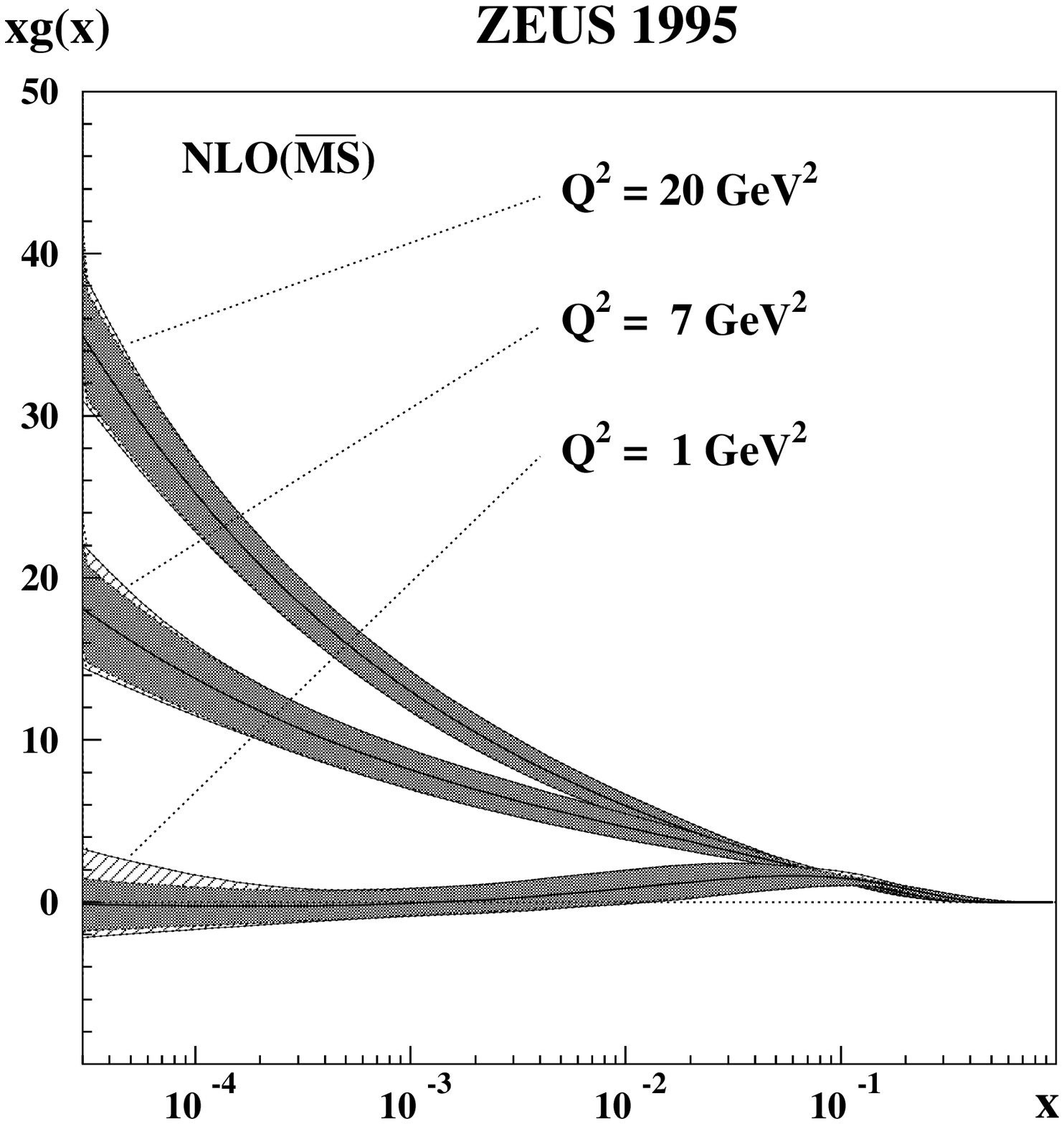,bbllx=0pt,bblly=0pt,bburx=594pt,
bbury=842pt,width=82mm}}
\put(-175,183){\makebox(0,0)[l]{\fcolorbox{white}{white}{
\textcolor{white}{aaaaa}}}}
\put(-107,183){\makebox(0,0)[l]{\fcolorbox{white}{white}{
\textcolor{white}{ZEUS 1995aaa}}}}
\put(-71,90){\makebox(0,0)[l]{\fcolorbox{white}{white}{\footnotesize ZEUS 1995}}}
\put(-145,9){\makebox(0,0)[l]{\small (c)}}
\put(18,-80){
\epsfig{file=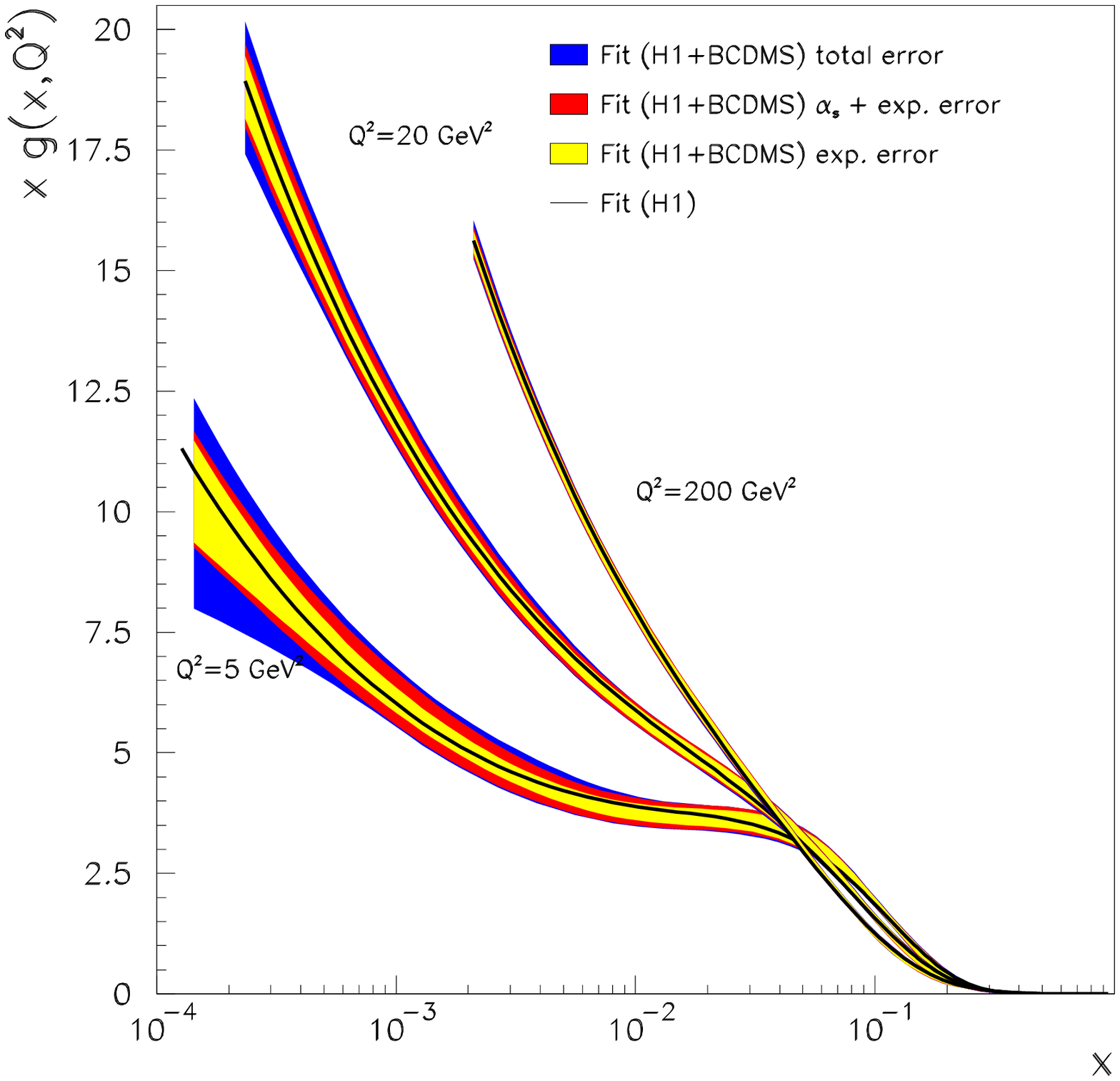,bbllx=0pt,bblly=0pt,bburx=594pt,
bbury=842pt,width=82mm}}
\put(65,9){\makebox(0,0)[l]{\small (d)}}
\put(135,120){\makebox(0,0)[l]{\footnotesize H1 1996-97}}
\put(-185,130){
\epsfig{file=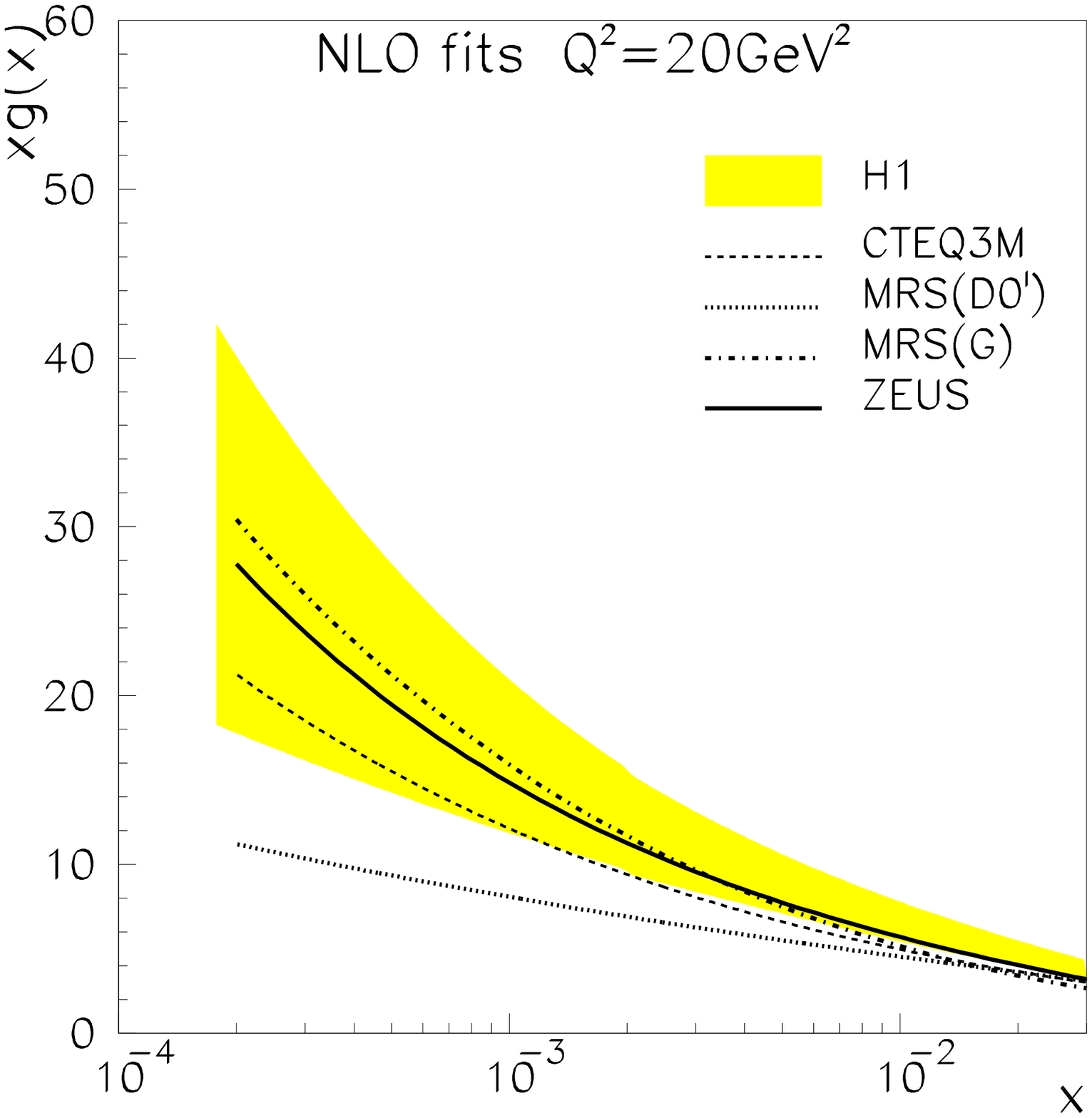,bbllx=0pt,bblly=0pt,bburx=594pt,
bbury=842pt,width=77mm}}
\put(-6,355.5){\makebox(0,0)[l]{\scriptsize 1993}}
\put(-145,211){\makebox(0,0)[l]{\small (a)}}
\put(25,130){
\epsfig{file=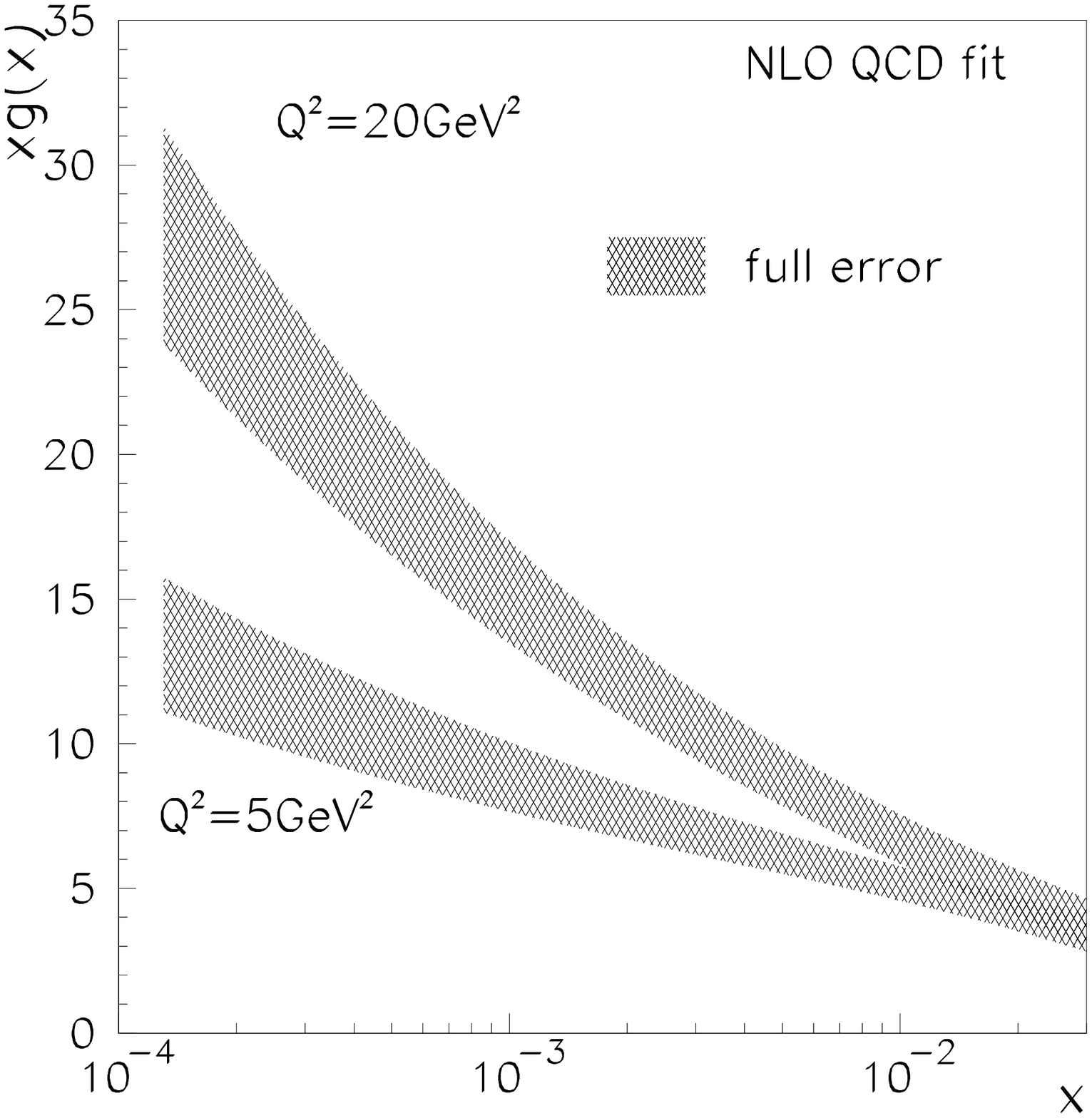,bbllx=0pt,bblly=0pt,bburx=594pt,
bbury=842pt,width=77mm}}
\put(165,339.5){\makebox(0,0)[l]{\fcolorbox{white}{white}{\footnotesize H1
1994}}}
\put(65,211){\makebox(0,0)[l]{\small (b)}}
\end{picture}
\end{center}
\caption{\label{fig:gluon}\small\sl {The time evolution of the
determination of the gluon density $xg(x)$ at HERA (a) the H1 result in
1993~\cite{h1gluon93} compared with the ZEUS result~\cite{zeusgluon93} and
parameterizations CTEQ\,3M~\cite{cteq3}, MRS(D0$^\prime$) and MRS(G),
(b) the H1 result in 1994~\cite{h1f294} and (c) the ZEUS result based on
their 1994 and 1995 data~\cite{zeusgluon95}, and (d) the most recent one from
H1 using data taken in 1996-1997.}}
\end{figure}

The gluon density for $Q^2=20\,{\rm GeV}^2$ obtained by H1~\cite{h1gluon93}
with the 1993 data is shown in Fig.\ref{fig:gluon}(a) in comparison 
with the result of ZEUS~\cite{zeusgluon93} and other determinations 
from the global analyses.
The early parameterization MRS\,D0$^\prime$, which did not use HERA data,
is clearly disfavored by the data as we have seen in Fig.\ref{fig:f2p_hera92}.
The other parameterizations which used the HERA $F_2$ data are in good
agreement with the results of H1 and ZEUS.
The error band represents the experimental errors with the statistical and
systematic errors added in quadrature. The experimental collaborations have
the advantage of knowing the correlations between their systematic errors so
that they could be properly treated~\cite{pz95}.
As the integrated luminosity increases, the accuracy of the gluon density
is further reduced by H1 (Fig.\ref{fig:gluon}(b)) using their 1994 $F_2$
data~\cite{h1f294} and by ZEUS (Fig.\ref{fig:gluon}(c)) using both 1994 and
1995 data~\cite{zeusgluon95}. The most precise determination from H1 is
shown in Fig.\ref{fig:gluon}(d). It is a result of a NLO DGLAP QCD fit to the
recent H1 $ep$ cross section data (sec.\ref{sec:h1f29697}) and
the BCDMS $\mu p$ data. An experimental precision (the inner error band) 
of 3\% at $x\simeq 10^{-3}$ and $Q^2=20\,{\rm GeV}^2$ has been reached 
for the first time. The gluon distribution has been obtained together with 
the strong coupling constant $\alpha_s$ (Sec.\ref{sec:alphas}). 
The effect of the uncertainty of the
latter is shown with the middle error band. The outer error bands represent
the uncertainties related to the QCD model and data range. For the first time,
the gluon distribution has been determined with the H1 inclusive data only. 
Both determinations are in good agreement.

\subsection{Uncertainty and future improvements} \label{ergluon}
For a wide range of theoretical and experimental applications, it is
important to know the range of uncertainties of the parton density functions.
However the task of deriving a reliable uncertainty is extremely complex and
difficult. The difficulty is related to many sources of uncertainty to be
considered, to name a few:
\begin{itemize}
\item The choice of experimental measurements and data sets. As an example,
the gluon density $xg(x)$ at $Q^2=25\,{\rm GeV}^2$ from the most recent
parameterizations MRST~\cite{mrst} and CTEQ5~\cite{cteq5} is compared in
Fig.\ref{fig:er_gluon_cteq}. 
\begin{figure}[htb]
\begin{center}
\begin{picture}(50,230)
\put(-145,-25){
\epsfig{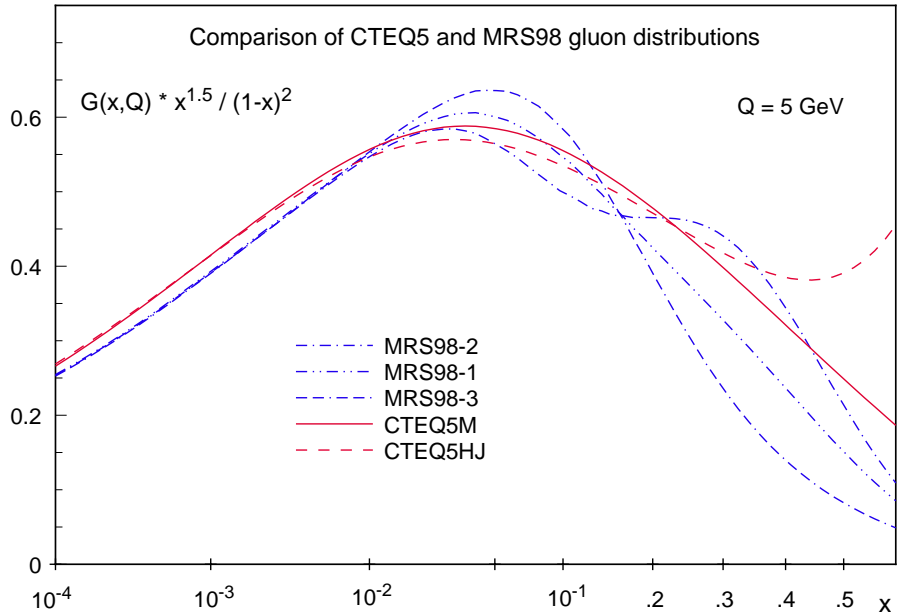}}
\end{picture}
\end{center}
\begin{flushright}
\begin{minipage}{13.5cm}
\caption{\label{fig:er_gluon_cteq}\small\sl {Comparison of the gluon density
functions from MRST~\cite{mrst} (the parameterizations MRS98-1,2,3
correspond respectively to the standard, high-gluon and low-gluon options)
with those from CTEQ5~\cite{cteq5} (the parameterizations CTEQ5M and CTEQ5HJ are
respectively the standard and large-gluon options, the latter one
being tailored to accommodate the Tevatron inclusive jet data, the high
$p_t$ tail of which was in excess with respect to the prediction based on
previous parton density parameterizations). The figure is from
Fig.19 of \cite{cteq5}.}}
\end{minipage}
\end{flushright}
\end{figure}
Both groups have used the HERA $F_2$ measurements
based on the data recorded in 1995 and before, which were the only available
constraint at low $x$ (see Fig.\ref{fig:gluon_constraints}). Consequently,
there is little difference at low $x$. On the other hand, the difference
is much larger at medium and large $x$,\footnote{The difficulty stems also from
the fact that the gluon density becomes very small towards large
$x$ (Fig.\ref{fig:gluon})} where in addition to the DIS
structure function data from fixed-target experiments, the prompt photon data 
from WA70~\cite{wa70} and E706~\cite{e706} were 
used in the analysis of MRST, while the inclusive jet data from 
CDF~\cite{cdfjet} and D0~\cite{d0jet} were used instead in the analysis 
of CTEQ5.
\item The experimental errors. The most precise structure function data are
dominated by the systematic errors. The non-trivial part of these are the
correlated systematic errors, which are not always available for the global
analyses.
\item The technical uncertainties due to the freedom in choosing
the initial scale $Q^2_0$ and the functional form at $Q^2_0$ with varying
number of free parameters, and the internal correlation
among different parton density functions and between different $x$ range.
\item The theoretical uncertainties due to effects such as the higher-order 
corrections, scale- and scheme-dependence, soft-gluon resummation, 
higher-twist effects, and nuclear (deuteron) corrections.
\item The evolution uncertainty related to the values of the strong coupling 
constant $\alpha_s$ and the strategies applied. The determination of 
$xg(x,Q^2)$ is strongly coupled with $\alpha_s$. One strategy is to determine 
$\alpha_s$, gluon and quark density functions together, an alternative is to 
take $\alpha_s$ and its uncertainty from other independent measurements.
\end{itemize}
These uncertainties are often internally correlated. However one thing is
clear, namely in order to reduce the uncertainties, both experimental and
theoretical efforts are needed. On the experimental side, more precise data
will directly reduce the experimental uncertainty. The precise data in
an extended kinematical range also verify whether the chosen functional form
is adequate within the conventional theoretical framework and test the validity
of the latter. On the theoretical side, one of the most urgent tasks is to
have a better understanding of the comparison between the measured prompt
photon data (E706) with the QCD prediction so that the uncertainty of
$xg(x,Q^2)$ at large $x$ can be reduced.

More independent (direct) measurements are highly desirable to either
provide more constraints or check whether new physics phenomena have been
artificially absorbed into the chosen functional form. As far as the gluon
density is concerned, several measurements could be used either as testing
ground or as a source of new constraints.

\begin{figure}[htb]
\begin{center}
\begin{picture}(50,290)
\put(-135,-80){
\epsfig{file=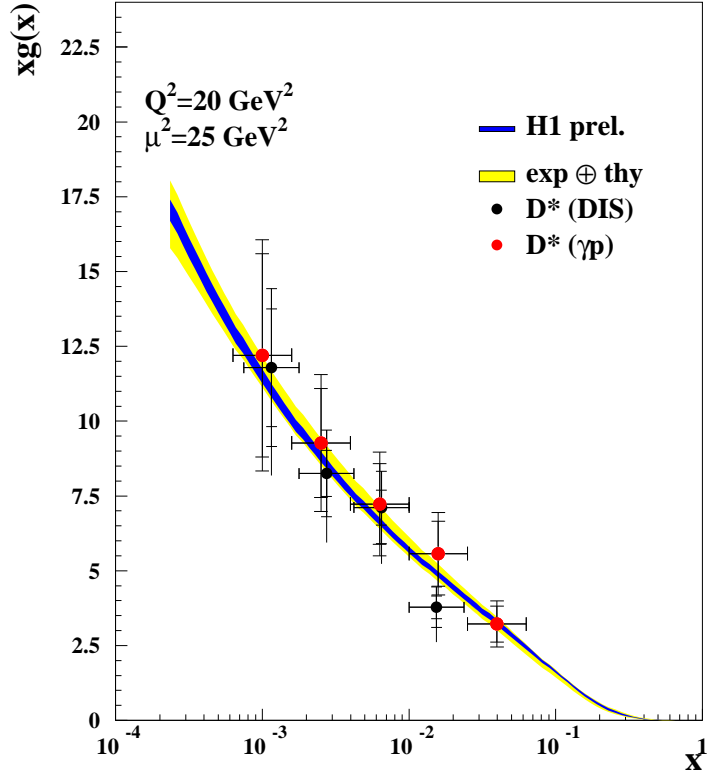,bbllx=0pt,bblly=0pt,bburx=594pt,
bbury=842pt,width=110mm}}
\end{picture}
\end{center}
\caption{\label{fig:charm_gluon}\small\sl {The gluon density at
$\mu^2=25\,{\rm GeV}^2$ determined from the open charm data collected by H1
in 1994-1996~\cite{h1charm_gluon} compared with the preliminary gluon
density at $Q^2=20\,{\rm GeV}^2$ based on the indirect scaling violation
data 1996-1997.}}
\end{figure}

One such measurement is the open charm production at HERA. 
The production of the charm quark at HERA proceeds in perturbative QCD 
almost exclusively via photon-gluon fusion, where the exchanged photon 
interacts with a gluon in the proton by forming a quark-antiquark pair 
($\gamma g\rightarrow c\overline{c}$, Fig.\ref{fig:bgf}). 
This holds both for DIS and for photoproduction where the exchanged photon 
is quasi real. Based on the data collected from 1994 to 1996, the H1
measurement~\cite{h1charm_gluon} is shown in Fig.\ref{fig:charm_gluon}.
The determination of the gluon density is in good agreement with the other 
determinations described above. The errors were however still too large 
to be competitive with the other determinations. 
Both high luminosity at HERA and
improved experimental setups (e.g.\ use of silicon vertex detectors) will
allow considerable improvements in the future.

Inelastic $J/\psi$ photoproduction at HERA has been suggested as a
measurement which could allow one to measure the gluon density. However it
appears that the perturbative calculation does not behave well in the limit
$p_T (J/\psi)\rightarrow 0$, and if the small $p_T$ region is excluded from
the analysis the predictions are not very sensitive to the small $x$
behavior of the gluon. Elastic (diffractive) $J/\psi$ production in DIS and
in photoproduction is more promising, since the cross section depends on
$(xg(x,\overline{Q}^2))^2$, where the scale of the process
is given by $\overline{Q}^2=(Q^2+M^2_{J/\psi})/4$ with $Q^2$ and
$M_{J/\psi}$ being respectively the virtuality of the photon and the rest
mass of the $J/\psi$~\cite{ryskin97}.
These data could give information on the gluon
distribution in the region $10^{-4}<x<10^{-2}$. At the present time the
theoretical framework for extracting the gluon distribution from these
measurements is still under development and the experimental precision of
the data is still fairly low.

From Eq.(\ref{eq:fl_qcd}), one sees that the longitudinal structure function
$F_L$ receives a direct contribution from the gluon, which dominates at low $x$.
Under certain conditions, a measurement of $F_L$ is almost a direct
determination of the gluon. Therefore it is important to measure the
longitudinal structure function at HERA. This will be the subject of the
next section.

\newpage
\section{Longitudinal structure function $F_L(x,Q^2)$} \label{sec:fl}
The longitudinal structure function is an important quantity to measure. The
earlier results on the smallness of $R$ supported the assignment of
half-integer spin for the quarks. A precise measurement allows an
independent check of the gluon density derived from the indirect scaling
violation of the structure functions.
The knowledge of $F_L$ is also needed to extract in a model independent way
the structure function $F_2$ from the measured cross section. 

\subsection{Current knowledge of $F_L(x,Q^2)$} \label{sec:fl_status}
Using Eqs.(\ref{eq:xs_tot})-(\ref{eq:r}) and neglecting the target mass
term, the differential cross section for one-photon exchange given in
Eq.(\ref{eq:xsnc_lowq}) can be rewritten in terms of $R$ and $\sigma_T$
as\footnote{In writing Eq.(\ref{eq:xs_r}), we have used the Hard
convention~\cite{hand63} $K=\nu=Q^2/(2xM)$ for the flux of the virtual photon
in analogy to the real photon.}
\begin{equation}
\frac{d^2\sigma(x,Q^2)}{dxdQ^2}=\frac{\alpha}{2\pi
xQ^2}Y_+\left[1+\varepsilon R(x,Q^2)\right]\sigma_T(x,Q^2)\,,
\label{eq:xs_r}
\end{equation}
where $Y_+=1+(1-y)^2$, $R=\sigma_L/\sigma_R$, and 
\begin{equation}
\varepsilon=\frac{2(1-y)}{Y_+}\,, \label{eq:epsilon_y}
\end{equation}
is the polarization of the virtual photon exchanged in the process.
Therefore for measuring $R$ at a given $(x,Q^2)$ point, it is necessary to
vary $\varepsilon$, i.e.\ $y$ or the center-of-mass energy squared $s$ 
because $y$ is related to $s$ as
\begin{equation}
y=\frac{Q^2}{xs}\,. \label{eq:y_s}
\end{equation}
The difficulty in
measuring $R$ or the longitudinal structure function $F_L$ is then directly
related to a good control of the relative normalization and systematic
errors at two different energies. This explains why measurements of
$F_L(x,Q^2)$ are so delicate, and why only a few fixed-target results
have been published.

The most extensive results from early fixed-target experiments were obtained
by BCDMS~\cite{bcdms}, and SLAC who has reanalyzed the data~\cite{r_slac},
covering respectively the kinematic range $0.07\leq x\leq 0.65$, $15\leq
Q^2\leq 50\,{\rm GeV}^2$ by BCDMS and $0.01\leq x\leq 0.9$,
$0.6\leq Q^2\leq 20\,{\rm GeV}^2$ by SLAC.
The data lie in a region where non-perturbative effects are likely to be
important. Most recently $R$ has been measured by a new SLAC experiment
E140X~\cite{newr_slac}, by CCFR~\cite{r_ccfr} and NMC~\cite{nmc97}, the
latter reached a lowest $x$ value at 0.002, which is still one or two orders 
of magnitude larger than the relevant kinematic region at HERA.

\begin{figure}[htbp]
\begin{center}
\begin{picture}(50,295)
\put(-150,-115){
\epsfig{file=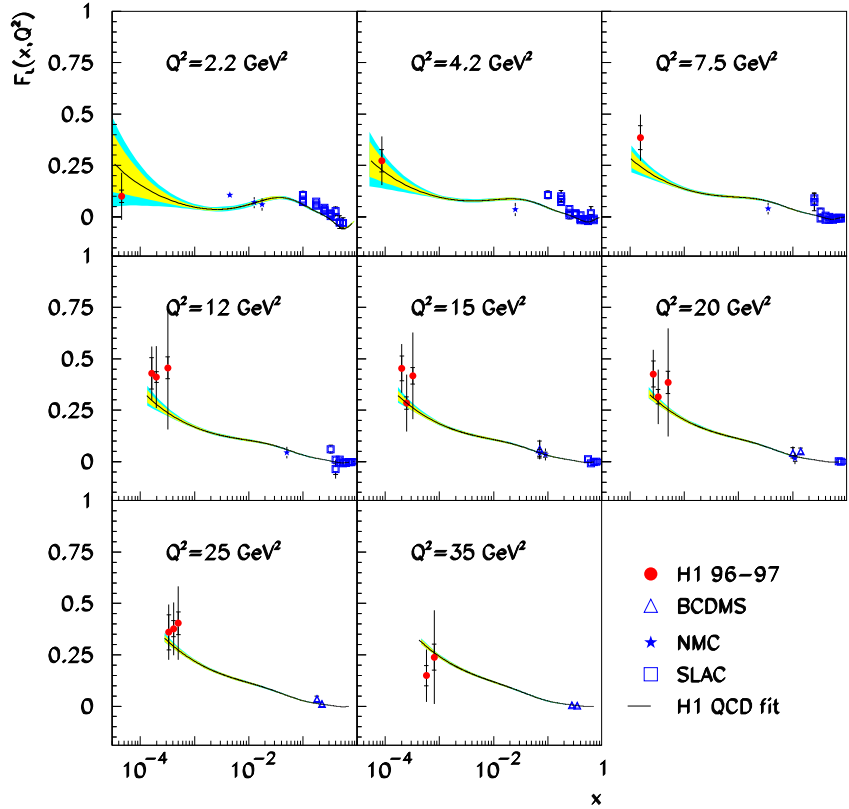,bbllx=0pt,bblly=0pt,bburx=594pt,
bbury=842pt,width=130mm}}
\end{picture}
\end{center}
\caption{\label{fig:fl_com}\small\sl {The extracted longitudinal structure
function $F_L(x,Q^2)$ by H1~\cite{h1f29697} in comparison with the direct
measurements by charged lepton-nucleon fixed-target experiments
BCDMS~\cite{bcdms}, NMC~\cite{nmc97}, and SLAC~\cite{r_slac}. The inner
error bars denote the statistical error. The outer error bars represent the
systematic and model errors. The curves and error bands show the NLO QCD fit
to the recent H1 low $Q^2$ data~\cite{h1f29697} and the high $Q^2$ from
1994-1997~\cite{h1hiq9497} for $y<0.35$ and $Q^2\geq 3.5\,{\rm GeV}^2$.}}
\end{figure}

At HERA, as the integrated luminosity increases, the latest cross section data
become precise enough to see the sensitivity to $F_L$ at the high $y$ region 
(see Fig.\ref{fig:h1f29697}). Two methods have been used to extract
$F_L(x,Q^2)$ by H1. This first method~\cite{h1fl96}, used for
$Q^2>10\,{\rm GeV}^2$, assumes that the dominant
contribution to the cross section from $F_2$ can be described by perturbative
QCD by an evolution from lower $Q^2$ at low $y<0.35$, where the contribution
of $F_L$ to the cross section is small. The second method~\cite{h1f29697},
applied for $Q^2<10\,{\rm GeV}^2$, is derived from the cross section
derivative
\begin{equation}
\frac{\partial \sigma_{\rm r}}{\partial \ln y}=\frac{\partial F_2}{\partial
\ln y}-F_L 2y^2\frac{2-y}{Y^2_+}-\frac{\partial F_L}{\partial \ln
y}\frac{y^2}{Y_+}
\end{equation}
taken at fixed $Q^2$ for $y=Q^2/sx$. At high $y$ the second term
dominates~\cite{h1f29697}.
This is in contrast to the $F_L$ influence on the reduced cross sections
$\sigma_r$ where the contribution of $F_2$ dominates for all $y$. A further
important advantage of the derivative method is that it can be applied down
to very low $Q^2$ where the first method can no longer be applied as reliable 
assumptions on the QCD description of $F_2$ are prohibited.
The most recent extraction is shown in Fig.\ref{fig:fl_com} in comparison with 
the other direct measurements mentioned above. 
The extracted $F_L$ is in good 
agreement with the expectation from the NLO QCD fit. It should be emphasized 
however that such an extraction does not represent a real measurement 
but a determination under specific assumptions. 

\subsection{Feasibility study and a new method for a direct measurement of
$F_L$ at HERA} \label{sec:fl_isr}
A direct measurement can be achieved by running the collider with 
reduced beam energy~\cite{Cooper}, but this procedure has the obvious 
draw back that a significant running time is lost for high energy physics and 
that the collider is not operated in optimal conditions.
For the $F_L$ measurement itself, in addition to the relative normalization
uncertainty mentioned above, a major experimental problem is 
the photoproduction background, when a hadron is wrongly taken as the electron
candidate. 
 
Another method proposed by Krasny et al.~\cite{Krasny}
makes use of the radiative events that we have discussed in
Sec.\ref{sec:isr}, namely those DIS events 
in which a real photon has been emitted in the direction of 
the incident electron beam, which corresponds to an effective decrease 
of the beam energy.
The spectrum of measured photon energies induces, for given \x\ and \qsq\
values, a continuous distribution of the \y, \s\ and $\varepsilon$ variables.
The relations (\ref{eq:epsilon_y}) and (\ref{eq:y_s})
stay the same, but the value of \s\ changes from event to 
event, depending on the photon energy.

The advantages of this method are that it can be used in parallel with
normal data taking, that it avoids luminosity normalization problems,
and that the statistical and systematic precisions increase continuously
during data taking.
As the basic principle of the $R(x,\qsq)$ measurement is to perform
a linear fit of cross section, Eq.(\ref{eq:xs_r}), as a function of
$\varepsilon$. The $R(x,\qsq)$ is thus determined from the {\it slope} of the
cross section. This procedure is independent of the knowledge of $F_2$,
however, it requires a very large integrated luminosity
(of the order of 200 \pbinv) to provide a significant measurement of
\R\ (with a statistical precision of $20\%-50\%$)~\cite{Krasny}.

In Ref.\cite{favart_fl}, a new method was proposed, in which we have considered
to use the available measurements of $F_2$ and to exploit
the dependence on \R\ of the {\it shape} of the $\varepsilon$ distribution
itself for radiative events. This is possible because the structure function 
$F_2$ has been
measured very precisely at HERA, and in view of the good understanding of
DIS processes with initial state radiation as demonstrated by the 
H1~\cite{h1f294} (Sec.\ref{sec:isr}) and ZEUS~\cite{zeusrad94} analyses.

\begin{figure}[hbt]
\begin{center}
\begin{picture}(50,280)
\put(-145,-10){
\epsfig{file=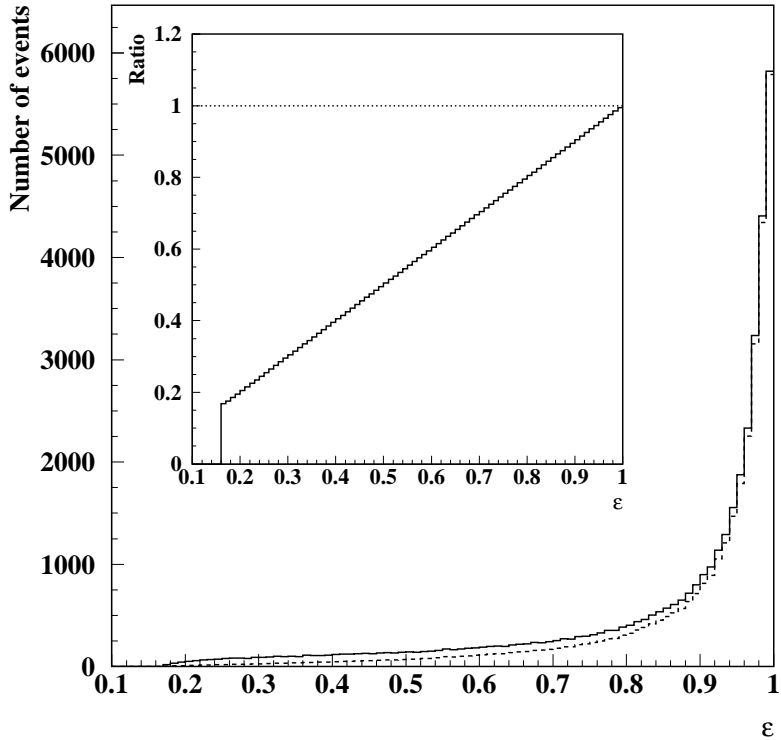,width=110mm}}
\end{picture}
\caption{Distributions of the $\varepsilon$ parameter of radiative events
generated according to Eq.(\ref{eq:cuts}) using the GRV parameterization of 
the structure function $F_2$ 
for $R =0$ (full line) and $R = \infty$ (dashed line), and for the kinematic 
domain covered by the bins shown in Fig.\ref{fig:flbins}. The ratio of the two 
distributions is shown in the inset. The electron and proton beam energy is
respectively 27.5\,GeV and 820\,GeV.}
\label{fig:epsil}
\end{center}
\end{figure}

\begin{figure}[htb]
\begin{center}
\begin{picture}(50,255)
 \put(-120,-10){
\epsfig{file=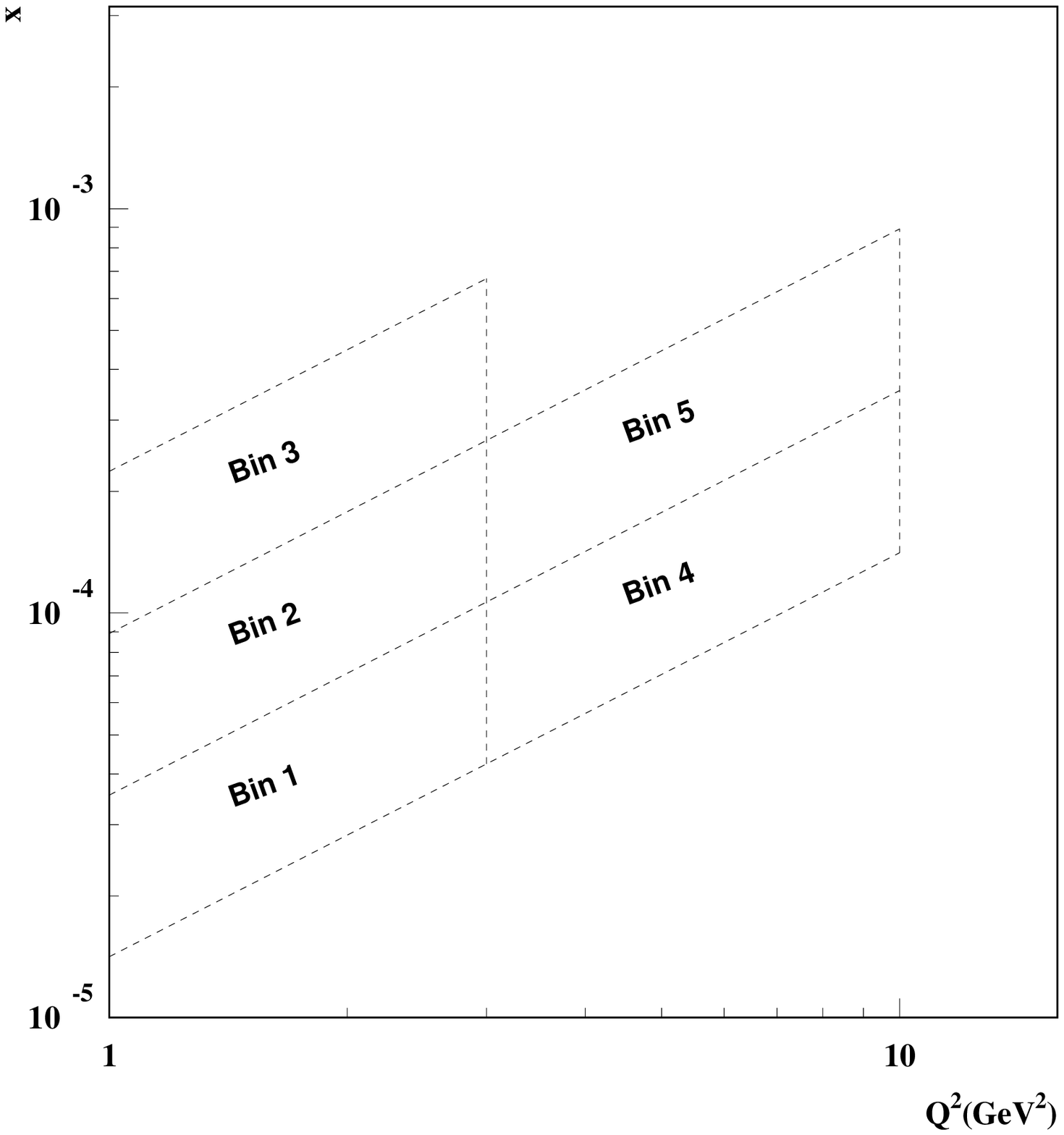,width=100mm}}
\end{picture}
\caption{Selected (\x,\qsq) bins.}
\label{fig:flbins}
\end{center}
\end{figure}

As an illustration, the $\varepsilon$ distribution is presented in 
Fig.\ref{fig:epsil} for $R=0$ (full line) and $R=\infty$ (dashed line),
for the full kinematic domain covered by the five bins in $x$ and $Q^2$
shown in Fig.\ref{fig:flbins}.
Three bins are designed for value of $Q^2$ around $2\,{\rm GeV}^2$
(with $x$ ranging from $4\cdot 10^{-5}$ to $2\cdot 10^{-4}$), and two bins for
$Q^2$ around $5\,{\rm GeV}^2$ (with $x$ ranging from $10^{-4}$ to 
$3\cdot 10^{-4}$).

\begin{figure}[htb]
\begin{center}
\begin{picture}(50,185)
\put(-170,-180){
\epsfig{file=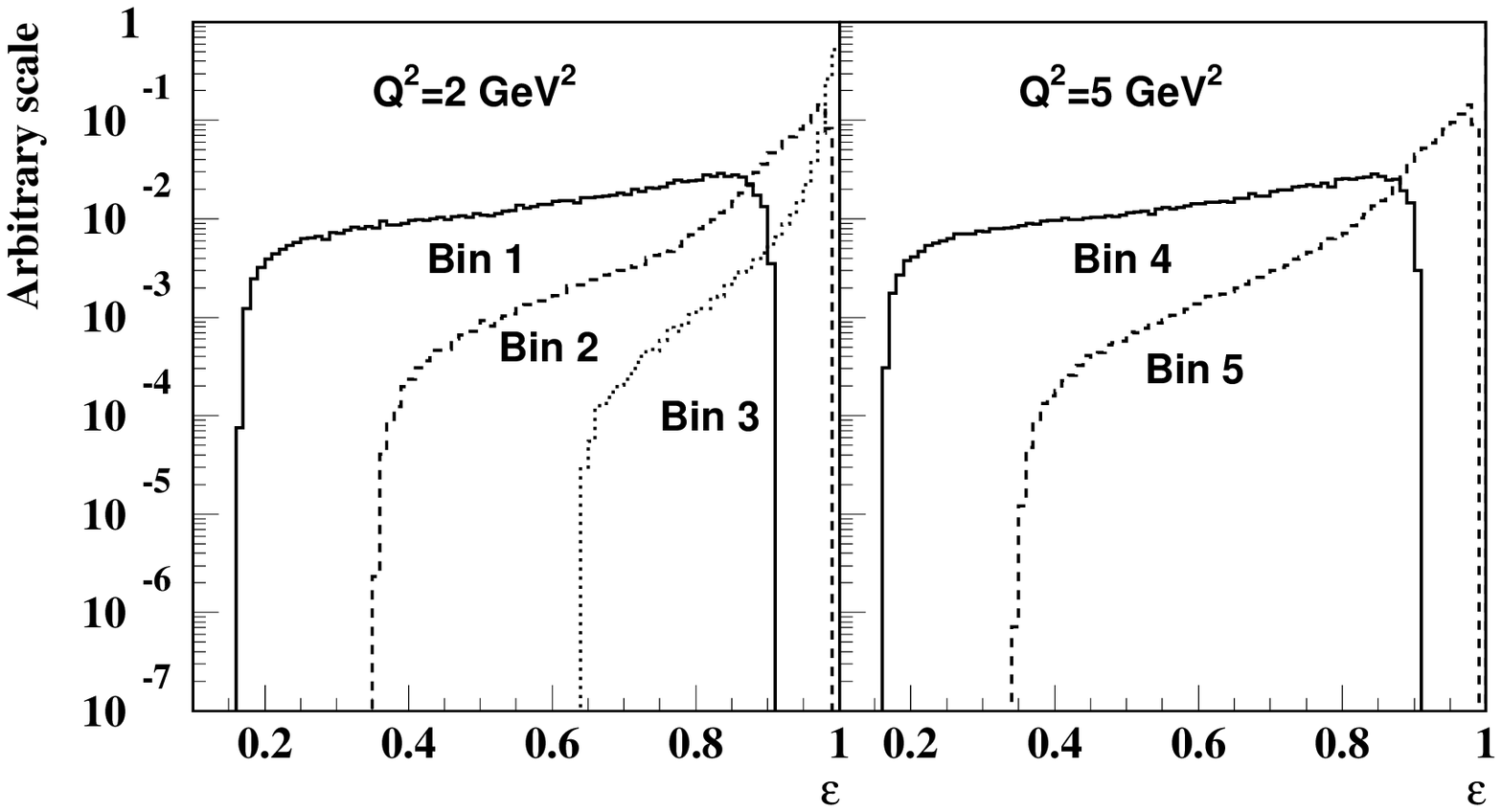,width=140mm}}
\end{picture}
\caption{Distributions of the $\varepsilon$ parameter in the selected bins,
for the GRV parameterization of the structure function $F_2$ and for $R = 0.5$.}
\label{fig:eps_4bins}
\end{center}
\end{figure}

In Fig.\ref{fig:eps_4bins}, the $\varepsilon$ distributions are presented 
for each of these 5 bins.
These distributions are obtained from a Monte Carlo simulation using the GRV
\cite{grv94} parameterization of the proton structure function $F_2$.
 
\begin{figure}[htbp]
\begin{center}
\begin{picture}(70,510)
\put(-175,-15){
\epsfig{file=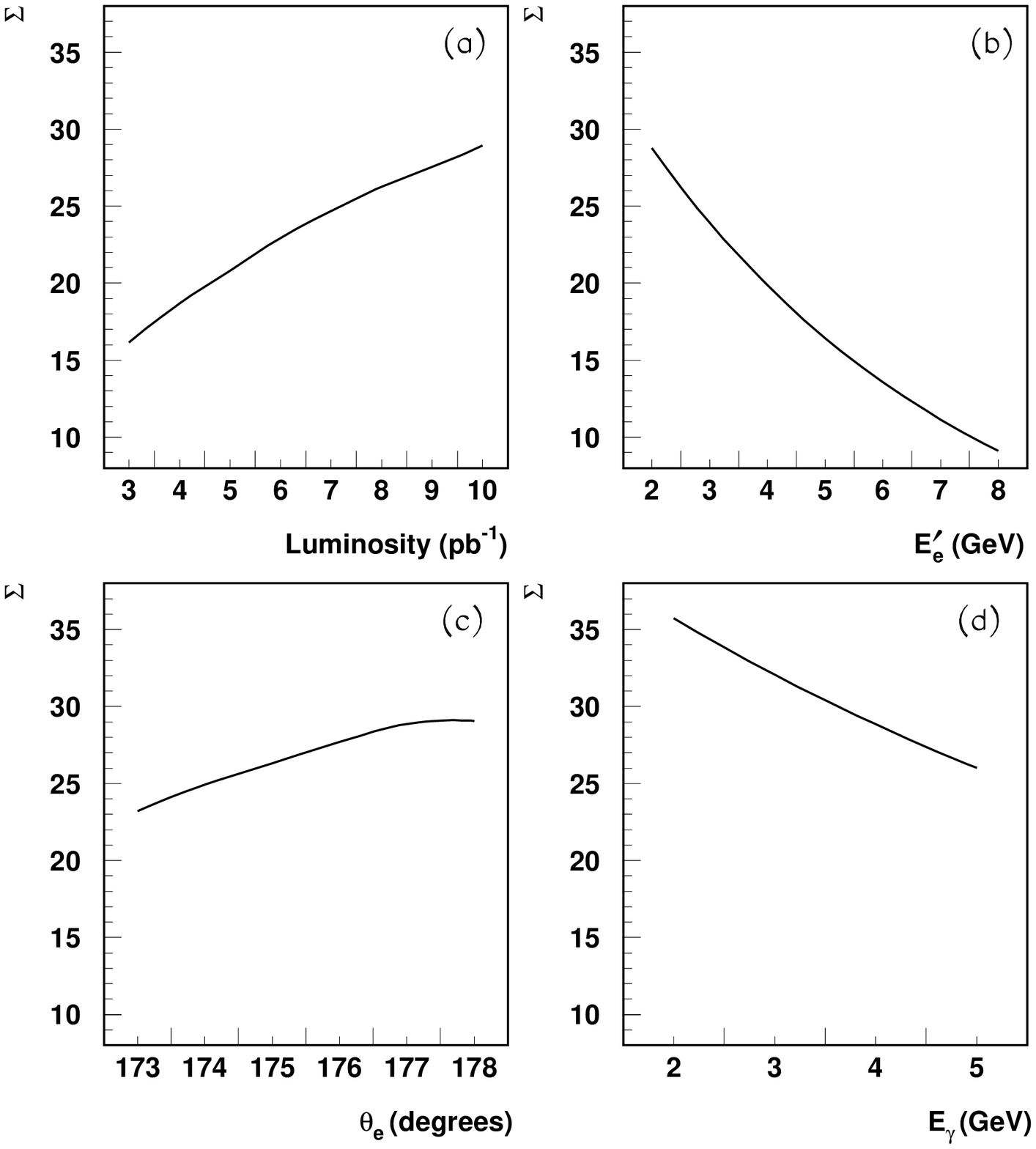,width=150mm}}
\end{picture}
\end{center}
\caption{Sensitivity $\Sigma$ as a function of
a) the integrated luminosity;
b) the electron energy threshold $E_e^{\prime}$;
c) the electron angular acceptance $\theta_e$;
d) the photon energy threshold $E_{\gamma}$,
for the GRV parameterization
under the experimental conditions specified in Eq.(\ref{eq:cuts})
and in the total kinematical region covered by the selected bins.}
\label{fig:sensit}
\end{figure}

The simulated integrated luminosity corresponds to 10 \pbinv, which is the 
luminosity expected from data at the time of the study. The electron and
proton beam energy is respectively 27.5\,GeV and 820\,GeV. The following 
kinematical cuts are applied for the feasibility study:
\begin{eqnarray}
& & E_e^{\prime}>2\, {\rm GeV}\,, \nonumber \\
& & \theta_e<177^{\circ}\,, \nonumber \\
& & E_{\gamma}>4\, {\rm GeV}\, ,
              \label{eq:cuts}
\end{eqnarray}
where $E_e^{\prime}$ and $\theta_e$ are the energy and the polar angle
(defined with respect to the proton beam direction) of the scattered electron.
As will be discussed later (see Fig.\ref{fig:sensit}), the sensitivity of
the method depends significantly on the lowest energies $E^\prime_e$ which
can be accepted. With the new H1 backward detectors, these are expected to
be below 5\,GeV and possibly as low as 2\,GeV, the value thus chosen here.
 
The \R\ dependence of the $\varepsilon$ distribution can be studied using the 
variable $\rhR$, defined as the ratio of the numbers of events with 
$\varepsilon$ smaller or larger than a chosen value $\varepsilon_0$:
\begin{equation}
\rhR=\frac{N(R;\varepsilon<\varepsilon_0)}{N(R;\varepsilon>\varepsilon_0)}\,.
              \label{eq:rho}
\end{equation}
 
\begin{figure}[htbp]
\begin{center}
\begin{picture}(70,510)
\put(-175,-15){
\epsfig{file=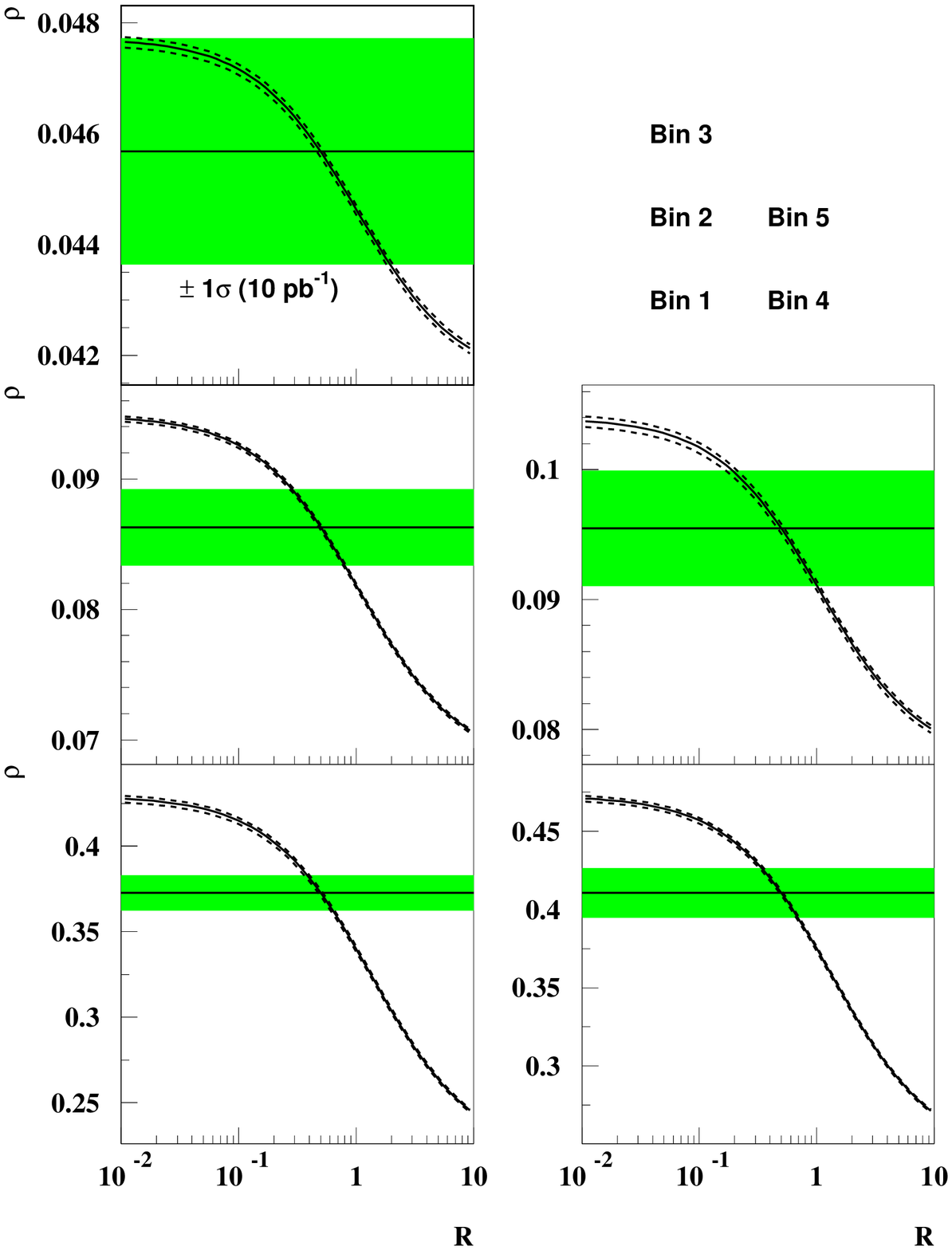,width=150mm}}
\end{picture}
\end{center}
\caption{The dependence of \rh\ on \R\ for the GRV parameterization
(full curves) and for the modified parameterizations described in the text
(dashed curves), in the selected bins.
The grey bands correspond to $\pm 1\sigma$ statistical errors for an
integrated luminosity of 10 \pbinv\ and for $R = 0.5$.}
\label{fig:rho_4bins}
\end{figure}

Fig.\ref{fig:rho_4bins} shows the \rhR\ dependence on $R$ in the selected 
(\x,\qsq)
bins, for the typical input value $R = 0.5$ as obtained by H1~\cite{h1fl96}
in a first analysis using one of the methods mentioned in
Sec.\ref{sec:fl_status}.
Since each bin covers a different $\varepsilon$ range,
the chosen optimal value $\varepsilon_0$ is bin dependent.
The dashed curves show the \rh\ distribution for an input structure function 
$F_2$ modified by $\pm 10 \%$ at $x = 10^{-4}$, the modification decreasing 
linearly to $\pm 5\%$ at $x = 10^{-2}$.
This corresponded to a conservative estimate of the uncertainty on $F_2$ at
the time of study. The uncertainty has been improved since. 
The grey bands correspond to the statistical precision of the \rh\ measurement
for an integrated luminosity of 10\,\pbinv.
 
\begin{figure}[htbp]
\begin{center}
\begin{picture}(50,185)
\put(-170,-180){
\epsfig{file=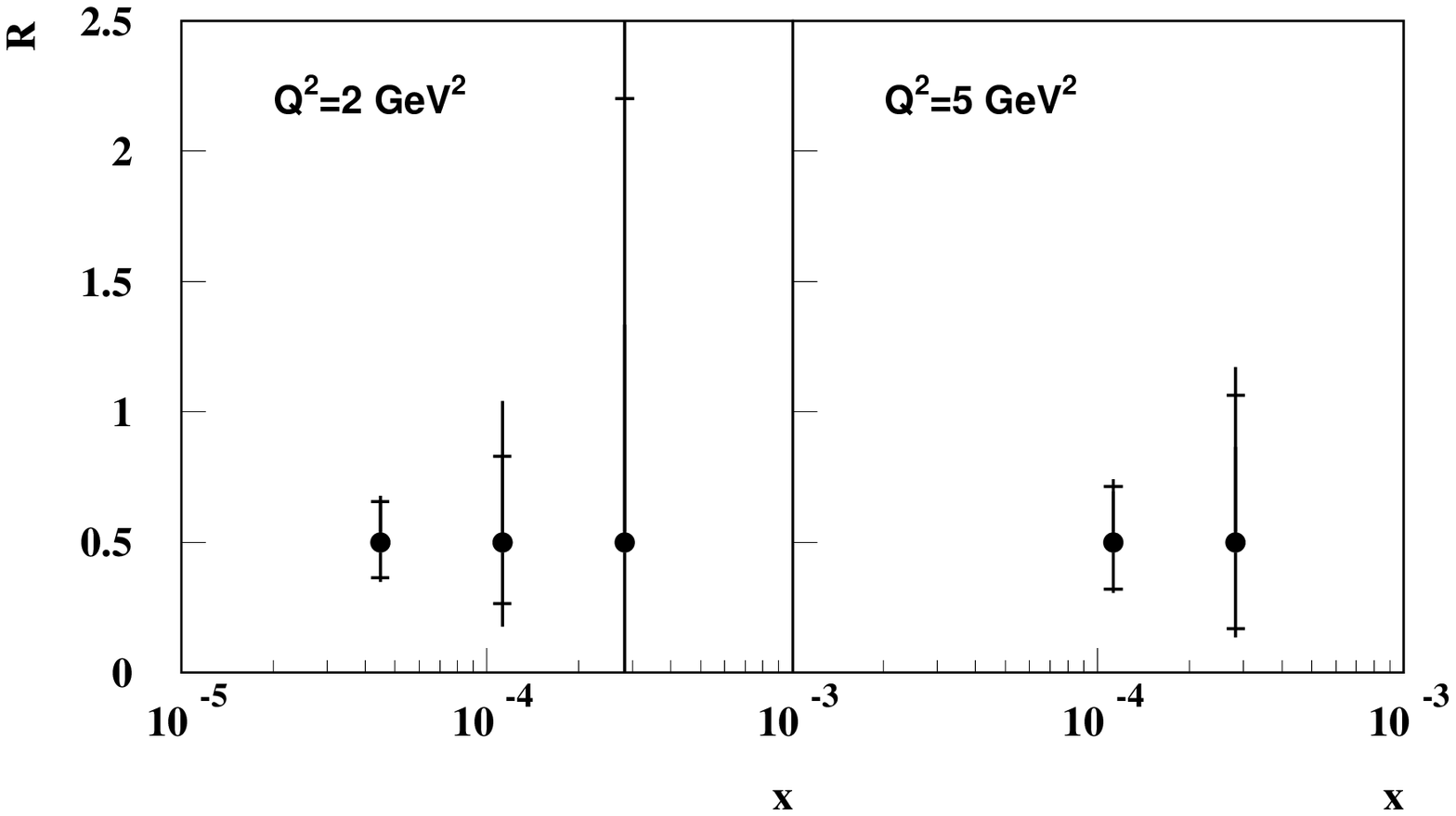,width=140mm}}
\end{picture}
\end{center}
\caption{Typical precision of the \R\ measurement in the selected bins, under
the experimental conditions specified in Eq.(\ref{eq:cuts}).
The inner error bars show the measurement precision for an
integrated luminosity of 10 \pbinv, taking also into account the uncertainty
on the structure function parameterization described in the text.
The outer error bars include, added in quadrature, the effects of the uncertainties
on $E_e^{\prime}$, $\theta_e$ and $E_{\gamma}$. }
\label{fig:R_4bins}
\begin{center}
\begin{picture}(50,185)
\put(-170,-180){
\epsfig{file=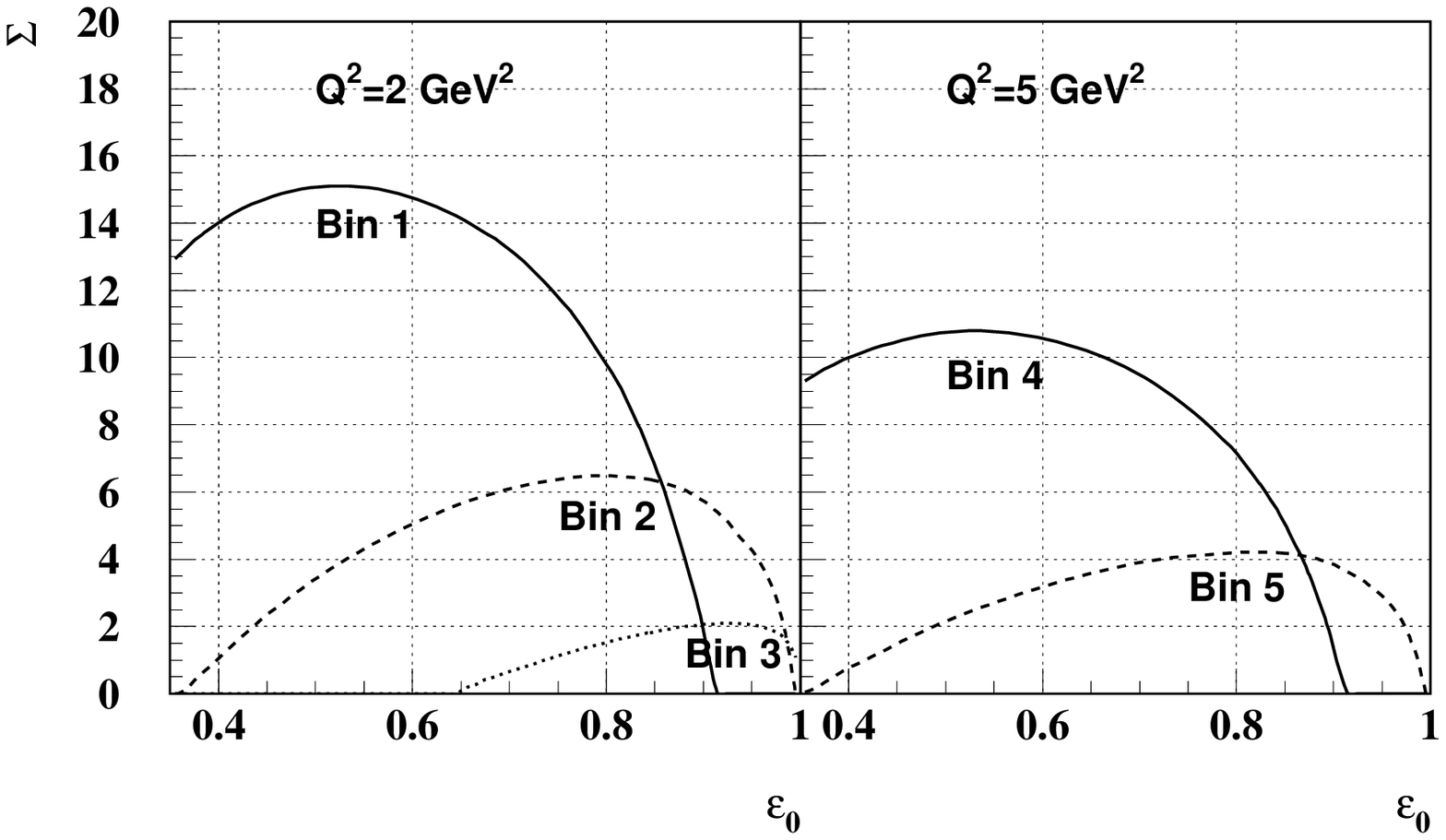,width=140mm}}
\end{picture}
\end{center}
\caption{Dependence of the sensitivity $\Sigma$
 on the $\varepsilon$ parameter in the selected bins,
for the GRV structure function.}
\label{fig:epsil0}
\end{figure}

The measurement of \R\ is deduced from the intersection of the grey bands with
the spread of curves describing the \rh\ dependence of the input structure
function $F_2$. The inner error bars in Fig.\ref{fig:R_4bins} show 
the statistical
precision of the \R\ measurement
for the cuts (\ref{eq:cuts}), an integrated luminosity of 10 \pbinv\ and 
the quoted uncertainty on $F_2$.
 
To estimate the sensitivity of the proposed method to several experimental
parameters, the variable $\Sigma$ is defined as
\begin{equation}
\Sigma(\varepsilon_0,{\cal L},F_2)=\frac{|\rhRz-\rhRi|}
                      {\sqrt{\srhRzsq+\srhRisq}}
              \label{eq:Sigma}
\end{equation}
for a given choice of $\varepsilon_0$, of the integrated luminosity ${\cal L}$
and of the input structure function $F_2$, \srhR\ being the statistical error
on \rh, estimated through the Monte Carlo simulation.
This variable quantifies the possibility of distinguishing between
the two extreme values of \R: $R = 0$ and $R = \infty$.
 
Fig.\ref{fig:epsil0} shows that the sensitivity $\Sigma$ for each
$(x,Q^2)$ bin is only weakly dependent on the
$\varepsilon_0$ value over a rather large domain in $\varepsilon_0$.
It is found that it also depends little on detector smearing effects.
 
On the other hand, as can be seen in Fig.\ref{fig:sensit}, the sensitivity 
$\Sigma$ is
strongly dependent on the detector acceptance conditions, in particular 
the electron energy threshold which is related to the $y$ and $\varepsilon$ 
ranges. For the same luminosity, the sensitivity is enhanced by a factor of 
2.1 for $E_e^{\prime}$ decreasing from 6 to 2 \gev.
A decrease on the photon energy threshold $E_{\gamma}$ also improves 
significantly the sensitivity.
The lowering of the electron energy threshold is a challenge for the HERA 
experiments because of the significant background from photoproduction 
interactions in which low energy hadrons are misidentified as the scattered 
electron.

Studies have been performed of the effects of experimental uncertainties,
which are in general bin to bin dependent.
The detector resolution was simulated using realistic smearing functions for
the H1 experiment; in addition, systematic uncertainties were taken into
account (1\% on $E_e^{\prime}$, 1\,mrad on $\theta_e$ and 1.5\% on
$E_{\gamma}$). The effects of these uncertainties, combined in quadrature with
the statistical errors and the effects of the uncertainty on the structure
function $F_2$, are displayed as the outer error bars on the \R\ measurements
of Fig.\ref{fig:R_4bins}.

The subtraction of the remaining photoproduction background is another
important source of systematic uncertainty, which affects mostly the lower x
bins. There, it was found to induce systematic errors of the same order as the
errors due to detector resolution.
One more source of systematic error will be the overlap of non radiative deep
inelastic events with bremsstrahlung events for which the photon is detected
in the photon detector and the scattered electron is not detected.
An electron tagger with a large energy acceptance is an important tool to
reduce this background.
Finally, as far as the uncertainty on the structure function $F_2$ is
concerned, it is observed in Fig.\ref{fig:rho_4bins} that it does not imply 
a large systematic uncertainty on \R.

Taking all these effects into account, we concluded that for an integrated
luminosity of 10 \pbinv\ a statistical precision of $\geq 30\%$ can be
achieved under the considered experimental conditions and the statistical 
errors dominate over the systematic errors in most of the chosen bins.
With increased statistics, a significant improvement of the measurement
precision is thus to be expected.
Detailed optimization studies are also expected to improve the measurement 
precision.

By now, we have collected by each HERA experiment a factor of ten more 
luminosity than what has been used in the feasibility study, therefore 
statistically there should be more than enough luminosity to provide a first 
direct $F_L$ measurement. The challenges are to achieve the best sensitivity 
with the lowest possible electron and photon energy thresholds and to keep the
experimental systematic uncertainties under control.
Currently the method is being used in an attempt to realize such a measurement.

\newpage
\section{The strong coupling constant $\alpha_s$}\label{sec:alphas}
The strong coupling constant $\alpha_s$, the only fundamental parameter of
the strong interaction sector of the Standard Model, has been measured in
last decades by a great variety of
processes~\cite{bethke00, davier98, stirling97}.
Here we briefly mention a few recent measurements of $\alpha_s(M_Z)$ 
based on DIS data, in particular those using the HERA data showing 
the impact of the precision measurement of the structure function 
data~\cite{h1f29697} (Sec.\ref{sec:h1f29697}).

One of the methods used on fixed-target data is the analysis of sum rules.
One such measurement is derived by using the nonsinglet structure function
data $xF_3$ in the Gross-Llewellyn Smith sum rule~\cite{gls_sr} 
which is known to order $\alpha^3_s$
\begin{equation}
\int^1_0dx\left[F^{\overline{\nu}p}_3(x,Q^2)+F^{\nu p}_3(x,Q^2)\right]=
3\left[1-a_s\left( 1
                   +3.58a_s
		   +19.0a^2_s-
\Delta HT\right) \right]
\end{equation}
where $a_s=\alpha_s/\pi$, and the higher-twist contribution $\Delta HT$ is
estimated to be $(0.09\pm 0.045)/Q^2$ in Refs.\cite{chyla92, braun87} and to
be somewhat smaller by Ref.\cite{dasgupta93}. The CCFR
experiment~\cite{ccfr_kim98}, combines their data with that from other
experiments~\cite{allasia98} and gives
$\alpha_s(\sqrt{3}\,{\rm GeV})=0.28\pm 0.035({\rm expt})\pm
 0.05({\rm syst})^{+0.035}_{-0.030}({\rm th})$.
The systematic uncertainty is dominated by the extrapolation of the integral
to the regions $x<0.01$ where no measurements exist, and to $x>0.5$ which
is substituted by $F_2$ data since it is more precise and has little
correlation to the poorly known gluon density at high $x$.
The error from higher-twist terms dominates
the theoretical uncertainty. If the higher-twist result of
Ref.\cite{dasgupta93} is used, the central value increases to 0.31 and
corresponds to ~\cite{pdg00}
\begin{equation}
\alpha_s(M_Z)=0.118\pm 0.011\,. \hspace{5mm} \mbox{(CCFR, $xF_3$ sum rule)}
\end{equation}

The original and still one of the most powerful quantitative tests of
perturbative QCD is the breaking of Bjorken scaling in DIS. The earliest and
many subsequent determinations of $\alpha_s$ in DIS were obtained by analyzing
the scaling violation of the structure function data. As an example, a
combined analysis of SLAC and BCDMS $F_2$ data in a $Q^2$ range from 0.5 to
260\,GeV$^2$ gives $\alpha_s=0.113\pm 0.005$~\cite{virchaux92}.
For several years, this result together with the one, 
$\alpha_s(M_Z)=0.111\pm 0.002({\rm stat})\pm 0.003({\rm syst})\pm 
0.004({\rm th})$, obtained by CCFR~\cite{ccfr93} based on
$xF_3$\footnote{The advantage of using the less precise structure function 
$xF_3$ instead of $F_2$ is that it has no correlation to the gluon density 
functions.} (and also $F_2$ at high $x$ for the same reason mentioned above
thus resulting in a statistically more precise measurement of $\alpha_s$)
is the most significant from DIS data.
However, these measurements are numerically smaller than typical values
obtained from $e^+e^-$ annihilation, $\alpha_s(M_Z)\sim 0.120$, thus raising
speculations about possible explanations. These speculations came to a halt
when the CCFR collaboration corrected their previous result, due mainly to
effects of a new calibration of the detector, to~\cite{ccfr97}
\begin{equation}
\alpha_s(M_Z)=0.119\pm 0.002({\rm expt})\pm 0.004({\rm th})\,.
\hspace{5mm}\mbox{(CCFR, $xF_3$)}
\end{equation}

Recent scaling violation analyses include
also HERA structure function data extending thus significantly the lever arm
in the $Q^2$ evolution. In Ref.\cite{ball96}, a NLO QCD fit is
performed using the early $F_2$ data of H1 from 1993~\cite{h1f293}
(Sec.\ref{sec:hiq2_93}) giving
\begin{equation}
\alpha_s(M_Z)=0.122\pm 0.004({\rm expt})\pm 0.009({\rm th})\,.
\hspace{5mm}\mbox{(BF, H1 $F_2$ 93)}
\end{equation}
The dominant part of the theoretical error is from the scale dependence;
errors from terms that are suppressed by $1/\ln(1/x)$ in the quark sector
are included~\cite{catani94} while those from the gluon sector are not.

In another analysis~\cite{santiago99}, where the 1994 $F_2$ data by
H1~\cite{h1f294} (Sec.\ref{sec:isr}) and ZEUS~\cite{zeusf294} are used
together with the fixed-target data from SLAC~\cite{slac92},
BCDMS~\cite{bcdms}, and E665~\cite{e66596} in a fit including all known NNLO
terms, the resulting $\alpha_s$ is
\begin{equation}
\alpha_s(M_Z)=0.1172\pm 0.0017 ({\rm expt})\pm 0.0017({\rm th})\,,
\hspace{5mm}\mbox{(SY, HERA $F_2$ 94)}
\end{equation}
where the theoretical error includes the uncertainties in the quark masses, 
higher-twist and target-mass corrections, and errors from the gluon
distributions. But the scale uncertainty is not explicitly taken
into account, and therefore following Ref.\cite{pdg00} the total error is 
increased to 0.0045.

The H1 collaboration have used their recent precise $F_2$ data together with 
the BCDMS data~\cite{bcdms} to obtain a precise determination of $\alpha_s$ 
as well as the gluon density~\cite{h1f29697} (Sec.\ref{sec:fit}):
\begin{equation}
\alpha_s(M_Z)=0.1150\pm 0.0017({\rm expt})^{+0.0011}_{-0.0012} ({\rm
model})\,, \hspace{5mm}\mbox{(H1, $F_2$ 96-97)}
\end{equation}
where the model error includes all uncertainties associated with the
construction of the QCD model used for the fit. In addition, a rather large
theoretical uncertainty resulting from the renormalization and factorization
scale choices should be added. This uncertainty amounts to about $\pm 0.005$
as is discussed in Refs.\cite{vogt96,blumlein96}.

\begin{figure}[htb]
\begin{center}
\begin{picture}(50,115)
\put(-230,-335){\epsfig{file=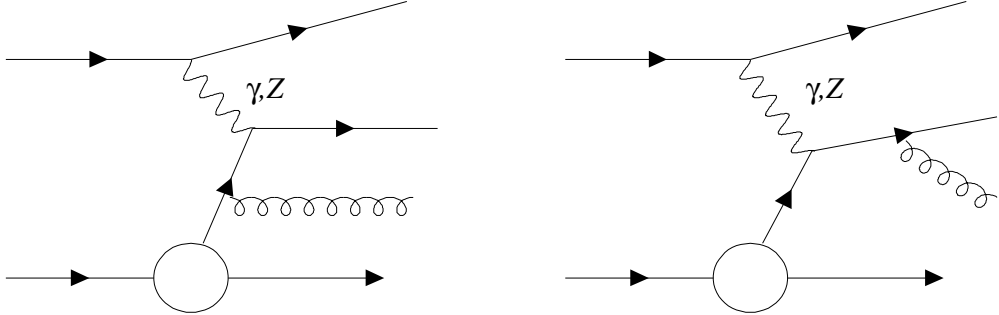,width=175mm}}
\end{picture}
\end{center}
\caption{\sl Diagrams of the QCD-Compton process in deep-inelastic
lepton-proton scattering.}
\label{fig:qcdc}
\end{figure}

At HERA, the strong coupling can also be determined in other ways. One such
measurement is obtained by comparing the rates for $(1+1)$ and $(2+1)$
jet\footnote{The notation ``$+1$'' refers to the proton remnant jet.}
processes. A final state of $(1+1)$ jet is produced at lowest order
in $\alpha_s$ in the $ep$ scattering process, while a $(2+1)$ jet final state
is produced at next order in $\alpha_s$ due to photon-gluon fusion
(Fig.\ref{fig:bgf}) and QCD-Compton processes (Fig.\ref{fig:qcdc}).
Unlike the similar measurements from the $e^+e^-$ experiments, the ones
in $ep$ scattering are not at a fixed energy scale thus allowing
the running of $\alpha_s$ to be tested in a single experiment (see 
Fig.\ref{fig:alphas_et} below).
The first determination by H1 was based on the data from 
1993~\cite{h1_alphas_jet93}. A similar determination by ZEUS was obtained 
from the 1994 data~\cite{zeus_alphas_jet94}.
In the meantime, the NLO QCD calculation have been improved such
that the prediction can be used to compare with experimental data in an
extended kinematic phase space~\cite{th_jet}.
Using the new prediction, the H1 collaboration has had an update using the
1994-1995 data\cite{h1_alphas_jet9495}. The latest result
was determined from the measured inclusive jet cross section
$d^2\sigma_{\rm jet}/dE_TdQ^2$ in the range $49<E^2_T<2500\,{\rm GeV}^2$ and
$150<Q^2<5000\,{\rm GeV}^2$ using the 1995-1997 $e^+p$
data~\cite{h1_alphas_jet9597}. Jets were
defined by the inclusive $k_T$ cluster algorithm~\cite{jet_kt} in the
Breit frame\footnote{In the Breit frame, the gauge boson exchanged in the $t$
channel is purely space-like with four momentum $q=\{0,0,0,-Q\}$
and collides head-on with a parton from the proton. In the leading process
($1+1$ jet) the incoming quark is back-scattered and no transverse energy is
produced. The high $E_T$ inclusive jet cross section receives only higher
order contribution and is therefore directly sensitive to the
strong coupling constant $\alpha_s$.} in the pseudorapidity
range $-1<\eta<2.5$ in the laboratory frame. The values determined in
different $E_T$ range are shown in Fig.\ref{fig:alphas_et} and are consistent
\begin{figure}[tb]
\begin{center}
\begin{picture}(50,250)
\put(-115,-25){\epsfig{file=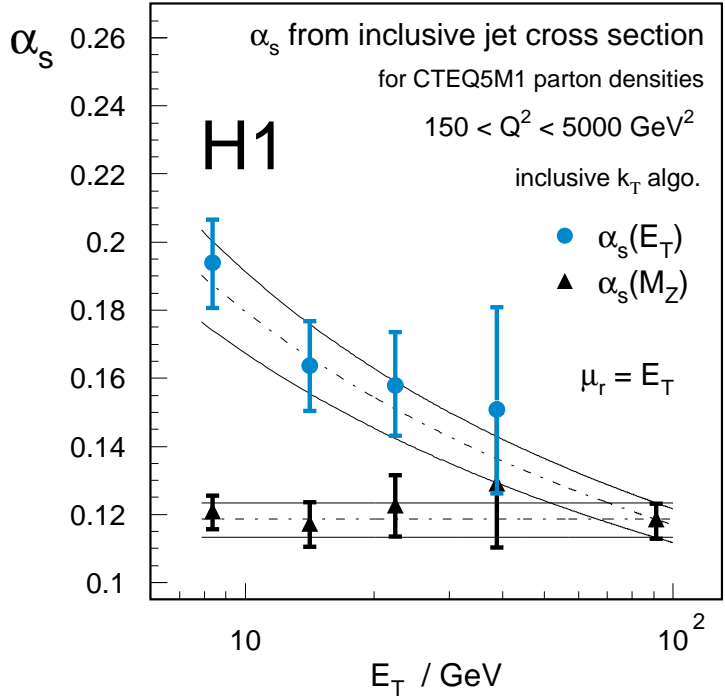,bbllx=0pt,
bblly=0pt,bburx=594pt,bbury=842pt,width=230mm}}
\end{picture}
\end{center}
\caption{\sl Values of $\alpha_s$ determined from the inclusive jet cross
section. The results are shown for each $E_T$ value (circles). The single
values are extrapolated to the $Z^0$ mass (triangles). The upper curves show
the expected scale dependence and the size of the error band indicates the
precision of the combined result.}
\label{fig:alphas_et}
\end{figure}
with the scale dependence predicted by the renormalization group equation.
The combined result choosing $E_T$ as the renormalization scale yields
\begin{equation}
\alpha_s(M_Z)=0.1186\pm 0.0030({\rm exp})\pm 0.0051({\rm th})\,. 
\hspace{5mm}\mbox{(H1, jet 95-97)}
\end{equation}
The most recent result from ZEUS was derived from the measured dijet rate 
in the range $470<Q^2<20\,000\,{\rm GeV}^2$ using the 1996-1997 
data~\cite{zeus_alphas_jet9697}
\begin{equation}
\alpha_s(M_Z)=0.1161^{+0.0039}_{-0.0047}({\rm exp})^{+0.0057}_{-0.0044}({\rm th})\,.
\hspace{5mm}\mbox{(ZEUS, dijet rate 96-97)}
\end{equation}
The dominant experimental error in both experiments is from the hadronic energy
scale uncertainty, which is expected to be significantly reduced according to
recent studies (Sec.\ref{sec:had_scale}). The theoretical error includes
uncertainties arising from scale choice, parton density functions, and
hadronization correction. 

\begin{figure}[htb]
\begin{center}
\begin{picture}(50,310)
\put(-170,-100){\epsfig{file=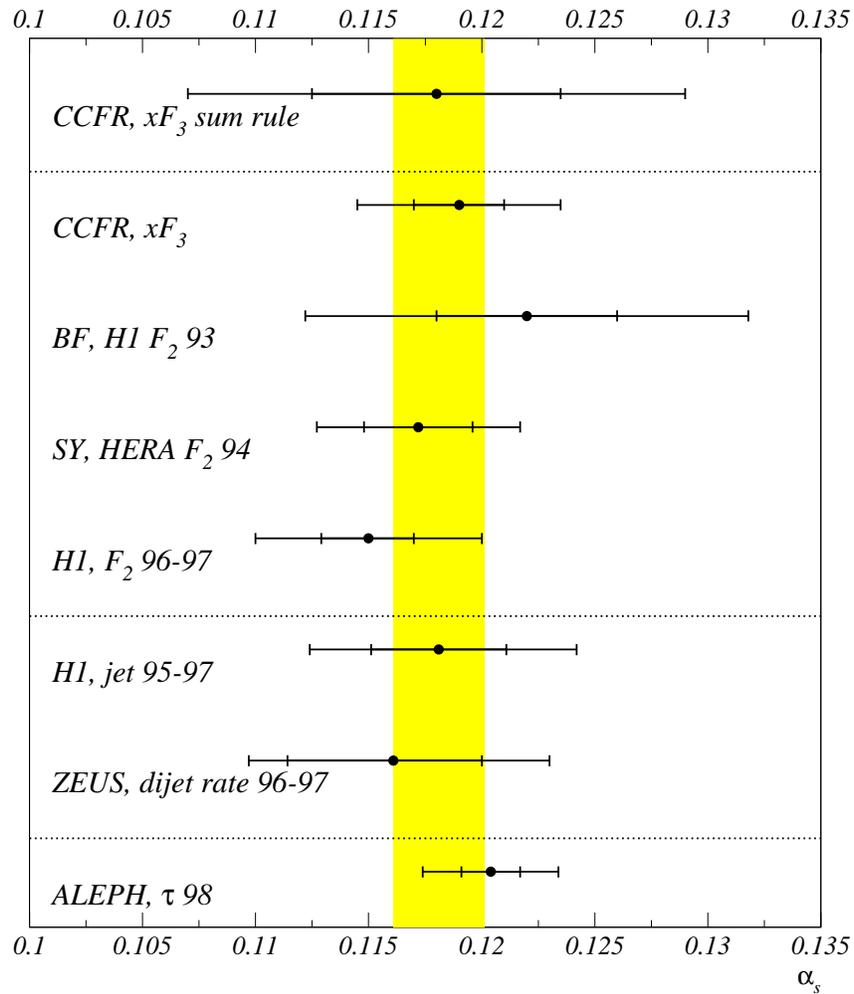,bbllx=0pt,bblly=0pt,
bburx=594pt,bbury=842pt,width=140mm}}
\end{picture}
\end{center}
\caption{\sl A comparison of a few selected $\alpha_s$ measurements from the
DIS data. Also shown are one of the most precise measurements from $\tau$
lepton decays by ALEPH~\cite{aleph_tau98} and the new world
average from \cite{pdg00} (the shaded band). The inner and full error bars
represent respectively the experimental and total errors. See text for the
label definitions and the references.}
\label{fig:alphas}
\end{figure}

These results are shown in Fig.\ref{fig:alphas} together with the new world
average value $\alpha_s=0.118\pm 0.002$~\cite{pdg00} and one of the most
precise measurements from $\tau$ lepton decays by ALEPH~\cite{aleph_tau98}. 
The comparison shows that the results from the DIS experiments are consistent 
among themselves and also with the other independent measurement. 
Furthermore, the DIS measurements are now becoming increasingly precise, 
in particular in terms of the experimental uncertainties. 
The future high statistics HERA data are expected to improve further the 
experimental precision. Similar improvement on the theoretical side is also 
expected when higher order QCD calculations for the structure functions and 
the jet rates will become fully available.

\chapter{Neutral and charged current interactions at high $Q^2$} 
\label{chap:hiq2}
In the previous chapter, the emphasis was put on the measurements of
structure functions at relatively low $Q^2$ and also low $x$, a new
kinematic domain unexplored by fixed-target experiments, and their impact on
the determination of parton (in particular gluon) density functions of the 
proton.
As the integrated luminosity increases, the new kinematic domain at high
$Q^2$ up to a few 10\,000\,GeV$^2$ has seen sufficient data to be studied
for the first time. Contrary to low $x$ HERA data, these data at high $Q^2$
cover a range of $x$ values which have been precisely measured by fixed-target
experiments. This allows the DGLAP evolution equations be tested over four
orders of magnitude in $Q^2$. Deviations from the standard DIS
expectations may be found as the largest $Q^2$ value attainable at HERA is
equivalent to a resolution power of one thousandth of the proton radius.
In fact, already with the data from years in 1994 to 1996, H1 and
ZEUS have reported a low statistics excess of events at high $Q^2$ and large
$x$. In this chapter, the analysis on three independent data samples taken
since 1994 will be discussed in some detail. The first $e^+p$ data sample,
corresponding to an integrated luminosity of 35.6\,pb$^{-1}$ and a 
center-of-mass energy of 300\,GeV (see Table \ref{tab:hera_year}), was taken
from 1994 to 1997. Both the $e^-p$ data of 1998-1999 and the $e^+p$ data of
1999-2000 are taken at a center-of-mass energy of 320\,GeV resulting from
the increased proton energy of 920\,GeV. The $e^-p$ data have an integrated
luminosity of 16.4\,pb$^{-1}$. The recent higher energy $e^+p$ data of
45.9\,pb$^{-1}$ correspond to the data taken until the beginning of June,
2000.\footnote{The data taking continued till the beginning of September 2000
when a long upgrade shutdown has taken place.}
In Sec.\ref{sec:hiq2_tech} a few relevant technical issues will be discussed 
first. The cross section results are given in Sec.\ref{sec:hiq2_xs}.

\section{Technical aspects} \label{sec:hiq2_tech}

The inclusive neutral (NC) and charged current (CC) cross section analysis
on the later $e^-p$ data of 1998-1999 and $e^+p$ data of 1999-2000 did not
differ significantly from the one on the earlier $e^+p$
data from 1994 to 1997 apart from a few technical improvements. Therefore in
the following discussion, no separation is given unless stated otherwise.

\subsection{Event selections and background studies} \label{sec:evtsel}

\subsubsection{NC event selection}

The selection of NC events is only briefly described here as it
did not differ in essence from the one on the earlier data of 1993 described 
in Sec.\ref{sec:hiq2_93} and was based on the characteristic feature of the
events; an identified scattered electron~\cite{bruel_thesis} in the liquid 
argon (LAr) calorimeter and the hadronic final state which is mainly
measured in the LAr calorimeter as well. 
In fact the main selection cuts were only slightly modified with respect to
the one used in the earlier analysis:
\begin{itemize}
\item A reconstructed event vertex $|z_{\rm vtx}-z_0|<35$\,cm. As mentioned
 already in the previous analyses (Secs.\ref{sec:hiq2_93} and \ref{sec:isr}), 
 this requirement suppresses efficiently the non-$ep$ background contributions.
\item An energy $E^\prime_e$ of the scattered electron measured in the LAr 
 calorimeter greater than 11\,GeV. Above this threshold the trigger was fully 
 efficient and a small fraction of inefficient regions was excluded by applying
 fiducial cuts~\cite{beate_thesis,burkard_thesis}.
\item An inelasticity $y_e$ lower than 0.9. This cut becomes more restrictive
 than the cut on $E^\prime_e$ for $Q^2\geq 907.5\,{\rm GeV}^2$.
\item A longitudinal momentum conservation verifying $\Sigma_e+\Sigma_h$
 greater than 35\,{\rm GeV}. For an ideal detector without energy loss, the
 sum is expected to be $2E_0$.
\end{itemize}
The last three requirements minimize the size of radiative corrections and 
reduce the background from photoproduction.

The dominant remaining background contribution was from photoproduction events
where a hadronic final state faked the scattered electron. For the analysis
on the 1994-1997 $e^+p$ data, it was subtracted statistically according to the
simulated photoproduction Monte Carlo events generated using the {\sc
pythia}~\cite{pythia} generator. The size of the subtraction was controlled
with a subsample in which the genuine scattered electron is tagged in the
small angle electron tagger at $z=-33$\,m away from the interaction point.
This method suffers however from a large uncertainty (30\%) on the estimated
background events to be subtracted. For the analysis on the new data recorded 
since 1998, a new method was applied relying on the fact that the track charge
associated to the electron candidate in the photoproduction event could be
either negative or positive instead of being negative as expected from the
$e^-p$ collisions. Therefore, the background can be subtracted with two times 
of positive charged tracks based on the data itself thereby reducing
considerably the uncertainty in the subtraction.

\subsubsection{CC event selection}

\begin{figure}[btp]
\begin{center}
\begin{picture}(50,325)
\put(-245,-50){\epsfig{file=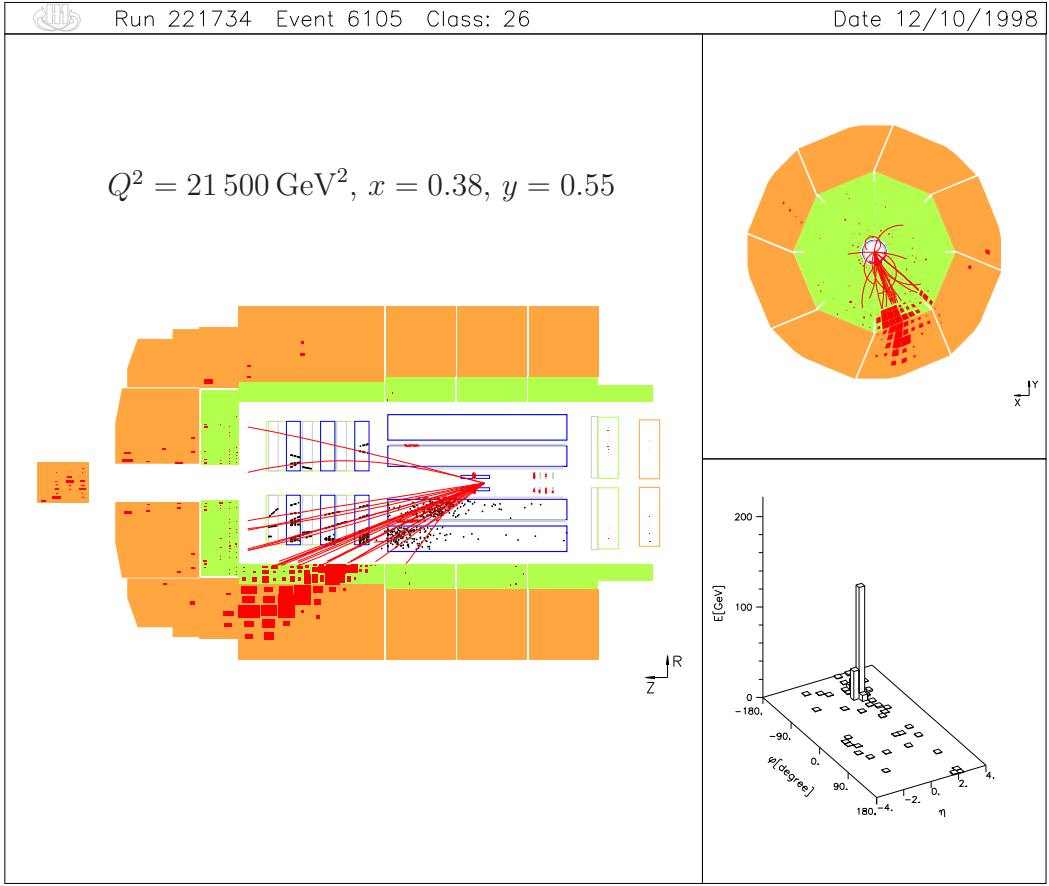,bbllx=0pt,bblly=0pt,
bburx=594pt,bbury=842pt,width=142mm,angle=90}}
\put(-150,275){\makebox(0,0)[l]{\fcolorbox{white}{white}{\textcolor{white}{AAAAAAAAAAAAAAAAA}}}}
\put(-135,250){\makebox(0,0)[l]{$Q^2=21\,500\,{\rm GeV}^2$, $x=0.38$, $y=0.55$}}
\put(-105,220){\makebox(0,0)[l]{\fcolorbox{white}{white}{\textcolor{white}{AAAAAAAAAAAAAAAAAAAA}}}}
\end{picture}
\end{center}
\caption{\sl A typical charged current event measured by the H1 detector.}
\label{fig:cc_r221734}
\end{figure}

\begin{figure}[htb]
\begin{center}
\begin{picture}(50,315)
\put(-155,-25){\epsfig{file=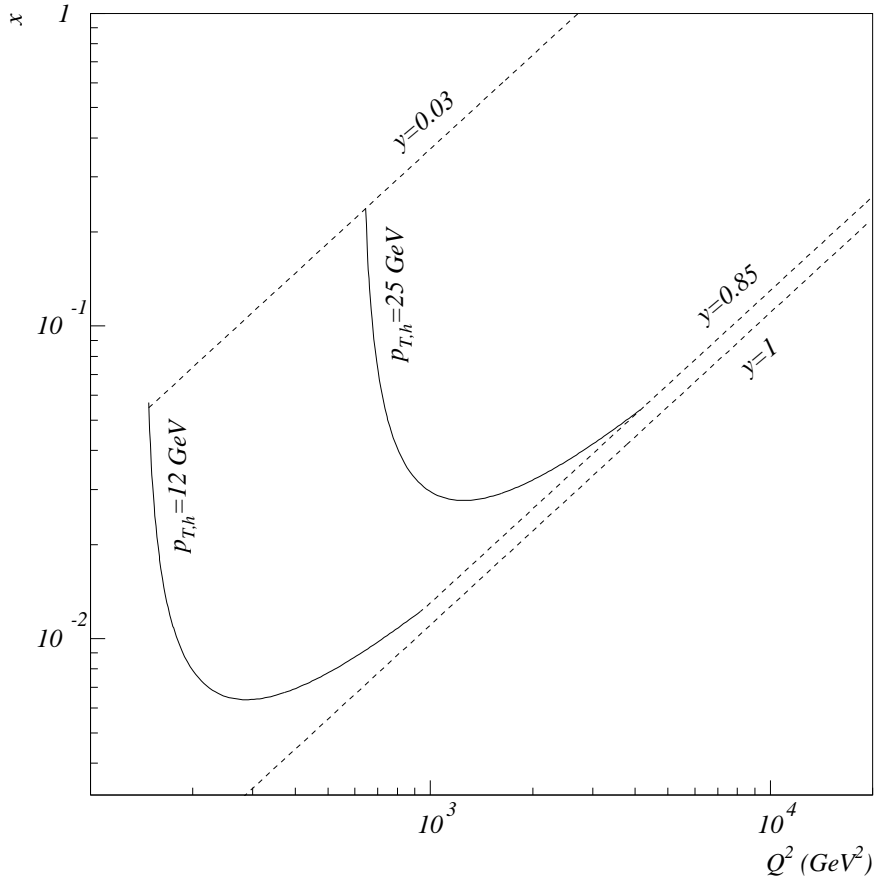,width=130mm}}
\end{picture}
\end{center}
\caption{\sl The extended kinematic region when the cut $p_{T,h}>25$\,GeV is
lowered to $p_{T,h}>12$\,GeV. Also indicated are the HERA kinematic limit 
$y=1$ and the actually analyzed y range $0.03<y<0.85$.}
\label{fig:kine_cc}
\end{figure}

Contrary to the NC selection, the selection of CC events represents 
a major challenge.
The CC events are characterized by a missing transverse momentum $p^{\rm
miss}_T\equiv p_{T,h}$ due to the undetected neutrino in the final state
(see Fig.\ref{fig:cc_r221734} for a typical CC event).
The main selection cut was
\begin{equation}
p_{T,h}>12\,{\rm GeV}\,,
\end{equation}
which was much lower than the cut $p_{T,h}>25$\,GeV
applied in the earlier studies~\cite{h1cc93,h1cc94,h1cc9394}
based on the 1993-1994 data. While the kinematic coverage has been
significantly extended to the lower $Q^2$ (Fig.\ref{fig:kine_cc}), 
the acceptance to the background events were opened up at the same time.
In addition to the $p_{T,h}$ cut and the same vertex requirement as for the NC
analysis, the CC analysis was limited in the inelasticity range $0.03<y<0.85$ 
as also indicated in Fig.\ref{fig:kine_cc} in order that the kinematic
quantities which can only be reconstructed from the hadronic final state are
reasonably precise and that the trigger efficiency remains reasonably large
(Sec.\ref{sec:cctrig}).

\begin{figure}[htb]
\begin{center}
\begin{picture}(50,270)
\put(-230,-100){\epsfig{file=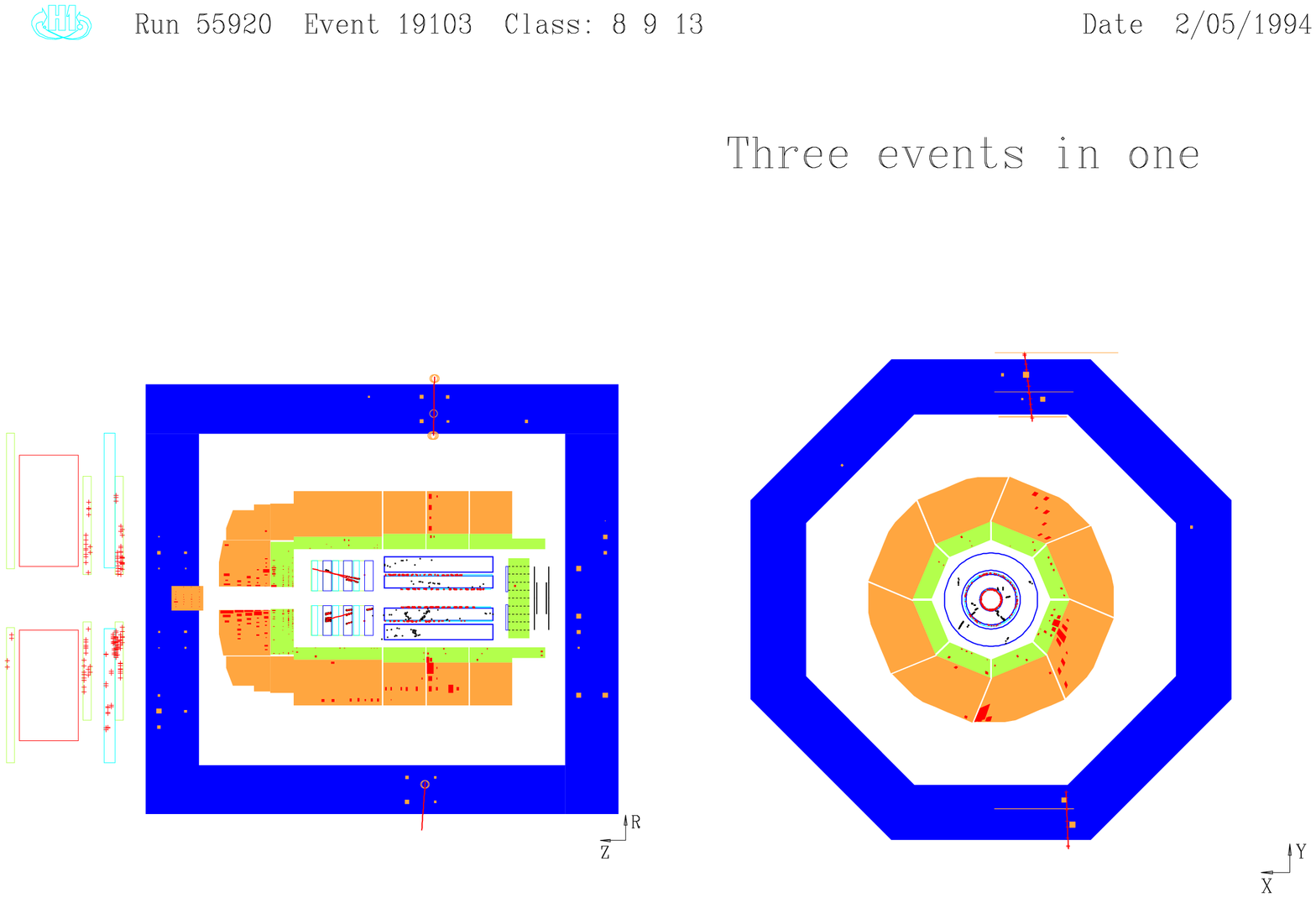,bbllx=0pt,bblly=0pt,
bburx=594pt,bbury=842pt,width=130mm,angle=90}}
\end{picture}
\end{center}
\caption{\sl A pile-up event from three background sources: halo muon,
cosmic ray, and beam-gas/wall interaction.}
\label{fig:bg-pile-up}
\end{figure}

Three background sources contribute in the remaining non-$ep$ background
events. These sources are halo muons, cosmic rays and beam-gas and beam-wall
interactions. Since the rate of these background events are a few orders of
magnitude higher than the CC rate (Table \ref{tab:xs_rates}), there is a
high probability that these events pile up among themselves or with other
$ep$ events. An example of such pile-up events is shown in
Fig.\ref{fig:bg-pile-up}.

One efficient rejection was achieved by applying the timing informations
provided by both the central jet chambers (CJC) and the LAr calorimeter.
The CJC timing, $t_{\rm CJC}$, was determined from tracks which cross the sense
wire planes, and has a resolution around 2\,ns.
The LAr timing~\cite{pieuchot95}, $t_{\rm LAr}$, was
determined from energy deposits in the trigger readout, which consists of 512
big towers\footnote{A small fraction of the big towers are formed from the
backward calorimeter BEMC/SPACAL and the forward PLUG calorimeter.}, 
a coarser structure compared to the highly segmented cell granularity.
The resolution of $t_{\rm LAr}$ is about 10\% of a bunch crossing (96\,ns).

The other powerful rejection used
a set of topological filters {\sc qbgfmar}~\cite{qbgfmar},
which was based on the characteristic signature
of cosmic and halo muon events. Part of the filters were developed for the
earlier analysis described in Sec.\ref{sec:hiq2_93}. The efficiency of the
background filters was determined to be well above 95\% based on the pseudo-CC
sample (see Sec.\ref{sec:cctrig})
and was in good agreement with the MC simulation.

The residual background 
events were further reduced by other track based criteria. Two examples are
\begin{itemize}
\item {\bf Track-Cluster link:} For the pile-up events, the charged tracks
contributing to the event vertex have in general no correlation with the 
energy deposit in the LAr calorimeter. These events were efficiently 
suppressed by requiring a minimum distance of 0.5 in the $\phi$ and 
pseudo-rapidity $\eta$ plane between the tracks and two largest energy
deposits in transverse momentum.
\item {\bf Track multiplicity:} The track multiplicity of a beam-gas or
beam-wall event can be large. An important fraction of these tracks do
not originate from a common vertex. These events were rejected by requiring
that there were 10 or more tracks originating far away from the event 
vertex by more than 35\,cm along $-z$ direction.
\end{itemize}
The final contamination of non-$ep$ background events in the CC sample
was determined by visual scanning to be 3.7\% for the 1994-1997 $e^+p$ 
analysis and was improved by a factor of two for the new data and
subsequently removed from the CC sample. The visual scanning technique 
as part of the final selection procedure was important because it revealed
not only often the need for further improvements in the background rejection 
but also occasionally interesting (exotic) events. In fact, the first events
of observed isolated lepton events with large missing transverse momentum were
found in this way.

The dominant $ep$ background sources originated from NC events and
photoproduction events due to the finite detector resolution and the limited 
geometrical acceptance (including the crack region in $\phi$ and $z$).
Nevertheless, in the transverse plane the energy flow is expected to
be more isotropic in the photoproduction and NC events than in CC events. 
This is quantified by the ratio $V$ of two variables $V_{ap}$ and $V_p$
defined~\cite{martin_thesis} as:
\begin{eqnarray}
V_{ap}=-\sum_i \frac{\vec{P}_{T,i}\cdot \vec{P}_{T,h}}{P_{T,h}} &
{\rm for} & \vec{P}_{T,i}\cdot \vec{P}_{T,h}<0 \label{eq:v_ap}\\
V_p=\sum_i \frac{\vec{P}_{T,i}\cdot \vec{P}_{T,h}}{P_{T,h}} &
{\rm for} & \vec{P}_{T,i}\cdot \vec{P}_{T,h}>0 \label{eq:v_p}
\end{eqnarray}
standing respectively for the transverse energy flow antiparallel and
parallel to the direction of the transverse momentum of the event. The sum
in (\ref{eq:v_ap}) and (\ref{eq:v_p}) extends over all hadronic final state
situated in the hemisphere which is opposite to or along $\vec{P}_{T,h}$.
For the analysis on the 1994-1997 $e^+p$ data, the background was
significantly suppressed by requiring
\begin{equation}
V=\frac{V_{ap}}{V_p}<0.15\,.\label{eq:v}
\end{equation}

\begin{figure}[htbp]
\begin{center}
\begin{picture}(50,180)
\put(-185,-240){\epsfig{file=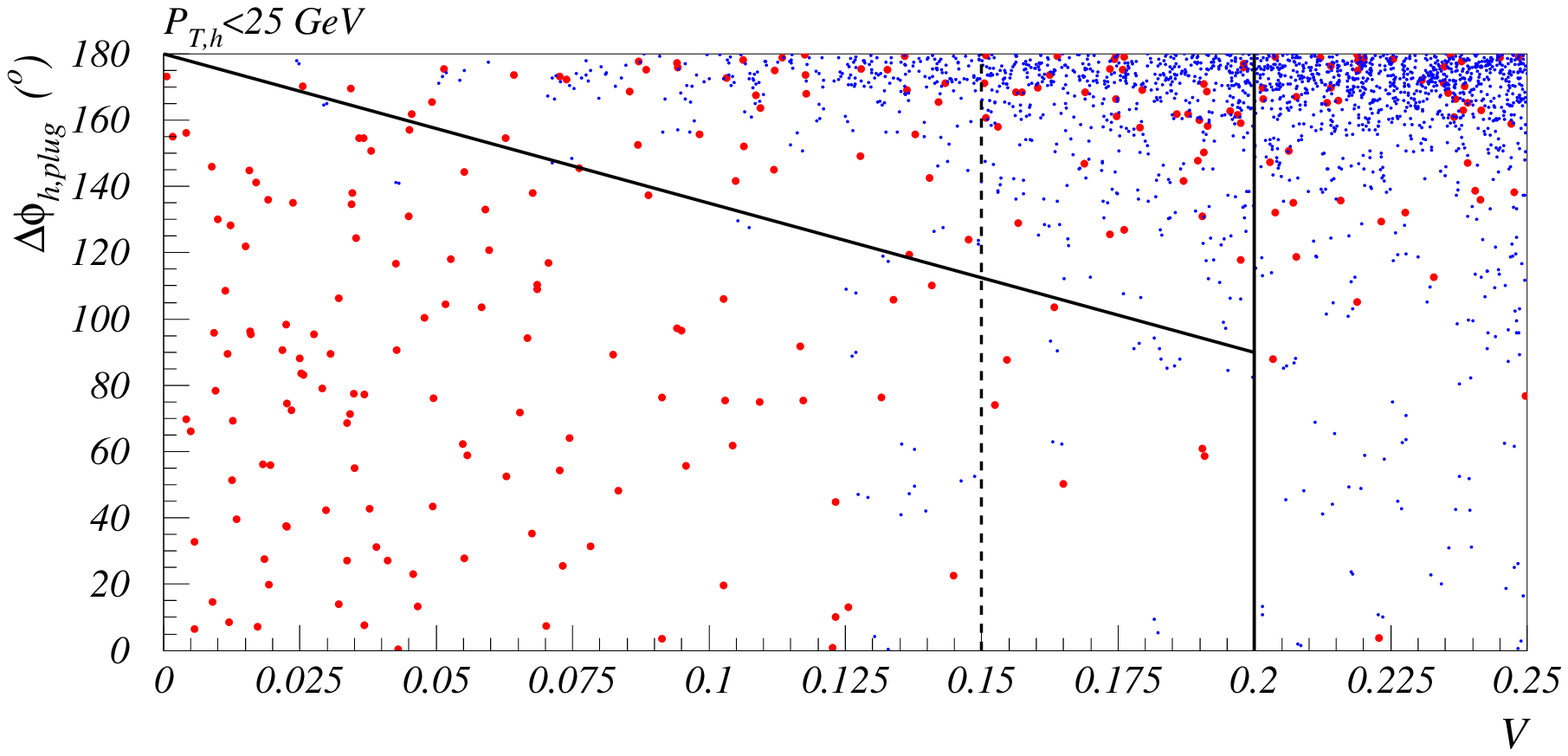,bbllx=0pt,bblly=0pt,
bburx=594pt,bbury=842pt,width=165mm}}
\end{picture}
\end{center}
\caption{\sl Distribution of $\Delta \phi_{h,{\rm PLUG}}$ versus
$V$ (see (\ref{eq:dphi}) and (\ref{eq:v}) for definitions) for CC candidates
in the real data (large dots) and for the simulated photoproduction
Monte Carlo events
(small dots) which represent a factor of six larger integrated luminosity than
in data. The full lines indicate the new cut applied for two example
$P_{T,h}$ values at 12\,GeV and $\geq 25\,{\rm GeV}^2$ in the $e^-p$ analysis 
while the dashed line for the old cut applied in the $e^+p$ analysis.}
\label{fig:dphi_v}
\begin{center}
\begin{picture}(50,190)
\put(-205,-345){\epsfig{file=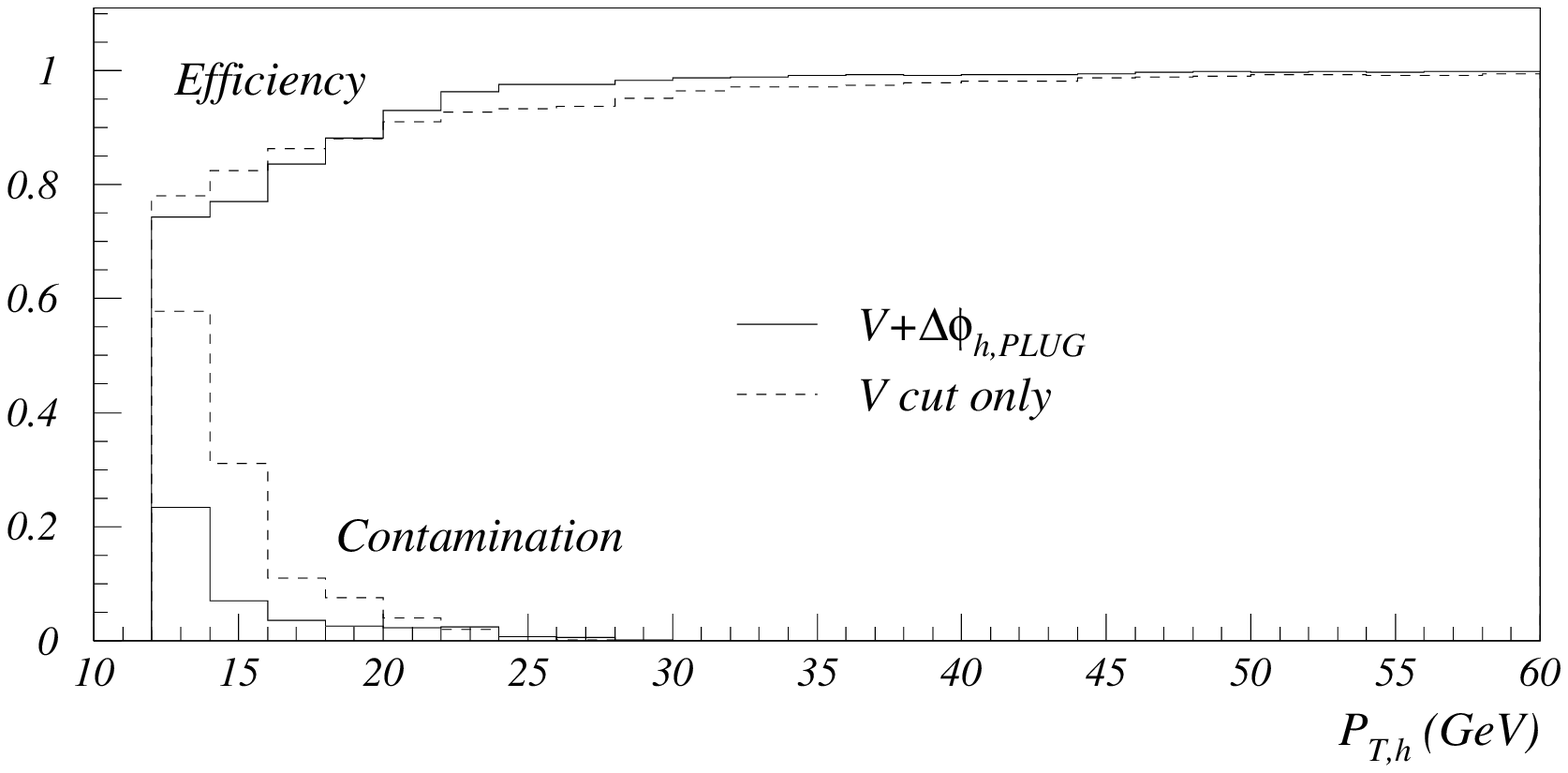,bbllx=0pt,bblly=0pt,
bburx=594pt,bbury=842pt,width=165mm}}
\end{picture}
\end{center}
\caption{\sl Improvement on the CC selection efficiency and the suppression of
the photoproduction contamination as a function of $P_{T,h}$ for the 1998-1999
$e^-p$ analysis with the $V$ and $\Delta \phi_{h,{\rm PLUG}}$ cuts (full
lines) compared with the $V$ cut only (dashed lines) used in the 1994-1997 
$e^+p$ analysis.} \label{fig:eff_antigp}
\end{figure}

For the analysis of the new data, the photoproduction rejection
was improved by introducing a new variable $\Delta \phi_{h,{\rm
PLUG}}$~\cite{mehta}:
\begin{equation}
\Delta \phi_{h,{\rm PLUG}}\equiv |\phi_h-\phi_{\rm PLUG}|\,,\label{eq:dphi}
\end{equation}
where $\phi_h$ and $\phi_{\rm PLUG}$ are the $\phi$ angles of
the hadronic final state measured respectively in the LAr calorimeter and in
the forward PLUG calorimeter. While the $\Delta \phi_{h,{\rm PLUG}}$
distribution is nearly flat for the $CC$ events, the photoproduction events
are mainly distributed at large angles as illustrated in Fig.\ref{fig:dphi_v}
for $P_{T,h}<25$\,GeV. 
In order to optimize the efficiency and the background contribution, a
$P_{T,h}$ dependent cut for $P_{T,h}<25$\,GeV is applied as
\begin{equation}
\Delta\phi_{h,{\rm PLUG}}<90+\frac{180-90}{V(P_{T,h})-0.2}(V-0.2)\,,
\end{equation}
with
\begin{equation}
V(P_{T,h})\equiv 0.2\left(\frac{P_{T,h}-12}{25-12}\right)^2\,
\end{equation}
and $\Delta\phi_{h,{\rm PLUG}}$ in degrees and $P_{T,h}$ in GeV.
At the same time, the $V$ cut is extended from 0.15 to 0.2.
Consequently the
photoproduction contamination has been further suppressed by a factor of two
for an efficiency which is hardly affected at $P_{T,h}<25$\,GeV and better
at higher $P_{T,h}$ values (Fig.\ref{fig:eff_antigp}).
The final photoproduction contribution, which was statistically subtracted
from the data, was estimated to be 2.9\% and 1.4\%
respectively for the 1994-1997 $e^+p$ and 1998-1999 $e^-p$ analysis. The
largest contribution is located at low $Q^2$ and high $y$ and did not exceed 
1\% in most of the kinematic region studied.

Part of the NC DIS background contribution was suppressed by the $V$ cut. The
remaining contribution was further reduced by requiring that if an isolated
lepton candidate was identified, neither the lepton in the LAr nor
the associated track should satisfy $|\phi-\phi_h|>120^\circ$, where $\phi$
and $\phi_h$ were respectively the $\phi$ angle of the lepton 
(or of the track) and of the hadronic final state.
The final NC background was estimated for both $e^+p$ and $e^-p$ analyses
to be around 0.7\% of the total selected samples. Again as for the
photoproduction background the largest contribution was 
distributed at the low $P_{T,h}$ region and was negligible for
most of the kinematic region under consideration.

The final CC samples comprised about 700 events for 1994-1997 $e^+p$ as well
as 1998-1999 $e^-p$ analyses\footnote{The integrated luminosity of the $e^+p$
1994-1997 data was about a factor of three larger than that of the $e^-p$
1998-1999 data. On the other hand,
the cross sections are the other way around so that the total selected number
of events were about the same for the two data samples.}, and 1000 events
for the 1999-2000 $e^+p$ analysis.

\subsection{Alignment and electron angle measurement} \label{sec:align_theta}
For the NC analysis, one of the key quantities needed for the kinematic
reconstruction is the polar angle of the scattered electrons. The polar
angle is determined both by the trackers and the position measured in 
the high granularity calorimeter together with the event vertex. The two
determinations are complementary:
\begin{itemize}
\item When the trackers are fully operational, the former determination
gives a better precision than the latter one. On the other hand, it suffers
from occasional inefficiencies of part of the tracking system\footnote{In
order to maximize the statistical precision of the cross section measurements,
which is still limited at high $Q^2$, the data sample used includes data in 
which part of the trackers could be temporarily off.} 
and needs a delicated modeling of the fraction of 
inefficient regions and its time dependence in the Monte Carlo simulation.
\item The latter determination though intrinsically less precise is more
stable as a function of time. However, it is achieved when the calorimeter is
well aligned to the tracking system.
\end{itemize}

\begin{figure}[htbp]
\begin{center}
\begin{picture}(50,180)
\put(-195,-345){\epsfig{file=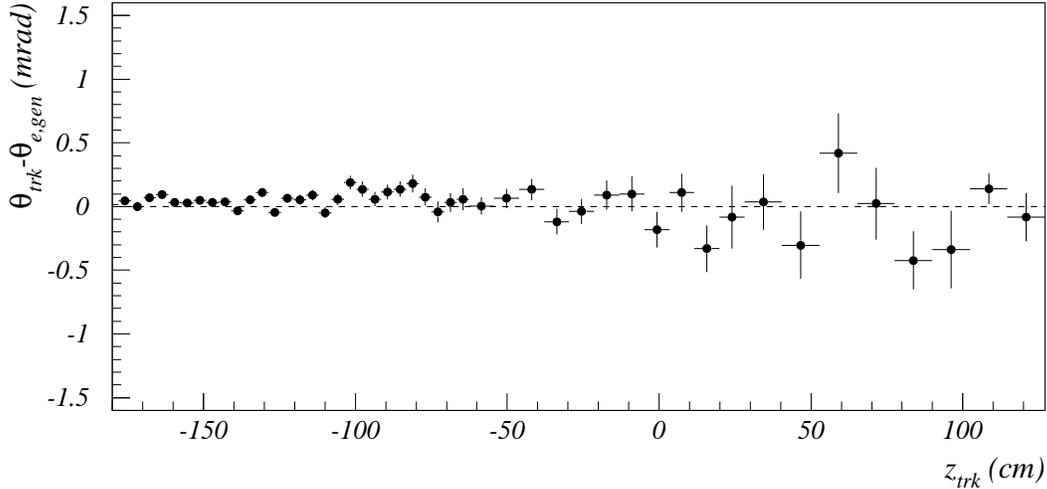,bbllx=0pt,bblly=0pt,
bburx=594pt,bbury=842pt,width=165mm}}
\end{picture}
\end{center}
\caption{\sl The polar angle difference between the generated scattered
electron and the associated track as a function of the $z$ impact of the
track.} \label{fig:trk_gen}
\end{figure}

The alignment was performed in this analysis based on a subsample where the
track associated to the scattered electron was well measured by the central
tracking device (i.e.\ both the CJC and the inner and outer $z$
chambers were operational). Under such condition, the polar angle of the
track $\theta_{\rm trk}$ gave a good reference for the angle of the scattered
electron $\theta_e$. This is illustrated by Fig.\ref{fig:trk_gen} where
the angle $\theta_{\rm trk}$ is found in good agreement with
$\theta_{e,{\rm gen}}$, the generated angle of the scattered electron,
independent of the impact position $z_{\rm trk}$ of the track on the front
of the calorimeter.
An octagon
around the $z$ axis with a minimum 
radius of 105\,cm was taken to approximate the front plane of 
the calorimeter. The track impact point was defined on this octagon after 
having been extrapolated from the event vertex
$(x_{\rm vtx}, y_{\rm vtx}, z_{\rm vtx})$ taking into account its curvature
measured by the tracker within the magnetic field. The impact position of 
the corresponding cluster measured in the LAr calorimeter on the same octagon 
was obtained, however, with a straight line between the barycenter of 
the cluster and the event vertex.

Defining $(x_{\rm LAR}, y_{\rm LAr}, z_{\rm LAr})$ as the center of the LAr
calorimeter in the coordinator system of CJC with additional parameters
$\alpha$, $\beta$, and $\gamma$ being respectively for anti-clock-wise
rotations around the $x$, $y$ and $z$ axes, the cluster barycenter
$(x_e, y_e, z_e)$ of the scattered electron measured in the LAr calorimeter
can be written in the CJC system as\footnote{Since the rotation angles are
expected to be small, the following approximations $\cos x\simeq 1$ and
$\sin x\simeq x$ can thus be made. The $z$ component $z_e$ in data 
should be replaced by $z^{\rm cold}_e$ which is related to $z_e$ by $z^{\rm
cold}_e=23.67+(z_e-23.67)(1.0-0.0027)$. This correction arises because
during the final step of data processing the LAr geometrical parameters
were mistakenly taken to correspond to a warm environment instead of the cold
LAr in reality.}
\begin{eqnarray}
& & x^\prime_e=x_e-x_{\rm LAr}+ \beta z_e- \gamma y_e\,, \nonumber \\
& & y^\prime_e=y_e-y_{\rm LAr}-\alpha z_e+ \gamma x_e\,, \\
& & z^\prime_e=z_e-z_{\rm LAr}+\alpha y_e- \beta x_e\,. \nonumber
\end{eqnarray}

\begin{figure}[htbp]
\begin{center}
\begin{picture}(50,300)
\put(-195,-145){\epsfig{file=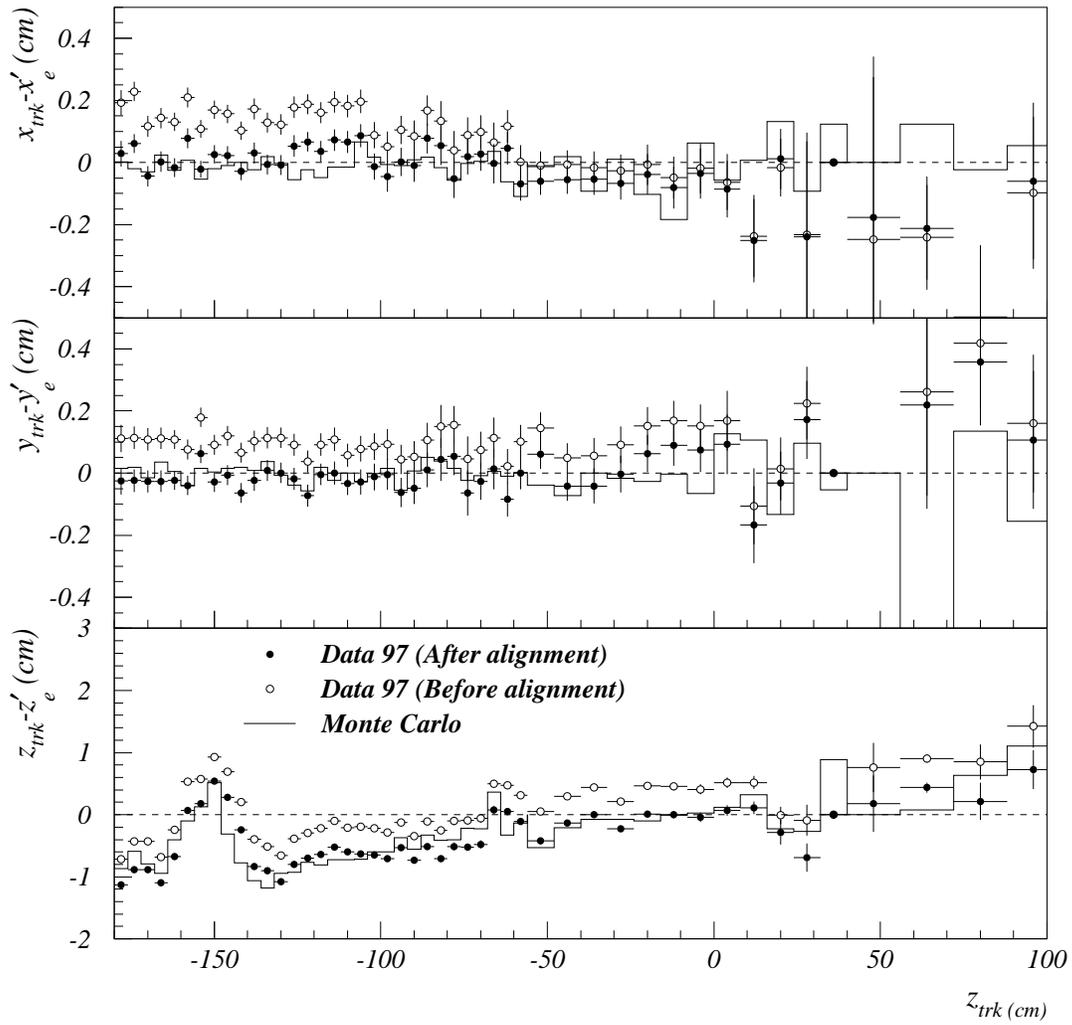,bbllx=0pt,bblly=0pt,
bburx=594pt,bbury=842pt,width=165mm}}
\end{picture}
\end{center}
\caption{\sl The difference of the impact point $x$, $y$ and $z$ 
between track and cluster as a function of $z$ impact of the track 
before and after the alignment.}
\label{fig:align_z}
\end{figure}

The parameters were determined by minimizing the sum of the difference
$a_{\rm trk}-a^\prime_e$ with $a=x$, $y$ and $z$.
The resulting parameters are shown in Table \ref{tab:align} for the 1997
$e^+p$, 1998-1999 $e^-p$, and 1999-2000 $e^+p$ data.
As an example, Fig.\ref{fig:align_z} shows how the alignment on the 1997
$e^+p$ data improves the difference $a_{\rm trk}-a^\prime_e$.
Before the alignment, there was a significant rotation around the $y$ axis
which resulted in a tilt in the $x-z$ plane. Around the other axes, there
were no important rotation, but the shifts in $y$ and in particular in $z$
were sizable. As far as the measurement of the scattering angle $\theta_e$ 
is concerned, the quantity $z_{\rm trk}-z^\prime_e$ is most relevant, 
which as shown in Fig.\ref{fig:align_z} is in better agreement 
with the Monte Carlo simulation after the alignment.
The overall shape, which stems partly from the pad structure and 
partly from the energy distribution on the pads from which the barycenter 
can be defined, cannot of course be improved by the alignment. But this
effect can be and has been corrected for based on the Monte Carlo.
\begin{table}[tb]
\begin{center}
\begin{minipage}{14.1cm}
\caption{\sl Parameters describing the relative alignment between the LAr
calorimeter and the central tracking system for the 1997 $e^+p$,
1998-1999 $e^-p$, and 1999-2000 $e^+p$ data. 
} \label{tab:align}
\end{minipage}
{\footnotesize \begin{tabular}{|c|r|r|r|r|r|r|}
\hline
Year & $x_{\rm LAr}$ (mm) & $y_{\rm LAr}$ (mm) & $z_{\rm LAr}$ (mm) &
$\alpha$ (mrad) & $\beta$ (mrad) & $\gamma$ (mrad) \\\hline
1997 $e^+p$ & $-0.5\pm 0.2$ & $-1.5\pm 0.2$ & $-4.2\pm .08$ &
$0.6\pm 0.1$ & $-1.7\pm 0.1$ & $0.7\pm .06 $ \\
98-99 $e^-p$ & $0.6\pm 0.2$ & $-0.5\pm 0.2$ & $-1.7\pm .07$ &
$-0.5\pm 0.1$ & $-0.0\pm 0.1$ & $2.4\pm .06$ \\
99-00 $e^+p$ & $0.9\pm 0.1$ & $ 0.4\pm 0.1$ & $-2.0\pm .03$ &
$-0.6\pm .05$ & $-0.2\pm .05$ & $0.9\pm .03$ \\\hline
\end{tabular}}
\end{center}
\end{table}

For the 1994-1997 $e^+p$ data analysis, the final angle of the scattered
electron was determined from the track when
it was well measured by the central tracking system ($\theta_e>35^\circ$)
and in other cases from cluster measured in the calorimeter. The fraction of
latter determination was about 40\% and 100\% respectively for
$\theta_e>35^\circ$ and $\theta_e<35^\circ$. The corresponding systematic
uncertainty from the tracker and the calorimeter was respectively 1\,mrad
and 3\,mrad. For the analyses of other data samples, the angle was
determined from the cluster position in the calorimeter together with the
event vertex due to a reduced efficiency of the central trackers.

\subsection{Electron energy measurement} \label{sec:em_scale}
For the cross section measurements to be shown below, the energy measurement
of the scattered electron has been improved by an {\sl in situ} calibration
by using the increased event sample and the overconstrained kinematic
reconstruction.

Several complementary methods and event samples have been used:
\begin{itemize}
\item {\bf The double angle method based on NC DIS events (DIS DA):} As
shown in (\ref{eq:eda}), the energy of the scattered electron can be
predicted from the polar angles of the electron $\theta_e$ and of the
hadronic system $\theta_h$. The resolution of $\theta_h$ is sufficiently
good when one uses the subsample at $y_\Sigma<0.3$ ($y_\Sigma<0.5$) for
$80^\circ \lesssim \theta_e \lesssim 153^\circ$ ($40^\circ \lesssim \theta_e
\lesssim 80^\circ$). In this region, the event statistics is large enough to
improve the energy measurement locally in finely segmented $z$ and $\phi$
grid defined by the impact position of the electron track on the LAr
calorimeter.
\item {\bf The elastic QED Compton events and exclusive two photon $e^+e^-$
pair production (QED Compton/$e^+e^-$):} Due to the limited number of events
from NC DIS process in the forward region ($\theta_e<40^\circ$), two
additional event samples are used to check and obtain the calibration
constant together with the DIS sample. These samples are the elastic QED
Compton events and the exclusive two photon $e^+e^-$ pair production. Contrary
to the DIS events, these samples provide two leptonic final states for the
calibration when both are measured in the LAr calorimeter.
Furthermore since these events cover a large energy range, 
they allow the energy linearity be studied.
\item {\bf The $\omega$ method based on NC DIS events ($\omega$ DIS):} This
is an approximate kinematic method~\cite{omega} which assumes that the
relative error in the measured $\Sigma_h$ and $P_{T,h}$ is the same 
(i.e.\ $\delta\Sigma_h/\Sigma_h=\delta P_{T,h}/P_{T,h}$). 
This method was also based on the NC DIS sample and therefore it was not 
independent of the DA method. 
On the other hand, the $\omega$ method is less sensitive to the effects of 
initial state QED radiation than the DA method.
\end{itemize}

\begin{figure}[htbp]
\begin{center}
\begin{picture}(50,220)
\put(-155,-35){\epsfig{file=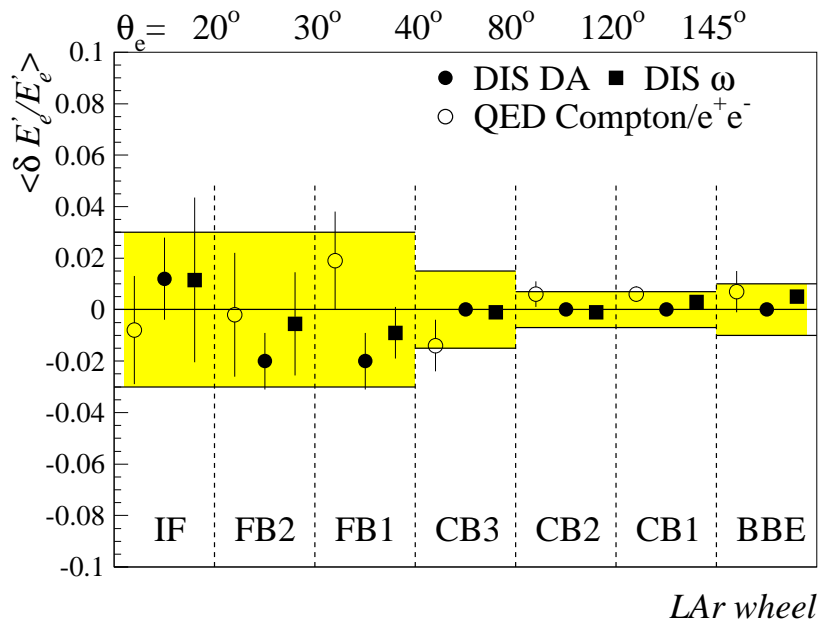,bbllx=0pt,bblly=0pt,
bburx=594pt,bbury=842pt,width=130mm}}
\end{picture}
\end{center}
\caption{\sl Comparison of the electromagnetic energy scale as determined by
different methods and event samples. Shown is the mean fractional energy
shift of the different method from the absolute energy scale, $\langle
\delta E^\prime_e/E^\prime_e\rangle$. The shaded error band shows the
systematic uncertainty on the energy scale quoted for the cross section
measurement, which varies from 0.7\% to 3\%, depending on the location of
the electron in the detector.}
\label{fig:ecal_summary}
\end{figure}

The resulting electromagnetic energy scale as determined for the 1994-1997
$e^+p$ data analysis is shown in Fig.\ref{fig:ecal_summary} together with 
the quoted systematic uncertainty as indicated by the error band.

For the analyses on the new data, the energy scale was checked using
the newly available event sample. Small variation within the quoted
systematic uncertainty was found and taken into account in the new analysis.

\subsection{Hadronic energy measurement} \label{sec:had_scale}
A precise hadronic energy measurement is crucial for all measurements
relying on the hadronic final state for the kinematic reconstruction. In
fact many of these measurements are now limited by the uncertainty of the
hadronic energy scale, which was estimated to be about 4\%.
For instance, such an uncertainty dominates the experimental error of
the $\alpha_s(M^2_Z)$ measurement of $\alpha_s(M^2_Z)$ based on the inclusive 
jet cross section~\cite{h1_alphas_jet9597} (Sec.\ref{sec:alphas}).

For the cross section measurements presented here, a detailed study was
carried out~\cite{h1note571} in order to improve the uncertainty.
While the precision of the electron energy calibration was limited
in the forward region due to the limited statistics, the hadron energy
calibration does not have such a limitation as the hadronic final state
covers rather uniformly all the calorimeter region. The calibration method
is based on the transverse energy conservation, namely the transverse
momentum of the hadronic final state $P_{T,h}$ should be equal to that
of the scattered electron $P_{T,e}$, the latter can be again predicted with
the double angle method as $P_{T,{\rm DA}}$.

A clean and unbiased calibration sample is defined as follows:
\begin{itemize}
\item An identified scattered electron having $P_{T,e}>10$\,GeV,
\item The predicted $P_{e,{\rm DA}}$ is not strongly affected by
the initial state QED radiative events by requiring\footnote{As mentioned
previously, a cut on $\Sigma_e+\Sigma_h$ is also very efficient in
suppressing the radiative effect. On the other hand, since the sum includes
a contribution which is directly proportional to the energy scale of 
the hadronic final state, such a cut may bias the calibration.}
$P_{T,e}/P_{T,{\rm DA}}>0.88$,
\item One and only one reconstructed jet based on the cone
algorithm~\cite{cone} with a cone radius of
1\,rad in the $\phi$ and pseudorapidity $\eta$ plane and a minimum
transverse momentum of 4\,GeV. In addition, in order that the calibration
factor to be determined is relevant for the LAr calorimeter, the jet has
to be well contained in the LAr calorimeter by requiring that there is no
more than 1\% energy or transverse energy deposit in either the backward
calorimeter or the tail-catcher. The jet is also required to be well
away from the beam pipe in the forward direction $(\theta_{\rm jet}>7^\circ)$
so that the calibration would not correct for the beam energy loss.
\end{itemize}

The calibration is then performed in two steps. In the first step, the
relative difference in $P_{T,h}/P_{T,{\rm DA}}$ between data and the Monte
Carlo simulation is determined in finely segmented $\phi$ and
$\theta^\prime_{\rm jet}$\footnote{The quantity $\theta^\prime_{\rm jet}$
differs from $\theta_{\rm jet}$ in that the former is calculated always with
a fixed vertex position at zero instead of with the measured event vertex
as in the latter. In this way, the angle $\theta^\prime_{\rm jet}$ is well
defined independent of the vertex position.} grid to account for possible 
detector effects which may be year dependent. In the second step, an
absolute hadronic energy scale is achieved both in data and simulation by
imposing the transverse momentum balance between the scattered electron and
the hadronic final state.

\begin{figure}[htbp]
\begin{center}
\begin{picture}(50,370)
\put(-202.5,-145){\epsfig{file=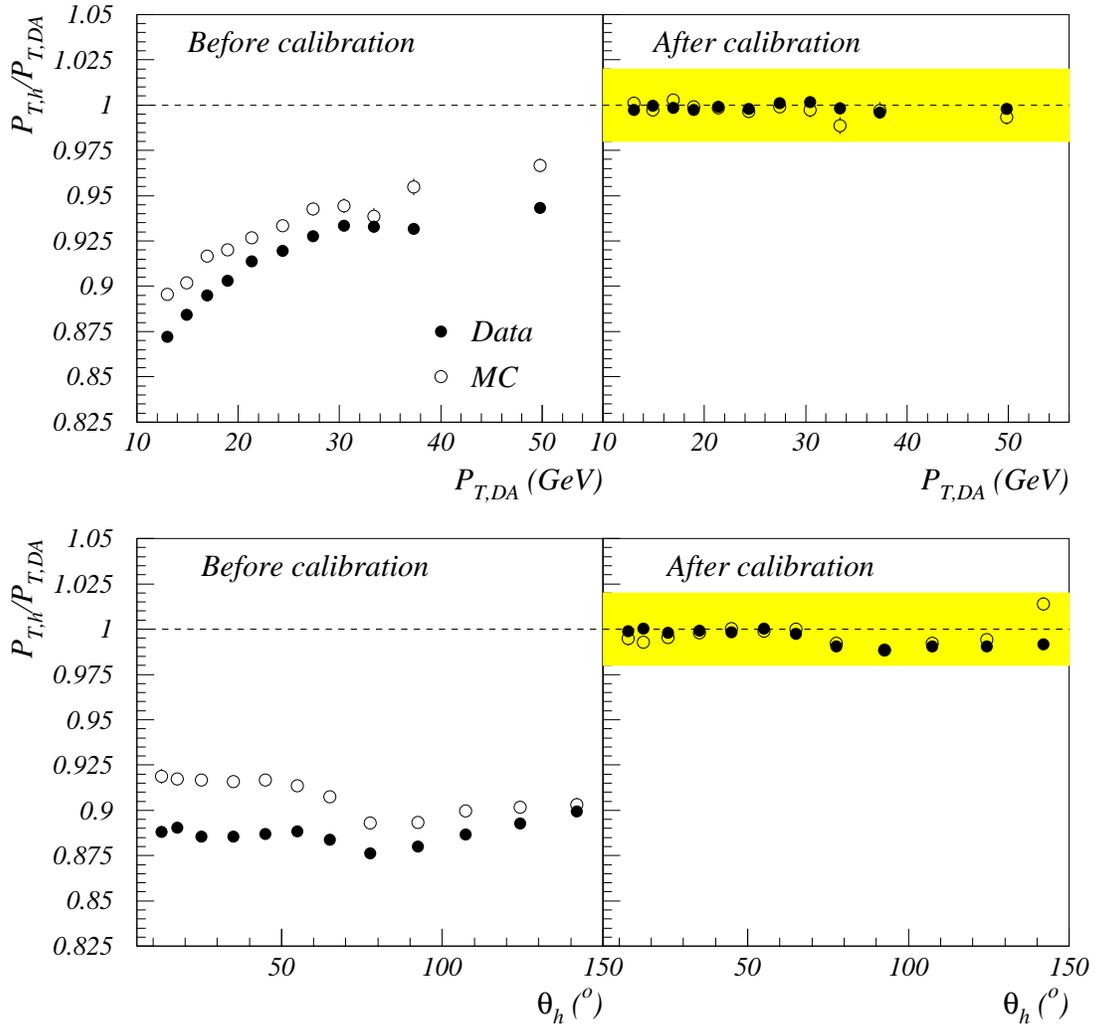,bbllx=0pt,bblly=0pt,
bburx=594pt,bbury=842pt,width=165mm}}
\end{picture}
\end{center}
\caption{\sl The transverse momentum of the hadronic final state, $P_{T,h}$,
over that of the scattered electron, $P_{T,{\rm DA}}$, as predicted with 
the double angle method, as function of $P_{T,{\rm DA}}$ and the inclusive 
angle of the hadronic final state $\theta_e$ before and after the calibration.}
\label{fig:pt_the}
\end{figure}

Fig.\ref{fig:pt_the} shows the performance of the calibration based on a
large inclusive event sample which was selected with less restrictive cuts
as was used for defining the calibration sample. Before the calibration,
the hadronic scale was found to be significantly lower than expected and
there was in addition a strong $P_{T,h}$ dependence. The difference between
data and the simulation was about 4\%. After the calibration, a precision of
2\% was achieved for the first time in H1 both on the hadronic scale and the
systematic difference between data and simulation. 

The performance of the calibration was further checked on a few other aspects:
\begin{itemize}
\item {\bf Hadronization model dependence:} The check concerns whether there is
 any dependence of the calibration on the usage of two different 
 hadronization models in the Monte Carlo event simulations. One model uses
 the color dipole model in its {\sc ariadne} implementation and the other
 model uses the matrix element plus parton shower model (MEPS) as implemented
 in {\sc lepto}. While some differences were found in each calibration step,
 the overall calibration procedure was found to be independent of the Monte
 Carlo models as expected because of the imposed balance in transverse 
 momentum.
\item {\bf Independent event sample:} While the calibration correction
 factors were obtained using the clean one jet sample, the performance of the
 calibration was checked with an independent sample in which two or more jets
 are reconstructed. 
\item {\bf Other jet reconstruction algorithms:} The calibration
 factors obtained from the cone algorithm were applied to jets which were
 reconstructed with two other jet algorithms to check the dependence on the
 jet reconstruction.
\end{itemize}
In all cases, the results obtained were within the quoted systematic
uncertainty. In addition, the absolute energy calibration also improved the
relative energy resolution $\sigma(P_{T,h})/P_{T,h}$ by 5\%
at $P_T\simeq 35$\,GeV and 15\% at $P_T\sim 12$\,GeV~\cite{h1note571}. 

\begin{figure}[htbp]
\begin{center}
\begin{picture}(50,360)
\put(-195,-225){\epsfig{file=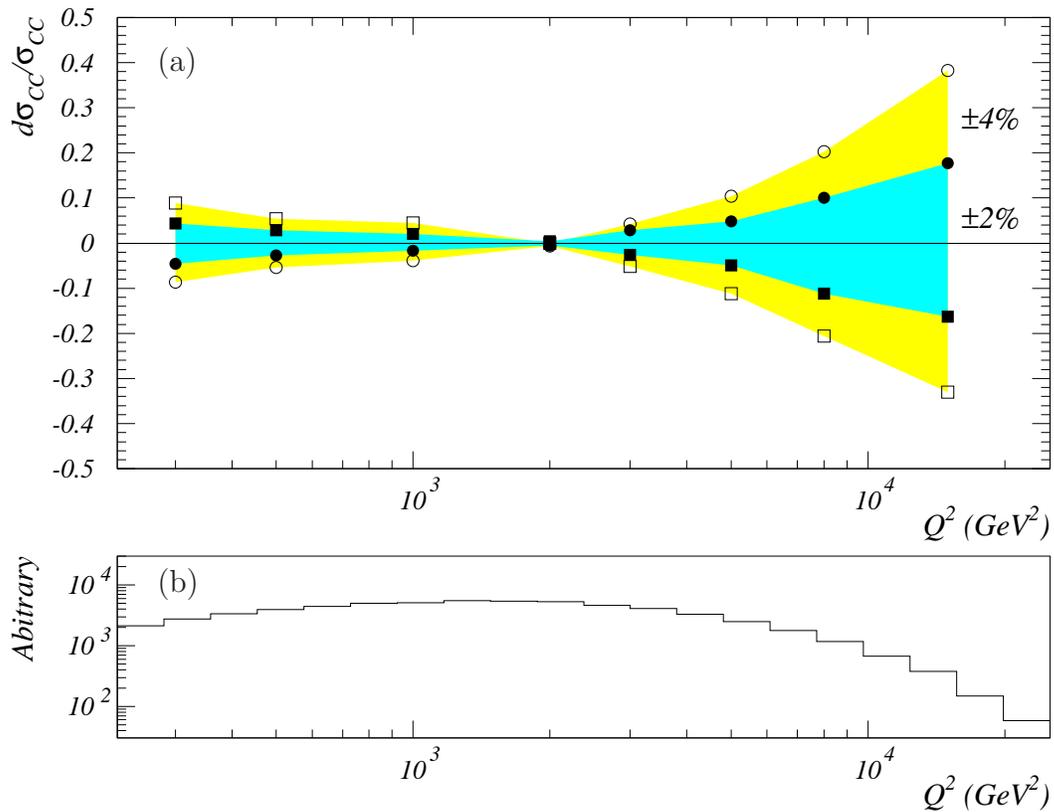,bbllx=0pt,bblly=0pt,
bburx=594pt,bbury=842pt,width=165mm}}
\put(-120,265){(a)}
\put(-120,68){(b)}
\end{picture}
\end{center}
\caption{\sl (a) The resulting uncertainties on the measurement of
the $e^+p$ CC cross section from the improved hadronic energy scale uncertainty
of $\pm 2$\% (dark sharded area enclosed with full points) in comparison with 
an uncertainty of $\pm 4$\% (light shared area enclosed with open points). 
(b) The selected event distribution as a function of $Q^2$ upon
which the form of the uncertainty depends.}
\label{fig:syst_hs}
\end{figure}

The reduced systematic uncertainty on the hadronic energy scale has 
an important impact on analyses relying on the hadronic final state. 
One example is the measurement of the charged current cross sections.
This is illustrated in Fig.\ref{fig:syst_hs}(a).
The magnitude and turn-over behavior at around 2000\,GeV$^2$ depend on the
selected event distribution shown in Fig.\ref{fig:syst_hs}(b). At the
intermediate $Q^2$, the event distribution is rather flat and the energy
scale uncertainty on the cross section is smallest, as the event
migration from one side compensates the migration from the other side. 
At lower $Q^2$
the drop in the event distribution is completely due to the the selection cuts,
while at higher $Q^2$ the selection efficiency is very high and therefore 
the shape of the event distribution is intrinsically due to the falling 
cross section as $Q^2$ increases.

\subsection{Calorimetric noise suppression} \label{sec:noise}

One of the problems in the analyses relying on
the liquid argon calorimeter (LAr) is the distorted measurement of the
kinematic variables at the low $y$ region~\cite{h1note571}. The problem is
related to the fact that at
low $y \lesssim 0.05$, hadrons are produced in the forward direction.
The $z$ component $p_{z,h}$ is comparable with the energy $E_h$ 
thereby resulting in 
a small value of $\Sigma_h\equiv E_h-p_{z,h} (\lesssim 2.76\,GeV)$. 
Any background energy located in the
barrel and backward calorimeter if taken as part of the hadronic final state
will thus distort significantly the total $\Sigma_h$ (the more backward 
a given energy $E_h$ is deposited, the larger the resulting $E_h-p_{z,h}$).
Background energies arise primarily from the following sources:
\begin{itemize}
\item {\bf Electronic noise:} the noise inherent in the electronics of the
calorimeter readout,
\item {\bf Backscattering:} the contribution from the secondary scattering of
final state particles,
\item {\bf Pile-up events:} the contribution from the accidental coincidence 
of a low $y$ event with a background process, e.g.\ a beam halo or 
a cosmic ray event,\footnote{The LAr time sensitivity of several microseconds
allows energy deposits of these background events to pile up on real physics
events.}
\item {\bf Hadron/photon separation:} the large angle radiative photons 
measured in the LAr are misinterpreted as part of the hadronic final state 
if they are not explicitly identified,
\item {\bf Hadron/electron separation:} for neutral current events, a wrong
separation between the scattered electron and the hadronic final state can
result in a biased measurement for the $E_h-p_{z,h}$ measurement. 
This could happen when the
scattered electron is misidentified. The imperfect cluster algorithm
can give rise to multiple clusters for the scattered electron
in particular when it hits a $\phi$ crack between two octants,
or a $z$ crack between two wheels, or the overlapping region between
the BBE and the backward calorimeter SPACAL. In such a case, part of the
electron energy may be wrongly attributed to hadrons.
Since the electron is located predominantly at large polar angle
in the backward region of the calorimeter, a fraction of its energy if 
assigned to hadrons may dominate over the small $\Sigma_h$ in the forward 
region, thus bias the hadronic measurement.
\end{itemize}
The last source, which affects only neutral current events, 
can be eliminated by improving either the current electron
identification programs or the clustering algorithm.
The radiative contribution
is more difficult to deal with for an inclusive analysis.
In order to study the other noise contributions, these last two contributions 
are explicitly removed or suppressed from Monte Carlo samples used. 

\begin{figure}[htb] 
\begin{center}
\setlength{\unitlength}{1mm}
\begin{picture}(70,115)
\put(-37.5,-40){
\epsfig{file=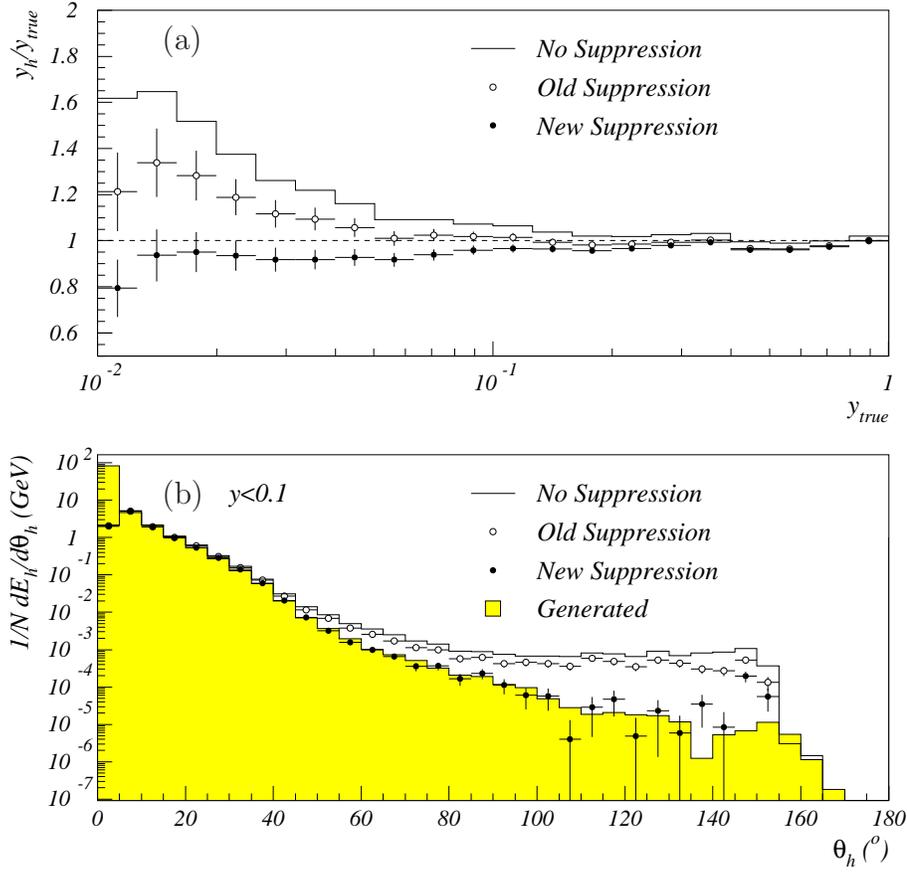,
bbllx=0pt,bblly=0pt,bburx=594pt,bbury=842pt,width=14cm}}
\put(-8,107){(a)}
\put(-8,46.5){(b)}
\end{picture}
\caption {\sl A comparison of the measured $y_h$ as a function of $y$ (a),
and of the measured $E_h$ as a function of $\theta_h$
at low $y<0.1$ (b) based on the neutral current MC files.}
\label{fig:noise_sup_com}
\end{center}
\end{figure}

Part of the electronic noise has been suppressed during the online
and the offline reconstruction~\cite{vladimir288}. The remaining
background contribution was further suppressed in the past by various noise
suppression algorithms. The basic idea of one of these algorithms, to be
called ``old suppression''~\cite{fscomb} in the following, is to identify and
subsequently suppress isolated low energy deposits (clusters)
in the LAr calorimeter. After the suppression, a significant improvement is
achieved in particular for $y$ down to $\sim 0.05$
(Fig.\ref{fig:noise_sup_com}(a)). For the region at smaller $y$, 
there is still an important bias to the measured $y_h$. 

A closer look at the energy distribution as a function of $\theta_h$
(Fig.\ref{fig:noise_sup_com}(b)) for $y<0.1$ shows that there is an energy
surplus towards the large angles. A detailed study~\cite{note580} revealed
that two distinct types of background energies contribute: energetic deposits
affect a small fraction of the sample and less energetic ones affect the
majority of the events. The first type originates from the accidental
coincidence of a DIS event with a halo or a cosmic muon event. These
background muons can produce rather energetic showers resulting in very
large distortions in the measurement of $y_h$ (up to an order of magnitude
with respect to $y_{\rm true}$). Two specific algorithms were thus developed
to identify the background energy patterns and subsequently suppress them
from the energy measurement. The second type is due to other
contributions such as the residual electronic noise and the backscattering
contribution. To suppress these background energies, a higher energy threshold
was found necessary~\cite{note580}. After the ``new suppression'',
the improvement is clearly demonstrated not only on the average value of
the measured $y_h$ (Fig.\ref{fig:noise_sup_com}(a)) but also in terms of
the energy distribution over the detector (Fig.\ref{fig:noise_sup_com}(b)).
It should be pointed out that the measured $E_h$ and $y_h$
do not have to coincide with the generated quantities as the absolute
calibration discussed in the previous section is not applied for this
comparison. At very low $y$, the energy loss in the beam pipe becomes 
increasingly important (see the energy distributions at small angles in 
Fig.\ref{fig:noise_sup_com}(b)), one expects $y_h$ to be smaller than 
$y_{\rm true}$.

\begin{figure}[htbp]
\begin{center}
\begin{picture}(50,180)
\put(-210,-345){\epsfig{file=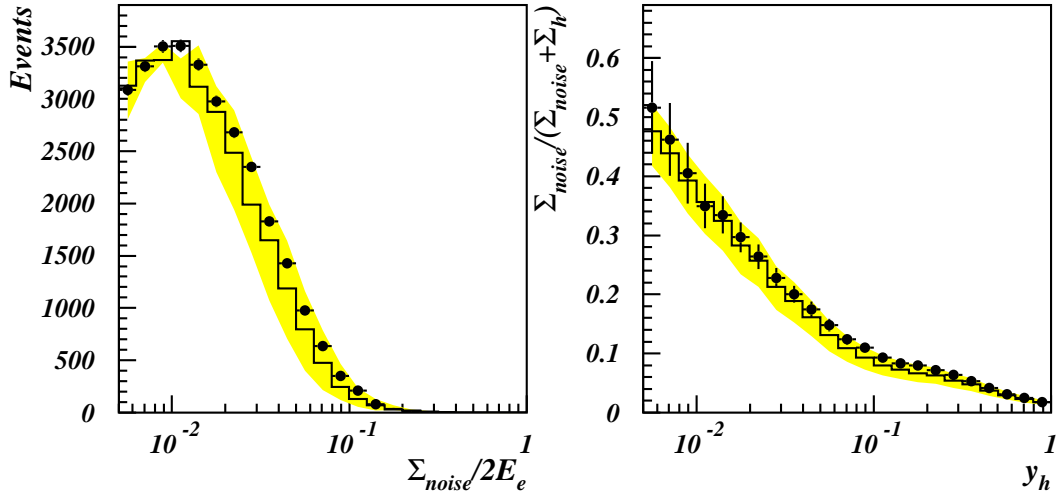,bbllx=0pt,bblly=0pt,
bburx=594pt,bbury=842pt,width=165mm}}
\end{picture}
\end{center}
\caption{\sl Comparison of (a) the event distribution as a function of
$\Sigma_{\rm noise}/2E_e$, (b) the relative contribution of the suppressed
$\Sigma_{\rm noise}$ to the total measured
$\Sigma (=\Sigma_h+\Sigma_{\rm noise})$ as a function of $y_h$.
The data (points) are the NC events from the $e^-p$ data taken in 1998-1999,
while the histograms show the corresponding Monte Carlo simulation. The
shaded error bands represent the quoted systematic uncertainty ($\pm 25\%$)
on the suppressed noise contribution.}
\label{fig:noise_com}
\end{figure}

It has been checked~\cite{note580} based on the simulated MC samples that
what has been suppressed were indeed the various noise contributions discussed
above and there was no evidence that signal energies were affected.
The suppressed noise in the data is finally compared with that in 
the simulation using the NC $e^-p$ data taken in the years from 1998 to 1999 
as an example.
First in Fig.\ref{fig:noise_com}(a), the event distribution as a function of
$\Sigma_{\rm noise}/2E_e$ 
($\Sigma_{\rm noise}=E_{\rm noise}-p_{z,{\rm noise}}$)
is compared. In Fig.\ref{fig:noise_com}(b), the relative contribution of
the suppressed $\Sigma_{\rm noise}$ to the total measured $\Sigma$ including 
$\Sigma_{\rm noise}$ is shown as a function of $y_h$, which does not include 
the contribution from the suppressed noise. The shaded bands show the quoted
systematic uncertainty ($\pm 25\%$) on the noise suppression. Within the
uncertainty quoted, the data are well described by the simulation.

\subsection{Trigger efficiency for charged current events}
\label{sec:cctrig}
The trigger efficiency for CC events is one of the most
critical parts among the various efficiencies needed for the cross section
measurements as it was not automatically simulated in the Monte Carlo 
production. 

The logics designed for triggering CC events are based mainly on the missing 
transverse energies provided by the LAr calorimeter and on the timing 
information from either the proportional chambers or the LAr calorimeter.
From the detector point of view, the hadronic final state of a CC event is the
same as that of a NC event. Therefore, the CC trigger efficiency as well as 
other CC efficiencies can be determined from high statistics NC events from 
real data when all information associated to the scattered electron is removed.
This is the so called pseudo-CC sample\footnote{These events were further 
reweighted to the CC cross section such that they behave kinematically just 
like real CC events.} mentioned already in Sec.\ref{sec:evtsel}.

\begin{figure}[htbp]
\begin{center}
\begin{picture}(50,180)
\put(-200,-345){\epsfig{file=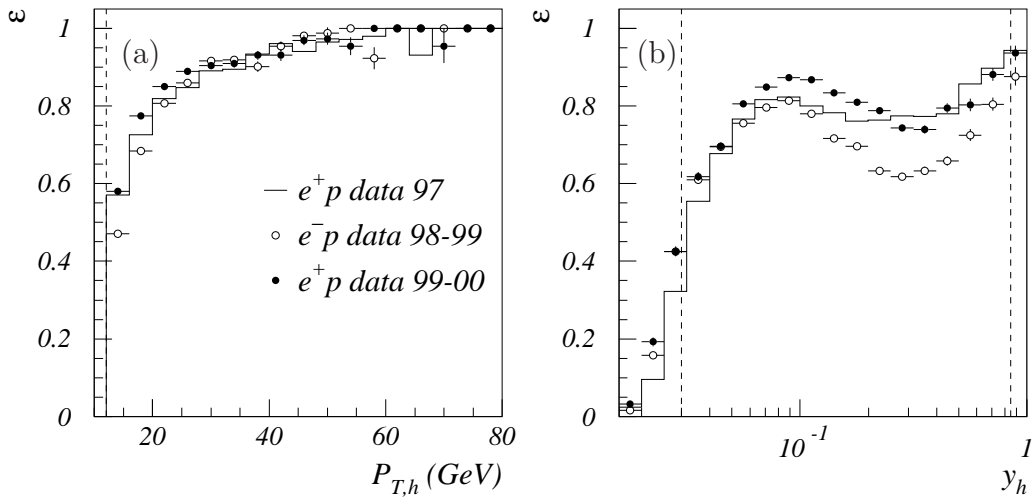,bbllx=0pt,bblly=0pt,
bburx=594pt,bbury=842pt,width=165mm}}
\put(-130,145){(a)}
\put(65,145){(b)}
\end{picture}
\end{center}
\caption{\sl Comparisons of CC trigger efficiencies from three different run
periods as a function of (a) the missing transverse momentum $P_{T,h}$ and
(b) the kinematical variable $y_h$. The selection cuts $P_{T,h}>12$\,GeV and
$0.03<y_h<0.85$ are indicated with the dashed lines.}
\label{fig:trig_cc}
\end{figure}

The CC trigger efficiencies for three different run periods are compared in
Fig.\ref{fig:trig_cc} as functions of the missing transverse momentum
$P_{T,h}$ and the kinematic variable $y_h$.
The lower cuts on these two quantities in the analysis are respectively 
12\,GeV and 0.03 (Sec.\ref{sec:evtsel}). This ensures that the trigger 
efficiency is always above $\sim 50\%$. During the shutdown between 
1997 and 1998, part of the preamplifiers with large capacity were upgraded by
low-noise amplifiers.
With this upgrade~\cite{h1note470} and other hardware
modifications~\cite{h1note559}, the trigger efficiency is expected to be
improved as indeed seen at the low $y$ region. Unfortunately, during the first
part ($\sim 64\%$) of the $e^-p$ data taking period, the missing transverse
energies provided from the LAr trigger towers were modified to correspond to
an earlier bunch crossing instead of the nominal one thereby resulting in a
significant efficiency loss. The problem was fixed in the second part of the
data taking. The efficiency shown in Fig.\ref{fig:trig_cc} for the $e^-p$
data is the averaged value of the two parts.
The inefficiency at the low $y$ region occurs because two big towers close to
the forward beam pipe are not considered so far for the sum in the missing
transverse energies. Currently there has been efforts investigating the
possibility of reopening these trigger towers to improve the situation.

\section{Measurement of inclusive cross sections at high $Q^2$} 
\label{sec:hiq2_xs}
In this section, many cross section results will be given and compared 
with the standard DIS expectations. The cross sections are
obtained following the measurement procedure shown in Sec.\ref{sec:method}.
The results are presented in the form of reduced cross sections, which
are related with the double differential cross sections (Eq.(\ref{eq:xsnc})) 
as:
\begin{equation}
\tilde{\sigma}^\pm=\frac{Q^4x}{2\pi\alpha^2Y_+}\frac{d^2\sigma^\pm}{dxdQ^2}
=\tilde{F}_2(x,Q^2)-\frac{y^2}{Y_+}\tilde{F}_L(x,Q^2)\mp
\frac{Y_-}{Y_+}x\tilde{F}_3(x,Q^2)
\label{eq:redxs_hiq}
\end{equation}
where $\sigma^\pm$ stand for cross sections respectively for $e^+p$ and 
$e^-p$ collisions and $Y_\pm=1\pm (1-y)^2$ are the helicity functions.

Similarly the CC reduced cross sections are related with the double
differential cross sections (Eq.(\ref{eq:xscc}))\footnote{In the considered
kinematic range ($Q^2\gtrsim 300\,{\rm GeV}^2$), the proton mass term $M^2$ 
can be neglected.}
\begin{equation}
\tilde{\sigma}^\pm_{\rm CC}=\frac{2\pi
x}{G^2_F}\left(\frac{Q^2+M^2_W}{M^2_W}\right)^2\frac{d^2\sigma^\pm_{\rm
CC}}{dxdQ^2}=\frac{1}{2}\left(Y_+F^{\rm CC}_2 -y^2F^{\rm CC}_L \mp
Y_-xF^{\rm CC}_3\right) \label{eq:redxs_cc}
\end{equation}
and have the following
simple quark flavor decomposition when the LO structure function formula 
in Eqs.(\ref{eq:f2cce-})-(\ref{eq:f3cce+}) are used:
\begin{eqnarray}
& & \tilde{\sigma}^+_{\rm CC}=x\left[\left(\overline{u}+\overline{c}\right)+
(1-y)^2\left(d+s\right)\right] \label{eq:redxscce+} \\
& & \tilde{\sigma}^-_{\rm CC}=x\left[\left(u+c\right) +
(1-y)^2\left(\overline{d}+\overline{s}\right)\right]. \label{eq:redxscce-}
\end{eqnarray}

Before presenting these results, various systematic sources and their effects 
on the cross sections will be discussed.

\subsection{Systematic sources and resulting uncertainties on the cross
section measurements}
The uncertainties in the measurement lead to systematic errors on the cross
sections which can be point to point correlated or uncorrelated. All the
correlated systematic errors were checked to be symmetric to a good
approximation and are assumed so in the following. The correlated systematic
errors and main uncorrelated systematic errors of the NC and CC cross
sections are given below:
\begin{itemize}
\item An uncertainty of the energy of the scattered electron is
(see Fig.\ref{fig:ecal_summary})
$1\%$ if the $z$ position\footnote{The variation as a function of $z$ is
mainly a reflection of the statistical precision of the data samples with
which the uncertainty could be checked.} of its impact on the calorimeter is
in the backward part ($z < -145\,{\rm cm}$), $0.7\%$ in the CB1 and CB2 wheels
($-145 < z < 20\,{\rm cm}$), $1.5\%$ for $20 < z < 100\,{\rm cm}$ and
$3\%$ in the forward part ($z > 100\,{\rm cm}$).
These uncertainties are obtained by the quadratic sum of an uncorrelated
uncertainty and a point to point correlated uncertainty.
This correlated uncertainty comes mainly from the potential bias of
the calibration method and is estimated to be $0.5\%$
in the whole LAr calorimeter. The resulting correlated (uncorrelated)
systematic error on the NC cross sections is $\lesssim \ 3 \ (5)\%$ except for
the measurement at the two highest $x$ values~\cite{h1hiq9497}.
Due to the smaller luminosity for the 1998-1999 $e^-p$ data, the uncertainty
in the CB1 and CB2 wheels was verified to 1\% with the same uncertainty for
the rest of the calorimeter region~\cite{h1hiq9899}. 
For the high statistics $e^+p$ data of 1999-2000, the uncertainty of 
the electron energy scale is expected to be improved in particular in the
forward region. For the preliminary cross section results which were based on 
about 2/3 of the full data sample, the uncertainty was conservatively quoted 
to be the same as the $e^-p$ data~\cite{h1hiq9900}.
\item A correlated uncertainty of 1\,mrad on the electron polar angle, and
an uncorrelated uncertainty of 2.8\,mrad when the angle is determined with
the position measured in the calorimeter and the event vertex. The resulting
correlated (uncorrelated) systematic error is small, typically $\lesssim
1(2)\%$.
\item An uncertainty of 2\% on the hadronic energy in the LAr calorimeter
which is obtained from the quadratic sum of an uncorrelated systematic
uncertainty of 1.7\% and a correlated one of 1\% originating from the
calibration method and from the uncertainty of the reference scale of the
scattered electron. The resulting correlated systematic error increases at
low $y$, and is typically $\lesssim 4\%$ except at high $Q^2$ for the CC
measurements (see Fig.\ref{fig:syst_hs}). For the preliminary results of the
1999-2000 $e^+p$ data, a conservative uncertainty of 3\% was quoted. 
It should be noted that the same uncertainty on the hadronic energy measurement 
can result in different uncertainties on the cross section measurements. 
For CC cross sections at high $Q^2$, it is larger for the $e^+p$ data at
$\sqrt{s}\simeq 300\,{\rm GeV}$ and smaller for the $e^-p$ data at
$\sqrt{s}\simeq 320\,{\rm GeV}$ due to different $Q^2$ dependences of the 
cross sections (see Sec.\ref{sec:dsdq2}).
\item An uncertainty of 7(3)\% on the energy of the hadronic final state
measured in the SPACAL (tracking system\footnote{An improvement in the
energy resolution of about $10-20\%$, for events having a $P_{T,h}$ between
10 to 25\,GeV, is obtained by using a combination of the
momentum of low transverse momentum particles ($P_T<2\,{\rm GeV}$) measured
in the central tracking detector with the energy deposited by other
particles of the hadronic final state measured in the calorimeter.}). Their
influence on the cross sections is small compared to the uncorrelated
uncertainty of the LAr calorimeter energy.
\item A correlated uncertainty of 25\% on the energy identified as noise in
the LAr calorimeter (see Sec.\ref{sec:noise}). The resulting systematic
error is largest at low $y$, reaching $10-15\%$ at $x=0.65$ and $Q^2\leq
2000\,{\rm GeV}^2$ in the NC measurements, while it remains below 5\% for 
the CC measurements.
\item A variation of the anti-photoproduction cuts (see
Sec.\ref{sec:evtsel}). The resulting correlated systematic error reaches a
maximum of 12\% at low $x$ and $Q^2$ in the CC analyses.
\item An uncertainty of 30\% on the subtracted photoproduction background.
The resulting correlated error is always smaller than 5\% in the NC and CC
analyses and is further reduced in the analyses of the 1998-1999 $e^-p$ and
1999-2000 $e^+p$ data due to the improved subtraction method for NC and 
the additional anti-photoproduction cut for CC.
\end{itemize}
The other considered uncertainties giving rise to uncorrelated systematic
errors on the cross sections are:
\begin{itemize}
\item An error of 2\% (4\% at $y>0.5$ and $Q^2<500\,{\rm GeV}^2$) from the
electron identification efficiency in the NC analyses.
\item an error of 1\% from the efficiency of the track-cluster link
requirement in the NC analyses.
\item An error of 0.5 ($3-8$)\% from the trigger efficiency in the NC (CC)
analyses.
\item An uncertainty of 1(3)\% from the QED radiative corrections in the NC
(CC) analyses~\cite{beate_thesis}.
\item An error of 3\% from the efficiency of the non-$ep$ background filters
in the CC analyses.
\item An error of 2\% (5\% for $y<0.1$) from the efficiency of the event 
vertex reconstruction in the CC analyses.
\end{itemize}
The typical total systematic error for the NC (CC) double differential cross
section is about 4(8)\%. In addition a normalization error of 1.5\%, 1.8\%, and
1.7\% respectively for the 1994-1997 $e^+p$ data, the 1998-1999 $e^-p$
data and the 1999-2000 $e^+p$ data has to be taken into account, which is
not included in the results shown in the following subsections.

\subsection{Measurement and comparison of NC and CC reduced cross sections in
$e^+p$ collisions at two center-of-mass energies} \label{sec:redxs_e+}
The NC reduced cross sections obtained from 35.6\,pb$^{-1}$ of the $e^+p$ data
taken from 1994 to 1997 at the center-of-mass energy of 300\,GeV has recently
been published by H1~\cite{h1hiq9497}. The measurement covers a kinematic
region for $Q^2$ between 150 and 30\,000\,GeV$^2$, for $x$ between 0.0032 and
0.65, and for $y$ between 0.007 and 0.88. This kinematic range has
significantly extended the previous HERA measurements~\cite{h1f294} both in
$Q^2$ (from 5000 to 30\,000\,GeV$^2$) and towards higher $x$, with
measurements at $x=0.65$ for $Q^2$ between 650 and 20\,000\,GeV$^2$.
At $Q^2\lesssim 500\,{\rm GeV}^2$ the total error is dominated by the
systematic uncertainties in the energy scale and identification efficiency
of the scattered electron and by the uncertainty in the energy scale of the
hadronic final state. In this region the systematic error is typically 4\%.
At higher $Q^2$ the statistical error becomes increasingly dominant.

\begin{figure}[hp]
\begin{center}
\begin{picture}(50,550)
\put(-170,-33){
\epsfig{file=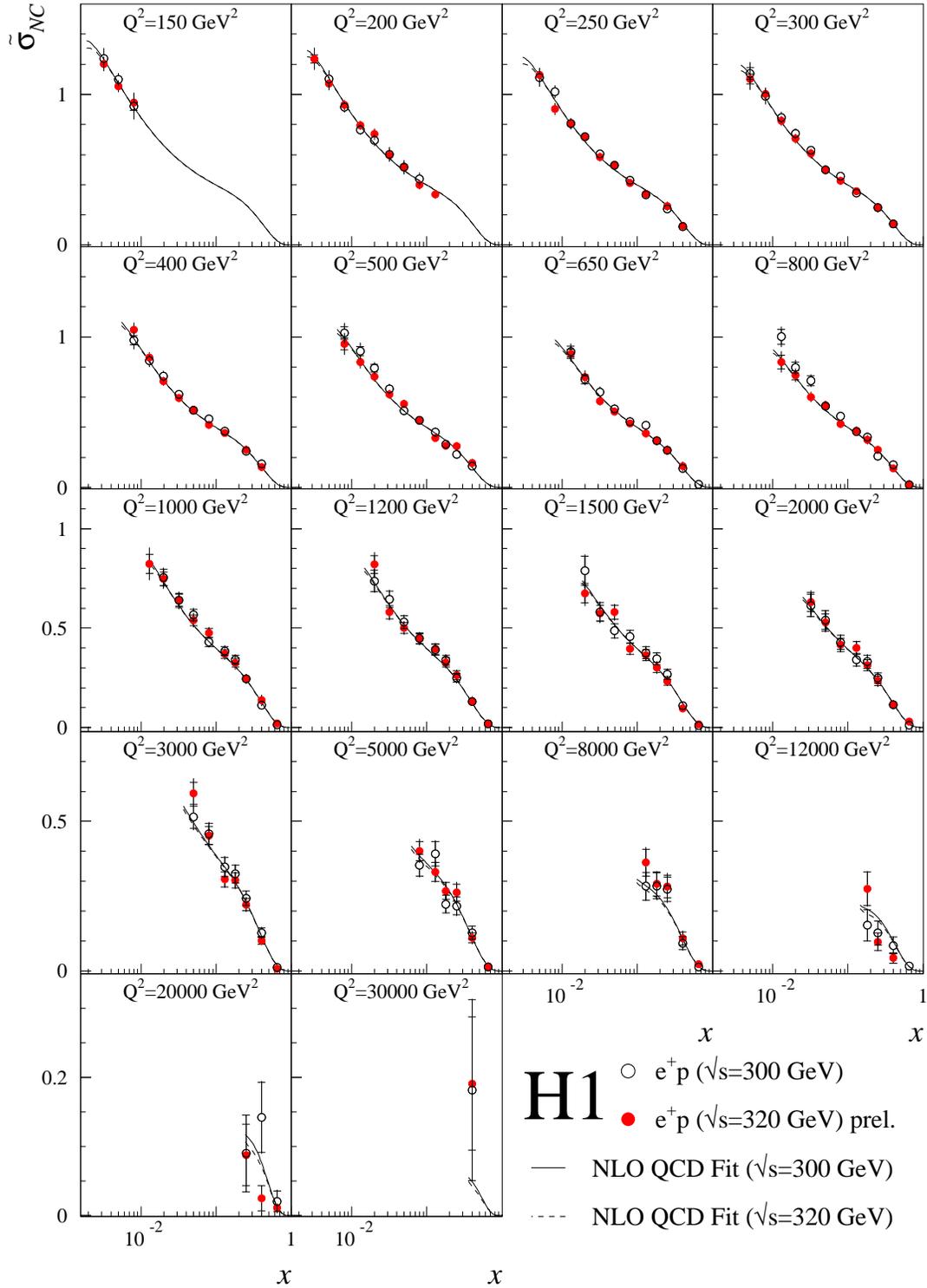,bbllx=48,bblly=30,bburx=550,
bbury=780,width=140mm}}
\end{picture}
\end{center}
\caption{\sl The $e^+p$ NC reduced cross section
$\tilde{\sigma}_{NC}$ is compared to the NLO QCD Fit~\cite{h1hiq9497}. 
Shown are the measurements from 1994-1999 data~\cite{h1hiq9497} and 
from 1999-2000 data~\cite{h1hiq9900} and the NLO QCD Fit predictions
for different center-of-mass energies. The inner error bars represent 
the statistical error, and the outer error bars show the total error.}
\label{fig:nce+_com2}
\end{figure}

The new $e^+p$ data of 45.9\,pb$^{-1}$ at a higher center-of-mass energy of
320\,GeV has been analyzed and the preliminary results have been made available
for the summer conferences~\cite{h1hiq9900}. The new data with slightly
improved statistical precision cover the same kinematic range as the
1994-1997 $e^+p$ data. The two independent measurements are compared
in Fig.\ref{fig:nce+_com2}
and found in good agreement. The cross sections rise strongly as $x$ decreases
as seen in the low $Q^2$ data, which can be interpreted as
the increasing contributions from the gluon and sea quarks at low $x$. 
The measurements
are well described by the standard DIS expectation from a NLO QCD
Fit~\cite{h1hiq9497} to the $e^+p$ 1994-1997 data\footnote{The agreement is
equally good when the input data used in the fit was limited to $Q^2\leq
150\,{\rm GeV}^2$~\cite{h1hiq9497}. The fit is any case independent of the new
$e^+p$ and $e^-p$ data taken since 1998.} as well as H1 structure 
function data at low $Q^2$~\cite{h1f294} and fixed-target data by 
NMC~\cite{nmc95} and BCDMS~\cite{bcdms89}. The difference in the 
center-of-mass energies is also indicated but remains small.

\begin{figure}[hp]
\begin{center}
\begin{picture}(50,400)
\put(-170,-110){
\epsfig{file=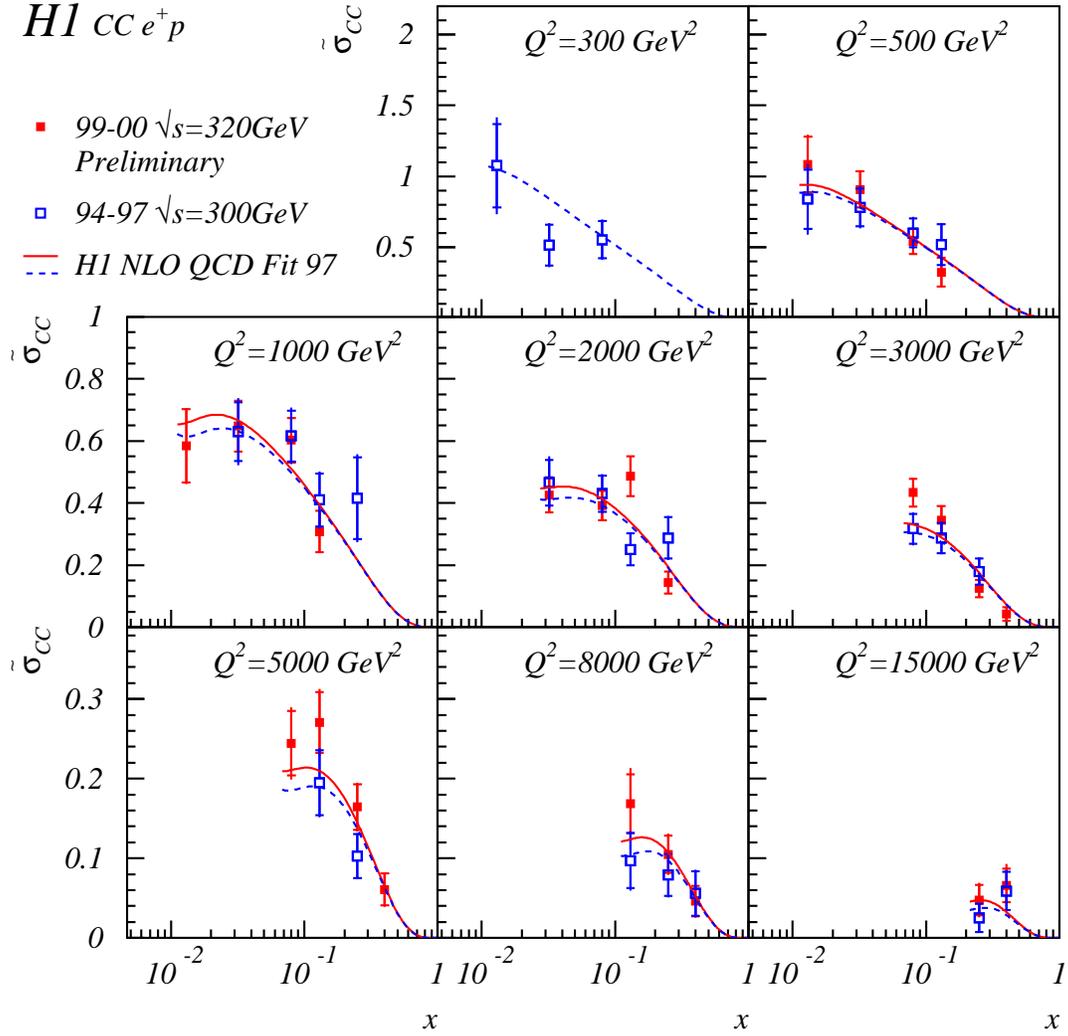,bbllx=48,bblly=30,bburx=550,
bbury=780,width=140mm}}
\end{picture}
\end{center}
\caption{\sl The CC $e^+p$ reduced cross section
$\tilde{\sigma}_{CC}$ is compared to the NLO QCD Fit~\cite{h1hiq9497}. 
Shown are the measurements from 1994-1999 data~\cite{h1hiq9497} and 
from 1999-2000 data~\cite{h1hiq9900} and the NLO QCD Fit predictions 
respectively for $\sqrt{s}=320$\,GeV (full curves) and 300\,GeV 
(dashed curves).
The inner error bars represent 
the statistical error, and the outer error bars show the total error.}
\label{fig:cce+_com2}
\end{figure}

The similar comparison for the CC reduced cross sections is shown in
Fig.\ref{fig:cce+_com2}.
The $e^+p$ 1994-1997 double differential CC cross sections
cover the kinematic region for $Q^2$ between 300 and 15\,000\,GeV$^2$ and 
for $x$ between 0.013 and 0.4~\cite{h1hiq9497}. 
The uncertainties on the measurements are dominated by the statistical errors.
The largest systematic errors come from the uncertainty on the energy scale
of the hadronic final state at high $Q^2$ and from the uncertainty on
the trigger efficiency at low $Q^2$ and high $x$. A similar measurement has
been published by the ZEUS experiment~\cite{zeuscc9497}. The new $e^+p$
data~\cite{h1hiq9900}, covering a similar range, are in good agreement with the
published data and both measurements agree also well with the Standard Model 
(SM) expectations based on the same NLO QCD Fit mentioned above.

Given the good agreement between the two data sets and in order to facilitate
the comparison of the $e^+p$ data with the $e^-p$ date in the later sections,
we have combined the $e^+p$ data at different center-of-mass energies
to the higher energy in the following way\footnote{The alternative way would
be to make the correction to the measured cross section at 300\,GeV from
the 1994-1997 $e^+p$ data and to combine the corrected cross section with
the one at 320\,GeV from the 1999-2000 $e^+p$ data with the standard
combination method using error weights. Indeed, the combined results obtained 
from both methods agree in the low $Q^2$ region where the data are precise.
However, the combined results can be substantially different in the region of
higher $Q^2$ and high $x$ where the statistical error of the data is large 
($>30\%$).
The method using the luminosity weights is believed to be less sensitive to 
the statistical fluctuation.}
\begin{equation}
\sigma_i=\frac{\sigma^{\rm meas}_{i,300}{\cal L}_{300}+\sigma^{\rm
meas}_{i,320}{\cal L}_{320}}{{\cal
L}_{300}\left(\sigma^{\rm th}_{i,300}/\sigma^{\rm th}_{i,320}\right)+{\cal
L}_{320}} \label{eq:xs_comb}
\end{equation}
where $\sigma^{\rm meas}_{i,\sqrt{s}}$ and $\sigma^{\rm th}_{i,\sqrt{s}}$
are respectively the measured and expected cross section. The statistical
error is determined correspondingly and the systematic error is assumed to
be fully correlated between the two measurements and 
taken from the new measurement.

\subsection{Measurement and comparison of NC and CC reduced cross sections
in $e^+p$ and $e^-p$ collisions} \label{sec:xs_e+_e-}
\begin{figure}[hp]
\begin{center}
\begin{picture}(50,480)
\put(-180,-70){
\epsfig{file=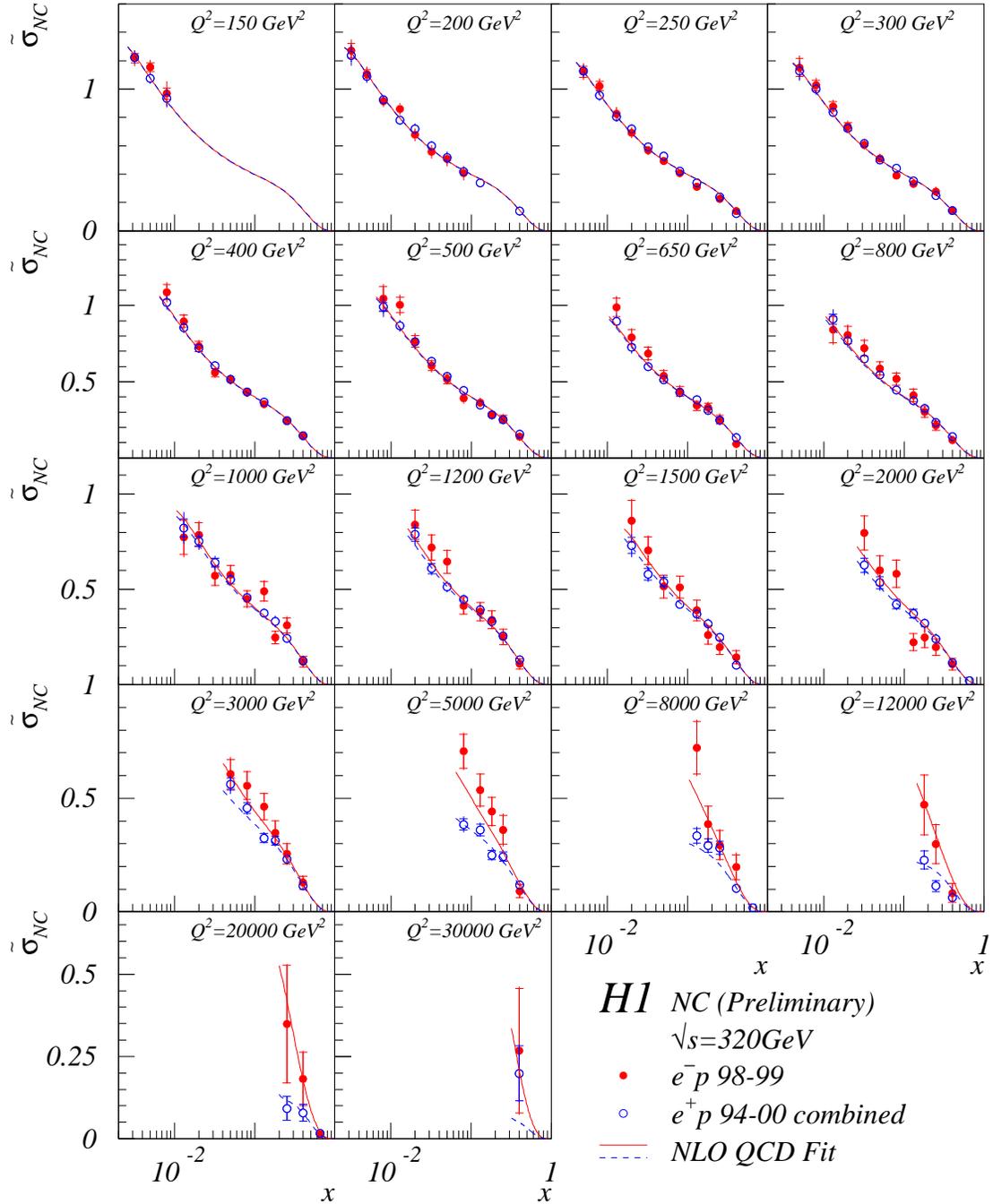,bbllx=48,bblly=30,bburx=550,
bbury=780,width=145mm}}
\end{picture}
\end{center}
\caption{\sl The combined $e^+p$ NC reduced cross section 
$\tilde{\sigma}_{NC}$ from 1994-2000 is compared to the $e^-p$ cross section 
from 1998-1999 and the corresponding expectations from the NLO QCD Fit.
The inner error bars represent the statistical error, and the outer error bars 
show the total error.}
\label{fig:nc_xq2_com}
\end{figure}

The combined $e^+p$ NC reduced cross sections are
compared with the results from the $e^-p$ 1998-1999
data~\cite{h1hiq9899} in Fig.\ref{fig:nc_xq2_com}.
For $Q^2\lesssim 1200\,{\rm GeV}^2$, where the cross section is dominated by
the contribution from one-photon exchange, the data are essentially
independent of the lepton-beam charges. With the increasing $Q^2$
the $e^-p$ cross section becomes larger than the $e^+p$ cross section.
The data thus give the clear evidence for the presence of
the structure function $xF_3$ (Eq.(\ref{eq:redxs_hiq})).
The standard DIS expectations from the NLO QCD Fit~\cite{h1hiq9497}, which was
partly constrained by the 1994-1997 $e^+p$ cross section data, are compared
with the measured cross sections. The fit is found to give a good
description of the $x,Q^2$ behavior of the data, though in some of the phase
space it has a slight tendency to be lower than the $e^-p$ cross sections.

\begin{figure}[hp]
\begin{center}
\begin{picture}(50,400)
\put(-170,-110){
\epsfig{file=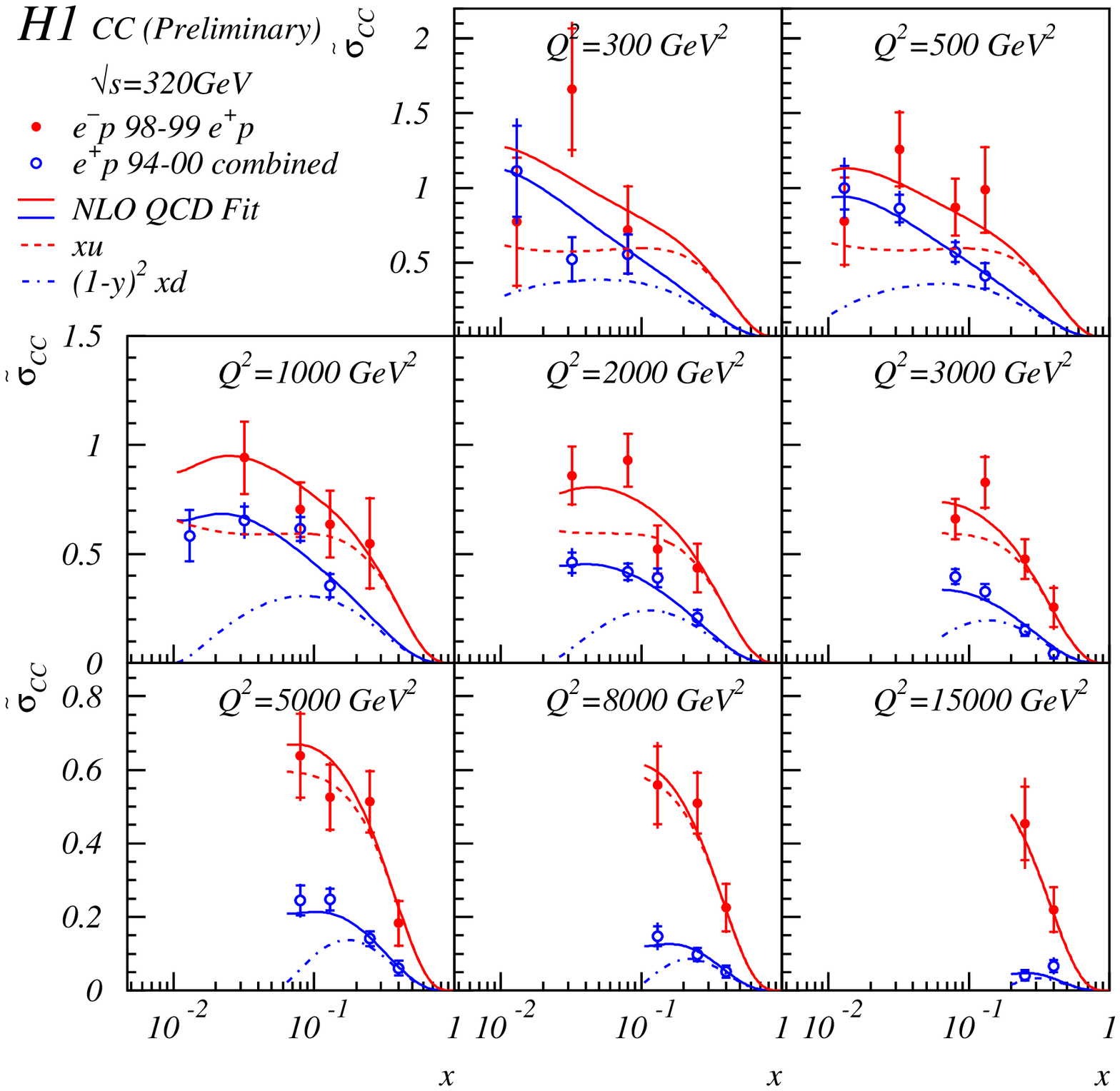,bbllx=48,bblly=30,bburx=550,
bbury=780,width=140mm}}
\end{picture}
\end{center}
\caption{\sl The combined $e^+p$ CC reduced cross section
$\tilde{\sigma}_{CC}$ from 1994-2000 is compared with the $e^-p$ cross section
from 1998-1999 and the corresponding expectations from the NLO QCD Fit. 
The inner error bars represent the statistical error, and the outer error bars
show the total error.}
\label{fig:cc_xq2_com}
\end{figure}

The combined $e^+p$ CC reduced cross sections are compared with the
results of the $e^-p$ 1998-1999 data~\cite{h1hiq9899} in 
Fig.\ref{fig:cc_xq2_com}.
Contrary to the NC data, the CC $e^-p$ data are everywhere different from
the $e^+p$ data. This is the particular feature of the CC interaction
mentioned already in Sec.\ref{sec:qpm}. At low $Q^2$, the difference is less
pronounced because the sea quark contribution is relatively important,
whereas at high $Q^2$ and high $x$, the difference can mostly be attributed
to the underlying quark flavor difference between $xu$ and $(1-y)^2xd$ as
indicated respectively with the dashed and dash-pointed lines in
Fig.\ref{fig:cc_xq2_com}. Therefore the $e^-p$ CC cross section at high
$Q^2$ and high $x$ can be used to constrain the $u$ valence quark while the
$e^+p$ CC cross section provides the constraint for the $d$ valence quark
(see Sec.\ref{sec:ud}). The expectations from the same NLO QCD Fit are found
to give a good description of the measured CC cross sections. 

\subsection{Measurement of the structure function $x\tilde{F}_3$ at high
$Q^2$}\label{sec:xf3}
\begin{figure}[hp]
\begin{center}
\begin{picture}(50,420)
\put(-195,-25){\epsfig{file=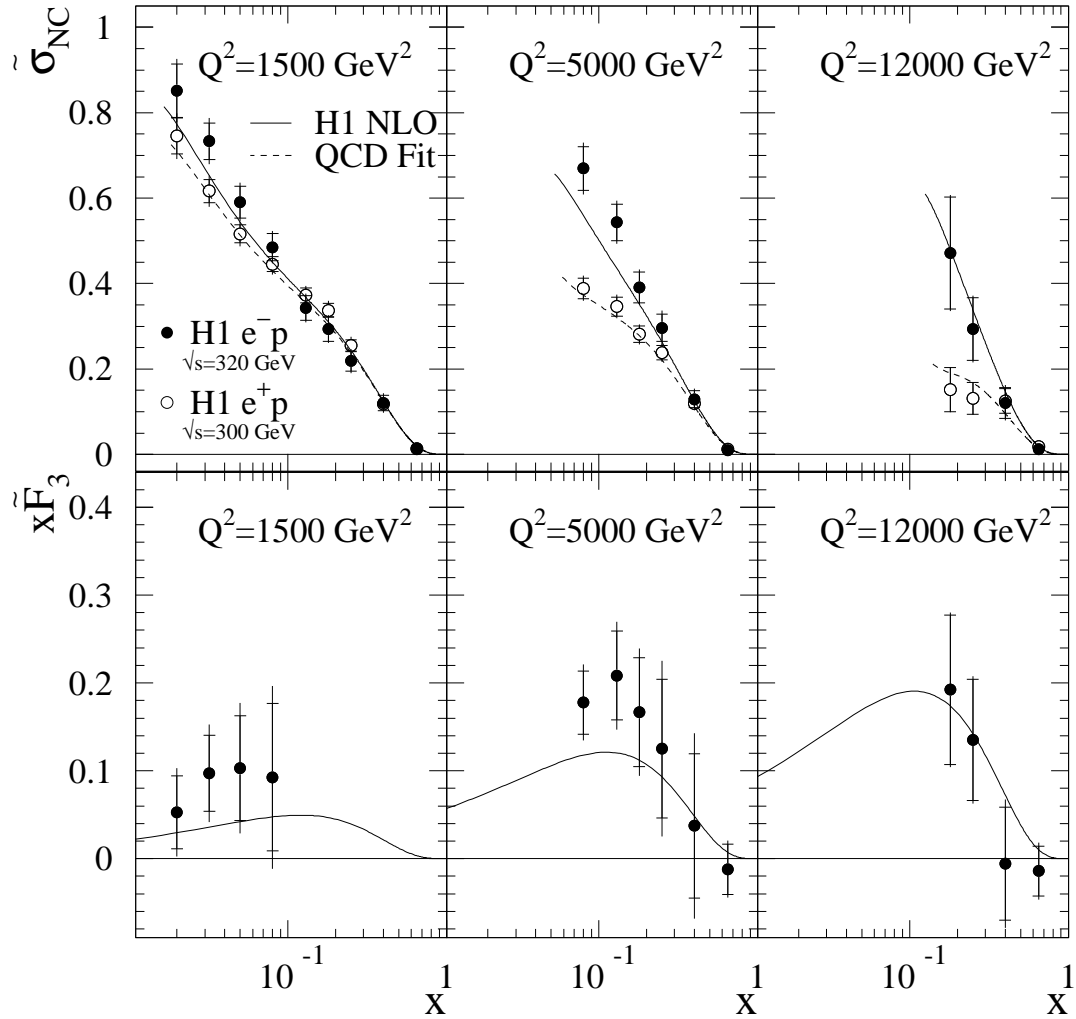,width=155mm}}
\end{picture}
\end{center}
\caption{\sl A measurement of the structure function
$x\tilde{F}_3$ by H1~\cite{h1hiq9899} as a function of $x$ for
three value sof $Q^2$ at 1500\,GeV$^2$, 5000\,GeV$^2$, and 12\,000\,GeV$^2$
based on the $e^+p$ 1994-1997 and
$e^-p$ 1998-1999 NC data. The inner error bars represent the statistical error,
and the outer error bars show the total error. The SM expectations 
for the reduced cross sections and $x\tilde{F}_3$ from the H1 NLO QCD 
Fit~\cite{h1hiq9497} are also shown.}
\label{fig:h1xf3}
\end{figure}

\begin{figure}[hp]
\begin{center}
\begin{picture}(50,360)
\put(-130,-80){
\epsfig{file=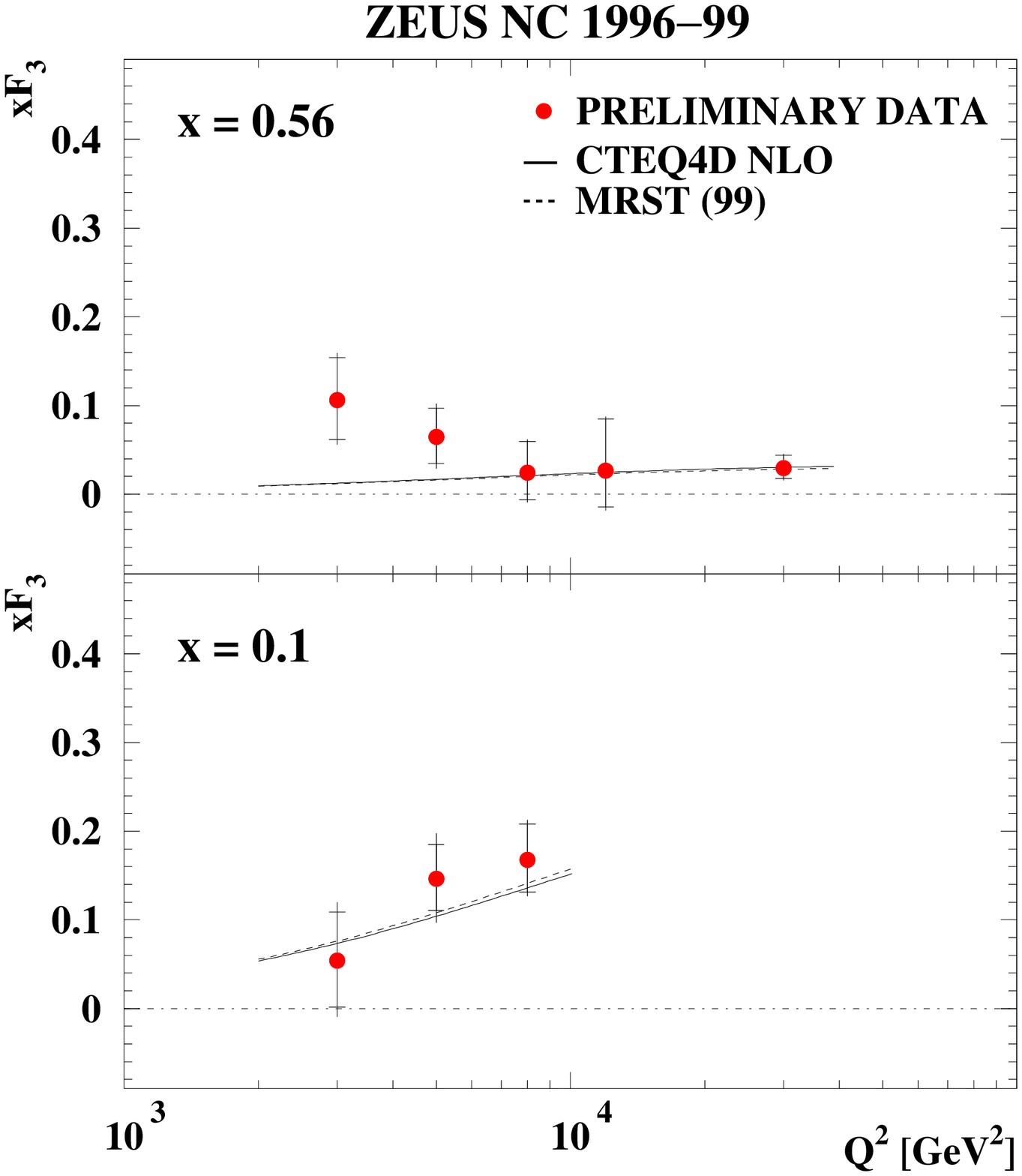,bbllx=48,bblly=30,bburx=550,
bbury=780,width=110mm}}
\end{picture}
\end{center}
\caption{\sl A measurement of the structure function
$x\tilde{F}_3$ by ZEUS~\cite{zeusnc9899} as a function of $Q^2$ for 
two values of $x$ at 0.56 and 0.1 based on the $e^+p$ 1994-1997 and
$e^-p$ 1998-1999 NC data. The inner error bars represent the statistical error,
and the outer error bars show the total error. The SM expectations 
from CTEQ4~\cite{cteq4} and MRST~\cite{mrst} are also shown.}
\label{fig:zeusxf3}
\end{figure}

Using the measured $e^+p$ and $e^-p$ double differential NC cross sections,
both ZEUS~\cite{zeusnc9899} and H1~\cite{h1hiq9899} have measured
the structure function\footnote{It should be pointed out that
the structure function $x\tilde{F}_3(x,Q^2)$ measured at HERA arises
from the contributions of the $\gamma Z^0$ interference and $Z^0$ exchange
with the dominant contribution from the former (Eq.\ref{eq:xf3_nc}), in which
the weak neutral current couplings enter. Therefore it is different from the
corresponding structure function $xF_3=x(u_v+d_v)$, the sum of the $u$ and
$d$ valence quarks (Eq.(\ref{eq:xf3_vn})), measured by neutrino-fixed-target
experiments.} $x\tilde{F}_3(x,Q^2)$ at the high $Q^2$ region.
The measurement was obtained using Eq.(\ref{eq:redxs_hiq}) by subtracting the
$e^+p$ cross section from the $e^-p$ cross section.
Since the data are taken at slight different center-of-mass energies,
a small correction\footnote{The correction, estimated with $\tilde{F}_L$ from
the NLO QCD Fit~\cite{h1hiq9497}, is about 10\% at the lowest $x$ and
negligible elsewhere.} from the longitudinal structure function is needed:
\begin{equation}
x\tilde{F}_3=\left(\frac{Y_{-,320}}{Y_{+,320}}+
\frac{Y_{-,300}}{Y_{+,300}}\right)^{-1}
\left[\left(\tilde{\sigma}^--\tilde{\sigma}^+\right)+
\tilde{F}_L\left(\frac{y^2_{320}}{Y_{+,320}}-
\frac{y^2_{300}}{Y_{+,300}}\right)\right]
\end{equation}
where the subscripts 300 and 320 denote the center-of-mass energies.
The H1 results are shown in Fig.\ref{fig:h1xf3} as a function of $x$ at
three values of $Q^2$ at 1500\,GeV$^2$, 5000\,GeV$^2$, and 12\,000\,GeV$^2$.
The data are in good agreement with the expectation based on the H1 NLO QCD
Fit~\cite{h1hiq9497} with a peak at $x\simeq 0.1$
reflecting the valence quark structure of the constituent quarks.
The ZEUS results are shown in Fig.\ref{fig:zeusxf3} as a function of $Q^2$
for two values of $x$ at 0.56 and 0.1.
The expectation of the SM, evaluated with both
the CTEQ4~\cite{cteq4} and the MRST~\cite{mrst} parton distributions, gives a
good description of the data.

\subsection{Helicity structure of the CC cross sections} \label{sec:phicc}
\begin{figure}[htbp]
\begin{center}
\begin{picture}(50,400)
\put(-172.5,-110){
\epsfig{file=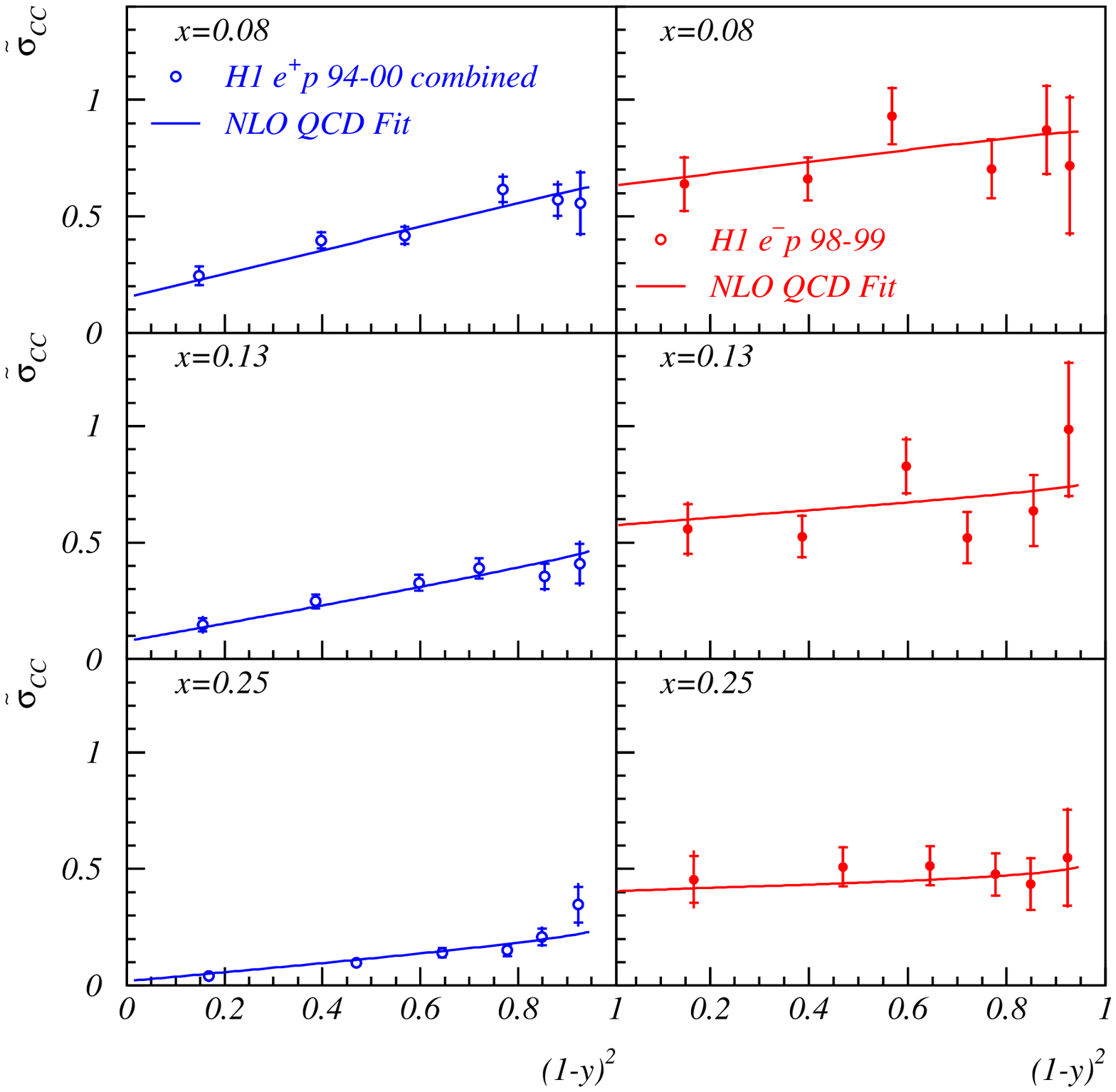,bbllx=48,bblly=30,bburx=550,
bbury=780,width=140mm}}
\end{picture}
\end{center}
\caption{\sl The CC reduced cross sections $\tilde{\sigma}_{\rm CC}$ from
the combined $e^+p$ 1994-2000 data (left) and from the $e^-p$ 1998-1999 data
(right) shown as a function of $(1-y)^2$ for three $x$ values 0.08, 0.13 and
0.25. The inner (outer) error bars represent the statistical (total) errors.
The curves are expectations from the NLO QCD Fit~\cite{h1hiq9497}.}
\label{fig:phicc}
\end{figure}

The measured double differential CC cross sections can be used to test the
helicity dependence of the electron-quark interaction in the region of
approximate Bjorken scaling, $x\sim 0.1$. In Fig.\ref{fig:phicc}, the
reduced cross sections\footnote{In Refs.\cite{h1hiq9497,h1hiq9899},
the so-called structure function term $\phi_{\rm CC}$ was introduced, which is
related to the reduced CC cross section by a weak correction factor
$\tilde{\sigma}=\phi (1+\delta_{\rm weak})$. However, since this correction is
small with respect to the uncertainty of the measurements,
we choose not to distinguish them here.}
from the combined $e^+p$ 1994-2000 data and the $e^-p$ 1998-1999 data
are shown as a function of $(1-y)^2$. This kinematic variable is directly
related to the scattering angle $\theta^\ast$ in the electron-quark
center-of-mass system through $\cos^2(\theta^\ast/2)=1-y$.
The $e^+p$ and $e^-p$ data show a different behavior. This is consistent
with the expectations from the NLO QCD Fit~\cite{h1hiq9497}. Indeed, from
Eqs.(\ref{eq:redxscce+}) and (\ref{eq:redxscce-}), one expects a small
isotropic distribution from positron-antiquark scattering
$(\overline{u},\overline{c})$ and a strong angular dependence on $(1-y)^2$
from positron-quark scattering $(d,s)$, while for the $e^-p$ interaction,
the isotropic distribution is larger $(u,c)$ and the angular dependence is
weaker $(\overline{d},\overline{s})$. In addition, both the isotropic
distribution in the $e^+p$ cross sections and the angular dependence in the
$e^-p$ cross sections increase as $x$ decreases, this is consistent with an
increasing sea quark contribution at lower $x$.

\subsection{Measurement and comparison of the $Q^2$ dependence of NC and CC
cross sections in $e^+p$ and $e^-p$ collisions}
\label{sec:dsdq2}
\begin{figure}[hp]
\begin{center}
\begin{picture}(50,500)
\put(-170,-60){
\epsfig{file=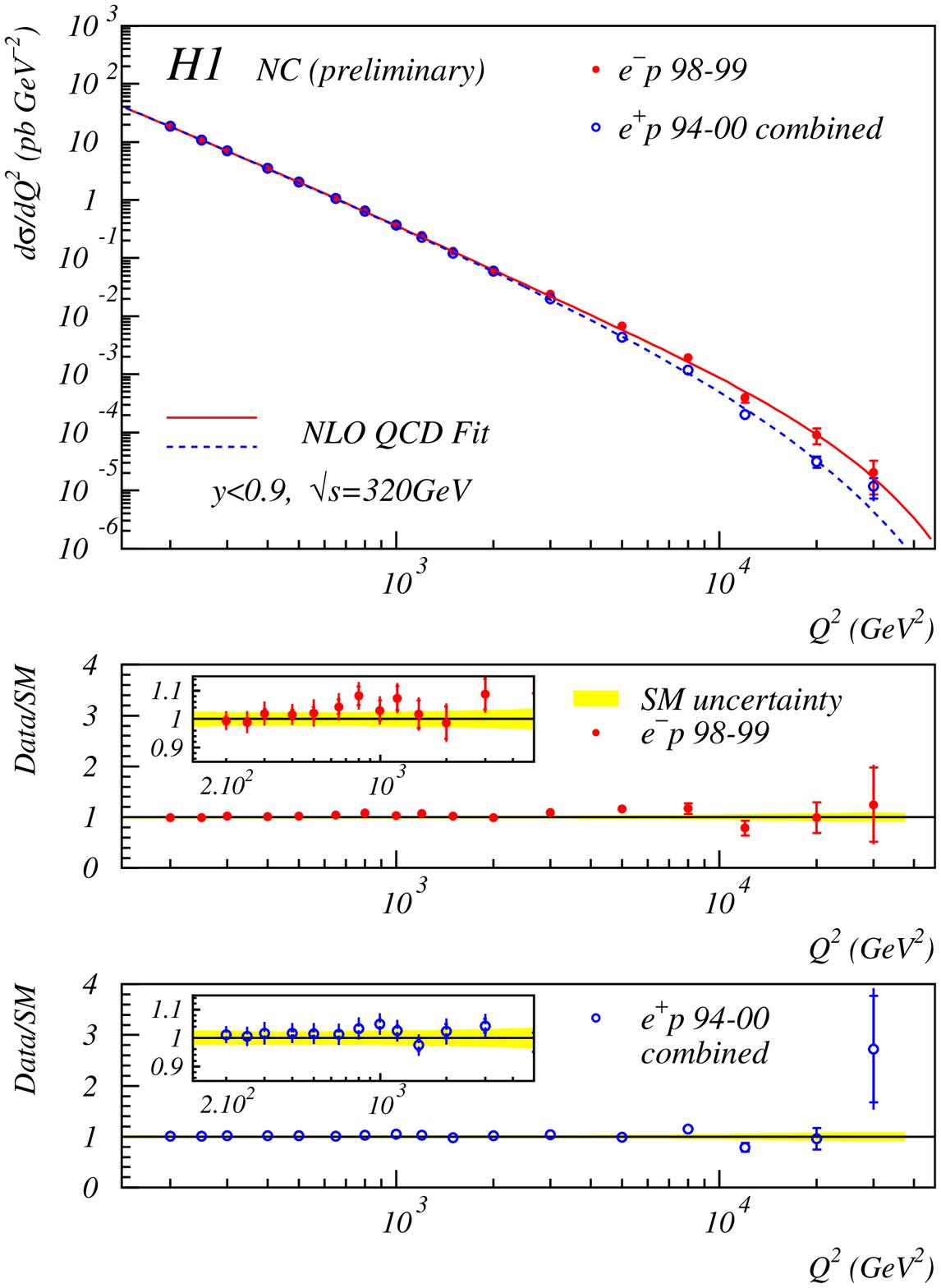,bbllx=48,bblly=30,bburx=550,
bbury=780,width=140mm}}
\end{picture}
\end{center}
\caption{\sl The combined $e^+p$ NC cross sections $d\sigma/dQ^2$
from 1994-2000 are compared with the $e^-p$
cross sections from 1998-1999 and the corresponding expectations from
the NLO QCD Fit. The inner error bars represent the statistical error, 
and the outer error bars show the total error. The shaded error bands
represent the uncertainties on the SM expectations of the $e^-p$
and $e^+p$ cross sections.}
\label{fig:nc_q2_com}
\end{figure}

\begin{figure}[hp]
\begin{center}
\begin{picture}(50,500)
\put(-170,-60){
\epsfig{file=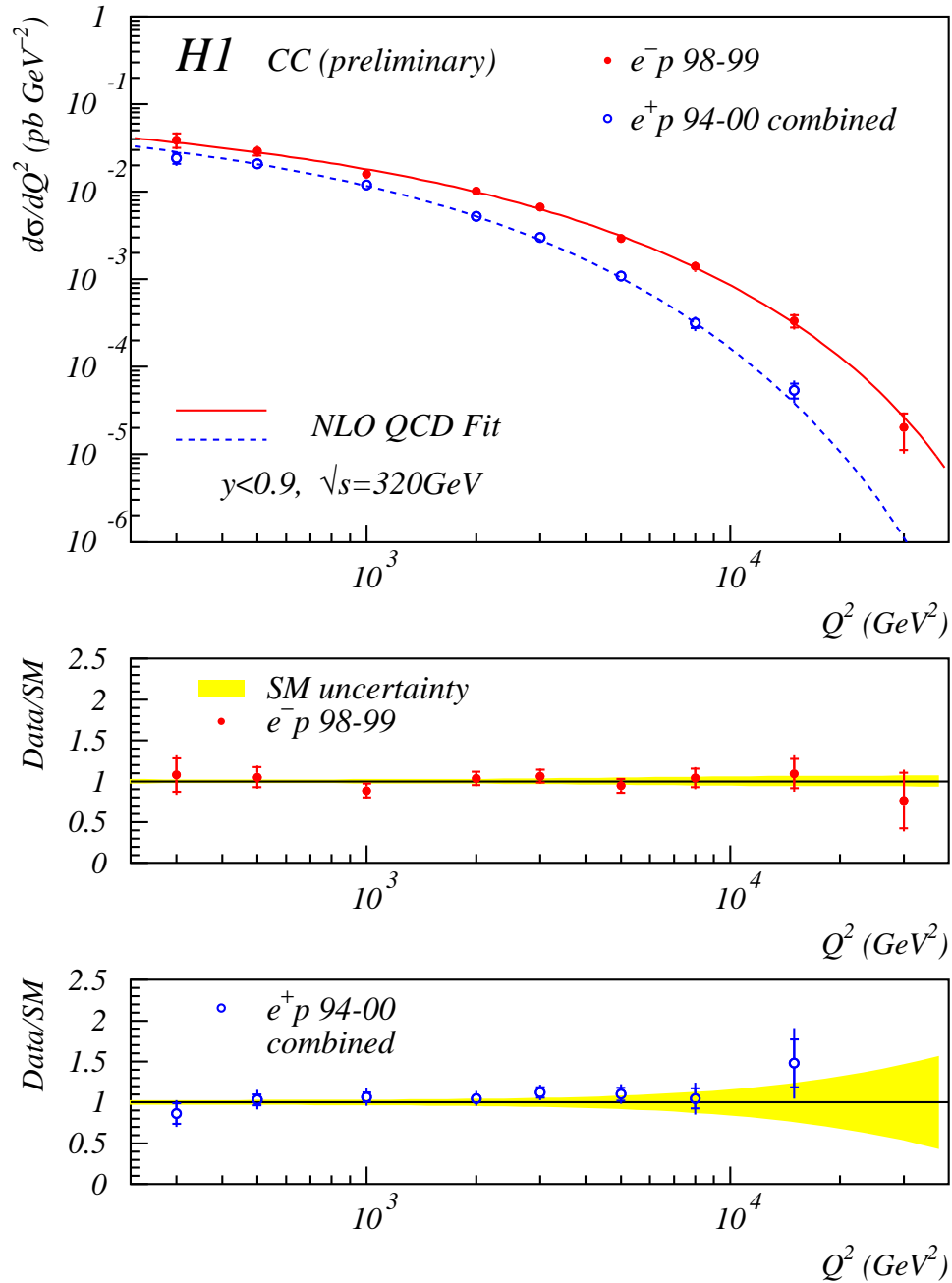,bbllx=48,bblly=30,bburx=550,
bbury=780,width=140mm}}
\end{picture}
\end{center}
\caption{\sl The combined $e^+p$ CC cross sections $d\sigma/dQ^2$
from 1994-2000 are compared with the $e^-p$
cross section from 1998-1999 and the corresponding expectations from
the NLO QCD Fit. The inner error bars represent the statistical error,
and the outer error bars show the total error. The shaded error bands
represent the uncertainties on the SM expectations of the $e^-p$ 
and $e^+p$ cross sections.}
\label{fig:cc_q2_com}
\end{figure}

The single differential cross sections $d\sigma_{\rm NC(CC)}/dQ^2$ have also
been measured by H1~\cite{h1hiq9497,h1hiq9900,h1hiq9899} and by
ZEUS~\cite{zeusnc9497,zeuscc9497,zeusnc9899,zeuscc9899}. 
The combined $e^+p$ NC and CC data are compared with the 
$e^-p$ data respectively in Figs.\ref{fig:nc_q2_com} and \ref{fig:cc_q2_com}.
The measurement of the NC cross sections spans more than two orders of
magnitude in $Q^2$. The cross sections fall with $Q^2$ by about
six orders of magnitude. The $e^-p$ NC data have a statistical precision
varying between 1.3\% and 58\% from low to high $Q^2$ and are dominated
by the systematic uncertainty for $Q^2\lesssim 1000\,{\rm GeV}^2$.
The combined $e^+p$ NC data have a better statistical precision
between 0.5\% to 38\%, which dominates over the conservatively
estimated systematic uncertainty only at $Q^2>8000\,{\rm GeV}^2$.
The ratio of the measurements with the
SM expectations, represented in the middle and lower figures
respectively for the $e^-p$ and the combined $e^+p$ data, show that the data
have no significant deviation from the SM DIS expectations,
albeit may have some structure at the very high $Q^2$ region in particular
for the combined $e^+p$ data. The SM uncertainty, which varies
$\sim 3\%$ to $\sim 8\%$ from the low to high $Q^2$ values shown, 
represents the uncertainty of the expectation due to the assumptions made in 
the fit, as well as the uncertainties of the experimental data entering the
fit~\cite{h1hiq9497}.

\begin{figure}[htbp]
\begin{center}
\begin{picture}(50,280)
\put(-178,-350){
\epsfig{file=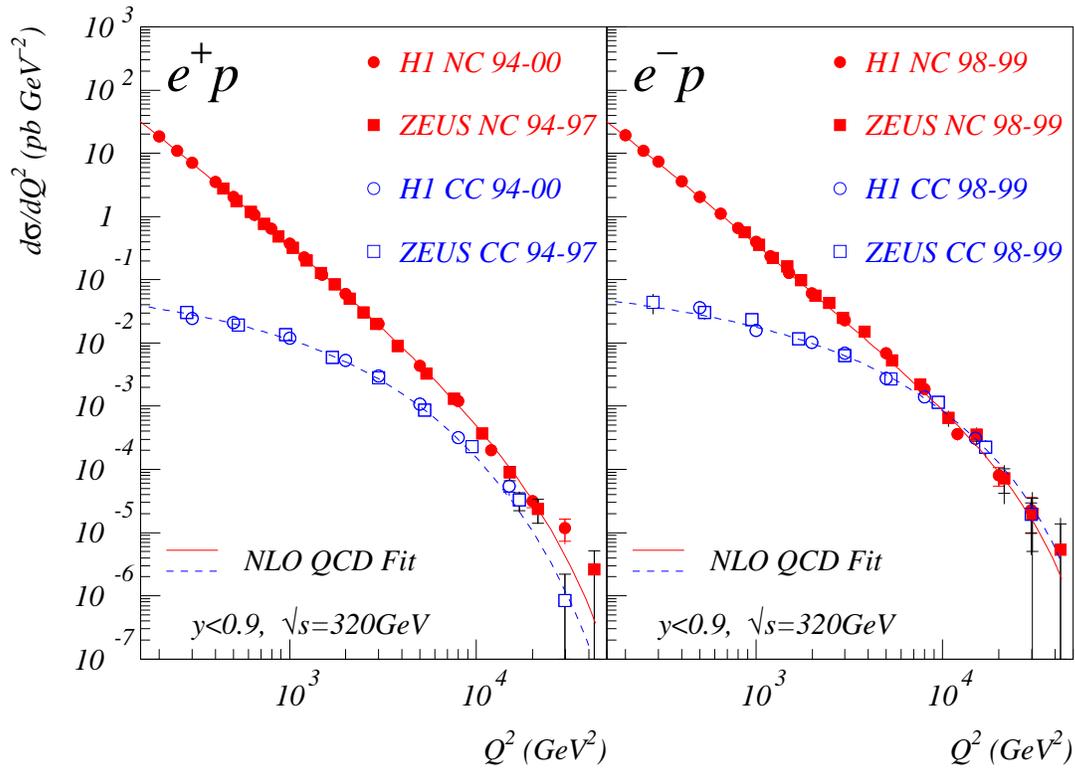,bbllx=48,bblly=30,bburx=550,
bbury=780,width=160mm}}
\end{picture}
\end{center}
\caption{\sl The NC and CC cross sections measured by
H1 and ZEUS for $e^+p$ (left) and $e^-p$ (right) collisions in comparison
with the corresponding SM expectations from the NLO QCD Fit~\cite{h1hiq9497}. 
The inner error bars represent the statistical error, and the outer error bars
show the total error.}
\label{fig:nccc_q2_h1zeus}
\end{figure}

The $e^-p$ CC data have a statistical precision between 7.5\% and 19.2\%
with the best precision at $Q^2\simeq 3000\,{\rm GeV}^2$. 
The combined high statistics $e^+p$ CC data have about the same precision 
at high $Q^2$ and marginally better precision at other $Q^2$ values 
due to the smaller $e^+p$ CC cross sections.
The cross section uncertainties are mostly dominated by the statistical errors.
The CC cross sections covering about the similar $Q^2$ range as the NC data
shown in Fig.\ref{fig:nc_q2_com}
only vary over about three orders of magnitude.
The difference is due to the dominant contribution at low $Q^2$ from the photon
exchange in the NC process.
At high $Q^2$, the NC cross sections are comparable with the CC cross
sections, as shown in Fig.\ref{fig:nccc_q2_h1zeus}, 
confirming electroweak unification of bosons exchanged in the $t$-channel.
The SM uncertainty is larger for the $e^+p$ data, as at high
$Q^2$ the cross section is dominated by the $d$ valence quark which is less
well constrained than the $u$ valence quark.

\subsection{Measurement and comparison of the $x$ dependence of NC and CC
cross sections in $e^+p$ and $e^-p$ collisions}
\label{sec:dsdx}
\begin{figure}[hp]
\begin{center}
\begin{picture}(50,460)
\put(-182,-93){
\epsfig{file=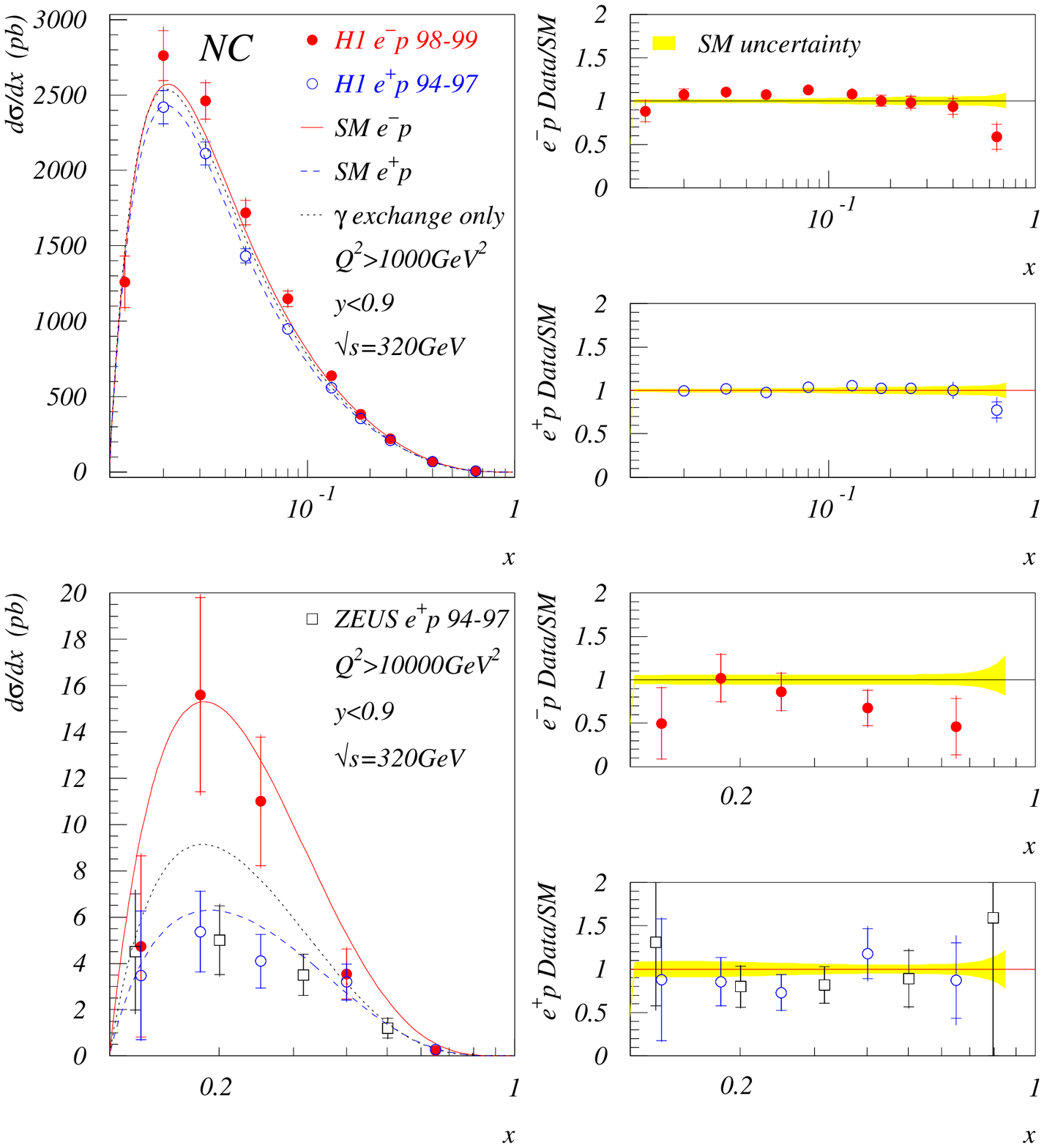,bbllx=48,bblly=30,bburx=550,
bbury=780,width=145mm}}
\end{picture}
\end{center}
\caption{\sl The $e^-p$ NC cross sections $d\sigma/dx$ from 1998-1999
in comparison with the $e^+p$ data from 1994-1997 and
the corresponding expectations from the NLO QCD Fit~\cite{h1hiq9497}. 
The dotted lines
represent the expectations from the pure photon exchange only.
The inner error bars represent the statistical error, 
and the outer error bars show the total error. The shaded error bands
represent the uncertainties on the SM expectations of the $e^-p$
and $e^+p$ cross sections.}
\label{fig:nc_x_com}
\end{figure}

The single differential cross sections $d\sigma_{\rm NC(CC)}/dx$ have also
been measured by H1~\cite{h1hiq9497,h1hiq9899} and by ZEUS~\cite{zeusnc9497}.
In Figs.\ref{fig:nc_x_com} the $e^-p$ NC cross sections from 1998-1999 are
compared with the $e^+p$ data from 1994-1997. In order to see the difference
due to the change in the lepton-beam charge, a small correction is
applied to the $e^+p$ data to account for the different center-of-mass
energies.
The H1 measurement for $Q^2>1000\,{\rm GeV}^2$ ($Q^2>10\,000\,{\rm GeV}^2$)
extends in $x$ from 0.02 to 0.65 (from 0.13 to 0.65). The cross sections rise
towards low $x$, a behavior seen already from the double differential cross
sections (Figs.\ref{fig:nce+_com2} and \ref{fig:nc_xq2_com}). The decrease
of the cross section at low $x$ edges are due to the kinematic cut $y<0.9$.
The SM expectations for $e^-p$ and $e^+p$ collisions
are compared with the data. While only a small difference between
the $e^-p$ and $e^+p$ data is seen for $Q^2>1000\,{\rm GeV}^2$, it is much
more pronounced for $Q^2>10\,000\,{\rm GeV}^2$ in accordance with the
constructive and destructive interference from the contribution of
$x\tilde{F}_3$ (see Fig.\ref{fig:weak_cor} and Sec.\ref{sec:xf3}).
The measured cross sections clearly disagree with
the expectation for pure photon exchange only, which is independent of the
lepton-beam charges, as indicated with the dotted curves in
Fig.\ref{fig:nc_x_com}.
The data are in good agreement with the SM expectations with a
tendency to be lower\footnote{The agreement between the $e^+p$ data and the
expectation from the NLO QCD Fit is better since the data were included in
the fit. The new $e^-p$ data, which indicate some difference with the fit, 
should bring additional constraints on the various parton distribution
functions.} than the expectation for the $e^-p$ data at
$Q^2>10\,000\,{\rm GeV}^2$, as well as at $x=0.65$ for all data in the full
$Q^2$ range. The implication will be discussed in Sec.\ref{sec:ud}.
Also shown are the SM uncertainties, which for $Q^2>1000\,{\rm
GeV}^2$ varies between $\sim 2.5\%$ at $x=0.02$ to $\sim 7\%$ at $x=0.65$
and degrades slightly for $Q^2>10\,000\,{\rm GeV}^2$.

\begin{figure}[htbp]
\begin{center}
\begin{picture}(50,210)
\put(-182,-325){
\epsfig{file=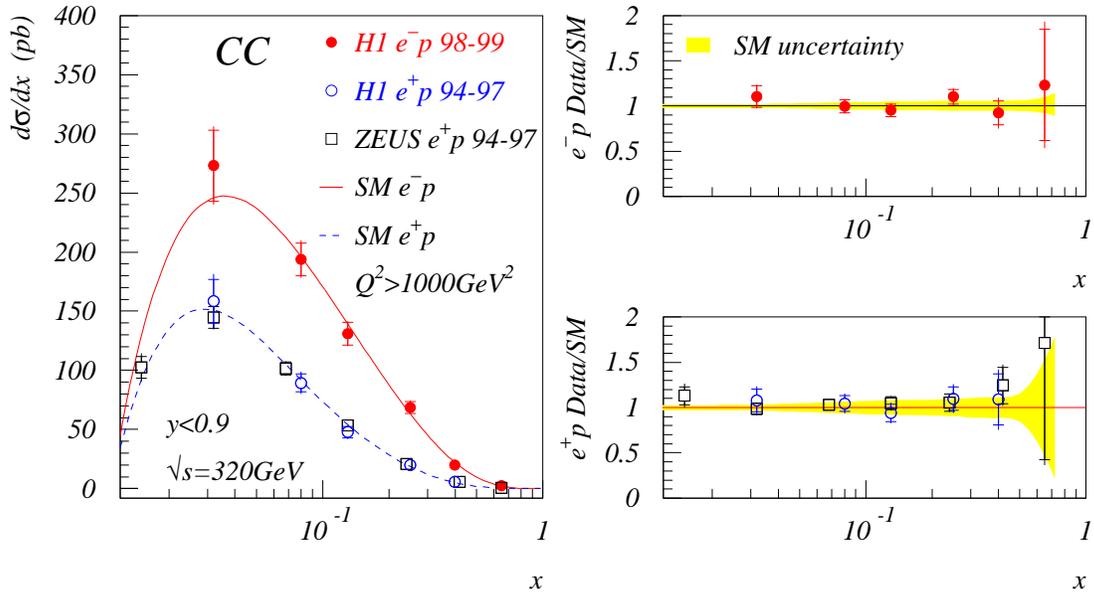,bbllx=48,bblly=30,bburx=550,
bbury=780,width=145mm}}
\end{picture}
\end{center}
\caption{\sl The $e^-p$ CC cross sections $d\sigma/dx$ from 1998-1999
in comparison with the $e^+p$ data from 1994-1997 and
the corresponding expectations from the NLO QCD Fit. 
The inner error bars represent the statistical error, 
and the outer error bars show the total error. The shaded error bands
represent the uncertainties on the SM expectations of the $e^-p$
and $e^+p$ cross sections.}
\label{fig:cc_x_com}
\end{figure}

The new $e^-p$ CC cross sections measured by H1 using data taken in
1998-1999~\cite{h1hiq9899} is compared in Fig.\ref{fig:cc_x_com}
with the $e^+p$ H1 data from 1994-1997~\cite{h1hiq9497}. 
As for NC, a small correction is applied to account for the
different proton beam energies. The $e^+p$ cross
sections from ZEUS~\cite{zeuscc9497} is also shown\footnote{Since the
original cross sections are measured for $Q^2>200\,{\rm GeV}^2$, a (large)
correction is needed, which amounts to 0.299, 0.594, 0.728, 0.783, 0.812,
0.823, and 0.817 respectively for $x=0.015, 0.032, 0.068, 0.13, 0.24, 0.42$,
and 0.65.}.
The larger $e^-p$ cross sections compared to the $e^+p$ cross sections reflect
again the different quark flavors probed by the $W^\pm$ bosons in the $e^\pm
p$ scatterings and the helicity structure of the CC interaction (see
Eqs.(\ref{eq:redxscce+}) and (\ref{eq:redxscce-})). 
Within the experimental errors, the data agree well with the
SM expectations represented by the NLO QCD Fit~\cite{h1hiq9497}.
The uncertainties of the SM expectations are shown by the
shaded error bands. As mentioned already in Sec.\ref{sec:dsdq2}, the
uncertainty, which becomes increasingly large towards high $x$, is
significantly larger for the $e^+p$ expectation than for $e^-p$.

\section{Valence quark distribution at high $Q^2$ and high $x$}
\label{sec:ud}
As it was demonstrated in Fig.\ref{fig:cc_xq2_com} (Sec.\ref{sec:xs_e+_e-}),
the CC cross sections at high $Q^2$ and high $x$ are dominated by the up and
down valence quarks ($xu_v$, $xd_v$) contributions. 
These data can therefore be used to constrain these valence quark densities 
at high $Q^2$ and high $x$. The NC cross sections, 
though they do not possess the same quark flavor discriminant
power as the CC cross sections, provide an additional constraint on $xu_v$ 
since its contribution to the NC cross section dominates over that of $xd_v$ 
due to the larger quark charge.

There has been a renewed interest on the behavior of the ratio $d/u$ at
high $x$ in recent years~\cite{yang98,kuhlmann}. 
The ratio $d/u$ is relatively well constrained at $x<0.3$ by 
DIS structure function data as well as the $W$ asymmetry measurement.
At higher $x$ ($0.3<x<0.7$), the constraint becomes weaker since it is only
available from DIS structure function data alone. These data are uncertain
because of
\begin{itemize}
\item {\bf Experimental systematic uncertainties:} The fixed-target data 
are often dominated by various systematic errors. It is not at all
trivial how these errors are treated properly in a QCD analysis.
\item {\bf Higher-twist contribution:} The fixed-target data are located
at relatively low $Q^2$ and high $x$. The higher-twist contribution is
expected to be important as it behaves as $1/[(1-x)Q^2]$ with respect to
the leading-twist contribution.
\item {\bf Large nuclear corrections:} As far as the $x$ valence quark is
concerned, it is constrained mainly by the deuteron data, of which the
nuclear binding correction can be large. This explains why the uncertainty
of the SM expectation on the $e^+p$ CC cross section (dominated
by the $d$ valence quark at high $x$) was larger than that on the $e^-p$ CC
and $e^\pm p$ NC cross sections (dominated by the $u$ valence quark).
\end{itemize}
At even higher $x$ ($x>0.7$) there is
no reliable data available and the model predictions for $u/d$ vary widely 
from 0 to about 0.2. 
In most of the conventional QCD analyses performed up to now, 
the $d/u$ ratio was assumed to approach zero as $x$ goes to one. It has
been argued recently that a large value around 0.2 is able to give a better 
description of existing data at high $x$~\cite{yang98}.

When these parton densities are evolved to high $Q^2$, one has to take into
account additional uncertainties:
\begin{itemize}
\item The uncertainty on the strong coupling constant $\alpha_s$ and higher
order perturbative corrections.
\item The so called ``feed-down'' uncertainty~\cite{stirling99}. 
The parton density, in the convolution on the right-hand side of the DGLAP 
equations (\ref{eq:qns_evol}) and (\ref{eq:qs_g_evol}), 
is sampled at $x^\prime\geq x$ (Eq.\ref{eq:conv_int}). 
The evolution is thus susceptible to the feed-down
of an anomalously large contribution at $x\simeq 1$. Such a contribution
could escape detection by the fixed-target measurements while still
influencing the evolution to the high $Q^2$ region.
\end{itemize}

The HERA data at high $Q^2$ as constraints on the valence quarks have thus
the advantage that most of the these potential problems are absent. In the
following two independent methods will be discussed to provide a 
determination of the valence quark distributions at high $Q^2$.

\subsection{Local extraction of valence quark densities} \label{sec:local_ext}
The fact that the contribution of the valence quark densities dominates in
the $e^\pm p$ NC and CC cross sections at high $x$ makes it possible to
extract $xu_v$ and $xd_v$ locally and directly from the measured cross
sections. The first extraction was performed by H1 for two values of $x$ at
0.25 and 0.4 using the $e^+p$ NC and CC cross sections from the 1994-1997
data:
\begin{equation}
xq_v(x,Q^2)=\sigma_{\rm
meas}(x,Q^2)\left(\frac{xq_v(x,Q^2)}{\sigma(x,Q^2)}\right)_{\rm th}
\end{equation}
where $\sigma_{\rm meas}(x,Q^2)$ is the measured NC or CC double differential
cross sections, and the second factor on the right-hand-side of the equation
is the theoretical expectation from the previous H1 fit~\cite{h1hiq9497}. 
Only those points where the $xq_v$ contribution is greater than 70\% of 
the total cross section are selected. The extracted parton densities are thus 
rather independent of the theoretical input as the uncertainty on the dominant
valence quark contribution and that of the corresponding cross section largely
cancel in the ratio.

With the new $e^-p$ 1998-1999 and $e^+p$ 1999-2000 data, similar extractions
can be made and are extended to $x=0.65$ for $xu_v$. In practice, the $d$
valence quark densities are determined from the combined $e^+p$ CC cross
sections discussed in Sec.\ref{sec:redxs_e+} (Eq.(\ref{eq:xs_comb})).
The $u$ valence quark densities are determined respectively from the
combined $e^+p$ NC, $e^-p$ NC, and $e^-p$ CC cross sections. The three
independent determinations of $xu_v$ are then combined. 
The resulting $xu_v$ and $xd_v$ represent an improved statistical precision of 
typically 50\% and up to 100\% at high $Q^2$ with respect to 
the first extraction.
The results are shown as the data points in Fig.\ref{fig:xuxd}.

\subsection{Valence quark densities from a new NLO QCD Fit}
\label{sec:newfit}
The valence quark distributions have also been determined from a new NLO QCD
Fit\footnote{The fit differs from the one in \cite{h1hiq9497} in that the
new NC and CC cross sections measured from the 1998-1999 $e^-p$ and 1999-2000
$e^+p$ data at high $Q^2$ shown in the previous sections 
are included as well as the new low $Q^2$ precision data from 1996-1997 
mentioned in Sec.\ref{sec:h1f29697}, and at the same time no fixed targed data 
are used. The fit also differs from the one in \cite{h1f29697} in that 
the latters uses only the $e^+p$ data at $\sqrt{s}=300$\,GeV and for 
$Q^2<3000\,{\rm GeV}^2$ so that the heavy flavors could be treated massively.
The emphasis also differs, which for this fit is more on the valence quarks at 
high $Q^2$ and high $x$ while in \cite{h1f29697} it is more on the gluon 
density at low $x$ and on the extraction of $\alpha_s$.}
using all cross sections measured by H1 at high $Q^2$. The fit was
performed with the NLO DGLAP evolution equations using the $\overline{\rm
MS}$ renormalization and factorization scheme and treating the heavy flavors
as massless quarks. Five quark and gluon components are parameterized in an
MRS-like form at $Q^2_0=15\,{\rm GeV}^2$:
\begin{eqnarray}
& & xu_v=A_ux^{B_u}(1-x)^{C_u}(1+a_u\sqrt{x})\\
& & xd_v=A_dx^{B_d}(1-x)^{C_d}(1+a_d\sqrt{x})\\
& & xg=A_gx^{B_g}(1-x)^{C_g}\\
& & x(\overline{u}+\overline{c})=A_1x^{B_1}(1-x)^{C_1}(1+a_1 x)\\
& & x(\overline{d}+\overline{s}+\overline{b})=A_2x^{B_2}(1-x)^{C_2}(1+a_2 x)
\end{eqnarray}
The number of chosen parameters for each component are dictated by the input
data sets and their precision. In order to have a reliable constraint on the
gluon and sea quarks, the new low $Q^2$ H1 data~\cite{h1f29697}
(Sec.\ref{sec:h1f29697}) are also included. The minimum $Q^2$ cut on the
data is $Q^2_{\rm min}=20\,{\rm GeV}^2$.
The parameters $A_u$ and $A_d$ are constrained by the valence quark counting
rules
\begin{equation}
\int^1_0u_vdx=2 \hspace{1cm} {\rm and} \hspace{1cm} \int^1_0d_vdx=1\,.
\end{equation}
The momentum sum rule allows the determination of one further normalization
parameter, taken to be $A_g$. The rest of the parameters are constrained by the
fit which uses the {\sc minuit} program~\cite{minuit} to minimize the
$\chi^2$ defined as~\cite{h1f29697}
\begin{equation}
\chi^2=\sum^{N_{\rm dataset}}_{j=1}\left[\sum^{N^{\rm data}_j}_{i=1}
\frac{\left\{f^{\rm data}_{i,j}-f^{\rm fit}_{i,j}\left[1-\nu_j\delta{\cal
L}/{\cal L}-\sum_k\delta_{i,k}(s_{j,k})\right]\right\}^2}{\sigma^2_{i,{\rm
stat}}+\sigma^2_{i,{\rm uncor}}}+\nu^2_j+\sum_ks^2_{j,k}\right]
\end{equation}
where the sums run over the data $i$, various data sets $j$ and sources $k$ of
correlated systematic uncertainties. The quantity $\nu_j$ stands for the
number of standard deviations for the relative normalization uncertainty of
the data set $j$, and $\delta_{i,k} (s_{j,k})$ the relative shift of the data
$i$ induced by a change by $s_{i,k}$ standard deviations of the correlated
systematic source $k$.

The results of the fit are presented in Table \ref{tab:chi2} in which the
$\chi^2$ is given for each data sets, together with their optimal
normalization factors.
\begin{table}[htb]
\begin{center}
\begin{minipage}{12.8cm}
\caption{\sl Results of the new NLO QCD Fit showing the used data sets and
their contribution to the $\chi^2$ and the optimal normalization factors.}
\label{tab:chi2}
\end{minipage}
{\small \begin{tabular}{|c|c|c|c|c|c|c|}
\hline
Data set & $e^+p$ NC & $e^+p$ CC & $e^-p$ NC & $e^-p$ CC & Low $Q^2$ & Total
\\\hline
data points & 134 & 29 & 115 & 26 & 65 & 369 \\
$\chi^2$ & 81.9 & 19.9 & 95.6 & 36.4 & 55.3 & 289.1 \\
normalization & 1.001 & 1.006 & 1.010 & 1.005 & 0.9777 & \\\hline
\end{tabular}}
\end{center}
\end{table}
The total $\chi^2$ per degree of freedom (dof) is 289.1/(369-16)=0.82. The
normalization factors are also well within the quoted luminosity uncertainty.
The parameters of the fit are given in Table \ref{tab:para}.
\begin{table}[htb]
\begin{center}
\begin{minipage}{8.9cm}
\caption{\sl Fitted parameters of the quark and gluon distributions at
$Q^2_0=15\,{\rm GeV}^2$. The parameters $A_u$, $A_d$
and $A_g$ are constrained by the sum rules.}
\label{tab:para}
\end{minipage}
\begin{tabular}{|c|c|c|c|c|}
\hline
$xq$ & $A_q$ & $B_q$ & $C_q$ & $a_q$ \\\hline
$xu_v$ & 1.71 & 0.463 & 4.93 & 12.25 \\
$xd_v$ & 4.90 & 0.872 & 5.89 & $-0.015$ \\
$xg$   & 3.39 & $-0.162$ & 11.67 & \\
$x(\overline{u}+\overline{c})$ & 0.140 & $-0.268$ & 8.31 & 3.54 \\
$x(\overline{d}+\overline{s}+\overline{b})$ & 0.140 & $-0.268$ & 13.54 &
22.7 \\\hline
\end{tabular}
\end{center}
\end{table}

\begin{figure}[htbp]
\begin{center}
\begin{picture}(50,420)
\put(-200,-140){\epsfig{file=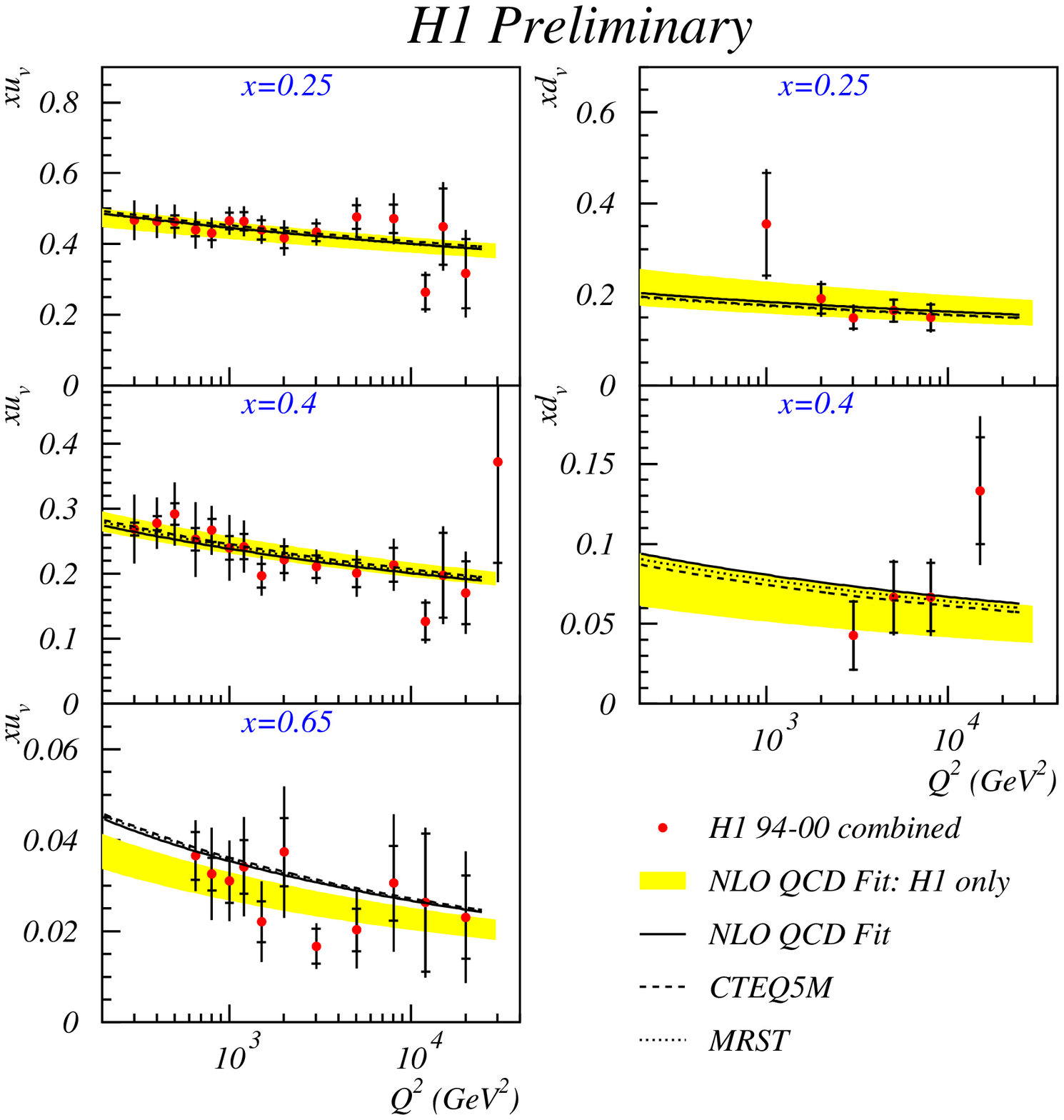,bbllx=0pt,bblly=0pt,
bburx=594pt,bbury=842pt,width=170mm}}
\end{picture}
\end{center}
\caption{\sl The valence quarks distributions $xu_v$ and $xd_v$ determined
both with the new NLO QCD Fit (shaded error bands, labeled as ``NLO QCD
Fit: H1 only'', see text for the difference with the ``NLO QCD Fit'' of
\cite{h1hiq9497}) using all H1 data at high $Q^2$
as well as the new low $Q^2$ data~\cite{h1f29697} and 
with the local extraction method (data points with 
the inner and full error bars showing respectively the statistical and total
errors) in comparison with other parameterizations which use fixed-target data
at low $Q^2$.}
\label{fig:xuxd}
\end{figure}

The resulting valence quark distributions $xu_v$ and $xd_v$ are shown in
Fig.\ref{fig:xuxd} labeled ``NLO QCD Fit: H1 only'' in comparison with
the results obtained from the local extraction method 
(Sec.\ref{sec:local_ext}),
and other parameterizations MRST~\cite{mrst}, CTEQ5~\cite{cteq5}, and the
previous H1 NLO QCD Fit~\cite{h1hiq9497}.

To conclude, the following points are ready to be listed:
\begin{itemize}
\item For the first time, the valence quark distributions at high $x$ are
constrained by the H1 experiment alone with an experimental precision
(shaded error bands in Fig.\ref{fig:xuxd}) varying between 6\% at $x=0.25$
and $x=0.4$ and $\sim 10\%$ at $x=0.65$ for $xu_v$ and $\sim 20\%$ for $xd_v$.
The quoted uncertainties correspond to the experimental errors only.
However, it has been studied with a few additional fits that 
other uncertainties are negligible:
\begin{enumerate}
\item The $Q^2_{\rm min}$ cut has been varied from 20\,GeV$^2$ up to
$200\,{\rm GeV}^2$. As expected, the valence quarks at high $x$ are
constrained by the high $Q^2$ data only.
\item The massless treatment of the heavy flavors has been replaced with the
massive one. The results are stable.
\item The uncertainty of $\alpha_s$ has been varied by $\pm 0.0017$,
corresponding to the experimental precision of the latest H1 
determination~\cite{h1f29697}. The resulting change on $xu_v$ and $xd_v$ is 
again small.
\end{enumerate}
\item The valence quark distributions determined with the new NLO QCD Fit
are in good agreement with the other parameterizations at all values of $x$
shown except for $xu_v$ at $x=0.65$, where the new valence density
is about $\sim 17\%$ lower than the other parameterizations 
with little dependence on $Q^2$ in the covered range.
These other parameterizations all used BCDMS data~\cite{bcdms89} to constrain
$xu_v$ at high $x$. However the discrepancy remains small, within
about two standard deviations. It does point to a larger ratio of $d/u$, a
preferable value as discussed by Yang and Bodek~\cite{yang98}. 
The only difference is that the larger ratio is achieved by a smaller $xu_v$ 
here instead of a larger $xd_v$.
It is therefore very important that the issue is clarified with the future
HERA high-precision data at high $Q^2$.
\item Finally, the results of the fit agree well with those obtained with the
local extraction method.
\end{itemize}

\newpage
\section{Electroweak tests at HERA} \label{sec:hera_ew}
So far in this report, we have concentrated on the tests and improvements on
the current knowledge of the strong sector of the SM by
implicitly assuming the validity of the electroweak sector. The electroweak
parameters are fixed to the world average values provided mainly by
$e^+e^-$ high precision experiments.

\begin{figure}[htbp]
\begin{center}
\begin{picture}(50,300)
\put(-225,-205){\epsfig{file=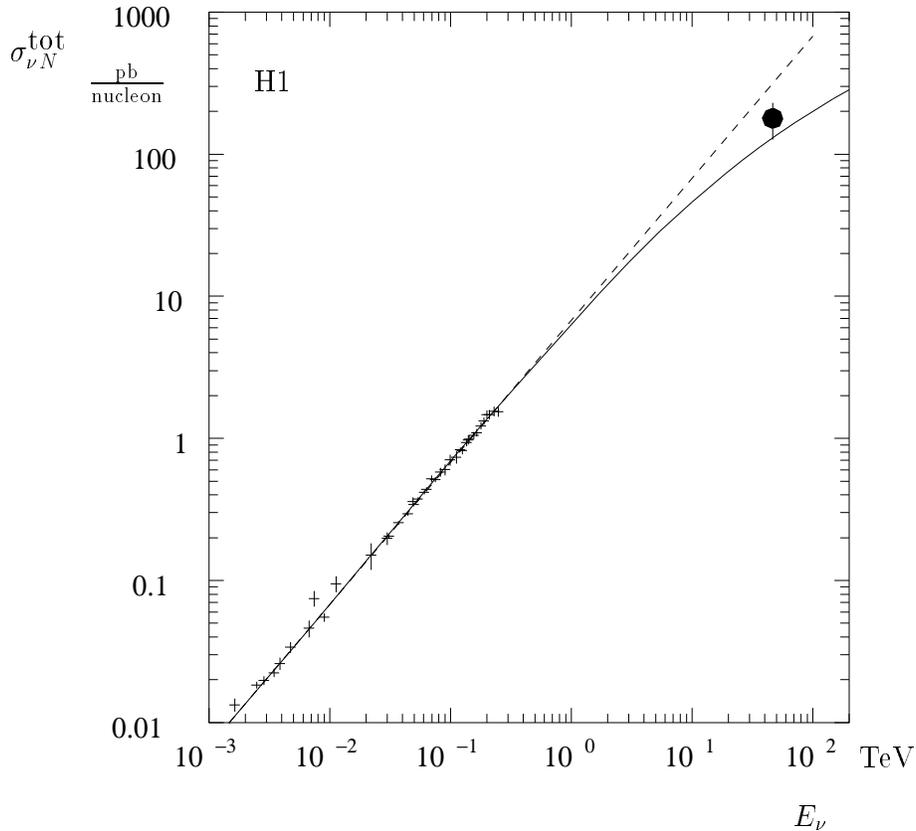,bbllx=0pt,bblly=0pt,
bburx=594pt,bbury=842pt,width=170mm}}
\end{picture}
\end{center}
\caption{\sl The energy dependence of the $\nu N$ cross section. The crosses
represent the low energy neutrino data~\cite{haidt88} while the full circle
refers to the CC result of the 1993 H1 data~\cite{h1cc93}, which for the
purpose of comparison, has been converted to a $\nu N$ cross section. The
experiment at HERA corresponds to an equivalent fixed-target energy of about
50\,TeV. The full line represents the predicted cross section including the
$W$ propagator. The dashed line is the linear extrapolation from low
energies.}
\label{fig:h1cc93}
\end{figure}

In principle, the electroweak sector can also be tested at HERA. A first
such test was performed by H1 with its first 14 CC candidates obtained from
0.35\,pb$^{-1}$ of data taken in 1993~\cite{h1cc93}. For the first time,
the CC cross section was observed to be damped at high energy due to the
propagator mass of the exchanged $W$ boson in the space-like regime
(Fig.\ref{fig:h1cc93}).

This first test was made on the total CC cross section. Additional tests are
now possible with differential CC cross sections. In fact,
the double differential CC cross section can generally be expressed in terms 
of a propagator mass $M_{\rm prop}$ of the exchanged particle:
\begin{equation}
\frac{d^2\sigma_{\rm CC}}{dxdQ^2}=\frac{G^2_F}{2\pi
x}\left(\frac{M^2_{\rm prop}}{M^2_{\rm prop}+Q^2}\right)^2\tilde{\sigma}_{\rm CC}
\end{equation}
where $\tilde{\sigma}_{\rm CC}$ is the reduced CC cross section defined in
Eq.(\ref{eq:redxs_cc}).
The normalization of the CC cross section is fixed
by the Fermi coupling constant $G_F$ while its $Q^2$ dependence by 
$M_{\rm prop}$.

Using the $e^+p$ data taken from 1994 to 1997, which represent an increased
integrated luminosity by two orders of magnitude with respect to the first
CC data from 1993, both H1 and ZEUS have made a fit of the propagator mass 
$M_{\rm prop}$ to their cross section data~\cite{h1hiq9497,zeuscc9497}.
The propagator mass obtained is respectively
\begin{eqnarray}
80.9\pm 3.3 ({\rm stat})\pm 1.7 ({\rm syst}) \pm 3.7 ({\rm pdf})\,{\rm GeV} 
& & {\rm H1}\\
81.4^{+2.7}_{-2.6}({\rm stat}) \pm 2.0 ({\rm syst}) ^{+3.3}_{-3.0} ({\rm
pdf})\,{\rm GeV} & & {\rm ZEUS}
\end{eqnarray}
where the last error is due to the uncertainty of the parton density
distributions (pdf) dominated by the $d$ valence quark.
From this aspect and
the fact that the $e^-p$ CC cross section at high $Q^2$ is an order of 
magnitude larger than the $e^+p$ cross section, a better precision is
expected from the $e^-p$ data. Indeed, with an integrated luminosity of
16.4\,pb$^{-1}$ for the 1998-1999 H1 $e^-p$ data, which is less than half
of that from the $e^+p$ data, a similar fit yields~\cite{h1hiq9899}
\begin{equation}
79.9\pm 2.2 ({\rm stat}) \pm 0.9 ({\rm syst}) \pm 2.1 ({\rm pdf})\,{\rm GeV}\,.
\end{equation}
The dominant experimental systematic error is from the hadronic energy scale
uncertainty, which was estimated to be $\pm 2\%$ for both the $e^+p$
and $e^-p$ analyses (Sec.\ref{sec:had_scale}). However, the
resulting uncertainty on the CC cross sections is smaller in $e^-p$
collisions than in $e^+p$ collisions due to the different $Q^2$ dependence
(Fig.\ref{fig:cc_q2_com}). In all these measurements, the pdf error is large
than the experimental systematic uncertainty demonstrating  
the importance and the necessity to improve our knowledge on the parton
distribution functions. 

These measurements are in good agreement with the combined results 
$M_W=80.427\pm0.046$\,GeV and $M_W=80.448\pm 0.062$\,GeV respectively from 
LEP\,II and $p\overline{p}$ colliders~\cite{gurtu_00}. 
To achieve such a precision at HERA would be difficult but is not impossible
at the future high luminosity run~\cite{brisson92}.
Nevertheless the fact that the $W$ mass measured at HERA in the space-like
regime agrees so well with those when it appears as a real boson with 
subsequent decays or as a virtual boson in the space-like regime indicates 
that there is little space left for other exotic contributions to the measured 
CC cross sections at HERA.

Other tests e.g.\ on the weak neutral current couplings of quarks $v_u$,
$a_u$, $v_d$, and $a_d$ are possible but again have to wait for the future high
luminosity runs with polarized beams~\cite{cashmore96}.

\newpage
\section{Searches for new physics beyond the Standard Model}
\label{sec:search}
From an early $e^+p$ data sample of 1994-1996 corresponding to an integrated
luminosity of 14.2\,pb$^{-1}$ and 20.1\,pb$^{-1}$ respectively for H1 and
ZEUS, both experiments have reported~\cite{h19496, zeus9496} an excess of
events at high $Q^2 (>15\,000\,{\rm GeV}^2)$ and at large masses
$M=\sqrt{xs}$, the center-of-mass energies of the electron-parton collision,
with respect to the SM DIS expectation for the NC process. H1
has also observed a less significant excess in the CC channel.
In 1997, the HERA machine was particularly successful and both experiments
have collected more $e^+p$ data than they had before.
Since then, new data from both $e^-p$ and $e^+p$ collisions have been
collected respectively in 1998-1999 and 1999-2000 at a higher center-of-mass
energy $\sqrt{s}\simeq 320$\,GeV. These data have been used extensively
to search for new physics phenomena beyond the SM, of which three examples
are briefly described in the following subsections.

\subsection{Search for leptoquarks at HERA} \label{sec:lq}
The $ep$ collider offers the unique possibility to search for $s$-channel
production of new particles which couple to lepton-parton pairs. Leptoquarks
are one such example. Leptoquarks are color triplet bosons which appear
naturally in various unifying theories beyond the SM such as
Grand Unified Theories~\cite{gut} and Superstring inspired $E_6$
models~\cite{ss_e6}, and in some Compositeness~\cite{compsite} and
Technicolor~\cite{technicolor} models. 

\begin{figure}[htb]
\begin{center}
\begin{picture}(50,80)
\put(-160,-270){\epsfig{file=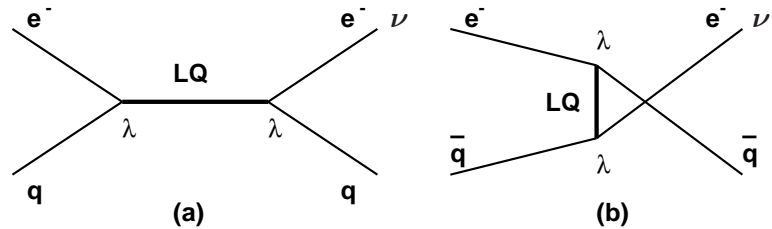,bbllx=0pt,bblly=0pt,
bburx=594pt,bbury=842pt,width=135mm}}
\put(38,62){\small\boldmath $\nu$}
\put(175,62){\small\boldmath $\nu$}
\end{picture}
\end{center}
\caption{\sl Diagrams (a) $s$-channel resonant production and (b)
$u$-channel exchange of a leptoquark (LQ) with fermion number $F=2$. Diagrams
involving a $F=0$ leptoquark are obtained from (a) and (b) by either
changing the lepton-beam charge or exchanging $q$ and $\overline{q}$.}
\label{fig:lq_dia}
\end{figure}

\begin{figure}[htbp]
\begin{center}
\begin{picture}(50,185)
\put(-160,-30){\epsfig{file=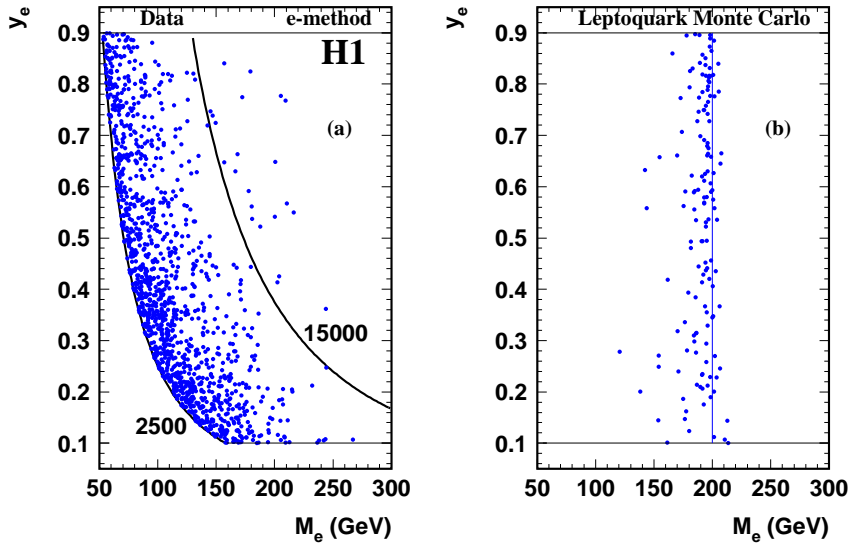,bbllx=0pt,bblly=0pt,
bburx=594pt,bbury=842pt,width=135mm}}
\end{picture}
\end{center}
\caption{\sl Kinematics in the $y_e-M_e$ plane of (a) the selected NC DIS
candidates from the $e^+p$ 1994-1997 H1 data (two isocurves at
$Q^2=2500\,{\rm GeV}^2$ and $15\,000\,{\rm GeV}^2$ are plotted as full
lines); (b) a scalar $F=0$ leptoquark of mass at 200\,GeV decaying into
$e+q$ for a coupling $\lambda=0.05$ generated with the {\sc lego} event
generator~\cite{lego}.}
\label{fig:y_mass_com}
\end{figure}

Leptoquarks are produced at HERA in the $s$-channel (Fig.\ref{fig:lq_dia}(a)). 
They appear also as exchanged bosons in the $u$-channel 
(Fig.\ref{fig:lq_dia}(b)).
Leptoquark states are classified according to the fermion number $F=L+3B$,
where $L$ is the lepton number and $B$ is the baryon number of the state.
Thus for $e^-p$ collisions, the fermion numbers are $F=2$ ($F=0$) for
electron fusing with a quark (antiquark) from the proton whereas for $e^+p$
collisions they are $F=0$ ($F=-2$) for positron fusing with a quark (antiquark)
from the proton. The $F=2$ leptoquarks would be produced with a larger cross
section in $e^-p$ collisions while the $F=0$ states are best searched for
in $e^+p$ collision data.
The unknown Yukawa coupling of the leptoquark to an $eq$ pair is
denoted by $\lambda$ in Fig.\ref{fig:lq_dia}. Leptoquarks coupling only to
first generation fermions give $e+q$ or $\nu+q^\prime$ final states leading to
individual events indistinguishable respectively from standard NC and CC DIS
events, which become backgrounds for the searches considered here.
Statistically, however, one expects for scalar leptoquarks produced
in $s$-channel a resonant peak in the mass distribution and an isotropical
$d\sigma/dy$ distribution where $y=(1+\cos\theta^\ast)/2$ with $\theta^\ast$
being the decay polar angle of the lepton relative to the incident proton
in the leptoquark center-of-mass frame. Events resulting from the production
and decay of vector leptoquarks would be distributed according to
$d\sigma/dy\propto (1-y)^2$. These $y$ spectra from scalar or vector
leptoquark production are markedly different from the distribution
($d\sigma/dy\propto y^{-2}$) expected at fixed $x$ for the dominant
$t$-channel photon exchange in NC DIS events. This is
illustrated in Fig.\ref{fig:y_mass_com}(a) for NC DIS events selected from
$e^+p$ 1994-1997 data and in Fig.\ref{fig:y_mass_com}(b) for a production of
a scalar leptoquark of mass at 200\,GeV, where the kinematic variables $y_e$
and $M_e=\sqrt{sx_e}$ were defined using the electron method
(Sec.\ref{sec:kinerec}). 

\begin{figure}[htbp]
\begin{center}
\begin{picture}(50,495)
\put(-185,225){\epsfig{file=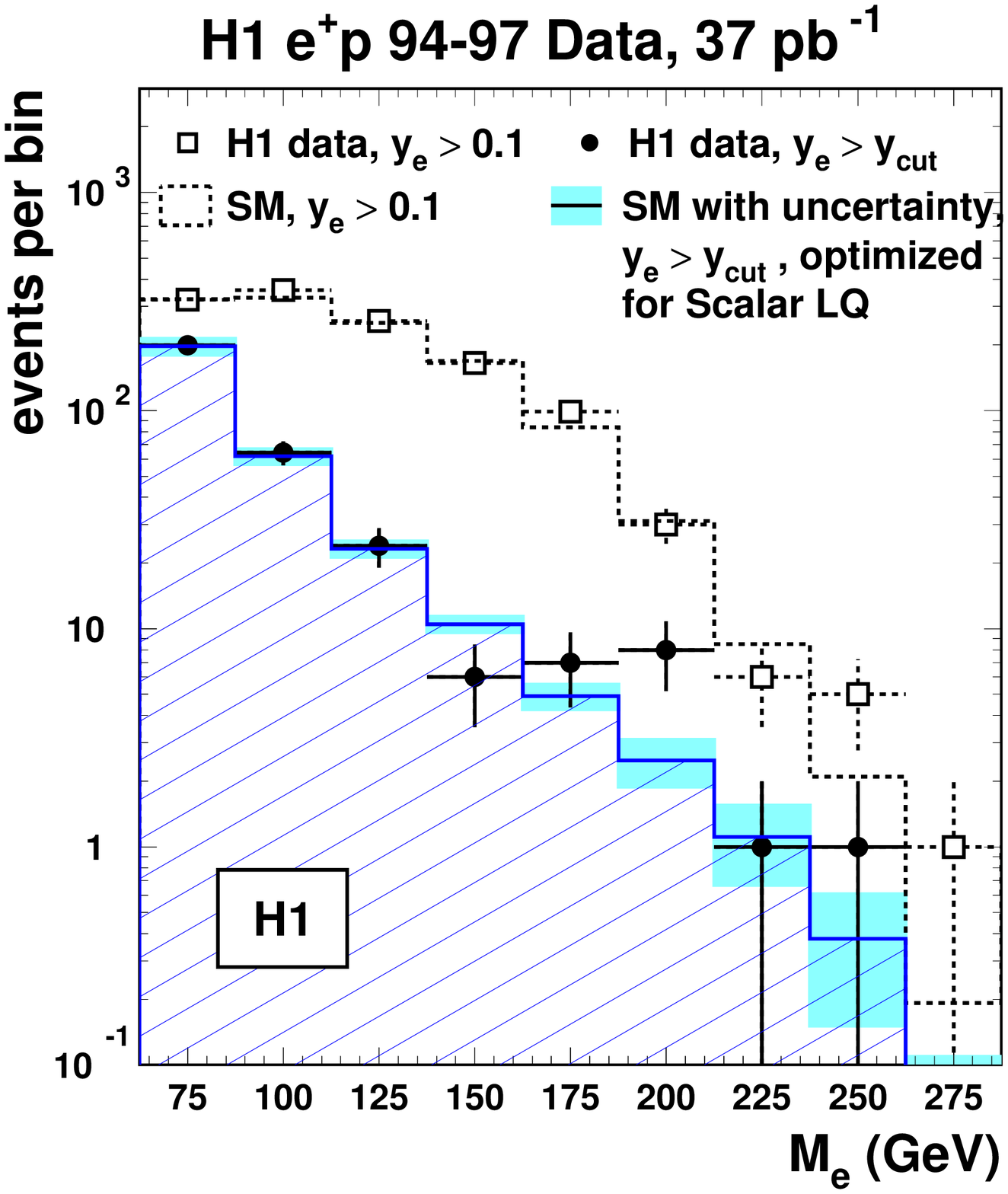,bbllx=0pt,bblly=0pt,
bburx=594pt,bbury=842pt,width=84mm}}
\put(-10,375){\makebox(0,0)[l]{(a)}}
\put(22,225){\epsfig{file=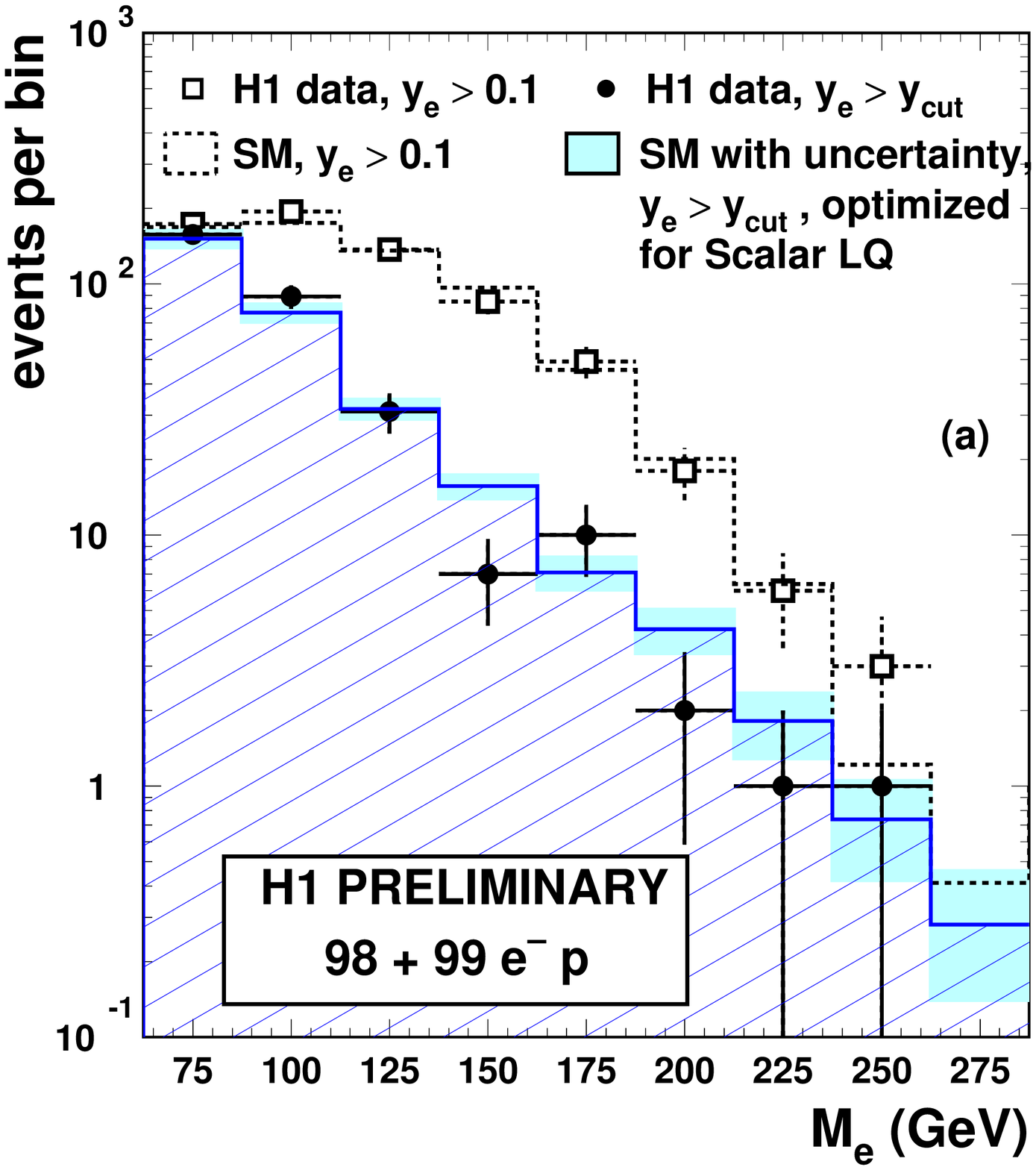,bbllx=0pt,bblly=0pt,
bburx=594pt,bbury=842pt,width=84mm}}
\put(202,372){\makebox(0,0)[l]{\fcolorbox{white}{white}{\textcolor{white}{kill}}}}
\put(198,375){\makebox(0,0)[l]{(b)}}
\put(67,457){\makebox(0,0)[l]{H1 $e^-p$ 98-99 Data, 14.4\,pb$^{-1}$}}
\put(-185,-20){\epsfig{file=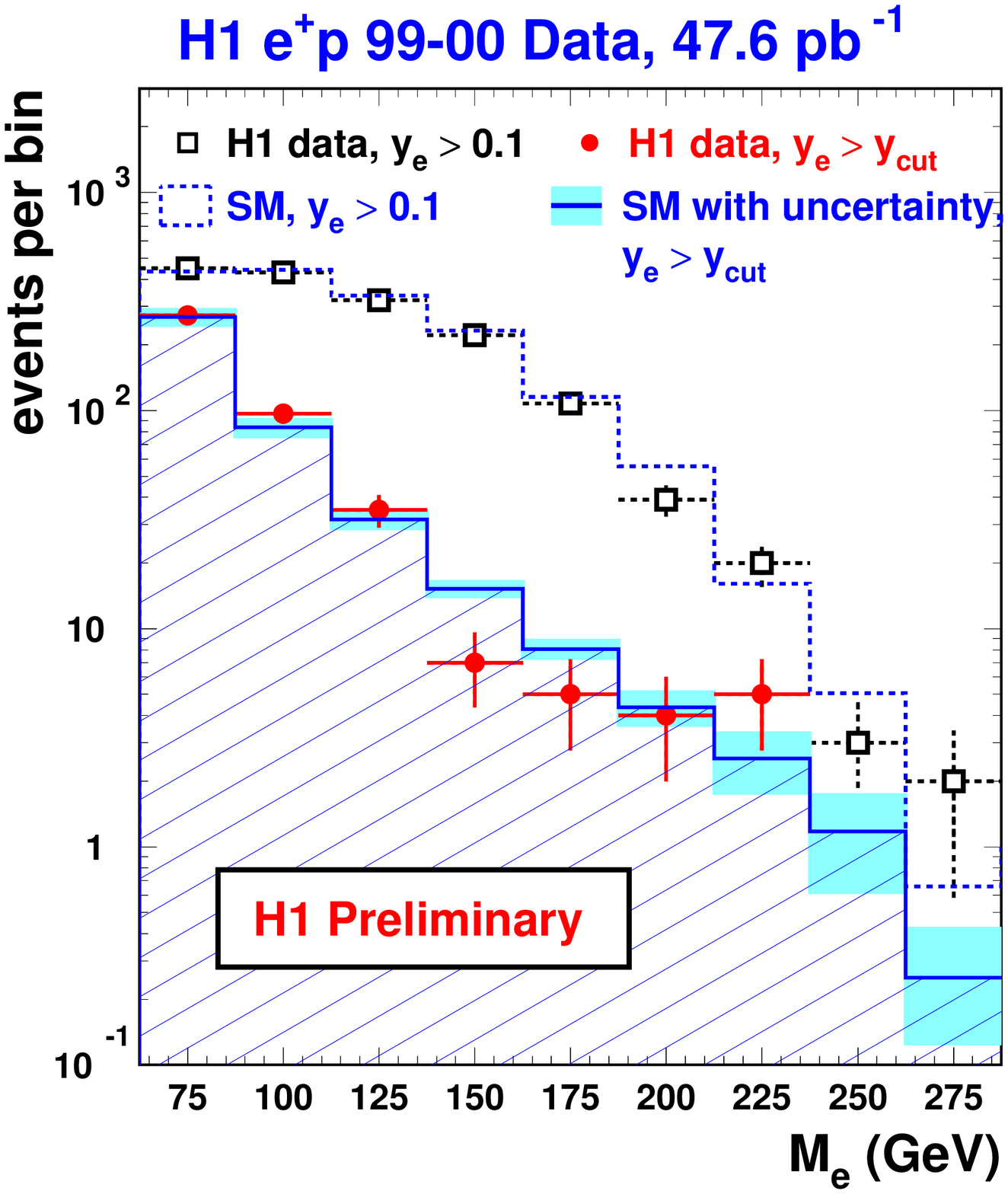,bbllx=0pt,bblly=0pt,
bburx=594pt,bbury=842pt,width=84mm}}
\put(-10,140){\makebox(0,0)[l]{(c)}}
\end{picture}
\end{center}
\caption{\sl Mass spectra for NC DIS-like final states selected from the $e^+p$
1994-1997 data (b), $e^-p$ 1998-1999 data (b), and $e^+p$ 1999-2000 data (c).
The data (symbols) are compared with DIS expectation (histograms) before and
after a mass dependent $y$ cut designed to maximize the significance of an
eventual scalar leptoquark signal. The greyed boxes indicate the $\pm 1
\sigma$ uncertainty of the statistical and systematic errors of the NC DIS
expectation.}
\label{fig:mass_slq}
\end{figure}

For this reason, a mass dependent $y$ cut is applied to optimize the signal
significance and background contribution. The resulting mass distributions
are shown in Fig.\ref{fig:mass_slq} for all $e^\pm p$ data taken by H1 
since 1994.
The excess around 200\,GeV in the $e^+p$ 1994-1997 data is not confirmed by 
the new $e^-p$ 1998-1999 and $e^+p$ 1999-2000 data.

\begin{figure}[htbp]
\begin{center}
\begin{picture}(50,335)
\put(-140,5){\epsfig{file=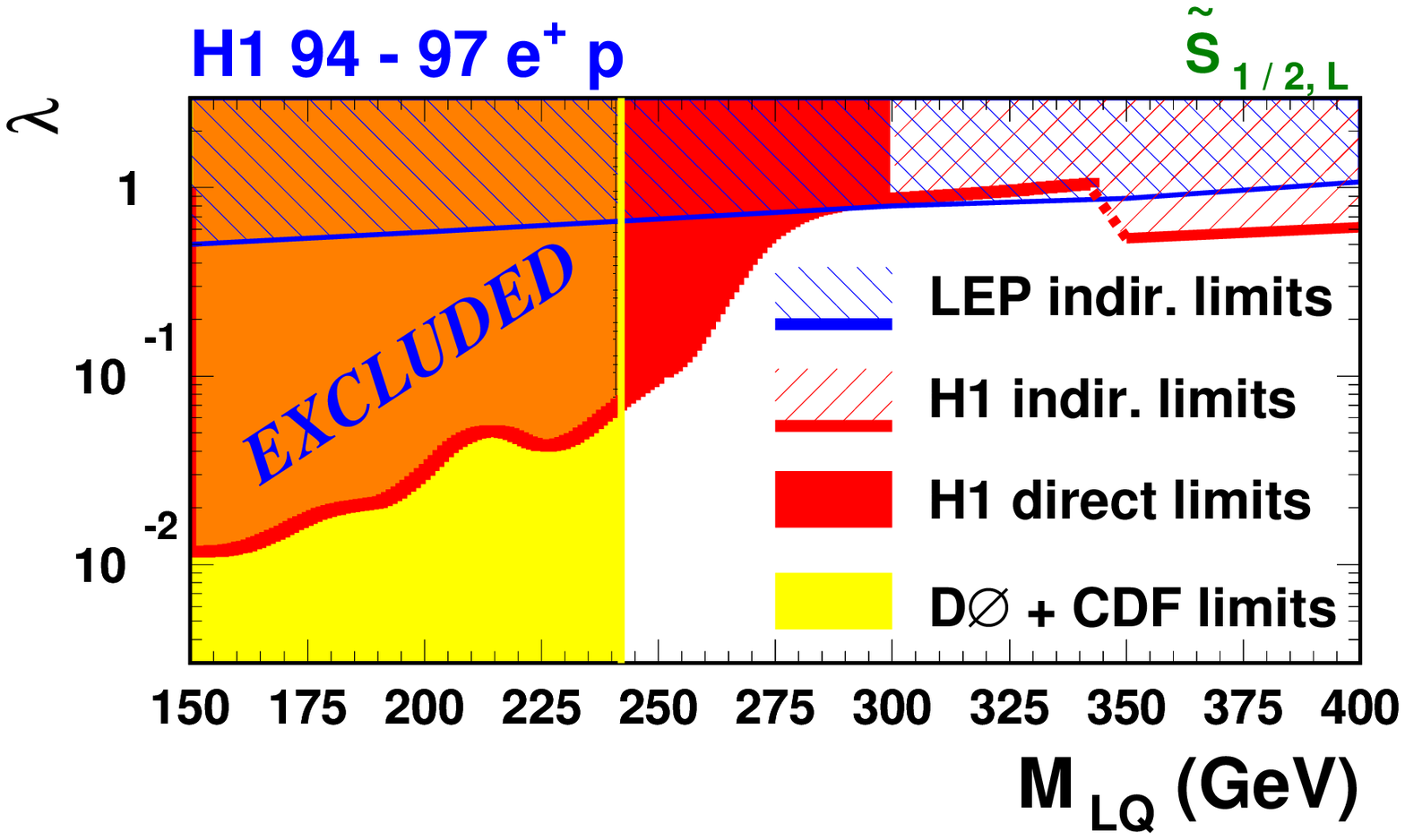,bbllx=0pt,bblly=0pt,
bburx=594pt,bbury=842pt,width=120mm}}
\put(-140,-180){\epsfig{file=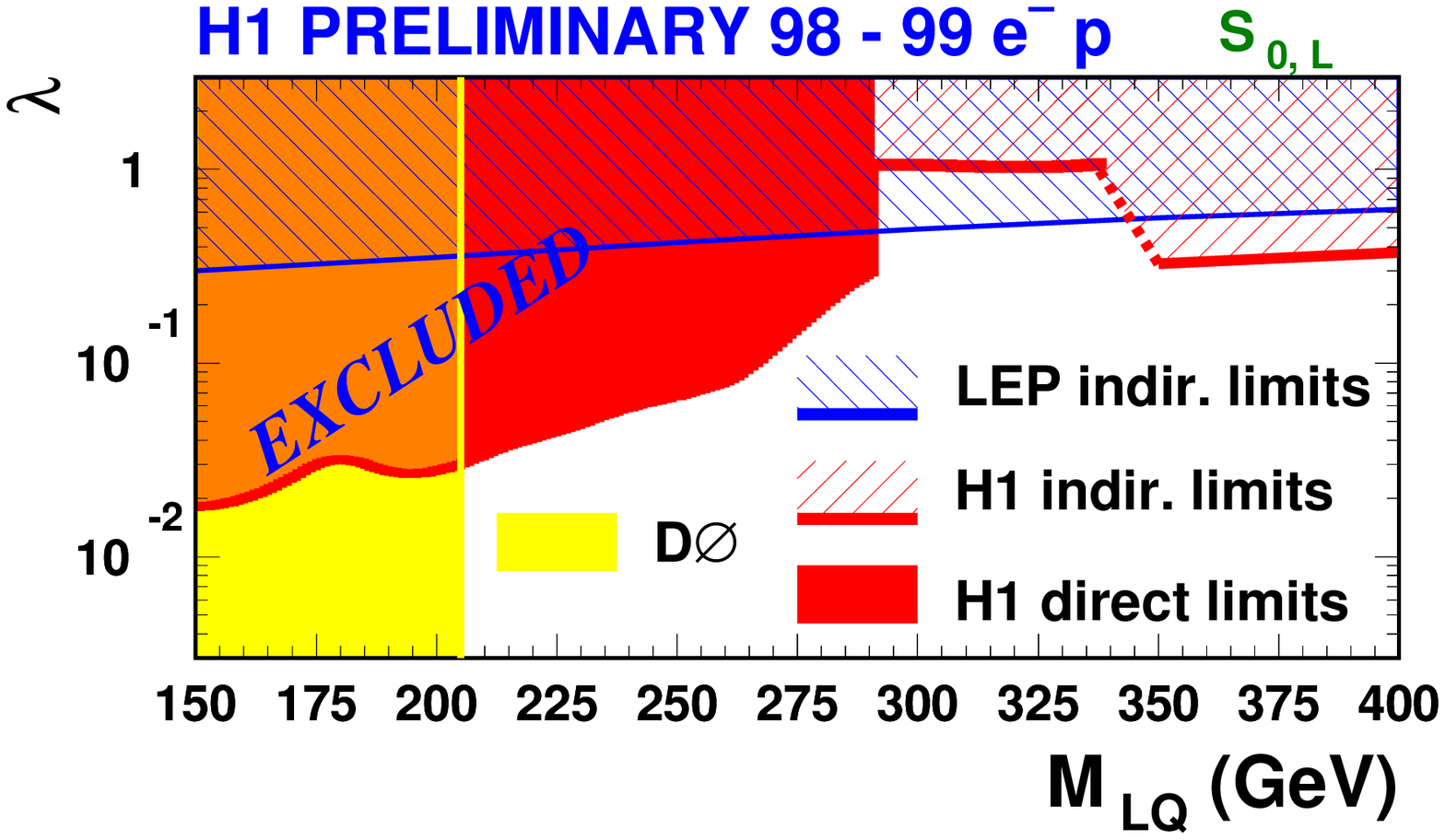,bbllx=0pt,bblly=0pt,
bburx=594pt,bbury=842pt,width=120mm}}
\end{picture}
\end{center}
\caption{\sl Exclusion limits at 95\% confidence level on the Yukawa
coupling $\lambda$ as a function of the leptoquark mass for (top) a scalar
leptoquark with $F=0$ and (bottom) a scalar leptoquark with $F=2$ described
by the Buchm\"uller-R\"uckl-Wyler model~\cite{brw}. Greyed and hatched
domains are excluded. The H1 indirect limits are derived in the analysis of
contact interactions described in Sec.\ref{sec:ci}.}
\label{fig:lq_limit}
\end{figure}

Since no significant evidence for a leptoquark signal has been observed,
constraints on the Yukawa coupling of leptoquarks were
derived~\cite{h1lq9497, h1lq9899, h1lq9400}.
Two examples are shown in Fig.\ref{fig:lq_limit} for a scalar leptoquark
with $F=0$ ($\tilde{S}_{1/2,L}$) from the $e^+p$ 1994-1997 data and a scalar
leptoquark with $F=2$ ($S_{0,L}$) from the $e^-p$ 1998-1999 data in
comparison with the indirect limits from H1~\cite{h1ci9497, h1ci9400}
(Sec.\ref{sec:ci}) and LEP~\cite{lq_opal}, as well as limits from the
Tevatron experiments~\cite{lq_tevatron}. At LEP~\cite{lq_opal, lq_l3,
lq_aleph}, sensitivity to a high-mass leptoquark arises from effects of
virtual leptoquark exchange on the hadronic cross section.
In both examples shown, the H1 limits extend beyond the present reach of
other colliders. Moreover the constrains obtained from the $e^-p$ data on $F=2$
leptoquarks are more stringent than those from the previous higher
statistics $e^+p$ data on $F=0$ states, due to the enhanced energy in the
center of mass of the $ep$ collision. Similar limits on $F=0$ leptoquarks 
have also been obtained by ZEUS using the $e^+p$ 1994-1997 
data~\cite{lq_zeus9497}.

\subsection{Search for supersymmetry at HERA} \label{sec:susy}
Supersymmetry (SUSY) is one of the most likely ingredients
for a theory beyond the SM. In particular the Minimal
Supersymmetric extension of the Standard Model (MSSM) describes as
well as the SM all experimental data, and in addition it offers among
its appealing consequences solutions for the cancellation of quadratic
divergences occurring in the scalar Higgs sector of the SM and models beyond
the SM.

SUSY relates fermions to bosons and predicts for each SM particle a partner
with spin differing by half a unit. For example selectrons
$\se_L$, $\se_R$ are scalar partners of electrons $e_L$, $e_R$, 
and similarly squarks
$(\su_L, \sd_L)$, $\su_R$, $\sd_R$ are the partners of
up and down quarks. Two Higgs doublets with vacuum expectation values
$v_2, v_1$ are necessary to generate masses for up-type quarks $(v_2)$ and
for down-type quarks and charged leptons $(v_1)$. The partners of the
gauge bosons $W^\pm, Z^0, \gamma$ and the two Higgs doublets are called
gauginos and higgsinos. They can mix and form two charged mass
eigenstates $\chi^\pm_{1,2}$ (charginos) and four neutral mass eigenstates
$\chi^0_{1,2,3,4}$ (neutralinos).

Since supersymmetric particles are not observed at the masses of their
SM partners, SUSY must be broken. In the MSSM, this breaking is achieved
by adding extra mass parameters $M_2$ and $M_1$ for the $SU(2)$ and $U(1)$ 
gauginos. Thus the masses of charginos and neutralinos depend on $M_1, M_2,
\tan\beta\equiv v_2/v_1$ and the higgsino mass parameter $\mu$.

\begin{figure}[htbp]
\begin{center}
\begin{picture}(50,200)
\put(-130,-80){\epsfig{file=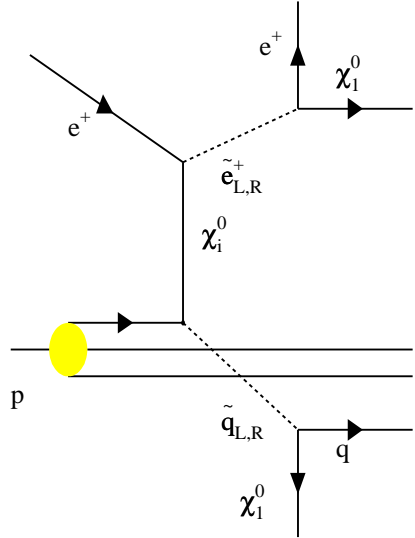,width=95mm}}
\end{picture}
\end{center}
\caption{Selectron-squark production via neutralino
exchange and the subsequent decays into the lightest supersymmetric
particle $\chi^0_1$.}
\label{fig:mssm}
\end{figure}

$R$-parity ($\rp$), defined as $\rp\equiv (-1)^{3B+L+2S}$, is a multiplicative
quantum number which distinguishes particles ($\rp=+1$) from SUSY particles
($\rp=-1$). Here $B$, $L$, and $S$ denote respectively baryon number, lepton
number, and spin of a particle. In SUSY models with $R$-parity conservation,
supersymmetric particles can only be produced in pairs and the
lightest supersymmetric particle (LSP), which is generally assumed to be
$\chi^0_1$, is stable. At HERA the dominant $\rp$-conservation process is
the production of a selectron and a squark via a $t$-channel exchange of
a neutralino $ep\ra \se\sq X$ (Fig.~\ref{fig:mssm}).
The $\se$ and $\sq$ can then decay
into any lighter gaugino and their SM partners. The decay involving
$\chi^0_1$ gives an experimentally clean signature of missing transverse
energy plus an electron and a hadronic system. Such a search has been
performed by H1~\cite{h1_mssm} using 6.38\,pb$^{-1}$ of $e^+p$ data from 
1994-1995, and by ZEUS~\cite{zeus_mssm} using 46.6pb$^{-1}$ of $e^+p$ data 
from 1994-1997. 

\begin{figure}[htbp]
\begin{center}
\begin{picture}(50,300)
\put(-175,125){\epsfig{file=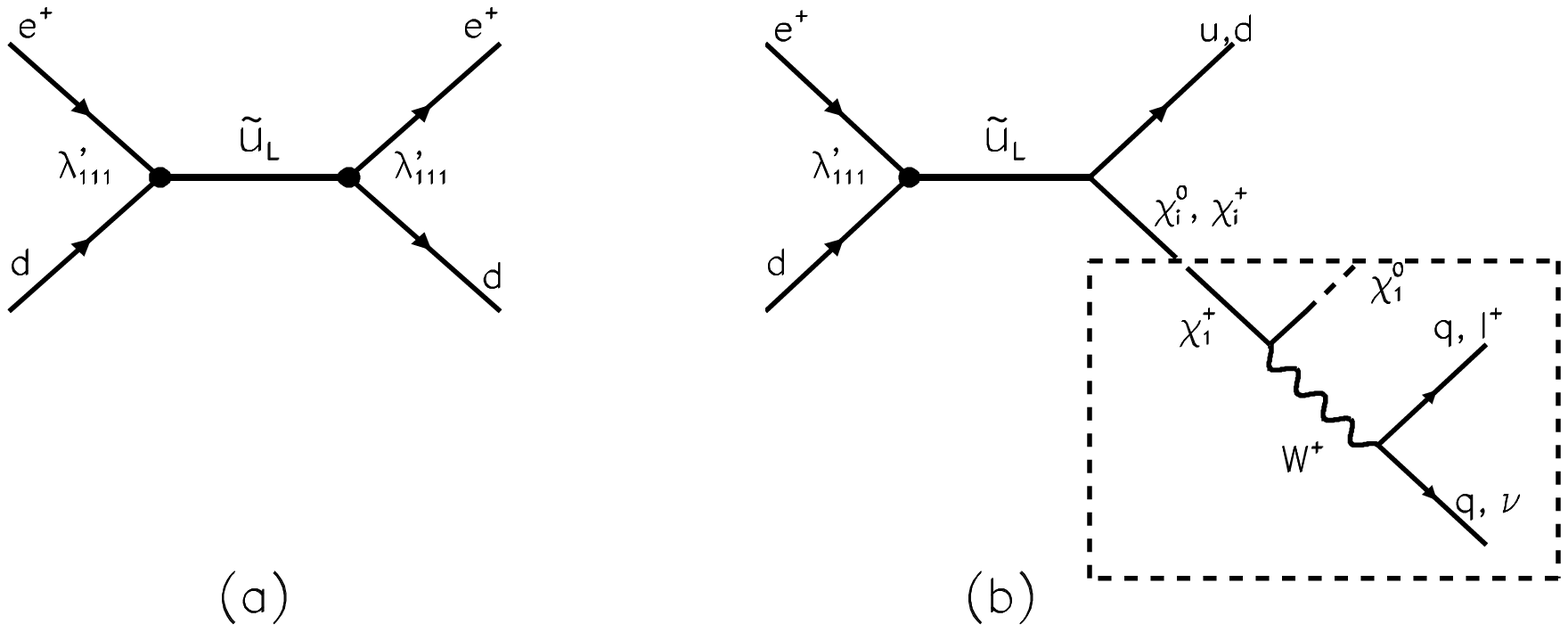,width=140mm}}
\put(-175,-35){\epsfig{file=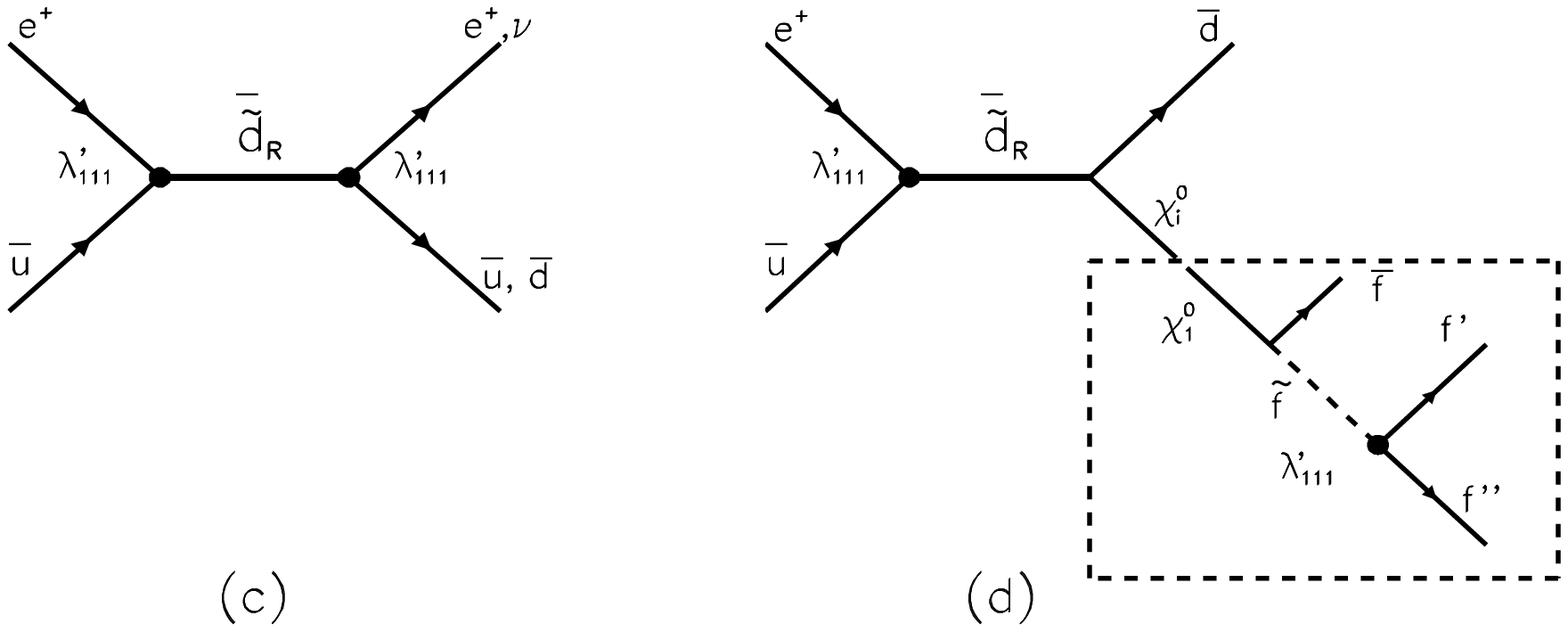,width=140mm}}
\end{picture}
\end{center}
\caption{Lowest order $s$-channel diagrams for first generation squark 
production at HERA followed by (a), (c) $\rpv$ decays and (b), (d) 
gauge decays. In (b) and (d), the emerging neutralino or chargino might
subsequently undergo $\rpv$ decays of which examples are shown in the 
dashed boxes for (b) the $\chi^+_1$ and (d) the $\chi^0_1$.}
\label{fig:rpvdia}
\end{figure}

The most general SUSY theory which preserves gauge invariance of the SM
allows, however, for $R_p$ violating ($\rpv$) Yukawa couplings $\lambda,
\lambda^\prime, \lambda^{\prime\prime}$ between one scalar squark or slepton
and two SM fermions:
\begin{equation}
W_{\not\! R_p}=\lambda_{ijk}L_iL_j\overline{E}_k+
\lp_{ijk}L_iQ_j\overline{D}_k+
\lambda^{\prime\prime}_{ijk}\overline{U}_i\overline{D}_j\overline{D}_k.
\label{eq:rpv}
\end{equation}
where $i,j,k=1,2,3$ are generation indices, $L_i(Q_i)$ are the lepton (quark)
$SU(2)_L$ doublet superfields and
$\overline{E}_i(\overline{D}_j,\overline{U}_j)$ are the electron
(down and up quark) $SU(2)_L$ singlet superfields.
Of particular interest for HERA are the $\rpv$ terms
$\lp L_iQ_j\overline{D}_k$ as HERA provides both
leptonic and baryonic quantum numbers in the initial state.
The resonant squarks at HERA are thus singly produced (in contrast to
the $\rp$-SUSY) in the $s$-channel (Fig.\ref{fig:rpvdia}) with masses up to 
the kinematic limit of $\sqrt{s}\simeq 300-320\,{\rm GeV}$. 
From the theoretical understanding of unification, there is no clear 
preference between $\rp$-conservation and $\rpv$, it is thus mandatory 
that the latter possibility is also sought experimentally~\cite{dreiner}. 

Compared to the squarks in $\rp$-SUSY, the squarks in \rpv-SUSY can have
additional decay modes decaying via Yukawa coupling into SM fermions 
as illustrated with two examples in Fig.\ref{fig:rpvdia}(b),(d).
Moreover, the LSP (again assumed to be $\chi^0_1$), which is no longer
stable, decays via $\lambda^\prime_{1jk}$ into a quark, an antiquark and a
lepton.

In cases where both production and decay occur through a
$\lambda^\prime_{1jk}$, the squarks in \rpv-SUSY behave as scalar leptoquarks
(Sec.\ref{sec:lq}) and the constraints obtained on
leptoquarks are also applicable for squarks. There is however an interesting
difference, namely the exclusion phase space
covered by HERA data can be relatively larger for squarks than for leptoquarks.
This is because the mass constraints of 242\,GeV (205\,GeV) from 
Tevatron~\cite{lq_tevatron} correspond to a branching ratio $\beta_{eq}$ of the
new particle into $e+q$ of 1(0.5), which can be naturally small in \rpv-SUSY
framework given the competition with gauge decay modes of the squarks.

With $e^+p$ collisions, HERA is best sensitive to couplings 
$\lp_{1j1}$ among the nine possible couplings 
$\lp_{1jk}$, where mainly $\tilde{u}^j_L$ squarks are
produced via processes involving a valence $d$ quark. On the contrary,
the $e^-p$ data will allow to better probe couplings
$\lp_{11k}$ and $\tilde{d}^k_R$ squarks.

Depending on whether the produced squarks undergo a $\rpv$ decay or 
a gauge decay, there are many different final state event topologies,
e.g.\ (a) a lepton plus a jet, (b) a neutrino plus a jet, (c) a right
sign lepton plus multijets, and (d) a wrong sign lepton plus multijets.
Topologies (a) and (b) are indistinguishable from NC and CC DIS
events respectively as for leptoquarks.
Topology (b) has only a low sensitivity with the $e^+$ beam since the produced
squarks $\tilde{d}^{k\ast}_R$ couples to a sea quark $\overline{u}$ from 
the proton (Fig.~\ref{fig:rpvdia}(c)), the density of which is small 
at high $x$.
The main SM background for topology (c) is also from NC DIS where QCD 
radiation leads to multijets.
Topology (d) has such a striking final state that it is essentially 
background free.

\begin{figure}[htb]
\begin{center}
\begin{picture}(50,230)
\put(-190,-15){\epsfig{file=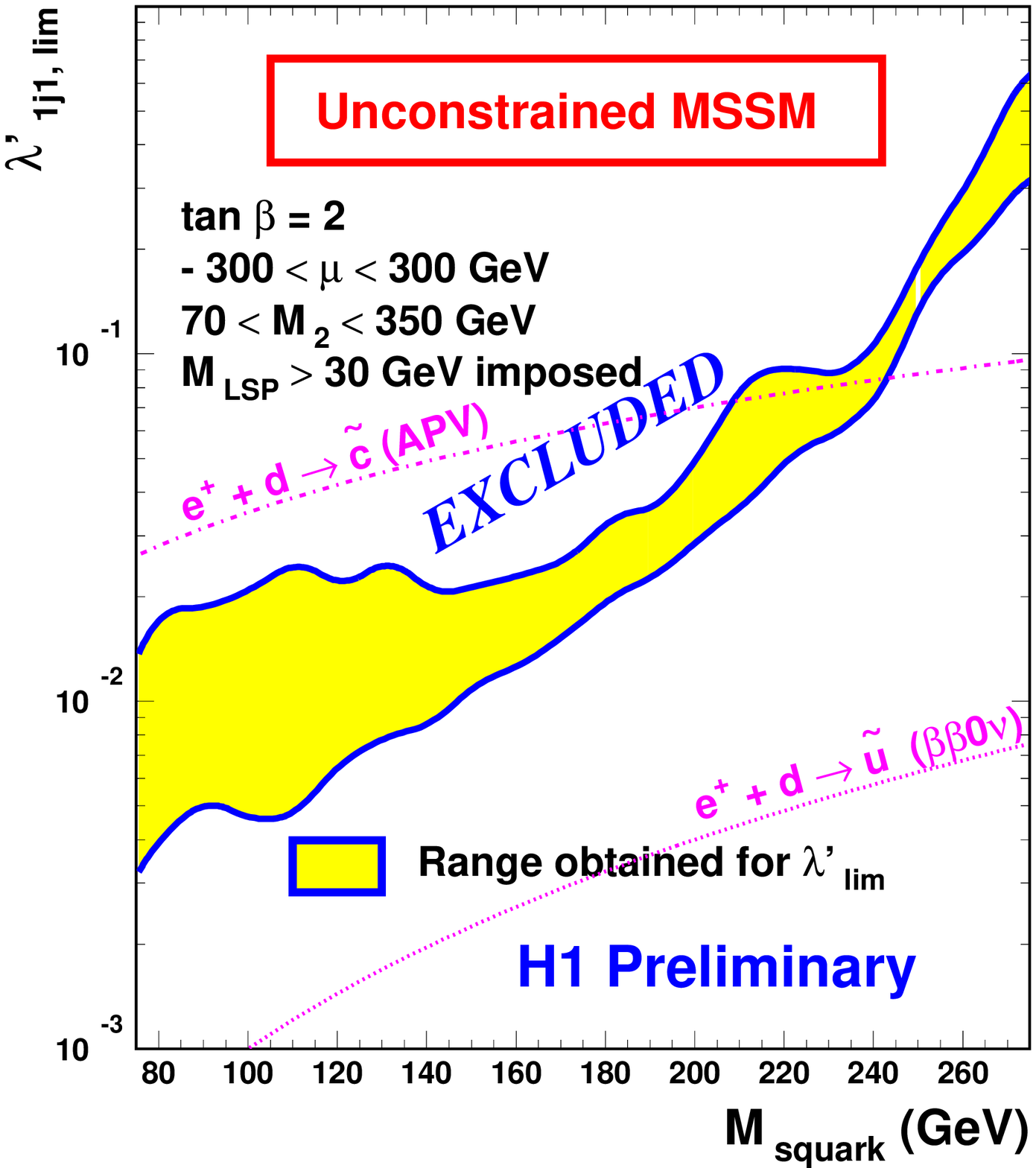,width=85mm}}
\put(15,-30){\epsfig{file=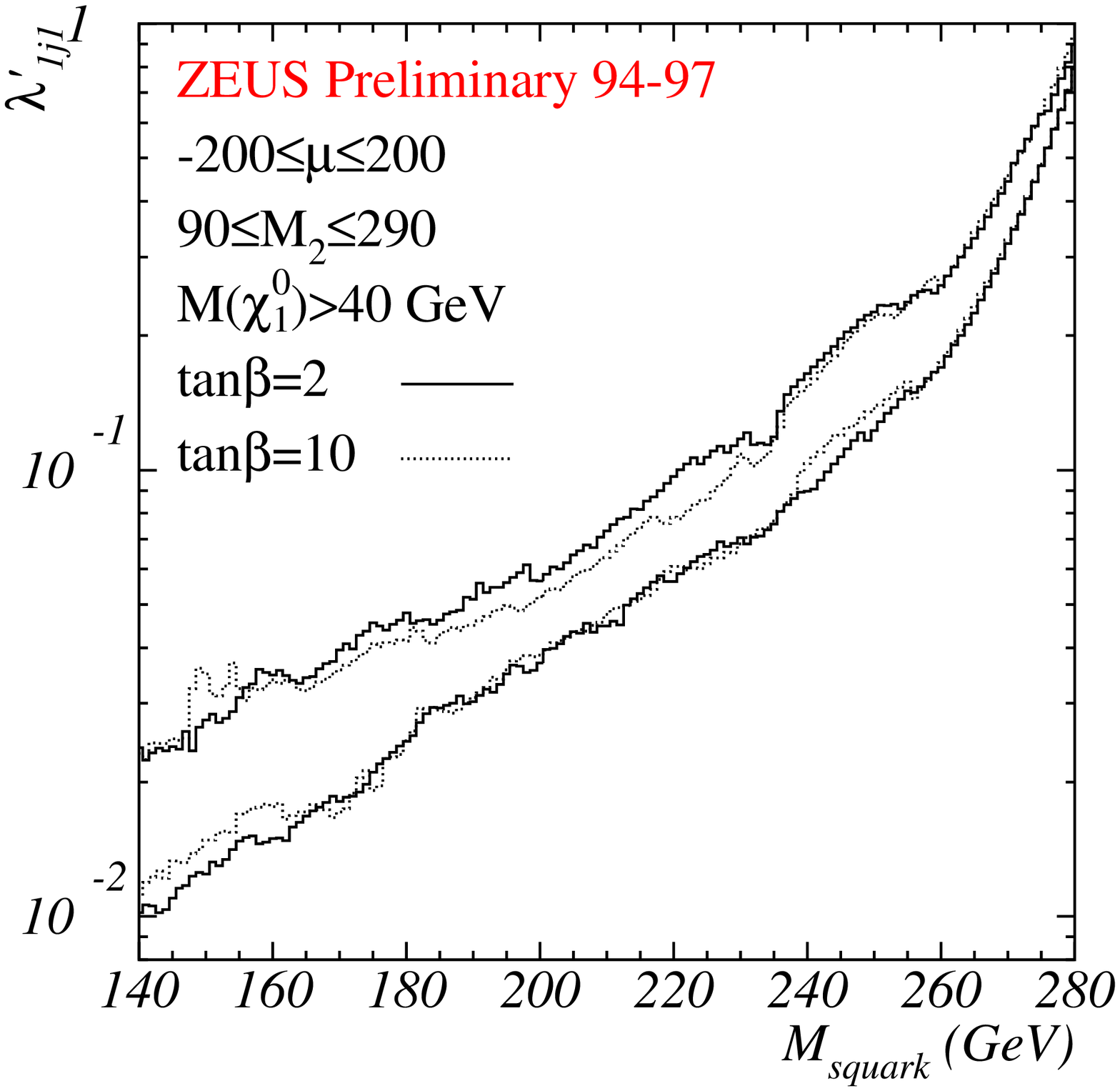,bbllx=0pt,
bblly=0pt,bburx=594pt,bbury=842pt,width=80mm,height=138mm}}
\put(-5,105){(a)}
\put(190,105){(b)}
\end{picture}
\end{center}
\caption{(a) Exclusion upper limits at $95\%$ confidence level by
H1~\cite{h1_squark9497} on the coupling
$\lambda^\prime_{1j1}$ as a function of the squark mass for $\tan\beta=2$,
in the ``phenomenological'' MSSM. For each squark mass, a scan on the MSSM
parameters $M_2$ and $\mu$ has been performed and the largest (lowest) value
for the coupling limit is shown by the upper (lower) curve. (b) The similar
limits from ZEUS~\cite{zeus_squark9497} for $\tan\beta=2$ as well as for
$\tan\beta=10$ (dotted lines).}
\label{fig:squark_limits}
\end{figure}

Under the assumption that only one of the Yukawa couplings
$\lp_{1j1}$ dominates\footnote{This is not unreasonable as in the SM the
top quark Yukawa coupling is almost a factor of 40 larger than the bottom
Yukawa coupling.}, mass dependent upper limits on these couplings
are derived by combining all topologies. Combining
all contributing channels with different topologies improves the sensitivity
considerably, up to a factor of $\sim 5$ at lowest masses, compared to what
would be achieved using only the contribution from topology (a).
The results~\cite{h1_squark9497, zeus_squark9497} are shown in 
Fig.\ref{fig:squark_limits}.
The H1 limits are compared in Fig.\ref{fig:squark_limits}(a) to the most
stringent indirect limits. The production of a $\tilde{u}$ squark via a
$\lambda^\prime_{111}$ coupling is very severely constrained by the
non-observation of neutrinoless double beta decay~\cite{dbdecay} as shown by
the dotted curve. The most sever indirect limit on the coupling
$\lambda^\prime_{121}$, which could allow for the production of squarks
$\tilde{c}$, comes from Atomic Parity Violation~\cite{dreiner, apv} and is
indicated by the dashed curve in Fig.\ref{fig:squark_limits}(a). For squark
masses below $\sim 240\,{\rm GeV}$ HERA limits significantly improve this
indirect constraint on $\lambda^\prime_{121}$ by a factor of up to $\sim 3$.
In Ref.\cite{h1_squark9497}, further limits are derived both in a constrained
version of the MSSM by assuming a universal mass parameter $m_0$ for all
sfermions at very high scale and in the framework of the Minimal Supergravity
Model, a more constrained SUSY model. The limits in the latter model extend
beyond the domain covered by other collider experiments~\cite{d0_msugra}
especially at large $\tan\beta$.


\subsection{Search for contact interactions} \label{sec:ci}
A broad range of hypothesized non-SM processes at mass scales beyond the
HERA center-of-mass energy, $\sqrt{s}=300-320\,{\rm GeV}$, can be
approximated in their low-energy limit by effective four-fermion $eeqq$
contact interactions, analogous to the effective four-fermion interaction
describing the weak force at low energies~\cite{fermi}.
Here three sources are briefly discussed, conventional contact interactions
arising from a substructure of the fermions involved, 
the exchange of a new heavy particle (e.g.\ leptoquark), 
low scale quantum gravity effects which may be
observable at HERA via the exchange of gravitons coupling to SM particles
and propagating into extra spatial dimensions.

New currents or heavy bosons may produce indirect effects through the exchange
of a virtual particle interfering the $\gamma$ and $Z^0$ fields of the SM.
In the present of $eeqq$ contact interactions which couple to a specific
quark flavor ($q$), the SM Lagrangian ${\cal L}_{\rm SM}$ received the
following additional terms~\cite{ci_th, haberl}:
\begin{equation}
{\cal L}={\cal L}_{\rm SM}+\sum_{a,b=L,R} 
\left[ \eta^q_s    (\overline{e}_a e_b) (\overline{q}_a q_b)
+      \eta^q_{ab} (\overline{e}_a \gamma_\mu e_b) 
                   (\overline{q}_a \gamma^\mu q_b)
+      \eta^q_T    (\overline{e}_a \sigma_{\mu\nu} e_b) 
                   (\overline{q}_a \sigma^{\mu\nu} q_b) \right]
\end{equation}
where the sum runs over the left-handed ($L$) and right-handed ($R$) fermion
helicities. The three terms correspond to respectively the scalar, vector, and
tensor interactions. Only the vector term is considered at HERA since
strong limits beyond the HERA sensitivity have already been placed on the
other terms~\cite{haberl, altarelli_ci97, barger_ci98}. The effective
coupling coefficients $\eta^q_{ab}$ is defined as
\begin{equation}
\eta^q_{ab}=\epsilon \left(\frac{g}{\Lambda^q_{ab}}\right)^2
\end{equation}
where $\epsilon=\pm 1, 0$ with $\pm 1$ representing constructive and 
destructive interference with the SM currents, $g$ is the overall coupling 
strength, and $\Lambda^q_{ab}$ is the effective mass scale.
In the study of possible fermion compositeness or substructure,
$g$ is conventionally chosen to be $\sqrt{4\pi}$, whereas for study of
the virtual leptoquark, these parameters are related to the leptoquark mass 
$M_{\rm LQ}$ and the coupling $\lambda$:
\begin{equation}
\frac{g}{\Lambda}=\frac{\lambda}{M_{\rm LQ}}\,.
\end{equation}

Based on the NC cross sections described in Sec.\ref{sec:dsdq2},
the lower bounds on the scale parameters $\Lambda^\pm$ obtained by
H1~\cite{h1ci9497} using the $e^+p$ 1994-1997 data range between
1.3\,TeV and 5.5\,TeV at 95\% confidence level for various chiral structures.
The limits from ZEUS~\cite{zeusci9497} are similar and range from 1.7\,TeV to
5\,TeV. The most restrictive range from both experiments is for the
$VV$ model, where all contact terms enter with the same sign.
Combining with the new $e^-p$ 1998-1999 and $e^+p$ 1999-2000 data,
the H1 limits have been substantially extended and amount to
$1.6-9.2$\,TeV~\cite{h1ci9400}.
The HERA results of direct searches for $(eq)$ compositeness are thus
competitive with those of other experiments at LEP~\cite{ci_opal, ci_lep} and
Tevatron~\cite{ci_tevatron}.

The same data have been used by H1 to derive lower limits on the ratio
$M_{\rm LQ}/\lambda$ for the exchange of virtual leptoquarks for range
between the center-of-mass energy $\sqrt{s}$ and up to
1.7\,TeV~\cite{h1ci9400}. These measurements extend thus the direct
leptoquark searches at HERA to high masses $M_{\rm LQ}>\sqrt{s}$. Two
examples have been shown for $\tilde{S}_{1/2, L}$ and $S_{0,L}$ 
in Fig.\ref{fig:lq_limit}. The most stringent limits, however, are those for
vector leptoquarks with coupling to up quarks.

It has been recently suggested that gravitational effects may become strong
at subatomic distances and thus measurable in collider
experiments~\cite{arkani}. In such a scenario, which may be realized in
string theory, gravity is characterized by a scale $M_s\sim {\cal O}({\rm
TeV})$ in $4+n$ dimensions. The extra spatial dimensions $n$ are restricted
to a volume associated with the size $R$ and the scale in $4+n$ and the
ordinary 4 dimensions are related by
\begin{equation}
M^2_P\sim R^nM^{2+n}_S\,,
\end{equation}
where $M_P\sim 10^{19}\,{\rm GeV}$ is the Planck mass. An exciting
consequence would be a modification of Newton's law at distance $r<R$, where
the gravitational force would rise rapidly as $F\propto 1/r^{2+n}$ and
become strong at the scale $M_S$. Experimentally, gravity is essentially not
tested in the sub-millimeter range~\cite{long} and scenarios with $n>2$
extra dimensions at large distances $R\lesssim 100\,\mu$m are conceivable.

In models with large extra dimensions the spin 2 graviton propagates into
the extra spatial dimensions and appears in the 4-dimensional world as a
spectrum of massive Kaluza-Klein excitations with masses $m^{(j)}=j/R$,
including the zero-mass state. The graviton fields $G^{(j)}_{\mu \nu}$
couple to the SM particles via the energy-momentum tensor $T^{\mu\nu}$
\begin{equation}
{\cal L}_G=-\frac{\sqrt{8\pi}}{M_P}G^{(j)}_{\mu \nu}T^{\mu\nu}\,.
\end{equation}
Summation over the whole tower of Kaluza-Klein states $j$ with masses up to
the scale $M_S$ compensates the huge $1/M_P$ suppression and results in an
effective contact interaction coupling~\cite{giudice}
\begin{equation}
\eta_G=\frac{\lambda}{M^4_S}
\end{equation}
where $\lambda$ is the coupling strength of order unity.

\begin{figure}[htbp]
\begin{center}
\begin{picture}(50,420)
\put(-120,-15){\epsfig{file=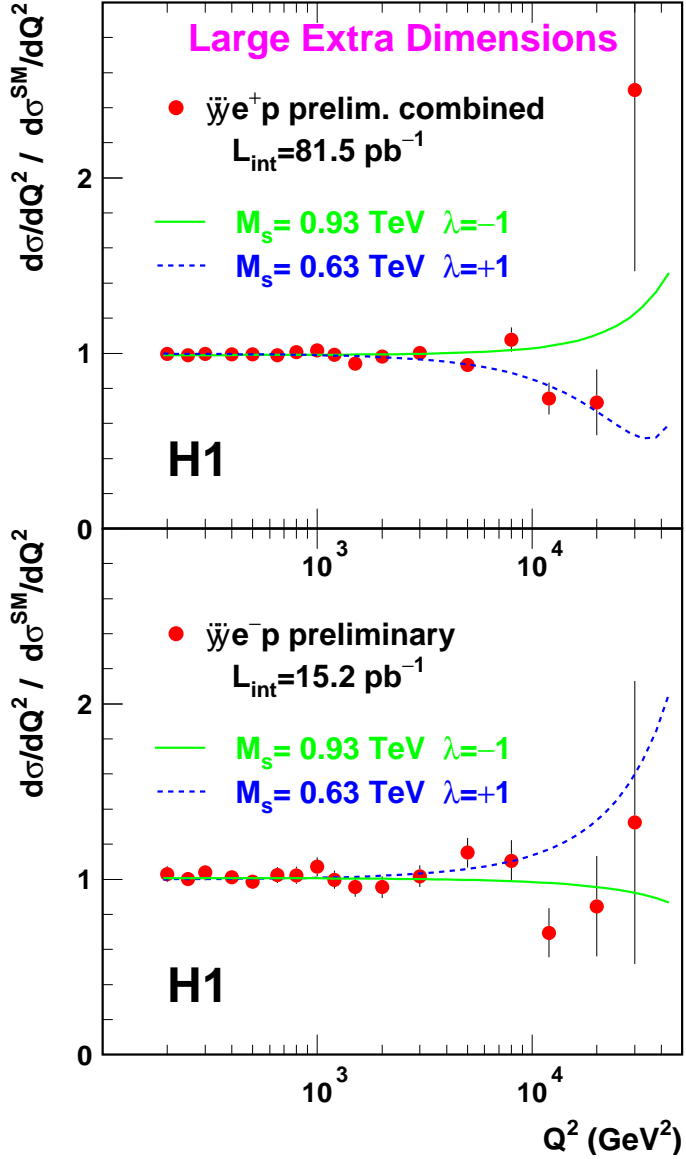,width=100mm}}
\end{picture}
\end{center}
\caption{NC cross sections $d\sigma/dQ^2$ normalized to the SM expectation
for the combined $e^+p$ and the $e^-p$ scattering data. The curves show the
effects of graviton exchange in large extra dimensions given by a common fit
to the scale $M_S$ (95\% confidence level lower limits) with coupling
$\lambda=1$ and $\lambda=-1$. The errors represent statistics and
uncorrelated experimental systematics.}
\label{fig:extradim}
\end{figure}

Lower limits from H1 on the scale parameter $M_S$ are derived again from fits
to the NC cross sections (Sec.\ref{sec:dsdq2}). For the $e^-p$ data stronger
bounds are obtained for positive coupling than for negative coupling. The
opposite behavior is observed in $e^+p$ scattering. Both lepton-beam charges
thus complement each other and a combined analysis~\cite{h1ci9400} of all
$e^\pm p$ data yields limits on $M_S$ of 0.63\,TeV for positive coupling
$\lambda=1$ and 0.93\,TeV for negative coupling $\lambda=-1$ as illustrated
in Fig.\ref{fig:extradim}. Similar investigations of virtual graviton
exchange in $e^+e^-$ annihilation provide comparable 
limits~\cite{ci_opal, extra_l3}.

\chapter{Summary and outlook} \label{chap:summary}
The $ep$ collider HERA has opened up a completely new kinematic domain
unexplored in the previous fixed-target experiments. The results of the
first measurement of the proton structure function at small $x$ down to 
$0.5\times 10^{-4}$ came as a surprise for many: the measured $x$ dependence 
of the structure function for a fixed momentum transfer $Q^2$ revealed 
a strong rise towards low $x$ reflecting 
an increasing number of slow partons being probed within the proton. 
This first measurement was based on about
25\,nb$^{-1}$ of data taken in 1992. One year later, the data have been
increased by more than a factor of 10. The data have allowed 
the structure function measurement to be extended to values of $Q^2$ 
beyond those measured at fixed-target experiments.
By making use of the hard initial radiative events, it was possible to extend
the kinematic region to lower $Q^2$ and high $x$. 
With more data and the improved apparatus, the precision
measurements are realized at HERA reaching a
statistical precision below 1\% and a systematic error of typically 3\%.

The structure function measurement of 1993 has allowed a first determination
of gluon density at low $x$ based on the NLO DGLAP evolution equations.
A strong-rise behavior toward low $x$ is also found and is
believed to be responsible for the observed behavior of the structure function.
The subsequently improved measurements of the structure function have
resulted in an ever-increasing knowledge of the gluon distribution 
at low $x$. Today the gluon density is better known at low $x$ than at 
other ranges, a situation completely different to several years ago, before 
the HERA start.
The latest precision measurement at low $Q^2$ and $x$ by H1 has not only 
provided a best determination of the gluon but also together with the BCDMS 
data one of the most precise measurements of the strong coupling constant 
$\alpha_s$.

The increased integrated luminosity has also allowed to explore the very high 
$Q^2$ region up to $\sim 30\,000\,{\rm GeV}^2$.
With the earlier data from 1994 to 1996, an excess of events at high $Q^2$
and high $x$ was reported by H1 and ZEUS. With new data
available taken for different lepton-beam charges and higher
center-of-mass energy, no excess was confirmed.
The NC and CC cross sections have been measured
both double differentially and single differentially in $x$ and $Q^2$.
The comparison of the $e^\pm p$ NC cross sections has revealed the presence of
the $\gamma Z^0$ interference contribution at high $Q^2$ and has allowed 
a first measurement of the structure function $x\tilde{F}_3(x,Q^2)$ at
high $Q^2$.
The comparison of the $e^\pm p$ CC cross sections has shown a huge
difference in the cross sections reflecting the different quark flavors
probed by the $W^\pm$ bosons in the $e^\pm$ scatterings.
The comparable NC and CC cross sections at high $Q^2$ has confirmed
electroweak unification in the space-like regime.

The measurement of the NC cross sections covers a $Q^2$ range of about
four orders of magnitude, over which the cross sections fall with $Q^2$ by
more than eight
orders of magnitude, a behavior well described by the Standard Model
expectation based mainly on low $Q^2$ data providing thereby a nontrivial
test of the validation the QCD evolution. The Standard Model expectations
also describe all other measurements with two possible exceptions:
\begin{itemize}
\item The NC cross sections at $Q^2>20\,000\,{\rm GeV}^2$ and $x$ around
 0.4 are higher than the expectation in both $e^+p$ data samples taken at
different center-of-mass energies. The excess remains however statistically
not very significant.
\item The NC cross sections at $x=0.65$, the largest $x$ value measured, are
for $e^+p$ and $e^-p$ at essentially all $Q^2$ lower than the expectation
which is mainly based on the BCDMS data.
\end{itemize}
A NLO QCD fit using all high $Q^2$ data shows the data at $x=0.65$ are about
17\% lower than the expectation and there remains a discrepancy of about
two standard deviations.
The fit has demonstrated that the quark distributions at high $x$ and
high $Q^2$ are constrained by the HERA data alone
with an experimental precision of $\sim 6\%$ at $x=0.25$ and
$x=0.4$ and $\sim 10\%$ at $x=0.65$ for the $u$ valence quark and about 20\%
for the $d$ valence quark. The results are in good agreement with that
obtained by the local extraction method. These determinations are free from
any nuclear correction in contrast to the current knowledge on the $d$ valence
quark which was mainly obtained from fixed-nuclear-target data at low $Q^2$.

The high $Q^2$ data have also been used to test the electroweak sector of 
the Standard Model. The derived propagator mass of the $W$ boson in the
space-like regime agrees well with other direct or indirect measurements of
a real or virtual boson in the time-like regime. The determination at HERA is
however already limited by the uncertainty of the parton density
distributions demonstrating the importance and necessity to improve our
knowledge on these parton distributions for future precision
measurements.

On the same data, various searches have been performed for seeking both 
resonance production of leptoquarks and squarks in $R$-parity violating 
supersymmetric models, and contact interactions of a substructure of 
the involved fermions, exchange of a new virtual heavy particle,
as well as low scale quantum gravity effects. No significant deviations from
the Standard Model expectations were
found and various limits have been derived. These limits either extend
beyond or are comparable with those provided by indirect measurements or by 
direct searches at other colliders.

At high $Q^2$, the precision of essentially all current measurements are
still limited by the statistical error. In order to improve the precision, to 
reach even higher $Q^2$ and to fully explore the discovery potential of HERA, 
it is important to have as much data as possible and in a relatively short 
period of time with stable running conditions to minimize additional 
systematic uncertainties. This can only be achieved with a luminosity upgrade.
The upgrade has started since the beginning of September 2000 and 
the new physics run of the HERA phase two is foreseen in September 2001. 
With this upgrade program for both the machine and the detectors together 
with the polarized beams, HERA, being one of few high energy machines running 
in the next years, will have an exciting and bright future.

\chapter*{Acknowledgements}

This is a report prepared for the `habilitation' defense on Dec.\ 1, 2000 
in front of the jury members: V.~Brisson, M.~Davier (chairman), J.\ Gayler,
F.\ Le Diberder, and W.\ J.\ Stirling.

I wish to thank all my colleagues from the H1 and ZEUS experiments on whose
research this report is based in particular U.~Bassler, G.~Bernardi, J.\ C.\
Bizot, V.\ Brisson, P.~Bruel, J.\ Cao, A.\ Courau, C.\ Diaconu, L.~Favart, 
M.~Fleischer, A.~Glazov, B.~Heinemann, M.~H\"utte, M.\ Jacquet, M.\ Jaffre, 
S.\ Kermiche, M.\ Klein, M.~W.~Krasny, Ch.~Leverenz, S.~Levonian, P.~Marage, 
A.~Mehta, T.~Merz, A.~Panitch, C.\ Pascaud, E.~Perez, J.~P.~Phillips,
G.~Radel, E.~Rizvi,
V.~Shekelyan, Y.~Sirois, C.\ Vall\'ee and F.\ Zomer for their collaborations 
during various stages in the past eight years. I thank V.~Brisson also for
her constant encouragement during these years.

I am grateful for V.\ Brisson, J.\ Gayler, F.\ Le Diberder and C.\ Pascaud
for their helpful comments on an earlier version of the report.

\end{document}